\newcommand{\True}{True}                % Basic T/F token. Don't mess with this
\newcommand{\ListofTables}{False}	% True:include list of tables
\newcommand{\ListofFigures}{False}	% True:include list of figures
\newcommand{\IncludeCR}{False}		% True:include copyright notice
\newcommand{\IncludeDed}{False}		% True:include dedication page
\newcommand{\IncludeAck}{True}		% True:include acknowledgements page
\newcommand{\IncludeApp}{False}		% True:include appendixes
\newcommand{\IncludeBio}{True}		% True:include biography page
\newcommand{\IncludeSig}{False}		% True:include signature page
\newcommand{\IncludeGen}{False}		% True:include general audience abstract

\newread\InStream
\newcount\Done
\openin\InStream=gradinfo.tex
\read\InStream to \FullName
\read\InStream to \ThesisType
\read\InStream to \DegreeType
\read\InStream to \GradMonth
\read\InStream to \GradYear
\read\InStream to \Department
\read\InStream to \Advisor
\closein\InStream

\documentclass[10pt,twoside]{report} %3
\usepackage{amssymb}
\usepackage{amsmath}
\usepackage{epsfig}
\usepackage{thesis}
\def\SingleSpace{\baselineskip=12.5pt} %2,3
\def\DoubleSpace{\baselineskip=12.5pt} %2,3
\begin{document}
\pagestyle{plain}
\pagenumbering{roman}

%%%% Title Page
\SingleSpace
\thispagestyle{empty}
\vspace*{63pt}
\SingleSpace
\ReadThesisTitle
\vspace*{5.5 cm}
\centerline{By}
\vspace*{.5 cm}
\centerline{\uppercase\expandafter{\FullName}}
\vspace*{5.5 cm}
\centerline{A \uppercase\expandafter{\ThesisType}\ PRESENTED TO THE 
            GRADUATE SCHOOL}
\centerline{OF THE UNIVERSITY OF FLORIDA IN PARTIAL FULFILLMENT}
\centerline{OF THE REQUIREMENTS FOR THE DEGREE OF}
\centerline{\uppercase\expandafter{\DegreeType}}
\vspace*{1 cm}
\centerline{UNIVERSITY OF FLORIDA}
\vspace*{1 cm}
\centerline{\GradYear}

\clearpage
%\pagebreak

%%%% Copyright (optional)
\ifx\IncludeCR\True
   input{copyrightpage.tex}
  \pagebreak
\fi

%%%% Dedication (optional)
\ifx\IncludeDed\True
  \input{dedicationpage.tex}
  \pagebreak
\fi

%%%% Acknowledgments (optional)
\ifx\IncludeAck\True
  \addcontentsline{toc}{chapter}{ACKNOWLEDGMENTS}
  \vspace*{63pt}
\centerline{ACKNOWLEDGMENTS}
\vspace{.5in}

%%%%%% Acknowledgements go here
\DoubleSpace

The first person I would like to acknowledge is my supervisor, Stan Dermott,
not only for his advice, encouragement, and influence on my work, but also for 
enticing me over to Florida in the first place --- it has been a tremendous
experience.
Equally I would like to thank my supervisor at Queen Mary and Westfield
College, Carl Murray, for inspiring me to continue with a career in Astronomy.
Thanks are also due to the Fulbright commission for supporting me in my first
year in Florida.

This dissertation is essentially an extension of the work that Stan Dermott and
his Solar System Dynamics group have accomplished over the past 15 or so
years in their efforts to model the zodiacal cloud.
It is fair to say that this dissertation could not have been achieved
without this foundation.
I wish to thank all members of the group, past and present, especially
Keith Grogan, for countless useful discussions.

Another person to whom I am indebted is Charlie Telesco, both for his
direct input to the dissertation through our work on the HR 4796 disk,
and for providing me with the opportunity to go observing with his
Infrared Astrophysics group.
The perspective I gained from this observing experience has been an essential
part of my Ph.D. education.
I acknowledge all of the members of the OSCIR observing team, especially
Robert Pi\~{n}a and Scott Fisher, for taking the time to teach me the
observing lore.

On a personal level, I would like to thank my parents for nourishing the gift
of life in me.
Most of all, I want to thank my fianc\'{e}e, Maxine, with whom I have shared
the last 8 years, and whose love and support throughout I could not have lived
without.
This dissertation is dedicated to her.

\SingleSpace

\clearpage
%  \pagebreak
\fi

%%%% Table of contents
\SingleSpace
\tableofcontents
\clearpage
%\pagebreak

%%%% List of tables (optional - need file thesis.lot)
\ifx\ListofTables\True
  \DoubleSpace
  \addcontentsline{toc}{chapter}{LIST OF TABLES}
  \listoftables
  \pagebreak
\fi

%%%% List of figures (optional - need file thesis.lof)
\ifx\ListofFigures\True
  \addcontentsline{toc}{chapter}{LIST OF FIGURES}
  \listoffigures
  \pagebreak
\fi

%%%% Abstract
\addcontentsline{toc}{chapter}{ABSTRACT}
\vspace*{66pt}
\SingleSpace

\centerline{Abstract of \ThesisType\ Presented to the Graduate School}
\centerline{of the University of Florida in Partial Fulfillment of the}
\centerline{Requirements for the Degree of \DegreeType}
\vspace*{2 cm}
\ReadThesisTitle
\vspace*{.5 cm}
\centerline{By}
\vspace*{.5 cm}
\centerline{\FullName}
\vspace*{.5 cm}
\centerline{\GradMonth\ \GradYear}
\vspace*{1 cm}
\centerline{Chairman: \Advisor \hspace*{\fill}}
\centerline{Major Department: \Department \hspace*{\fill}}

%%%% Body of Abstract goes here
\DoubleSpace

%% This abstract must contain fewer than 350 words.
Recent advances in astronomical instrumentation have led to a vast increase
in our knowledge of the environments of nearby stars.
In particular, we are now able to image the thermal emission from the disks
of dust around main sequence stars that may be the fossil remnants of
planetary formation.
These observations imply that the distribution of dust in the debris disks
is neither smooth nor symmetrical;
e.g., mid-infrared images of the disk of dust around the young
A0V star HR 4796A show two lobes of emission, one of which may
be $\sim 5$\% brighter than the other.
The observed structure of the debris disk in the solar system, i.e., the
zodiacal cloud, also contains asymmetries:
it has an offset center of symmetry, it is warped, and there is an asymmetric
ring of dust co-orbiting with the Earth.
Since the zodiacal cloud's asymmetries have been shown to be signatures of
the gravitational perturbations of the solar system's planets, it is hoped
that it may be possible to indirectly detect extrasolar planetary systems by
their signatures in debris disk observations.

This dissertation uses the physical processes that affect the evolution of
debris material in the solar system to create a generalized model for the
evolution of circumstellar debris material.
It then shows how planetary perturbations affect that evolution, thereby
causing the signatures of planets seen in the structure of the zodiacal cloud.
This model can be used to provide a quantitative interpretation of debris disk
observations, and the necessary modeling techniques are demonstrated by their
application to observations of the HR 4796 disk.
As well as determining the large scale structure of the HR 4796 disk, 
the modeling shows how a small body ($>10M_\oplus$) in the HR 4796 system
that is on an orbit with an eccentricity larger than 0.02 could be the cause
of the observed brightness asymmetry.
The modeling also shows that the disk's mid-IR emitting particles are hotter
than black body (and therefore small), and discusses whether they are in
the process of being blown out of the system by radiation pressure.

\SingleSpace

\clearpage
%\pagebreak

%%%% Start of chapters (double space text)
\addcontentsline{toc}{startchapter}{\ShiftHack CHAPTERS}
\DoubleSpace
\pagenumbering{arabic}
\pagestyle{myheadings}
%%%%%%%%%%%%%%%%%%%%%%%%%%%%%%%%%%%%%%%%%%%
\chapter{INTRODUCTION}
%%%%%%%%%%%%%%%%%%%%%%%%%%%%%%%%%%%%%%%%%%%
\label{c-intro}

%%%%%%%%%%%%%%%%%%%%%%%%%%%%%%%%%%%%%%%%%%%
\section{Motivation}
\label{s-motivation}
One of the most consistent human endeavors since the dawn of time has been
to understand the nature of the universe in which we live.
We are constantly questioning how the universe, and especially life on Earth,
came to be as it is.
Essentially this is a search for an understanding of our place in the
universe, which inevitably leads us to ponder whether life could
exist elsewhere in the cosmos.
Ever since the middle of the sixteenth century when the Copernican heliocentric
model of the universe superseded the geocentric view, this search has been a
humbling experience as we begin to grasp the gargantuan scale of the universe.
The observational data obtained over the previous 450 years paints a
picture of the universe as we see it today that places the Earth in a far more
insignificant role than was previously imagined ---
we now know that not only is the Earth not the center of the universe,
but neither is the Sun, since it is just one of an estimated 400 billion
stars that make up the Milky Way, which is itself just one galaxy out
of at least $\sim 80$ billion others in the universe.

However much we have learned about the nature of the universe, the
search for life outside the confines of the Earth has so far proved fruitless. 
Even in the corner of the universe about which we have the most
information, the solar system, the debate continues as to whether there is
(or indeed has been) life in some form or another on Mars.
The ocean under the frozen ices of Jupiter's moon Europa and the surface of
Saturn's moon Titan also provide potential habitats for life in the solar
system, possibilities that remain to be tested.
However, the solar system is but a small portion of the universe.
Since stars are supposed to have formed in the same manner as the
Sun, and the solar system's planets are thought to be a byproduct of that
formation process, this has led us to speculate that some of the $O(10^{22})$
stars in the universe could also have planets orbiting them, some of which
may (past, present, or future) harbor life.
Thus, it has been one of the holy grails of recent astronomy to find
habitable planets (i.e., ones that could potentially support life)
orbiting stars other than the Sun.

%%%%%%%%%%%%%%%%%%%%%%%%%%%%%%%%%%%%%%%%%%%%%
\section{The Search for Extrasolar Planets}
\label{s-search}
From an observational point of view, the direct detection of habitable
planets around even the closest stars is beyond our current technological
capabilities (\cite{back98}; \cite{wa98}).
There are two reasons for this.
First, technological constraints mean that until recently telescopes did
not have the resolving power to see the regions around the stars where it
was expected that planets would have formed.
Even now, the regions close to the star ($\sim 1$ AU) where we anticipate
that terrestrial planets would form are obscured from view by the
diffraction halo of the stellar image.
Second, planets by definition do not have an internal source of energy,
which means that they are not very bright ---
any light observed from them is either reflected or reprocessed starlight.
Thus, planets are far dimmer than the photospheric emission from star about
which they are orbiting and are too dim to be detected directly at present
(\cite{wa98}).

This led to the development of some ingenious techniques to indirectly detect
the presence of planets around stars.
The most productive of these techniques uses the gravitational effect of
a planetary system on the position of a star relative to the barycenter
of the star/planet system.
This causes a wobble in the motion of the star relative to the Sun.
This wobble can be detected in three ways:
astrometric detection, by measuring the change in position of a star on the
sky (e.g., \cite{gate87});
doppler shift detection, by measuring the variation in the radial velocity
of star (e.g., \cite{mq95}; \cite{mb98});
by measuring variations in the time arrival of periodic signals, such as
those from pulsars (e.g., \cite{wf92}).
There are also indirect detection methods that rely on the occurrence of
specific events such as:
the transit of a planet in front of a star that causes a reduction in the
brightness of the star (e.g., a planet was recently discovered around the
G0V star HD 209458 using this technique, \cite{hmbv00});
and gravitational lensing of a background star by a foreground
star/planet system.
A complete review of all techniques is given in NASA's Road Map for the
Exploration of Neighboring Planetary Systems (\cite{beic96}).

%%%% Marcy's figure of the currently known extrasolar planets
\begin{figure}
  \begin{center}
    \begin{tabular}{c}
      \epsfig{file=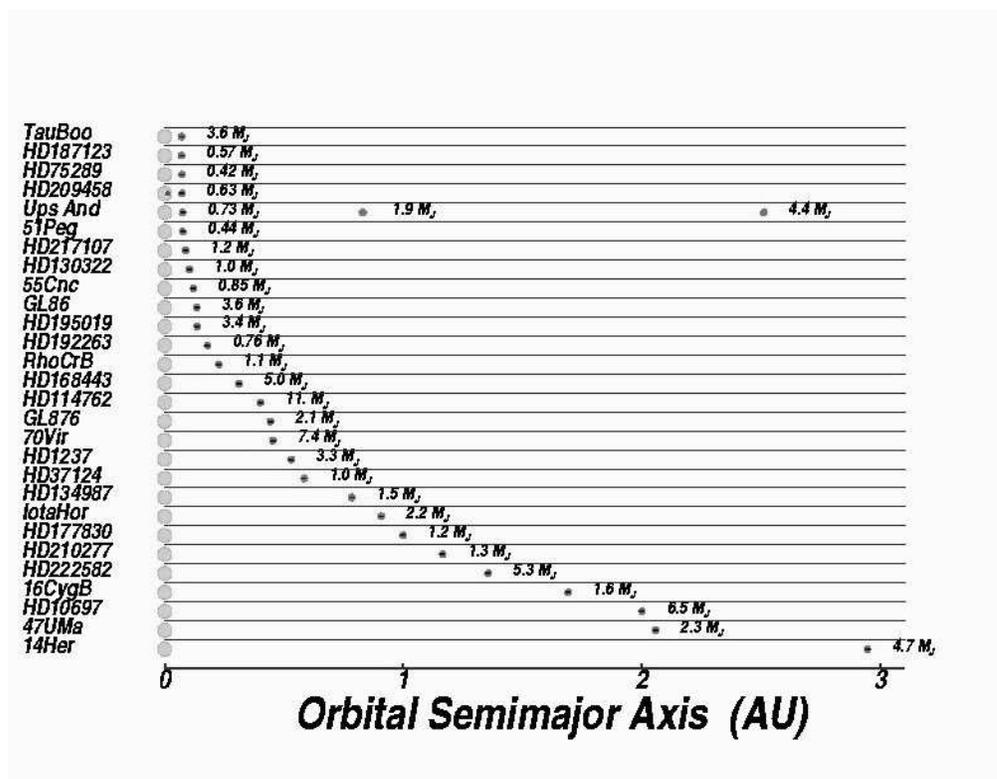,height=4in}
    \end{tabular}
  \end{center}
  \caption{Known extrasolar planet candidates around Sun-like stars
  (\cite{mbvf99}).}
  \label{figmarcyplanets}
\end{figure}

The first extrasolar planetary system that was detected was that around
a billion year old neutron star, the 6.2 ms pulsar PSR 1257+12 (\cite{wf92}).
This was followed three years later by the doppler shift detection of
a 0.5 Jupiter mass planet orbiting 0.05 AU from the G2IVa star 51 Peg
(\cite{mq95}).
There have since been many other detections (not all of them confirmed),
mostly using the doppler shift technique (e.g., \cite{mb98});
there are currently 27 confirmed detections of extrasolar planets
(see Fig.~\ref{figmarcyplanets}), with the number growing every month.
However, none of these detections are of habitable planets.
The doppler shift technique is biased towards detecting very massive planets
(to get a wobble that is large enough to detect) that are also very close to
the star (to detect the variation on a short enough timescale).
All of the planets detected using this technique are larger than
0.4 Jupiter masses and most are found to orbit closer than 0.5 AU to the star
(see Fig.~\ref{figmarcyplanets}).
This means that this technique excludes the detection of systems like the
solar system which we know to contain life.
The planets that were detected around PSR 1257+12 are terrestrial mass planets.
However, they are unlikely to harbor life due to the intensity of
radiation from the pulsar.

These detections have shown us that a solar planetary system-like configuration
is not the only possible planetary system configuration, and they demand
an explanation of how this diversity of planetary systems arose.
Essentially we want to be able to answer questions like:
How did these planetary systems form?
Is a planetary system a likely outcome of stellar evolution?
What are the other potential planetary systems like and what else is in
the systems (such as asteroids and comets)?
What are the implications for life in these systems?
Since a planetary system is the end state of a long process that starts
with the formation of a star from a cloud of gas and dust, this ultimately
comes down to understanding how stars and their environments
form and evolve.
The following sections describe our current understanding of
the physics behind the formation and evolution of both stars
and their environments.

%%%%%%%%%%%%%%%%%%%%%%%%%%%%%%%%%%%%%%%%%%%
\section{Paradigm of Star Formation}
\label{s-paradigms}

%%%% Figure of Star Formation from Shu, Adams, and Lizano 1987
\begin{figure}
  \begin{center}
    \begin{tabular}{c}
      \epsfig{file=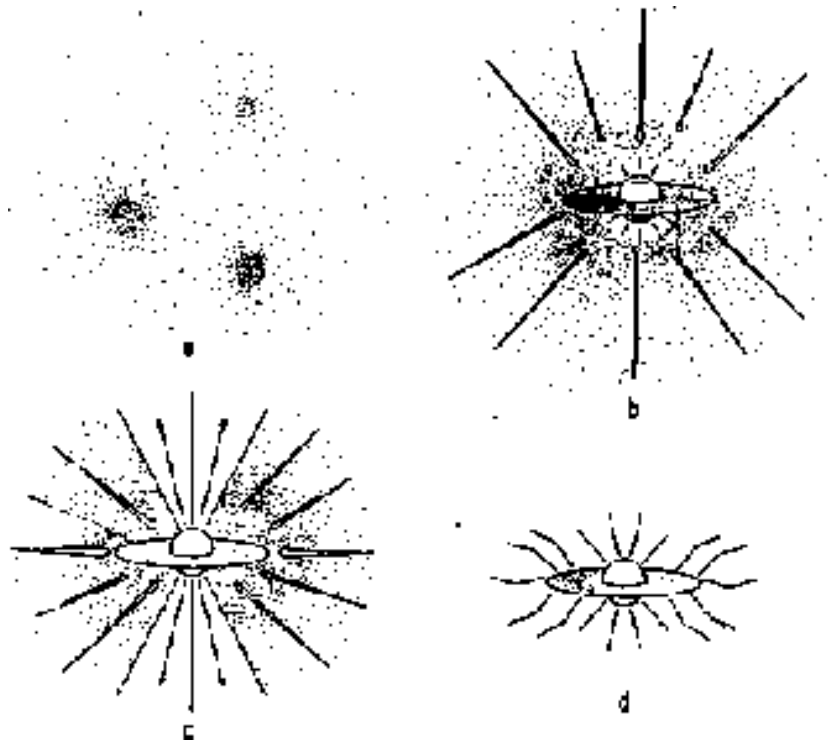,height=4in}
    \end{tabular}
  \end{center}
  \caption{The four stages of star formation (\cite{sal87}).
  (\textbf{a}) Clumps of gas and dust within molecular clouds collapse to
  form slowly rotating cloud cores.
  (\textbf{b}) A protostar with a surrounding nebula disk forms at the center
  of the cloud core collapsing from inside-out.
  (\textbf{c}) A stellar wind breaks out along the rotational axis of the
  system, creating a bipolar outflow.
  (\textbf{d}) The infall terminates, revealing a newly formed star with a
  circumstellar disk.}
  \label{figsal87fig7}
\end{figure}

Star formation is relatively well-understood.
Spectral energy distributions (SEDs) and images of young (pre main-sequence) 
stars, as well as observations of star forming regions provide a broad
paradigm that describes how stars like the Sun formed from molecular clouds
(e.g., \cite{sal87}).
This paradigm is outlined here;
for further information on the nuances of this process, the reader is
referred to the literature.
Shu et al. (1987) describe four stages in the formation of a star from
a molecular cloud (see Fig.~\ref{figsal87fig7}).
In the first stage (Fig.~\ref{figsal87fig7}a), clumps of gas and dust in
the cloud collapse to form slowly rotating cloud cores.
When the mass of the clump exceeds a critical stability limit, the second
stage is reached (Fig.~\ref{figsal87fig7}b).
This is characterized by dynamical collapse of the clump from
inside-out to form a central protostar with a surrounding accretion disk,
both of which are deeply embedded within an infalling envelope of dust
and gas.
In the third stage (Fig.~\ref{figsal87fig7}c), a stellar wind driven by
convection in the protostar (which has by now started to burn deuterium)
breaks out along the channels of weakest resistance, along the rotational
axis of the system, causing collimated jets and bipolar outflows.
The protostar is still accreting matter in this stage from the disk.
Eventually, the infall terminates when the stellar wind has managed to
sweep out the full $4\pi$ sr, leaving the fourth stage of the
evolution (Fig.~\ref{figsal87fig7}d):
a star (e.g., a T-Tauri star) with a surrounding nebula disk.
This disk then decays (possibly through the formation of planets or
other stellar companions) as the star evolves onto the main sequence.

It is the latter stages that we are most interested in, since it is in
these stages and in the nebula disk that the process of planetary formation
is thought to occur.
A variety of observations of nebula disks can provide important clues as
to the physical conditions of these systems at the onset of planetary formation.
Most inferences about the existence and nature of disks around young
stars comes from observations of their SEDs.
A strong infrared excess above that expected to arise from the stellar
photosphere is an indication of thermal emission from dust grains in
the disk that are heated by illumination from the star.
There are several classes of SED, all of which can be accounted for using
current models of the radiative transfer in nebula disks with different
structures (see, e.g., \cite{beck99}).
It turns out that to fit the SEDs, most of the disks need to have a cut-off
both close to (a few stellar radii) and far from ($O$(100 AU)) the star,
and to be flared.
Some SEDs require an additional component to the star and disk (e.g.,
a remnant halo from the molecular cloud).
Recently, there has been striking confirmation of these models of nebula
disk structure from images of the disks (\cite{bswk96}; \cite{pbss99});
e.g., near-IR images of HH30 (\cite{bswk96}), a star in the third stage of its
evolution, show reflected starlight from the jets perpendicular to a
dark disk seen edge-on with a gently flared shape, which causes illumination
of the top edge of the disk.

The fraction of young stars with infrared excesses shows that $\sim 50$\%
of young stars have disks (e.g., \cite{sns89}), and sub-millimeter
observations show that the mass of most disks is between 0.01 and
0.1$M_\odot$ (e.g., \cite{bscg90}).
The minimum mass needed in the solar nebula to create the
solar planetary system is $0.01-0.02 M_\odot$ (\cite{haya81}),
thus if planets are formed in disks, there appears to be enough mass in
the disks from which to make them (\cite{bs96}).
Observations of the fractions young stars with disks based on both their
near-IR and far-IR excesses show that both the inner and outer regions
of the disks dissipate between a few Myr and 10 Myr (see \cite{beck99}).
Thus, any planetary formation should take place within this timescale.
There is, however, little observational evidence of ongoing
planetary formation in the disks, although
authors have used SED observations to try to infer particle
growth in these disks (see, e.g., \cite{dmw98}; \cite{bhn99}).

%%%%%%%%%%%%%%%%%%%%%%%%%%%%%%%%%%%%%%%%%%%%%
\section{Paradigm of Planetary Formation}
\label{s-paradigmp}
The planetary formation process is more poorly understood than star
formation, since there is little observational evidence with which to test
the theories.
Most theories concentrate on describing how the current state of the solar
system --- i.e., the masses, orbits, bulk composition, and rotation of
the planets, and the distribution of comets, asteroids, and meteoroids
--- resulted from the supposed conditions in the primitive solar nebula
that are derived both from observations of other nebula disks, and from
the current state of the solar system (\cite{haya81}).
The theories also have to concur with information about the evolution of
the young solar system from current observations, e.g., of cratering
records.
The current paradigm that explains how the solar system formed from the
protosolar nebula is outlined here (for a good review, see
\cite{liss93}).

The first stage of planetary formation is the formation of the stable
protoplanetary disk, described in \S \ref{s-paradigms}, which takes
$\sim 10^5$ years from the onset of cloud collapse (\cite{sngo93}).
This disk is a mixture of gas and solid matter, which is a mixture of
interstellar grains and nebula condensates, both of which are small
(submicron-sized).
The composition of these condensates depends on the temperature and composition
of the gaseous disk (e.g., \cite{phbs94}; \cite{hs96}).
In the second stage, the microscopic grains settle towards the midplane of
the disk due to gravity.
During this sedimentation process, the grains collide and grow via pairwise
accretion, which accelerates the settling process (\cite{wc93}).
The mechanical and chemical processes related to grain agglomeration are
poorly understood, since they depend on sticking probabilities under
conditions that are difficult to simulate in the laboratory.
It is thought that the grains grow as fractal, or fluffy
aggregates (e.g., \cite{bwph98}) similar to the interplanetary particles
found in the solar system (\cite{gust94}).
Collisions between cm-sized aggregates, however, can provide some compaction
to the structure of the aggregates (\cite{dt97}).
Small particles are coupled to the gas in the disk, but the large particles
are not.
The gas drag on meter-sized bodies is such that the material that survives
to form planets completes the transition from cm- to km-size within
$\sim 10^4$ years, which is the settling timescale.

In the third stage, gravitational interactions between pairs of km-sized
planetesimals cause the planetesimals to conglomerate.
These interactions result in coagulation because the orbits of the planetesimals
have been circularized by gas drag resulting in low relative velocity
collisions. 
Runaway growth of the largest planetesimal in an accretion zone occurs
until protoplanets are dynamically isolated from one another.
These protoplanets can merge further if their orbits are affected
by either mutual gravitational interactions, or by tidal interactions with the
gas disk (\cite{gt80}).
Models of this phase of planetary formation show that accretion is not
complete until $\sim 10^8$ years, but that planets attain a sizeable
fraction of their mass within $\sim 10^7$ years (\cite{weth80}; \cite{liss93}).
A fourth stage of evolution occurs for the planetesimals that are large
enough to accrete substantial amounts of gas from the disk, thus forming
Jovian gas giant planets.
The existence of these giant planets, and their size and location, depend
on how fast they grow relative to the dissipation of the gas disk, and
their potential to halt (or enhance) their own growth via tidal
interactions with this disk (e.g., \cite{lp79}; \cite{pt99}).
The formation of gas giants and the observed dissipation of the
protoplanetary disks (\cite{beck99}) imply that protoplanetary cores
probably develop within the first $\sim 10^6$ years.
Current theories indicate that gas giants only form $> 4$ AU from the
star due to the abundance of ice grains there (\cite{boss95}), while
terrestrial rocky planets form inward of this.

While this paradigm is sufficient to explain the formation of the solar
system, including the formation of the minor bodies (\cite{liss93}),
it does not predict the formation of giant planets in orbits close to the
star like those detected in exosolar systems (e.g., \cite{mb98};
see Fig.~\ref{figmarcyplanets}).
To explain these observations, the paradigms have to include more complicated
physical processes.
For example, a common theory is that these giant planets actually formed at
$\sim 5$ AU from the star, but then suffered radial migration due to their
gravitational interaction with the disk (\cite{tbgl98}; \cite{pt99}).
It is evident, therefore, that the current theories of planetary formation
are far from finalized, and that further observations of developing and
developed planetary systems are needed to tell us how planets are really built.
It is interesting to speculate, however, that if the paradigms of
star and planetary formation hold true, then each star could
have its own planetary system, or at least a failed planetary system.

%%%%%%%%%%%%%%%%%%%%%%%%%%%%%%%%%%%%%%%%%%%%%
\section{Circumstellar Debris Disks}
\label{s-cdd}

%%%% IRAS fluxes of the 4 prototype debris disks from Walker & Wolstencroft 1988
\begin{figure}
  \begin{center}
    \begin{tabular}{rl}
      \epsfig{file=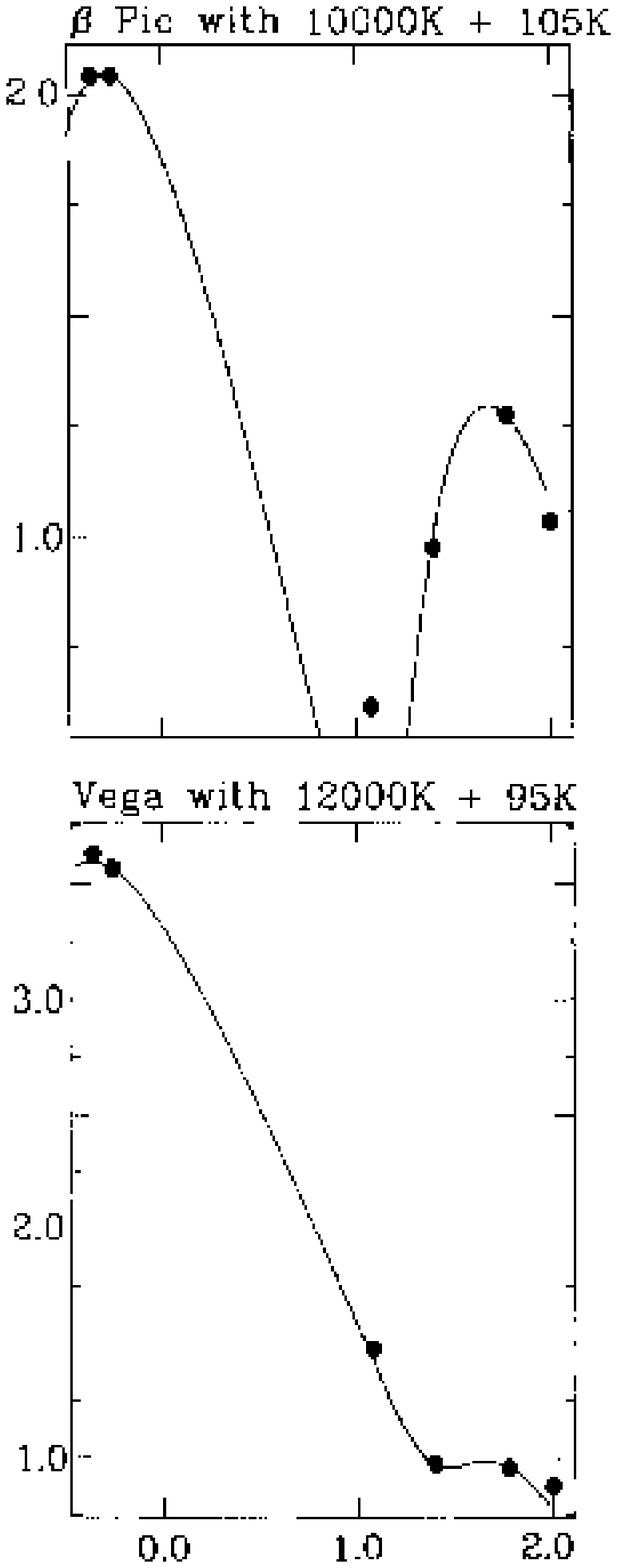,height=4in} &
      \epsfig{file=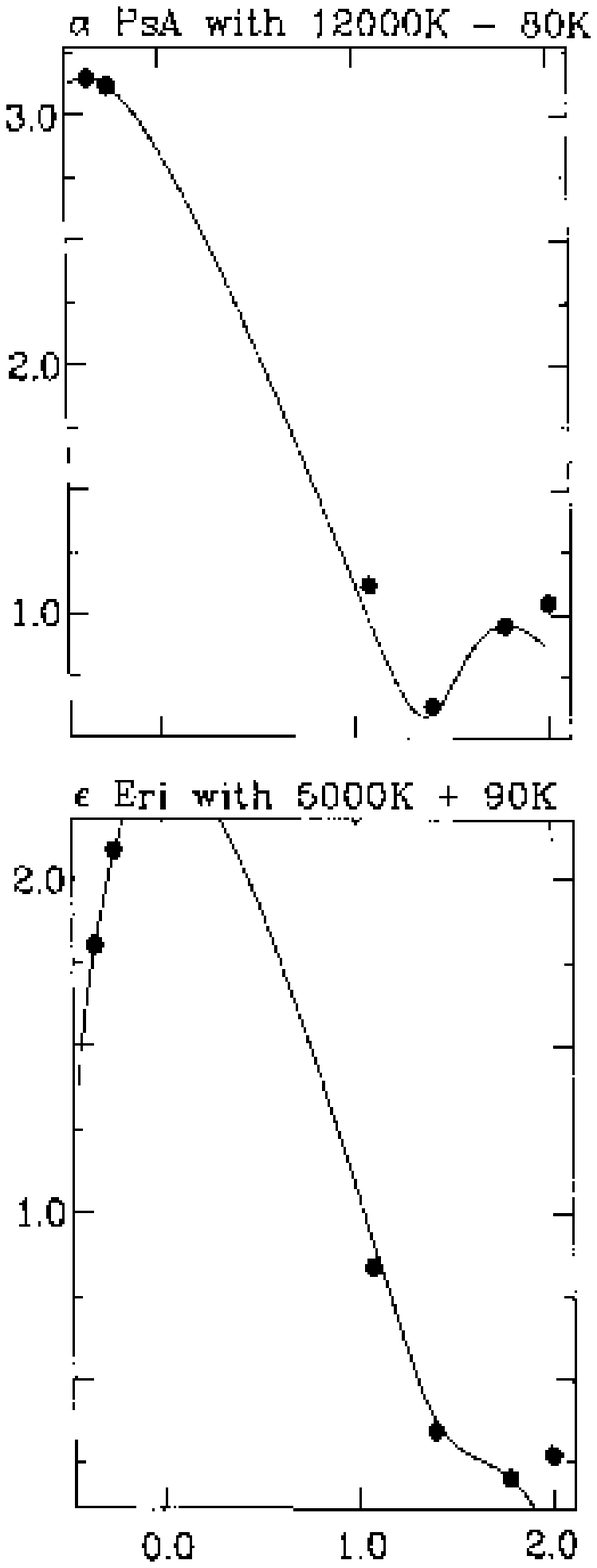,height=4in}
    \end{tabular}
  \end{center}
  \caption{Observed energy distributions and black body fits to the photospheric
  emission and the infra-red excess emission for the four prototype debris disk
  stars (\cite{ww88}).
  Log (flux density in Jy) is plotted against log(wavelength in $\mu$m) for
  $\beta$ Pictoris (top left), Fomalhaut (top right), Vega (bottom left),
  and $\epsilon$ Eridani (bottom right).}
  \label{figww88fig2a}
\end{figure}

Observations of young stars showed that protoplanetary disks dissipate
within $\sim 10^7$ years, supposedly due to planetary formation.
This process was thought to leave a bare main sequence star, possibly
with a planetary system orbiting it, and nothing but a very tenuous disk
of debris material that was unable to form planets, if any at all.
That all changed in 1983 when IRAS took routine calibration observations of
the A0V star Vega\footnote{The A0 is the star's temperature classification
and V is its luminosity classification.
An A0 star has $T_{eff} \approx 9500$ K and the V implies that the star is
on the main sequence}.
The SED of Vega, just like those of younger stellar objects, showed
a far-infrared excess above that expected from the stellar photosphere
(\cite{agbj84}; see Fig.~\ref{figww88fig2a}).
Limited spatial information about the 60 $\mu$m emission showed it to
come from a region of diameter $\sim 20$ arcseconds (160 AU at 8 pc,
the distance to Vega) around the star.
Further, the shape of the SED was shown to be consistent with thermal emission
from mm-sized material at 85 AU from the star that has been heated to
$85-95$ K by the stellar radiation field.
Subsequent IRAS observations of other main sequence stars in the solar
region revealed that many of them have IR excesses similar to that of Vega
(e.g., \cite{auma85}; \cite{ww88}).
After Vega, the next discoveries of infrared excesses were associated
with the stars $\beta$ Pictoris (A5V), Fomalhaut (A3V), and
$\epsilon$ Eridani (K2V+), and these 4 stars have been dubbed the
prototypes of this class (see Fig.~\ref{figww88fig2a}).
Recent analysis of the IRAS data found 108 stars with Vega-type excesses
(\cite{mb98b}).
Thus it was shown that cool ($50-125$ K) solid matter is not uncommon around
main sequence stars of a variety of ages and spectral types.

%%%% Figure of Beta Pictoris images from HST
\begin{figure}
  \begin{center}
    \begin{tabular}{c}
      \epsfig{file=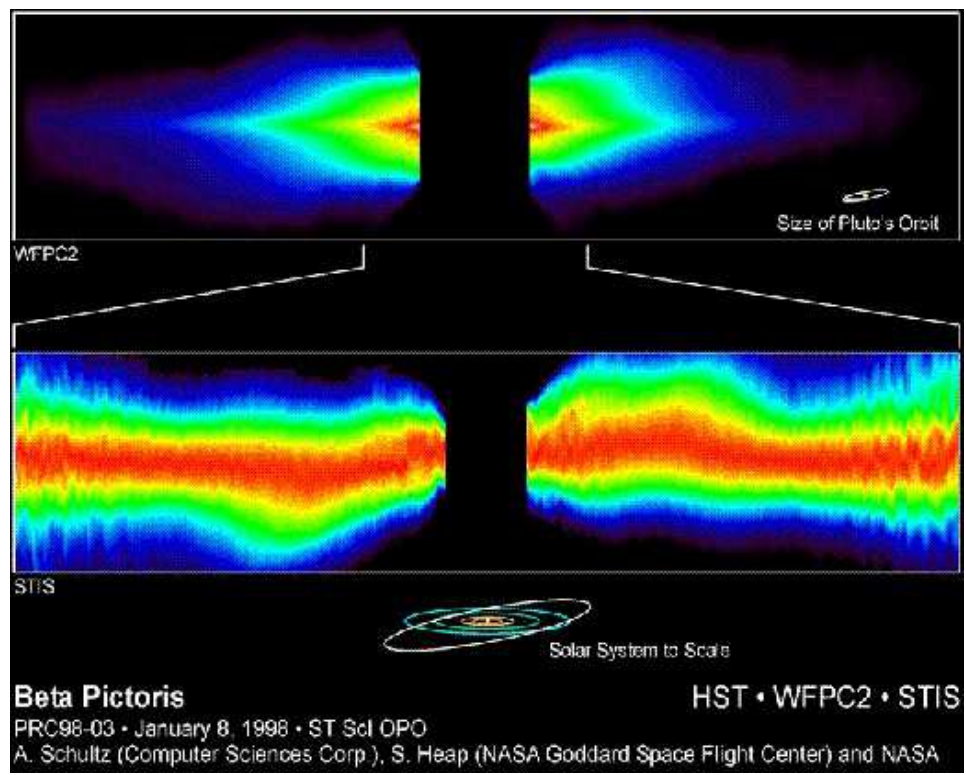,height=4in}
    \end{tabular}
  \end{center}
  \caption{Hubble Space Telescope visible-light views of the edge-on disk of dust
  around the star $\beta$ Pictoris (\cite{sh98}).}
  \label{figbpichst}
\end{figure}

Stunning confirmation that this excess emission comes from material
in orbit around the stars came soon after the IRAS observations with
ground-based optical imaging of $\beta$ Pictoris (\cite{st84};
see Fig.~\ref{figbpichst} for more recent images of the $\beta$ Pictoris disk).
Optical images taken using a coronographic spot to mask the bright stellar
emission showed scattered light from a disk of material seen
edge-on.
Our understanding of the Vega-type disks as it stood in 1993 was summarized
by Backman \& Paresce (1993) who concentrated on interpreting observations
of Vega, $\beta$ Pictoris, and Fomalhaut.
They concluded that the properties of these systems are very suggestive
of planetary systems.
The reasons they gave are:
the infrared emitting grains are larger than interstellar grains;
the material is orbiting, rather than falling into or away from, the star;
the material is arranged in disks in the stellar equatorial planes;
central regions similar in size to the solar planetary system are relatively
lacking in small grains;
and the rapid removal of the small grain population implies a reservoir
of larger undetected parent bodies.
There is, however, no direct evidence of planets in these systems.

Our understanding of the planetary formation process can easily
be invoked to explain these observations.
Consider the solar system.
Apart from the planets, there are several minor bodies orbiting the Sun
on stable orbits that can be explained as remnants of the planetary
formation process (\cite{liss93}):
the asteroids in the asteroid belt between 2 and 4 AU, that are
planetesimals that were inhibited from growth into a planet by
gravitational perturbations from Jupiter;
the Kuiper belt (see, e.g., \cite{jewi99}), which lies beyond the orbit
of Neptune (i.e., beyond 30 AU), is a reservoir of icy bodies that were
unable to form into planets due to the long accretion timescales there;
the short period comets, which have orbits that are more eccentric and
inclined than the asteroids, that are probably scattered Kuiper belt objects
(\cite{ld97});
and the Oort cloud of comets, which extends out beyond 1000 AU, which
are probably scattered planetesimals formed in the Saturn-Neptune region
(\cite{hm99}).
It is also known that in the early solar system there was probably a lot
more debris material that has since been either ejected from the system
(e.g., into the Oort cloud) or accreted onto the planets, since both
lunar and martian cratering records show that the first 800 Myrs
of the solar system was an era of heavy bombardment.
Collisions between debris material in regions in the Vega-type disks
that are analogous to the Kuiper belt in the solar system could provide an
ample reservoir of small particles with which to cause an infrared excess.

However, the planetary formation process is not well defined, and just
because emission is seen from Kuiper belt regions, one cannot automatically
assume that there is a solar-like planetary system interior to that
(\cite{kala98}) --- we could be seeing either failed planetary
systems, or systems quite unlike our own.
It is evident, therefore, that observations of this class of object can
reveal much information about the state of the planetary formation process
in these systems.
Combined with information about the age of the systems\footnote{Note that
age estimates for stars that are on the main sequence are notoriously imprecise
(\cite{ldlh99});
e.g., estimates for the age of $\beta$ Pictoris based on isochrone fitting
gave ages from zero age main sequence, 8 Myr, to a more mature 100 Myr
(the best current estimate based on a variety of techniques is 20 Myr;
\cite{bssc99}).}, this will help to refine the paradigm of planetary formation.
This fact has not gone unnoticed by the astronomical community and
there has been a wealth of observations (and consequent modeling) of
the stars found by IRAS to have Vega-type excesses.
The disk that has attracted the most attention is the $\beta$ Pictoris
disk, because until the 10 $\mu$m images of the K2 star SAO 26804
(\cite{ssgb95}), it was the only Vega-excess star with dust confirmed to
be confined to a disk geometry.
A variety of observational techniques spanning the electromagnetic
spectrum have provided a wealth of information about the $\beta$ Pictoris
system (for a recent review, see \cite{arty97}), but perhaps the most intriguing
result is that there is mounting suggestive evidence for planets in the disk.
Images of the disk in both optical (\cite{kj95}; \cite{sh98},
see Fig.~\ref{figbpichst}) and mid-IR (\cite{pla97})
wavebands show asymmetries in the outer disk, and a warp in the inner disk,
that have been shown by various authors to be possible consequences of
planets embedded in the disk (\cite{rsss94}; \cite{lsrg94}; \cite{mlpl97}).
A further indication of planets is the spectroscopic variability of the
star (\cite{lfv87}) which has been attributed to the
conversion of km-sized solid bodies into hot plasma near the star.
There is also evidence of photometric variability of $\beta$ Pictoris
that may be due to the passing of a giant planet in front of
the stellar image (\cite{ldvf95}).

One of the fundamental goals of the observing programs has been
to obtain resolved images of the material causing the IR excesses
observed by IRAS, since these images provide crucial information about the
spatial distribution of the emitting material.
The reason it took so long for disks other than $\beta$ Pictoris to be
resolved is that the other Vega-excess disks were either too dim or
too distant to be resolved with available technology.
The technology is now available and this field is going through a very
exciting time as disks are being resolved at a steady rate in
the near-IR (e.g., \cite{ssbk99}), the mid-IR (e.g., \cite{tfpk00};
\cite{ftpk99}), and the sub-mm (e.g., \cite{hgzw98}; \cite{ghmj98}).
The potential of these observations to reveal information suggestive
of planets in the systems, which was previously shown with the $\beta$
Pictoris disk, was confirmed with observations of an apparent asymmetry
in the disk around the A0V star HR 4796A (\cite{tfpk00}).
This asymmetry was shown to be a possible indicator of planets in the
system (\cite{wdtf00}).
Another link between these debris disks and planets was made when
a disk was observed around 55 Cancri (\cite{tb98}), which is one
of the stars with a planetary companion as inferred from radial velocity
measurements (see Fig.~\ref{figmarcyplanets}).

%%%%%%%%%%%%%%%%%%%%%%%%%%%%%%%%%%%%%%%%%%%%%
\section{The Zodiacal Cloud}
\label{s-zc}
\begin{figure}
  \begin{center}
    \begin{tabular}{c}
      \epsfig{file=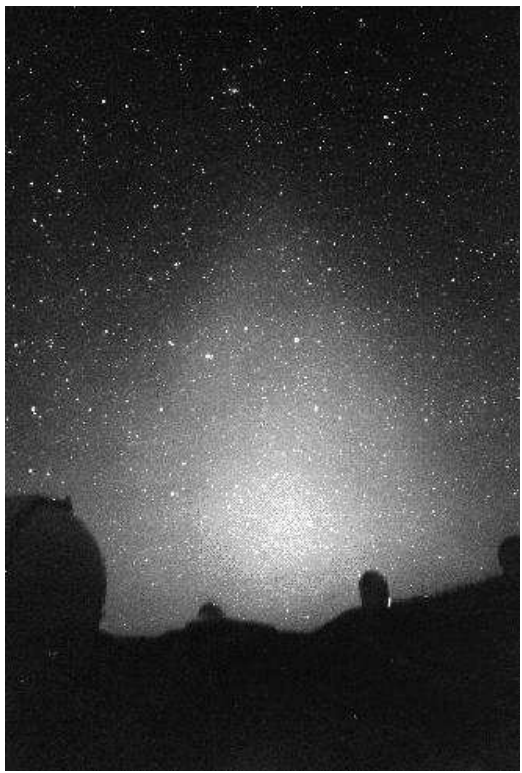,height=4in}
    \end{tabular}
  \end{center}
  \caption{A picture of the zodiacal light obtained at Mt. Mauna Kea, Hawaii
  (\cite{ishi99}).
  The zodiacal light is sunlight scattered by the interplanetary dust cloud
  (i.e., the zodiacal cloud).}
  \label{figzlight}
\end{figure}

At the moment, what is needed is a firm physical model with which
to interpret observations of circumstellar debris disks.
As with the paradigm for planetary formation process, it is most instructive
to start with the system about which we have the most information: the
solar system.
Not only does the solar system contain debris material, it also contains a
tenuous disk of dust, the zodiacal cloud.
The zodiacal cloud is visible to the naked eye from suitable viewing
locations as the zodiacal light, caused by the scattering of sunlight
by the dust particles (see Fig.~\ref{figzlight}).
From the shape and color of the zodiacal light, it was inferred that the
disk is comprised of 1-1000 $\mu$m dust particles extending from a
region beyond the asteroid belt into the Sun forming a disk with a
flattened lenticular shape (see, e.g., \cite{lg90}).
This disk is very tenuous;
it has an optical depth of $\sim 10^{-7}$ at 1 AU, and the total mass
of the disk is estimated to be $10^{16}-10^{17}$ kg, equivalent to the mass
of one large comet (\cite{lg90}).
Further evidence as to the nature of the dust particles in the disk came
from stratospheric dust collection experiments, impact craters on the
Moon, and deep-sea sediment records.
These confirmed the size range of the disk particles and showed some
evidence as to their elemental composition and morphology
(\cite{lg90}; \cite{gust94}).
Meteor observations showed that there is also a population of particles
in the disk that are larger than those responsible for the zodiacal light.

%%%% Zodiacal light picture

Because the zodiacal cloud resides in the solar system, we have a lot of
information available as to the possible sources of this dust.
The zodiacal cloud is generally accepted to be caused by a combination of
the grinding down of the asteroid belt (e.g., \cite{dnbh84};
\cite{dd97}) and the break-up of comets (\cite{slhl86}; \cite{ldx95}).
There is also a small proportion of interstellar grains (\cite{gdg96})
and there is an as yet undetermined amount of Kuiper belt material
(\cite{bds95}; \cite{lzd96}; \cite{gakg97}).
However, the exact contribution of each of these sources is still unknown.
The theory is that the small particles that are responsible for the
zodiacal light are continuously being created from these sources, and evolve
into the inner solar system due to radiation drag forces (\cite{bls79}).
This explains why the zodiacal cloud extends all the way in to the Sun,
while the orbits of the sources do not, and also accounts for the
detailed structure of the cloud (see next paragraph), as well as the size
and composition of the constituent particles.

Much of our information about the structure of the zodiacal cloud has
come from observations of the disk's thermal emission.
The infrared satellites IRAS and COBE mapped the large-scale structure
of the zodiacal cloud with unprecedented detail:
the observations have an extremely high resolution because of the proximity
of the cloud;
they also have a high signal-to-noise ratio because the zodiacal cloud is
the primary source of infrared flux in most directions except that of the
Galactic plane.
Such observations are particularly useful for studying the disk's structure,
since the observations change throughout the year as the Earth moves
through the disk.
These observations proved that the structure of the zodiacal cloud, which
was initially thought to be smooth, actually contains asymmetries:
low level bands of emission were found above the emission of the smooth
background cloud both above and below the ecliptic (\cite{lbgg84});
the center of symmetry of the cloud was shown to be offset from the
Sun (\cite{dghk99});
the cloud was shown to be warped (\cite{dghk99});
and an asymmetric ring of dust was discovered that co-orbits with the Earth
(\cite{djxg94}; \cite{rfwh95}).

To provide a successful interpretation of these observations, techniques
were developed that model the structure of the zodiacal cloud based on an
understanding of the dynamical evolution of the constituent particles from
their source in the asteroids and comets, to their sink at the Sun
(\cite{dgdg92}).
Such physically-based modeling techniques provided a quantitative method for
probing the structure of the zodiacal cloud, and were proven to be extremely
successful.
These techniques showed that the first asymmetry, the dust bands, are caused
by the break-up of individual asteroids (\cite{dnbh84}; \cite{gdjx97}),
while the remaining asymmetries are a direct consequence of the secular and
resonant gravitational perturbations of the solar planetary system (e.g.,
\cite{djxg94}; \cite{dghk99}); i.e., these last asymmetries are the planets'
signatures.

Thus, our understanding of the structure of the zodiacal cloud tells us
that if it were seen from afar, asymmetries in its observed structure could
be used to infer the existence of planets in the solar system.
Actually, even at a distance of 1 pc, the zodiacal cloud would be too dim and
its structure too close to the Sun to be detected with current technology
(\cite{back98}).
Observations of debris dust originating in the Kuiper belt, however, would be
on a much larger scale, and these too could be used to infer information
about the solar planetary system (\cite{lz99}).
The density of such a population is undetermined as yet and it may also be
too dim to detect.
Regardless of whether the solar debris disk is detectable, there are
debris disks that are dense enough to have been observed, and
our understanding of the structure of the zodiacal cloud shows that
if there are any planets in these systems then their signatures could
be imprinted on the observed disk structures.
Furthermore, the zodiacal cloud modeling techniques provide a method
for modeling those structures in a quantitative manner.

%%%%%%%%%%%%%%%%%%%%%%%%%%%%%%%%%%%%%%%%%%%%
\section{Aims and Outline}
\label{s-aims}
Just as recent technological accomplishments have unveiled a new era
in the study of circumstellar debris disks, so they have for the
fields of extrasolar planetary detection and the study of the
protoplanetary disks around young stars.
This means that we are amassing an unprecedented amount of information
about all stages of a star's evolution, and our concepts of that evolution
are continually being challenged, since we want to be able to fit all
observational evidence into a coherent picture of stellar evolution
that explains both the similarities and the dissimilarities between
different systems.
The broad theme that this dissertation aims to contribute to is
our quest to understand how stars and their environments evolve.
More specific questions that the dissertation aims to answer are:
What are the signatures of a planetary system in a circumstellar 
debris disk observation?
What can we infer from debris disk observations about the stellar environment?
Can we infer the existence of planets in the systems that might otherwise remain
undetected?

The observations that are dealt with in this dissertation are images
of a disk's thermal emission.
The physics behind these observations and the role of modeling in their
interpretation is discussed in Chapter \ref{c-obsn}.
Chapter \ref{c-structure} then gives a comprehensive discussion of the
physical processes that govern the evolution of a disk's particles
and uses this to provide a physical model for the evolution of
circumstellar debris disks.
This model is equally valid for the physics of the zodiacal cloud as
for the physics of most debris disks (within the limitations
of the model).
Chapter \ref{c-signatures} shows quantitatively how the presence of planets
in a disk affects its structure, causing the offset, warp and resonant
structures observed in the zodiacal cloud.
Chapter \ref{c-optprops} discusses possible models for the disk
particle's optical properties.
The theory of Chapters \ref{c-obsn} to \ref{c-optprops} can be combined
to model observed debris disk structure.
Such modeling is demonstrated by its application to the observations of
one particular disk, the HR 4796 disk. 
Chapter \ref{c-hrlitrev} gives a literature review of previous work on
the HR 4796 disk;
Chapter \ref{c-hrmodel} describes a model of the disk using these
techniques;
and Chapter \ref{c-interpretation} discusses its interpretation.
The dissertation is summarized in Chapter \ref{c-conclusions},
where recommendations for future work are also given.

%%%%%%%%%%%%%%%%%%%%%%%%%%%%%%%%%%%%%%%%%%%%
\chapter{DISSECTION OF A CIRCUMSTELLAR DISK OBSERVATION}
%%%%%%%%%%%%%%%%%%%%%%%%%%%%%%%%%%%%%%%%%%%%
\label{c-obsn}
The observed brightness of a circumstellar debris disk comes from two
sources:
starlight that has been absorbed by the disk particles and re-emitted as
thermal radiation (primarily at mid-IR, far-IR, and submillimeter
wavelengths at $\lambda > 5$ $\mu$m), and starlight that has been
scattered by the disk particles (primarily at optical and near-IR
wavelengths at $\lambda < 2$ $\mu$m).
This dissertation deals only with a disk's thermal emission.

%%%%%%%%%%%%%%%%%%%%%%%%%%%%%%%%%%%%%%%%%%%%%
\section{Disk Particle Thermal Emission}
\label{s-thermal}
A particle of diameter $D$, that is at a distance $r$ from a star, is heated
by the stellar radiation to a temperature $T$ that can be calculated from
the equilibrium between the energy that the particle absorbs and that
which it re-emits as thermal radiation.
This temperature depends on the particle's optical properties (\cite{gust94}):
\begin{equation}
  T(D,r) = [\langle Q_{abs} \rangle_{T_\star} /
            \langle Q_{abs} \rangle_{T(D,r)}]^{1/4}*T_{bb}, \label{eq:tdr}
\end{equation}
where this equation must be solved iteratively, since the particle's temperature
appears on both sides of the equation,
$\langle Q_{abs} \rangle_{T_\star}$ and $\langle Q_{abs} \rangle_{T(D,r)}$
are the particle's absorption efficiency averaged over the stellar
spectrum (which can be approximated as that of a black body radiating at
the star's effective temperature $T_\star$)
and the spectrum of a black body radiating at a temperature $T$,
and $T_{bb}$ is the equilibrium temperature of the particle if it were a
black body:
\begin{equation}
  T_{bb} = 278.3\sqrt{a_\oplus/r}*(4\sigma/A)(L_\star/L_\odot)^{1/4},
  \label{eq:tbb}
\end{equation}
where $T_{bb}$ is given in K, $\sigma/A$ is the ratio of the particle's
cross-sectional area to its surface area (e.g., spherical particles have
$\sigma = \pi D^2/4$ and $A = \pi D^2$, giving $\sigma/A = 1/4$),
and $L_\star$ and $L_\odot$ are the luminosities of the star and the
Sun.

If this particle is at a distance $R_\oplus$ from the Earth, the
contribution of its thermal emission to the flux density at a wavelength,
$\lambda$, received at the Earth is given by:
\begin{equation}
  F_\nu(\lambda,D,r) = Q_{abs}(\lambda,D)B_\nu[\lambda,T(D,r)]\Omega(D),
  \label{eq:flux}
\end{equation}
where $B_\nu$ is the Planck function, and $\Omega = \sigma/R_\oplus^2$ is
the solid angle subtended at the Earth by the cross-sectional area of the
particle.

%%%%%%%%%%%%%%%%%%%%%%%%%%%%%%%%%%%%%%%%%%%%%%%%
\section{Disk Structure Definition}
\label{s-structure}
A circumstellar disk consists of particles with a range of sizes,
compositions and morphologies.
Throughout this dissertation, however, disk particles are assumed to have
the same composition and morphology, and a particle's size is
characterized by its diameter, $D$.
Disk particles span a range of sizes from $D_{min}$, the smallest
(probably submicron-sized) particles sustainable for a given disk,
up to $D_{max}$, the largest (probably kilometer-sized) members of
the disk that were formed from the protoplanetary disk.
The spatial distribution of these particles can be defined by
$n(D,r,\theta,\phi)$, where $n(D,r,\theta,\phi)dD$ is the volume density
(number per unit volume) of particles in the size range
$D \pm dD/2$ at a location in the disk defined by $r$, the radial distance
from the star, $\theta$, the longitude relative to an arbitrary direction,
and $\phi$, the latitude relative to an arbitrary reference plane.
However, since it is a particle's cross-sectional area that is apparent in an
observation (eq.~[\ref{eq:flux}]), a disk's observable structure is better
defined in terms of $\sigma(D,r,\theta,\phi) = n(D,r,\theta,\phi)\sigma$,
the cross-sectional area per unit volume per unit diameter.

The definition of a disk's structure can be simplified by assuming
the size distribution of its particles to be independent of
$\theta$ and $\phi$:
\begin{equation}
  \sigma(D,r,\theta,\phi) = \bar{\sigma}(D,r)\sigma(r,\theta,\phi),
  \label{eq:sig2}
\end{equation}
where $\bar{\sigma}(D,r)dD$ is the proportion of the total cross-sectional
area of the disk at $r$ that is in particles in the size range $D \pm dD/2$,
and
\begin{equation}
  \sigma(r,\theta,\phi) = \int_{D_{min}}^{D_{max}} \sigma(D,r,\theta,\phi) dD
\end{equation}
is the spatial distribution of cross-sectional area of particles of all
sizes in the disk.

%%%%%%%%%%%%%%%%%%%%%%%%%%%%%%%%%%%%%%%%
\section{Line of Sight Brightness}
\label{s-los}
To determine the observed brightness of a circumstellar disk in, for example,
one pixel of an image of the disk, two components of the observation
must be defined:
the vector, $\mathbf{R}$, which extends from the observer to the disk, and
which describes how the line of sight intersects the disk in terms of
$r$, $\theta$, and $\phi$;
and the solid angle of the observation, $\Omega_{obs}$, where
$\Omega_{obs} = d_{pix}^2$ for pixels of width $d_{pix}$ radians.

Consider a volume element along this line of sight that is at a location in
the disk defined by $r,\theta,\phi$, and that has a length $\mathbf{dR}$;
the element volume is $dV = \Omega_{obs}R_\oplus^2\mathbf{dR}$.
The contribution of the thermal emission of the particles in this element
to the disk's brightness in the observation is given by:
\begin{eqnarray}
  dF_\nu(\lambda,r,\theta,\phi)/\Omega_{obs} & = &
    \int_{D_ {min}}^{D_{max}} Q_{abs}(\lambda,D)B_\nu[\lambda,T(D,r)]
    \sigma(D,r,\theta,\phi)dD \mathbf{dR}, 
    \label{eq:brightness} \\
    & = & P(\lambda,r)\sigma(r,\theta,\phi)\mathbf{dR}, \label{eq:fnu}
\end{eqnarray}
where equation (\ref{eq:fnu}) uses the simplification for the disk structure
given by equation (\ref{eq:sig2}), and
\begin{equation}
  P(\lambda,r) = \int_{D_ {min}}^{D_{max}} Q_{abs}(\lambda,D)
    B_\nu[\lambda,T(D,r)]\bar{\sigma}(D,r)dD. \label{eq:p}
\end{equation}
Thus, the brightness of this element is not affected by the solid angle
of the observation, neither is it affected by the distance of the element
from the Earth.

Equation (\ref{eq:p}) can also be written as:
\begin{equation}
  P(\lambda,r) = \langle Q_{abs}(\lambda,D)B_\nu[\lambda,T(D,r)]
    \rangle_{\sigma(D,r)}; \label{eq:p2}
\end{equation}
i.e., $P(\lambda,r)$ is simply $Q_{abs}(\lambda,D)B_\nu[T(D,r),\lambda]$
(which is a combination of the particles' optical properties)
averaged over the disk's cross-sectional area distribution, $\sigma(D,r)$.
Thus, $P(\lambda,r)$ can also be given by:
\begin{equation}
  P(\lambda,r) = Q_{abs}(\lambda,D_{typ})B_\nu[\lambda,T(D_{typ},r)],
    \label{eq:pdtyp}
\end{equation}
where $D_{typ}$ is the size of disk particle that characterizes
(and hence dominates) the disk's emission.
This characteristic particle size could be different in different
wavebands, as well as at different distances from the star, but always
lies in the range $D_{min} < D_{typ} < D_{max}$;
its value can be found by considering the relative contribution
of particles of different sizes to $P(\lambda,r)$.
Unless the optical properties of the particles prevent it, the
particles that dominate a disk's mid-IR ($\lambda = 10-20$ $\mu$m) emission
are also those that contribute most to its cross-sectional area, i.e., they are
those that dominate the disk's structure, $\sigma(r,\theta,\phi)$
(see, e.g., \S \ref{ssec-mpp}).

The total brightness of the disk in this observation is the integral of
equation (\ref{eq:brightness}) over $\mathbf{R}$.

%%%%%%%%%%%%%%%%%%%%%%%%%%%%%%%%%%%%%%%%%%%%%%%%%%%%%%%
\section{Observing Procedure: Real Images}
\label{s-procedure}
It is not in the remit of this dissertation to describe actual observing
procedures in detail (for an example of which, see \cite{tfpk00}).
This section merely aims to point out that real observations are far more
complicated than equation (\ref{eq:fnu}) might imply.
Consider an image of a disk that is made up of many pixels.
The disk's brightness along the line of sight, $\mathbf{R}$, of each pixel
is given by equation (\ref{eq:fnu}).
This describes a ``pure'' image of the disk emission;
i.e., the best image possibly attainable, albeit in unrealistic observing
conditions.

Consider this image (which is never actually observed) as it travels from
the disk to the detector.
First of all, it is joined by the photospheric emission from the
point-like star.
Before arriving at the Earth, the disk+star emission is subject to extinction
by the interstellar medium, and before arriving at the detector, it suffers
further extinction in the Earth's atmosphere, as well as in the telescope
and detector optics.
This extinction image is joined by an emission image, which includes the
extra emission along the path from the interstellar medium, the sky,
and the telescope and detector optics.
The combined extinction+emission image is smoothed by the observational
point spread function (PSF), a combination of the seeing conditions and the
telescope and instrumental optics, which, in the diffraction-limited case,
for an ideal instrument, can be approximated as gaussian smoothing with
$\textrm{FWHM} = \lambda/D$, where this $D$ is the diameter of
the telescope.
The observed image also contains random noise fluctuations from the
detector array, and suffers systematic variations from pixel to pixel due
to flat field effects (e.g., due to pixel responsivity or dust in the
instrument).

Luckily, observing procedures are able to recover a useful image of the
disk from the mess described in the last paragraph:
a procedure called chopping and nodding removes the sky and telescope
emission (although it cannot eliminate the random sky noise);
the calibration method of looking at nearby standard stars, the flux
densities of which are well-known, removes extinction errors;
the image of the star is subtracted by obtaining the PSF from observations
of point sources and using knowledge of the stellar flux density (usually
extrapolated from visual or near-IR observations assuming the photospheric
emission to fall off like a black body);
and the flat field is corrected for using observations of blank pieces of sky.
There are three potential worries about the image:
(1) The image still contains noise.
If this noise is above the level of disk emission, the disk emission cannot
be unambiguously detected.
Noise in an image can be reduced for the purposes of presentation or
analysis by further smoothing, although this results in a loss of spatial
resolution.
(2) The stellar image cannot be completely removed from the image,
because neither the PSF nor the stellar flux density are known with
infinite precision.
If the disk emission lies within the wings of the diffraction image of
the star, and it is not bright enough relative to the stellar image,
detection of the disk emission might again be ambiguous.
(3) Even if the disk is detected, the image obtained has still been
convolved with the PSF.
This effect can be removed by deconvolution techniques, but these do
not provide unique solutions.

%%%%%%%%%%%%%%%%%%%%%%%%%%%%%%%%%%%%%%%%%%%%
\section{Modeling Disk Images}
\label{s-modeling}
Thus, a disk observation has four components:
the disk's structure, the optical properties of the disk particles,
the orientation of the disk to the line of sight, and the observing procedure.
The first three combine to produce a ``pure'' image of the disk emission,
which is then modified by the observing procedure.
The observing procedure is designed such that its effect on the pure disk
image can be modeled (e.g., the PSF is found by taking observations of
point-like stars).
To model a disk observation, a model of the disk's structure and a model
of the optical properties of the particles are needed.
These models should accurately reflect the physics of the disk, and will
have some free parameters that relate to the physical properties of the disk.
These free parameters can be constrained by comparing the real observations
with observations of the pure model images at different orientations.
Thus the models can be shown either to be contradictory, in which
case they need revision, or consistent, in which cases they can be seen as
one possible solution, and their free parameters can be interpreted on
a physical basis.
It is usually the case when studying circumstellar debris disks that
there is too little information to constrain the models with any
absolute certainty;
i.e., it is easy to find a solution that fits the observation,
but harder to find the right one.
This is true even to some extent when studying the zodiacal cloud.

Observations of a disk in just one waveband cannot constrain any
information about the disk particles' optical properties.
Equation (\ref{eq:fnu}) shows us that the optical properties of the
disk particles only show up in the brightness of a disk in a given
waveband due to the radial dependence of $P(\lambda,r)$
(which is caused by the radial dependence of the temperature and
size distribution).
When modeling a disk image in just one waveband, this radial dependence
must be fixed if the model for the disk's structure is to be constrained.
The optical properties of a disk's particles can be constrained
using observations in different wavebands.
Ideally this is done using disk images in different wavebands, but it
is sometimes done using the disk's SED, with little or no information
about the spatial distribution of disk material.
The focus of this dissertation is on the interpretation of images of
disk structure.
The next two chapters are devoted to the development and discussion
of a model for debris disk structure, and a model for the disk particle
optical properties is discussed in Chapter \ref{c-optprops}.

%%%%%%%%%%%%%%%%%%%%%%%%%%%%%%%%%%%%%%%%%%%%
\chapter{THE STRUCTURE OF A DYNAMIC DEBRIS DISK}
%%%%%%%%%%%%%%%%%%%%%%%%%%%%%%%%%%%%%%%%%%%%
\label{c-structure}
This chapter outlines the theoretical framework upon which later
discussion of circumstellar disk structure (such as how planetary
perturbations affect this structure) is based.
A disk is a dynamic entity, the constituent particles of which
are undergoing constant dynamical and physical evolution.
\S\S \ref{s-gravity} - \ref{s-radiationf} give an extensive discussion
of the physical processes acting on disk particles.
\S \ref{s-categories} then summarizes our understanding of the dynamic disk
by showing how disk particles can be categorized according to the dominant
physical processes affecting their evolution.
\S \ref{s-diskevol} expands on this theme to give a tentative paradigm
for debris disk evolution.

If we are to make generalizations about the physical processes relevant to
debris disk evolution, then the zodiacal cloud is the best
example of a debris disk on which to base this understanding,
since its properties have been determined with a certain degree of
confidence.
This confidence stems from the wealth of observational, theoretical,
and physical evidence describing its present state, its evolutionary
history, and the physical environment of the system it is in.
The physical processes that are described in this section are those
thought to have dominated the evolution of the zodiacal cloud,
supposedly since the Sun reached the main sequence
(e.g., \cite{lg90}; \cite{gust94}).
These processes should serve as an adequate basis for an
understanding of the evolution of debris disks around
other main sequence stars that may be at different evolutionary stages.

There is, however, one obvious distinction between the zodiacal cloud
and exosolar dust disks.
The emission observed from the zodiacal cloud is dominated by that
from dust in the inner solar system (within 5 AU of the Sun), which
has its origins in the asteroid belt (\cite{dnbh84}; \cite{gdjx97}) and
the short period comets (\cite{slhl86}).
In contrast, the emission observed from exosolar dust disks is
dominated by that originating from dust in regions analogous to the
Kuiper belt in the solar system, i.e., $>30$ AU from the star
(e.g., \cite{bp93}).
Relatively little is known about the Kuiper belt, but, like the inner
solar system, it appears to be populated with many asteroid- or
comet-like objects that are probably the remnants of the solar
system's planetary formation phase (e.g., \cite{jewi99}).
Also, it is debatable, how far back in time a zodiacal cloud-like model
is applicable, since the evolution of very young disks could be dominated by
quite different physical processes such as gas drag, self-gravitational
perturbations, coagulation, or effects induced by planetary radial
migration.
Such a model should be valid back to the end of the planetary formation
process at $\sim 10^7$ years.

%%%%%%%%%%%%%%%%%%%%%%%%%%%%%%%%%%%%%%%%%%%%%%
\section{Gravity}
\label{s-gravity}
The dominant force acting on all but the smallest disk particles
is the gravitational attraction of the star:
\begin{equation}
  F_{grav} = GM_\star m/r^2, \label{eq:fgrav}
\end{equation}
where $G$ is the gravitational constant, $M_\star$ is the mass of the
star, $m$ is the mass of the particle, and $r$ is the distance of the
particle from the star.
All material is assumed to orbit the star on Keplerian (elliptical)
orbits, with other forces acting as perturbations to these orbits.
The orbit of a particle is defined by the following orbital elements:
the semimajor axis, $a$, and eccentricity, $e$, that define the radial
extent and shape of the orbit;
the inclination, $I$, and longitude of ascending node, $\Omega$, that
define the plane of the orbit (relative to an arbitrary reference plane);
and the longitude of pericenter, $\tilde{\omega}$, that defines the
orientation of the orbit within the orbital plane (relative to an arbitrary
reference direction).
The orbital period of the particle is given by:
\begin{equation}
  t_{per} = \sqrt{(a/a_\oplus)^3(M_\odot/M_\star)},
  \label{eq:tper}
\end{equation}
where $t_{per}$ is given in years, $a_\oplus = 1$ AU is the semimajor
axis of the Earth's orbit, and $M_\odot$ is the mass of the Sun.
At any instant, the location of the particle in its orbit is defined by the
true anomaly, $f$, where $f=0^\circ$ and $180^\circ$ at the pericenter and
apocenter, respectively, and its distance from the star, $r$, and its
velocity, $v$, are defined by:
\begin{eqnarray}
  r & = & a(1-e^2)/(1+e\cos{f}) \label{eq:rgrav}, \\
  v & = & \sqrt{GM_\star(2/r-1/a)}. \label{eq:vgrav}
\end{eqnarray}

Since at any given time a particle could be at any point along its orbit,
its contribution to the distribution of material in a disk can be
described by the elliptical ring that contains the mass of the particle
spread out along its orbit, the line density of which varies inversely
with the particle's velocity (eq.~[\ref{eq:vgrav}]).
Each disk particle has an orbit defined by a different set of orbital
elements, with a contribution to the spatial distribution of material in
the disk that can be described by a corresponding elliptical ring.
Thus, a disk's structure can be defined by the distribution of orbital
elements of its constituent particles, $n(D,a,e,I,\Omega,\tilde{\omega})$,
where $n(D,a,e,I,\Omega,\tilde{\omega})dDdadedId\Omega d\tilde{\omega}$
is the number of disk particles with sizes in the range $D \pm dD/2$,
and orbital elements in the range $a \pm da/2, e \pm de/2, I \pm dI/2$,
$\Omega \pm d\Omega/2, \tilde{\omega} \pm d\tilde{\omega}/2$.
A disk's orbital element distribution can be quantified in terms of
the evolutionary history of the system and the physical processes
acting on the disk's particles.
Using techniques such as those described in \S \ref{sssec-simul},
the resulting structure can then be compared with the disk's observable
structure, $\sigma(D,r,\theta,\phi)$, to link a disk observation with
the physics of the particles in that disk.
It is often convenient to discuss the dependence of the orbital
element distribution on the different parameters separately;
e.g., the distribution of semimajor axes, $n(a)$, is defined such that
$n(a)da$ is the total number of particles with orbital semimajor axes
in the range $a \pm da/2$.

%%%%%%%%%%%%%%%%%%%%%%%%%%%%%%%%%%%%%%%%%%%%%%
\section{Collisions}
\label{s-collisions}

%%%%%%%%%%%%%%%%%%%%%%%%%%%%%%%%%%%%%%%%%%%%%%
\subsection{Collisional Cascade}
\label{ss-cascade}
A typical disk particle is created by the break-up of a
larger parent body, either as the result of a collision with another
body, or simply by its disintegration.
This parent body could have been created by the break-up of an
even larger body, and the particle itself will most likely end up as a
parent body for particles smaller than itself.
This collisional cascade spans the complete size range of disk
material, excluding the planets\footnote{The planets are the result
of runaway accretion (\cite{liss93}) and are far larger than the
bodies impacting them.
This means that all collisions that they suffer result in accretion
of the impacting body onto the planet.
Collisions between planets, however, can be destructive, and this
may explain the origin of the Moon (see, e.g., \cite{came97}).},
and the particles that share a common ancestor are said to constitute
a ``family'' of particles.

The size distribution that results from this collisional cascade can be
found from theoretical arguments (\cite{dohn69}):
\begin{equation}
  n(D) \propto D^{2-3q}, \label{eq:nd}
\end{equation}
where $q=11/6$.
This distribution is expected to hold for disk particles that are large
enough not to be affected by radiation forces (\S \ref{s-radiationf}).
A disk with this distribution has its mass, $m(D) = n(D)m$,
concentrated in its largest particles, while its cross-sectional area,
$\sigma(D) = n(D)\sigma$, is concentrated in its smallest particles.
Collisions in such a disk are mostly non-catastrophic (see
eqs.~[\ref{eq:dimp}]-[\ref{eq:fcc}]), and a particle in this disk is most
likely to be broken up by a particle that has just enough mass
(and hence energy) to do so.
This in turn means that collisional fragments have velocities, and hence
orbital elements, that are almost identical to those of the original
particle;
i.e., in the absence of other forces, all members of the same family
have identical orbits.
Due to the interaction of the competing physical processes, the size
distribution of disk particles that are affected by radiation forces
is only really understood qualitatively (\S\S \ref{s-radiationf} and
\ref{ss-catsigs}).
Their distribution is particularly important, since, in general,
a disk's cross-sectional area (and hence its observable structure)
is concentrated in these smaller particles.

%%%%%%%%%%%%%%%%%%%%%%%%%%%%%%%%%%%%%%%%%%%%%%%
\subsection{Collisional Lifetime}
\label{ss-tcoll}
The importance of collisions in determining a particle's evolution
depends on its collisional lifetime.
Consider a collision between two disk particles, the larger of which
is denoted by the subscript 1, and the smaller by the subscript 2.
For this collision to be catastrophic, that is, for it to
result in the break-up of the larger particle, the impact energy of the
collision must be large enough both to overcome the tensile strength
of the larger particle, and to impart enough energy to the collisional
fragments to overcome its gravitational binding energy.
In the asteroid belt this limit means that a collision is only catastrophic
if $m_2/m_1 \geq 10^{-4}$ (\cite{dohn69}).
Since the impact energy of a collision is $\propto m_2 v_{rel}^2$,
assuming that exosolar disk particles have similar tensile strengths to the
solar system's asteroids, this limit can be scaled to exosolar disks
by the square of the ratio of the mean relative velocity of collisions in
the asteroid belt (at $\sim 3$ AU), $v_{rel} \approx 5$ km/s
(\cite{vedd98}), to that of collisions in the exosolar disk.
The mean relative velocity of collisions in exosolar disks can
be described by: 
\begin{equation}
  v_{rel}(r) = f(e,I)v(r), \label{eq:vrel}
\end{equation}
where $f(e,I)$ is some function of the disk particles' eccentricities and
inclinations, and $v(r) = 30\sqrt{(M_\star/M_\odot)(a_\oplus/r)}$ km/s
is the average velocity of particles at $r$ (eq.~[\ref{eq:vgrav}] with
$a$ replaced by $r$).
Thus, assuming $f(e,I) \approx 0.3$ as for collisions in the asteroid belt,
an exosolar disk particle of diameter $D \propto m^{1/3}$, would only
suffer a catastrophic collision if the other particle in the collision had a
diameter $\geq D_{cc}(D)$, where
\begin{equation}
  D_{cc}(D) = 0.03[(M_\odot/M_\star)(r/a_\oplus)]^{1/3}D.
  \label{eq:dimp}
\end{equation}

The collisional lifetime, i.e., the mean time between catastrophic
collisions, of a particle of diameter $D$, at a location in a disk
denoted by $r,\theta,$ and $\phi$, is the inverse of its catastrophic
collision rate (\cite{kess81}):
\begin{equation}
  t_{coll}(D,r,\theta,\phi) = [R_{coll}(D,r,\theta,\phi)]^{-1},
  \label{eq:tcol1}
\end{equation}
where
\begin{equation}
  R_{coll}(D,r,\theta,\phi) = \sigma_{cc}(D,r,\theta,\phi)v_{rel}(r),
  \label{eq:rcoll}
\end{equation}
$\sigma_{cc}(D,r,\theta,\phi)$ is the catastrophic collision
cross-section seen by the particle, and $v_{rel}(r)$ is the mean
encounter velocity of disk particles at $r$ (eq.~[\ref{eq:vrel}]).
Using the definition of a disk's structure given by equation
(\ref{eq:sig2}), this catastrophic collision cross-section is:
\begin{equation}
  \sigma_{cc}(D,r,\theta,\phi) = f_{cc}(D,r)\sigma(r,\theta,\phi),
\end{equation}
where
\begin{equation}
  f_{cc}(D,r) = \int_{D_{cc}(D)}^{D_{max}} (1+D/D^{'})^2
          \bar{\sigma}(D^{'},r) dD^{'}, \label{eq:fcc}
\end{equation}
and $D_{cc}(D)$ is the smallest particle with which a catastrophic
collision could occur (eq.~[\ref{eq:dimp}]).
  
However, unless $t_{coll} \ll t_{per}$, the particle's orbit takes
it through a range of $\theta$ and $\phi$ before a collision occurs
(there is also a variation of $r$ along the particle's orbit due to the
eccentricity of its orbit).
Thus, it is more appropriate to calculate the particle's collisional
lifetime using the mean catastrophic collision rate of the particles
in the size range $D \pm dD/2$ that are in the spherical shell of radius,
$r$, and width $dr$.
Consider an element of this shell that has a volume,
$dV = r^2drd\theta\cos{\phi}d\phi$.
The number of particles in the diameter range $D \pm dD/2$ in this
element is given by $n(D,r,\theta,\phi)dDdV$, and each of these particles
has a catastrophic collision rate given by equation (\ref{eq:rcoll}).
Integrating over the whole shell gives:
\begin{equation}
  t_{coll}(D,r) = \frac{ \int_{-I_{max}}^{+I_{max}}\int_0^{2\pi}
      \sigma(r,\theta,\phi) d\theta \cos{\phi} d\phi }
    { \int_{-I_{max}}^{+I_{max}}\int_0^{2\pi}
      [\sigma(r,\theta,\phi)]^2 d\theta \cos{\phi} d\phi
      \ast f_{cc}(D,r) v_{rel}(r)},
\label{eq:tcol3}
\end{equation}
where $I_{max}$ is the maximum inclination of the disk particles' orbits
to the reference plane.

Equation (\ref{eq:tcol3}) can be simplified by considering a cylindrical
shell, defined by $r,\theta,$ and $z$, rather than a spherical one.
An element of the cylindrical shell has a volume $dV = r dr d\theta dz$,
and the corresponding collisional lifetime of a particle in the shell is
given by equation (\ref{eq:tcol3}), but with $\phi$, $\cos{\phi}d\phi$ and
$\pm I_{max}$, replaced by $z$, $dz$, and $\pm h$, where
$h = r\sin{I_{max}}$.
Here we introduce the parameter $\tau_{eff}$, the disk's face-on effective
optical depth:
\begin{equation}
  \tau_{eff}(r) = \int_{-h}^{+h} \sigma(r,\theta,z) dz,
  \label{eq:taueffdefn}
\end{equation}
where the dependence on $\theta$ has been dropped since orbits sample the
full range of $\theta$.
This is not a true optical depth, since that would include a consideration
of the particles' extinction coefficients ($Q_{ext} = Q_{abs} + Q_{sca}$);
rather, it is the disk's face-on surface density of cross-sectional area,
which is equal to its true optical depth if its particles had $Q_{ext} = 1$.
Assuming that $\sigma(r,\theta,z)$ is independent of $z$, so that
$\int_{-h}^{+h} [\sigma(r,\theta,z)]^2 dz = 0.5\tau_{eff}(r)^2/h$,
and that the encounter speed is determined by the vertical motion of
particles in the disk, so that $f(e,I) \approx \sin{I_{max}}$,
equation (\ref{eq:tcol3}) can be simplified to:
\begin{equation}
  t_{coll}(D,r) = \frac{t_{per}(r)}{\pi f_{cc}(D,r) \tau_{eff}(r)},
\label{eq:tcol5}
\end{equation}
where $t_{per}(r)$ is the average orbital period of particles at $r$
(eq.~[\ref{eq:tper}] with $a$ replaced by $r$).

A disk's effective optical depth, $\tau_{eff}$, can be estimated
observationally from equation (\ref{eq:brightness}):
\begin{equation}
  \tau_{eff}(r) \approx (F_\nu(\lambda,r)/\Omega_{obs})/P(\lambda,r),
  \label{eq:taueff}
\end{equation}
where $F_\nu/\Omega_{obs}$ is the disk's face-on unsmoothed
brightness.
The disk's face-on unsmoothed brightness can be calculated either
from the observed brightness, making corrections to account for
both the disk's orientation, as well as for the PSF smoothing, or
from the disk's total flux density, $F_\nu(\lambda)$, and assuming
the disk material to be evenly distributed between $r \pm dr/2$:
\begin{equation}
  F_\nu(\lambda,r)/\Omega_{obs} = F_\nu(\lambda)C_fR_\star^2/rdr,
  \label{eq:fnuobs}
\end{equation}
where $C_f = 6.8 \times 10^9$ AU$^2$/pc$^2$/sr, and
$R_\star$ is the distance of the star from the observer.

%%%%%%%%%%%%%%%%%%%%%%%%%%%%%%%%%%%%%%%%%%
\subsubsection{Collisional Lifetime of Particles with the Most
Cross-sectional Area}
\label{sss-tcoll1}
Consider the particles in a disk that make up most of the disk's
cross-sectional area, i.e., those that have diameters close to $D_{typ}$
and that are expected to characterize the disk's mid-IR emission
(\S \ref{s-los}).
By definition, these particles are most likely to collide with each other
(a collision that would definitely be catastrophic), and so their collisional
lifetime can be found from equation (\ref{eq:tcol3}) using the
approximation:
\begin{equation}
  f_{cc}(D_{typ},r) \approx 4. \label{eq:fccdtyp}
\end{equation}
Applying this approximation to equation (\ref{eq:tcol5}) gives:
\begin{equation}
  t_{coll}(D_{typ},r) = t_{per}(r)/4\pi\tau_{eff}(r).
    \label{eq:tcol6}
\end{equation}
This is also the collisional lifetime derived by Artymowicz (1997) for
particles in $\beta$ Pictoris.

%%%%%%%%%%%%%%%%%%%%%%%%%%%%%%%%%%%%%%%%%
\subsubsection{Collisional Lifetime of a Disk's Large Particles}
\label{sss-tcoll2}
The collisional lifetime of particles of different sizes in a disk differ
only by the factor $f_{cc}(D,r)$.
This factor can be ascertained by making assumptions about the
disk particles' size distribution. 
Assuming that the size distribution of equation (\ref{eq:nd}) holds for
disk particles between $D_{min}$ and $D_{max}$, the
normalized cross-sectional area distribution is given by:
\begin{equation}
  \bar{\sigma}(D) = (3q-5)D^{4-3q}/D_{min}^{5-3q}.
  \label{eq:sigd}
\end{equation}
Substituting into equation (\ref{eq:fcc}) gives:
\begin{equation}
  f_{cc}(D) = \left( \frac{XD}{D_{min}} \right)^{5-3q}
    \left[ 1+\frac{6q-10}{(3q-4)X} + \frac{3q-5}{(3q-3)X^2} \right],
  \label{eq:fcc2}
\end{equation}
where $X = D_{cc}(D)/D$ for $D_{cc}(D) > D_{min}$, and $X = D_{min}/D$
for $D_{cc}(D) \leq D_{min}$;
the collisional lifetime of particles in a disk with this distribution
is a minimum for particles for which $D_{cc}(D) = D_{min}$.
The size distribution of particles in a real disk is more complicated than
equation (\ref{eq:sigd}), however, equation (\ref{eq:fcc2}) can be used to
give a crude approximation for the collisional lifetime of a disk's large
particles:
\begin{equation}
  t_{coll}(D,r) \approx t_{coll}(D_{typ},r)*(D_{cc}(D)/D_{typ})^{3q-5}.
  \label{eq:tcol7}
\end{equation}

Thus, the collisional lifetime of particles with $D > D_{typ}$ can be
considerably longer than that of equation (\ref{eq:tcol6}), and a disk's
largest particles, those for which $t_{coll}(D,r) > t_{sys}$, may not
have suffered any catastrophic collisions since they were first created;
such particles are primordial particles.
The cascades of very young disks may still contain a significant proportion
of primordial particles;
i.e., their cascades may not be fully evolved.

%%%%%%%%%%%%%%%%%%%%%%%%%%%%%%%%%%%%%%%%%%
\subsubsection{Other Considerations for a Particle's Collisional Lifetime}
\label{sss-tcoll3}
If there is a significant change in $r$ along a particle's orbit due to
the eccentricity of its orbit, this can be taken into account when calculating
its collisional lifetime.
By considering the washer-like disk of particles on orbits with $a$, $e$,
and random $\tilde{\omega}$ (\cite{syke90}):
$t_{coll}(a,e) = \pi [\int_{q}^{q'} (r/a)/[t_{coll}(r) \sqrt{(r-q)(q'-r})]dr]^{-1}$,
where $q=a(1-e)$ is this disk's inner edge, and $q'=a(1+e)$ is its outer
edge.

%%%%%%%%%%%%%%%%%%%%%%%%%%%%%%%%%%%%%%%%%%
\subsection{Evidence from the Zodiacal Cloud}
\label{ss-ccinzc}

%%%% Figure of Asteroid Belt Size Distribution
\begin{figure}
  \begin{center}
    \begin{tabular}{c}
      \epsfig{file=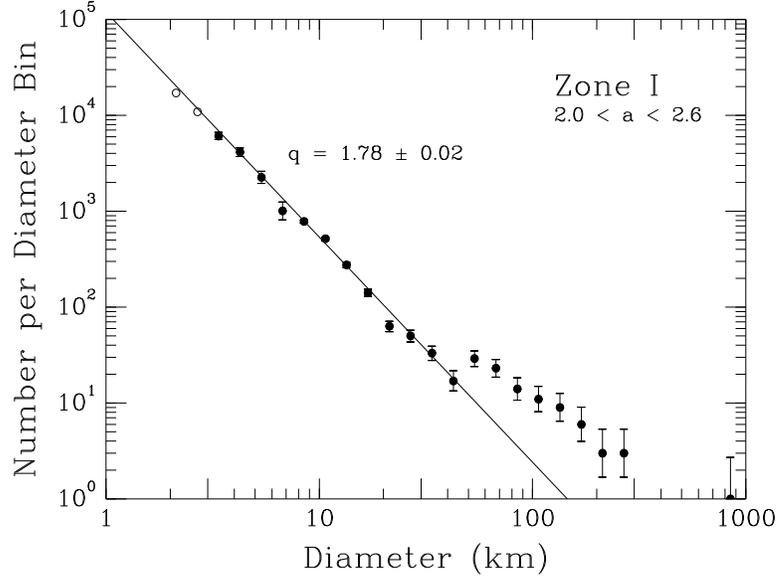,height=3in}
    \end{tabular}
  \end{center}
  \caption{Magnitude-frequency diagram of mainbelt asteroids with semimajor
  axes between 2.0 AU and 2.6 AU (\cite{dd97}).}
  \label{figzone1Dfit}
\end{figure}

\begin{figure}
  \begin{center}
    \begin{tabular}{c}
      \epsfig{file=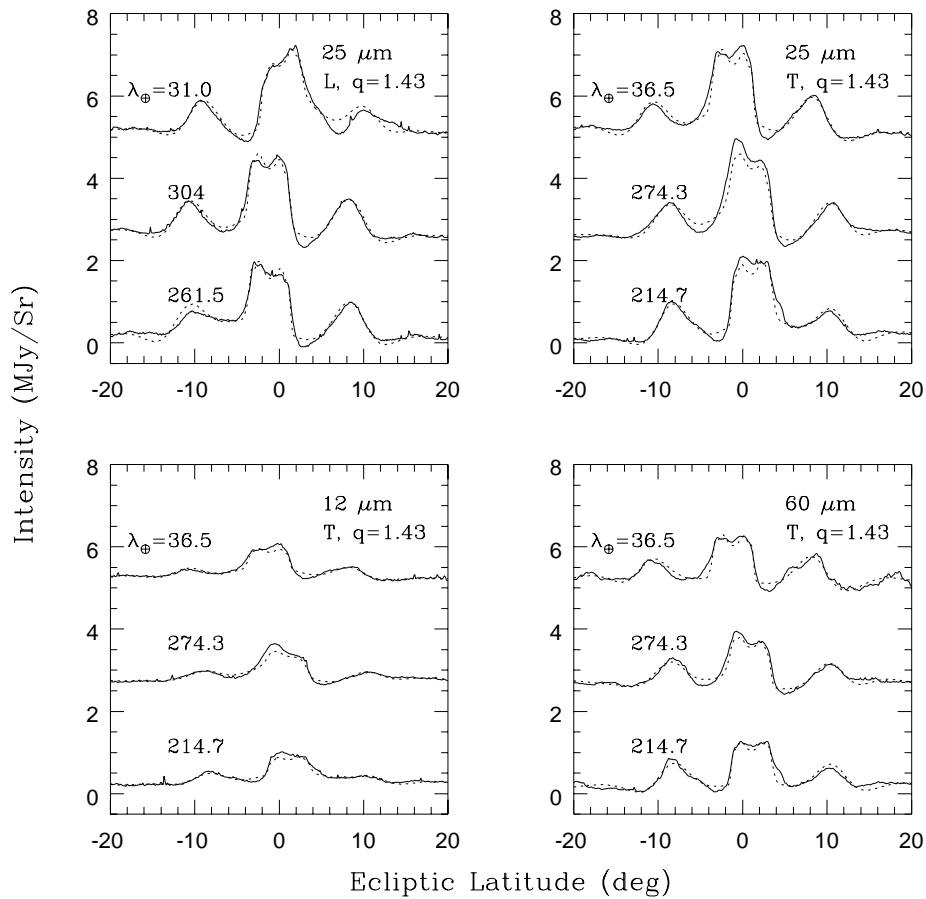,height=5in}
    \end{tabular}
  \end{center}
  \caption{Figure showing the fit between IRAS observations of the
  dust band residuals and a model of this observation (\cite{gdd00}).
  The observations are scans at a constant elongation angle in
  directions both leading (L) and trailing (T) the direction of the
  Earth's motion at three times throughout the year in the
  12, 25, and 60 $\mu$m wavebands.
  The smooth background emission has been subtracted by a Fourier
  filtering process.
  The model includes contributions from particles in the size
  range $1-100$ $\mu$m from the Eos, Themis, and Koronis asteroid
  families that have evolved into the inner system by P-R drag, and
  that have a size distribution that is described by $q=1.43$.}
  \label{figdbfit}
\end{figure}

The collisional cascade theory is well-supported by evidence from the
zodiacal cloud.
The size distribution of the largest ($D>3$ km) members of the zodiacal
cloud's collisional cascade, the observable asteroids, is well-approximated
by equation (\ref{eq:nd}) (\cite{dd97}; \cite{dgj98};
see Fig.~\ref{figzone1Dfit}).
The distribution of the very largest ($D > 30$ km) asteroids deviates from
this distribution, however, because of the transition from strength-scaling
to gravity-scaling for asteroids larger than $\sim 150$ m (\cite{dgj98}).
The size distribution of the zodiacal cloud's medium-sized
(1 mm $< D < 3$ km) members is also expected to follow equation
(\ref{eq:nd}) (\cite{dd97}), but there is no observational proof of this,
since these members are too faint to be seen individually, and too few to
be studied collectively (\cite{lg90}).
There is, however, proof that the zodiacal cloud's collisional cascade
extends from its largest members down to its smallest dust particles:
the dust band thermal emission features (\cite{lbgg84}) correspond to
those expected from the small ($1-1000$ $\mu$m) particles resulting
from the break-up, some time ago, of a few very large asteroids,
the largest fragments of which are still observable as the asteroids
in the Themis, Koronis, and, possibly, Eos families (\cite{dnbh84};
\cite{gdjx97}; \cite{gdd00}; see Fig.~\ref{figdbfit}).
The size distribution of the zodiacal cloud's smallest ($D < 1$ mm)
dust particles (e.g., \cite{lg90}; \cite{lb93}) can be explained
qualitatively (e.g., \cite{gzfg85}; see also \S \ref{s-radiationf}).

%%%% Figure showing fit between the dust band model and observations

Analysis of the collision rates of objects in the Kuiper belt
(\cite{ster95}) shows that a collisional cascade should exist here too.
There is also evidence to suggest that the Kuiper belt was once more
massive than it is today (\cite{jewi99}), meaning that in the past
collisions would have played a much larger role in determining its
structure than they do today, maybe even causing the supposed mass loss
(\cite{sc97}).
The size distribution of the observed Kuiper belt objects appears to
be slightly steeper than that in the inner solar system
($q > 11/6$, \cite{jewi99}), while observations have been unable, as yet,
to determine its dust distribution (\cite{bds95}; \cite{gakg97}), 

%%%%%%%%%%%%%%%%%%%%%%%%%%%%%%%%%%%%%%%%%%%
\section{Radiation Forces, $\beta$}
\label{s-radiationf}
For most of the collisional cascade, gravity can be considered to be
the only significant force acting on disk particles.
The smallest particles, however, are significantly affected by their
interaction with the photons from the star.

%%%%%%%%%%%%%%%%%%%%%%%%%%%%%%%%%%%%%%%%%%%
\subsection{Radiation Pressure}
\label{ss-rpressure}
Radiation pressure is the component of the radiation force that points
radially away from the star.
It is inversely proportional to the square of a particle's distance from
the star, and is defined for different particles by its ratio to the
gravitational force of equation (\ref{eq:fgrav}) (\cite{gust94}):
\begin{equation}
  \beta(D)  =  F_{rad}/F_{grav} =  C_r (\sigma/m) \langle Q_{pr}
    \rangle_{T_\star}(L_\star/L_\odot)(M_\odot/M_\star),  \label{eq:beta}
\end{equation}
where $C_r = 7.65\times 10^{-4}$ kg/m$^2$, $\sigma/m$ is the ratio of
the particle's cross-sectional area to its mass (e.g.,
$\sigma/m = 1.5/\rho D$ for spherical particles of density $\rho$), and
$\langle Q_{pr} \rangle_{T_\star} = \int Q_{pr}(D,\lambda)F_\lambda
d\lambda / \int F_\lambda d\lambda$ is the particle's radiation
pressure efficiency\footnote{A particle's radiation pressure
efficiency is related to its absorption and scattering efficiencies by
$Q_{pr} = Q_{abs} + Q_{sca}(1-\langle \cos{\theta}\rangle)$, where
$\langle \cos{\theta}\rangle$ accounts for the asymmetry of the
scattered radiation.} averaged over the stellar spectrum, $F_\lambda$.

An approximation for large particles is that
$\langle Q_{pr} \rangle_{T_\star} \approx 1$.
Thus, large spherical particles have: 
\begin{equation}
  \beta(D) \approx (1150/\rho D)(L_\star/L_\odot)(M_\odot/M_\star),
  \label{eq:beta2}
\end{equation}
where $\rho$ is measured in kg/m$^3$, and $D$ in $\mu$m.
This approximation is valid for particles in the solar system with
$D>20$ $\mu$m (\cite{gust94}).
For particles in exosolar systems, this limit scales with the wavelength
at which the star emits most of its energy,
$\lambda_\star \propto 1/T_\star$;
i.e., equation (\ref{eq:beta2}) is valid for spherical particles with
$D > 20(T_\odot/T_\star)$ $\mu$m, where $T_\odot = 5785$ K is the effective
temperature of the Sun.
Since $\beta \propto 1/D$, this means that the smaller a particle, the
larger its $\beta$.
This holds down to micron-sized particles, smaller than which $\beta$
decreases to a level that is independent of the particle's size
(\cite{gust94}).

The effect of radiation pressure is equivalent to reducing the mass of the
star by a factor $1-\beta$.
This means that a particle for which $\beta \neq 0$ moves slower
around the same orbit by a factor of $\sqrt{1-\beta}$ than one for which
$\beta = 0$ (eq.~[\ref{eq:vgrav}]).
It also means that daughter fragments created by the break-up of
a parent body move on orbits that can differ substantially from that
of the parent.
The reason for this, is that while the positions and velocities of a
parent and its daughter fragments are the same at the moment of break-up
(apart from a small velocity dispersion), their $\beta$ are different,
and so the daughter fragments move in effective potentials that are different
from that the parent moved in.
Daughter fragments created in the break-up of a parent particle that had
$\beta = 0$, and for which orbital elements at the time of the
collision were $a,e,I,\Omega,\tilde{\omega}$, and $f$, move in the
same orbital plane as the parent, $I^{'} = I$ and
$\Omega^{'} = \Omega$, but on orbits with semimajor axes, $a^{'}$,
eccentricities, $e^{'}$, and pericenter orientations,
$\tilde{\omega}^{'}$, that are given by (\cite{bls79}):
\begin{eqnarray}
  a^{'} & = & a(1-\beta) / \left[1 - 2\beta(1+e\cos{f})/(1-e^2)\right],
  \label{eq:aprime} \\
  e^{'} & = & (1-\beta)^{-1}\sqrt{e^2 + 2\beta e\cos{f} + \beta^2},
  \label{eq:eprime} \\
  \tilde{\omega}^{'} - \tilde{\omega} & = & f - f^{'} =
  \arctan{\left[ \beta \sin{f}/(\beta\cos{f} + e) \right]}. \label{eq:wprime}
\end{eqnarray}

%%%% Figure 1
\begin{figure}
  \begin{center}
    \begin{tabular}{c}
      \epsfig{file=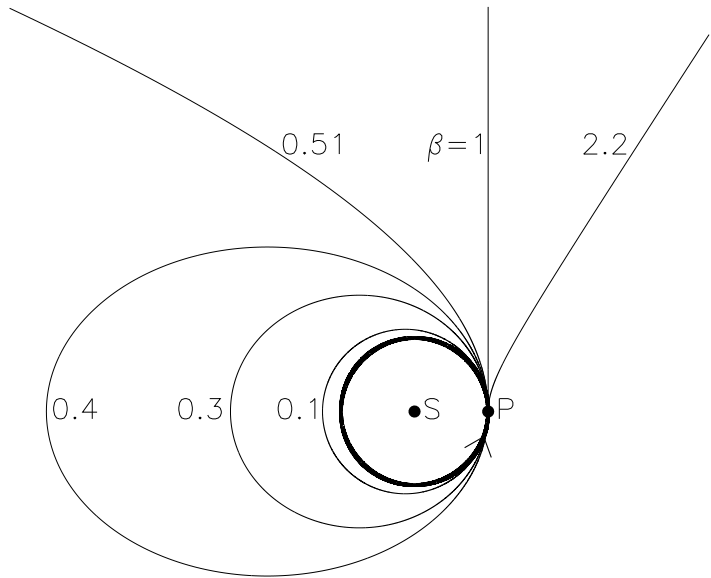,height=3in}
    \end{tabular}
  \end{center}
  \caption{Figure showing the new orbits of the fragments of a collision
  in which a large parent particle P, that was on a circular orbit
  around a star S, was broken up.
  Fragments of different sizes have different radiation pressure forces,
  characterized by a particle's $\beta$, acting on them, and so have
  different orbits:
  those with $\beta < 0.1$, the ``large'' particles, have orbits that are
  close to that of the parent;
  those with $0.1 < \beta < 0.5$, the ``$\beta$ critical'' particles, have
  orbits that have the same pericenter distance as the parent,
  but larger apocenter distances;
  and those with $\beta > 0.5$, the ``$\beta$ meteoroids'', have hyperbolic
  orbits.
  Since when they are created, the velocity vector of all fragments is
  perpendicular to the stellar direction, this is the point of their orbit's
  closest approach to the star.
  The orbits of fragments with $\beta = 0$, 0.1, 0.3, 0.4, 0.51, 1.0, and 2.2
  are shown here.
  The thick circular line denotes both the orbit of the parent particle
  and that of collisional fragments with $\beta = 0$.
  All particles are orbiting the star counterclockwise.}
  \label{fig1}
\end{figure}

Analysis of equations (\ref{eq:aprime})-(\ref{eq:wprime}) shows that
the orbits of the largest fragments, those for which $\beta < 0.1$,
are similar to that of the parent.
On the other hand, the smallest fragments, those for which
$\beta > 0.5(1-e^2)/(1+e\cos{f})$, have hyperbolic orbits ($e^{'} > 1$).
These particles are known as ``$\beta$ meteoroids''\footnote{Note that
particles with $\beta > 1$ are $\beta$ meteoroids even
if they were not created collisionally, since they ``see'' a negative
mass star.}.
Since $\beta$ meteoroids are lost from the system on the timescale
of the orbital period of the parent (eq.~[\ref{eq:tper}]), the diameter of
particle for which $\beta > 0.5$ essentially defines the lower end of
the collisional cascade.
However, there may also be a population of submicron particles that have
$\beta < 0.5$ (\cite{gust94}).
The intermediate-sized fragments, those for which $0.1 < \beta < 0.5$,
that we call ``$\beta$ critical'' particles, have orbits that differ
substantially from that of the parent.
However, the point of closest approach to the star of the orbits of
all daughter fragments, irrespective of their size, is the same as that of
the parent:
combining equations (\ref{eq:aprime}) and (\ref{eq:eprime}) gives the
pericenter distance of daughter fragments, $r_p^{'} = a^{'}(1-e^{'})$, as
\begin{equation}
  r_p^{'}/r_p = 1 + e(1-\cos{f}) + O(e^2). \label{eq:rp}
\end{equation}
The orbits of collisional fragments with different $\beta$ from a
parent particle that was on a circular orbit are shown in Fig.~\ref{fig1}.

%%%%%%%%%%%%%%%%%%%%%%%%%%%%%%%%%%%%%%%%%%
\subsection{Poynting-Robertson (P-R) Light Drag}
\label{ss-prdrag}
The component of the radiation force tangential to a particle's
orbit is called the P-R drag force.

%%%%%%%%%%%%%%%%%%%%%%%%%%%%%%%%%%%%%%%%%%
\subsubsection{Basic Equations}
\label{sss-preqns}
Like the radiation pressure force, the P-R drag force is
proportional to $\beta$.
It results in an evolutionary decrease in both the semimajor
axis and eccentricity of the particle's orbit (\cite{bls79}):
\begin{eqnarray}
  \dot{a}_{pr} & = & -(\alpha/a)(2+3e^2)/(1-e^2)^{3/2}
              = -2\alpha/a + O(e^2) \label{eq:adot} \\
  \dot{e}_{pr} & = & -(\alpha/a^2)2.5e/(1-e^2)^{1/2}
              = -2.5\alpha e/a^2 + O(e^2) \label{eq:edot},
\end{eqnarray}
where $\alpha = 6.24 \times 10^{-4} (M_\star/M_\odot)\beta$
AU$^2$/year.
P-R drag does not change the plane of the particle's orbit,
$\dot{I}_{pr} = \dot{\Omega}_{pr} = 0$;
neither does it affect the orientation of the particle's pericenter,
$\dot{\tilde{\omega}}_{pr} = 0$.
For a particle with zero eccentricity, equation (\ref{eq:adot})
can be solved to find the time it takes for the particle to spiral in from a
radial distance of $r_1$ to $r_2$:
\begin{equation}
  t_{pr} = 400(M_\odot/M_\star)[(r_1/a_\oplus)^2-(r_2/a_\oplus)^2]/\beta
  \label{eq:tpr},
\end{equation}
where $t_{pr}$ is given in years.

%%%%%%%%%%%%%%%%%%%%%%%%%%%%%%%%%%%%%%%%%%%%%
\subsubsection{P-R Drag without Collisions}
\label{sss-prnocoll}
Consider the daughter fragments created in the break-up of a parent body that
was on an orbit at a distance $r$ from the star.
The largest fragments are broken up by collisions before their orbits have
suffered any significant P-R drag evolution, while the smaller fragments,
for which the P-R drag evolution is faster, can reach the star (at which
point they evaporate) without having encountered another particle.
Particles for which P-R drag significantly affects their orbits in their
lifetime can be estimated as those for which their collisional lifetime
(eq.~[\ref{eq:tcol6}]; the use of this equation is justified in the next
paragraph) is longer than their P-R drag lifetime (eq.~[\ref{eq:tpr}]
with $r_2=0$), i.e., those for which $\beta > \beta_{pr}$, where
\begin{equation}
  \beta_{pr} = 5000\tau_{eff}(r)\sqrt{(r/a_\oplus)(M_\odot/M_\star)}.
    \label{eq:btr}
\end{equation}
For large spherical particles, this is also those for which $D < D_{pr}$,
where
\begin{equation}
  D_{pr} = [0.23/\rho \tau_{eff}(r)]\sqrt{(a_\oplus/r)(M_\odot/M_\star)}
    (L_\star/L_\odot), \label{eq:dtr}
\end{equation}
$\rho$ is measured in kg/m$^3$, and $D_{pr}$ in $\mu$m.

Consider the daughter fragments created in the break-up of an endless supply
of parent bodies that are on orbits with the same semimajor axis, $a_s$.
Ignoring collisional processes, the fragments with orbits that are
affected by P-R drag, those with $\beta > \beta_{pr}$, have their
semimajor axes distributed from $a=a_s$ to $a=0$ according to
(eq.~[\ref{eq:adot}]):
\begin{equation}
  n(a) \propto 1/\dot{a}_{pr} \propto a. \label{eq:napr}
\end{equation}
This corresponds to a volume density distribution that is roughly
inversely proportional to distance from the star\footnote{If the
particles had circular orbits, equation (\ref{eq:napr}) means
a spherical shell of width $dr$, the volume of which is $\propto r^2dr$,
would contain a number of particles that is $\propto rdr$
(see, e.g., \cite{gomt97}).}.
If the collisional processes leading to the size distribution of the
parent bodies, $n_s(D)$, still holds for the production of the P-R drag
affected particles, then their size distribution is given by:
\begin{equation}
  n(D) \propto n_s(D)/\dot{a}_{pr} \propto n_s(D)D.
  \label{eq:ndpr}
\end{equation}
If $n_s(D)$ can be given by equation (\ref{eq:nd}) with $q = 11/6$,
the cross-sectional area of a disk's P-R drag affected particles
is concentrated in the largest of these particles, while that
of its unaffected particles is concentrated in the smallest of these
particles;
i.e., most of a disk's cross-sectional area is expected to be concentrated
in particles with $D_{typ} \approx D_{pr}$, justifying the use of equation
(\ref{eq:tcol6}) for the collisional lifetime of these particles.

%%%%%%%%%%%%%%%%%%%%%%%%%%%%%%%%%%%%%%%%%%%%%%%%
\subsubsection{P-R Drag with Collisions}
\label{sss-prcoll}
If the P-R drag lifetime, $t_{pr}$ (eq.~[\ref{eq:tpr}]), is comparable to, or
shorter than $t_{coll}$, then the effect of P-R drag on a particle's
collisional lifetime must be accounted for;
e.g., a particle may evolve quickly out of a dense region where it has a short
collisional lifetime, into a less dense one where it can continue its evolution
without encountering other particles.
On average, particles from a parent at $r_{parent}$, survive
until they reach $r_{coll}$ where $\int_{r_{parent}}^{r_{coll}}
\frac{800(M_\odot/M_\star)r}{t_{coll}(r)\beta} dr = 1$.

The following paragraphs give a heuristic understanding of the structure of
a disk of particles that are evolving due to both P-R drag and collisions.
Consider a source of particles at $r_0$, where all of the particles produced
are of the same size, $D$, and consequently have the same P-R drag
force, $\beta$, affecting them, and all of which have circular orbits. 
If these particles evolve into the inner disk due to P-R drag and 
mutual collisions (which are assumed to destroy both particles involved in
the collision), then the distribution resulting from this
evolution can be found by considering the amount of material entering and
leaving an annulus at $r$ of width $dr$.
The steady state solution is that the amount entering the annulus due to
P-R drag is equal to that leaving due to P-R drag and that which is lost
by collisions (i.e., the continuity equation):
\begin{equation}
  d[n(r)\dot{r}_{pr}(r)]/dr = -N^{-}(r),
  \label{eq:continuity}
\end{equation} 
where $n(r)$ is the one dimensional number density (number of particles per
unit radius),
$\dot{r}_{pr}(r) = -2\alpha/r$ is their P-R drag evolution rate (from
eq.~[\ref{eq:adot}]),
and $N^{-}(r) = n(r)/t_{coll}(D,r)$ is the rate of collisional loss
of $n(r)$.

Taking the collisional lifetime of the particles from equation
(\ref{eq:tcol6}) and using the thin disk approximation that
$\tau_{eff}(r) = \sigma n(r)/2\pi r$, we find that
$N^{-}(r) = 2[n(r)]^2\sigma/rt_{per}(r)$.
Substituting into equation (\ref{eq:continuity}), and solving for the
variation of effective optical depth with distance from the star,
we find that
\begin{equation}
  \tau_{eff}(r)/\tau_{eff}(r_0) = \frac{1}{1+4\eta_0(1- \sqrt{r/r_0})},
  \label{eq:prwcoll}
\end{equation}
where $\eta_0 = 5000 \tau_{eff}(r_0)\sqrt{(r_0/a_\oplus)
(M_\odot/M_\star)}/\beta = \beta_{pr}/\beta$, and
$\tau_{eff}(r_0)$ can be calculated from the production rate of
these particles.
Note that $\eta_0 = 1$ corresponds to the situation when the
collisional lifetime of the particles (if they suffered no P-R drag
evolution) is equal to their P-R drag evolution time
(see eq.~[\ref{eq:btr}]).

%%%% Figure showing number density for disks with P-R drag and collisions
\begin{figure}
  \begin{center}
    \begin{tabular}{rccl}
      \textbf{(a)} & \hspace{-0.4in} \epsfig{file=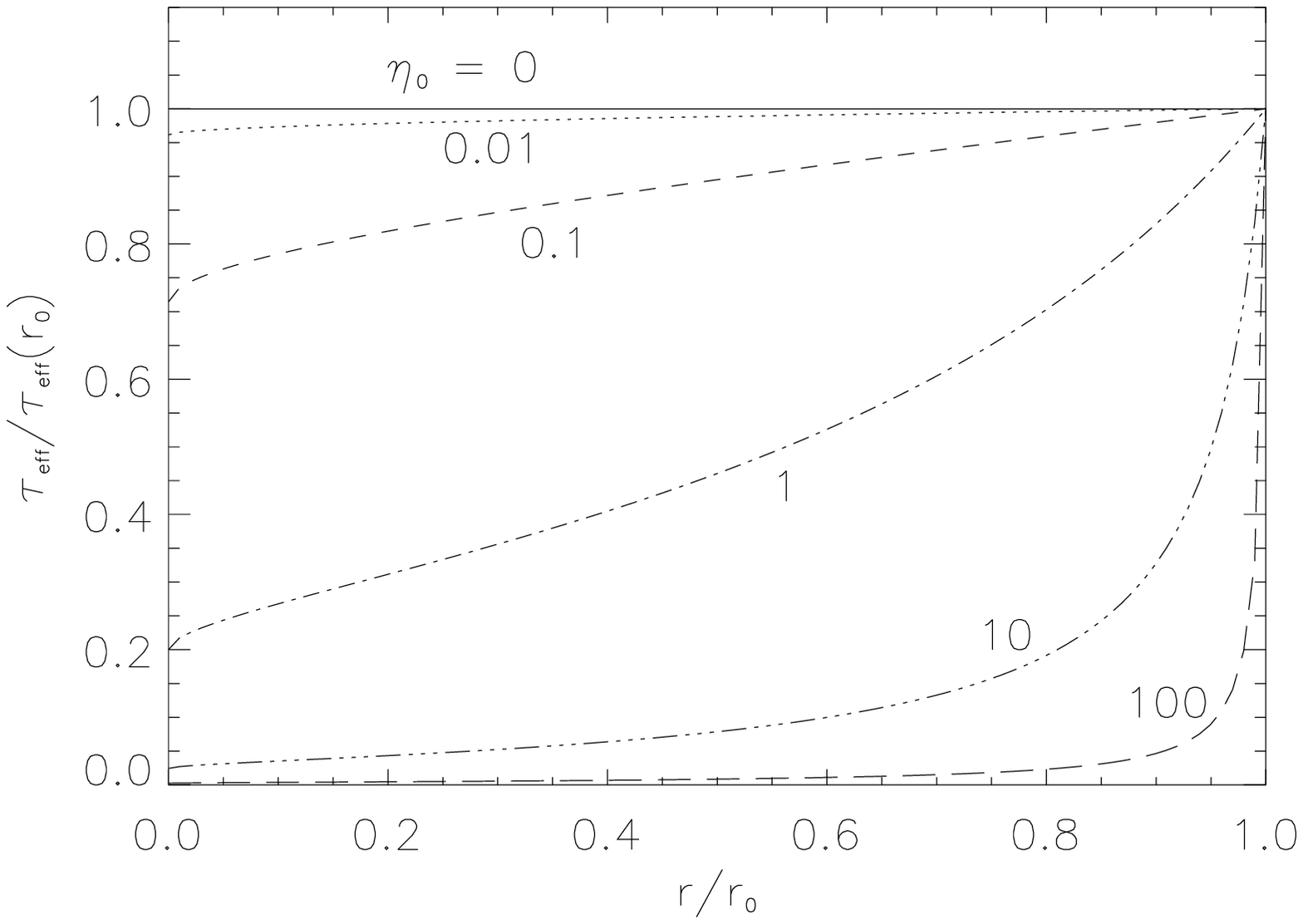,height=1.85in} &
      \epsfig{file=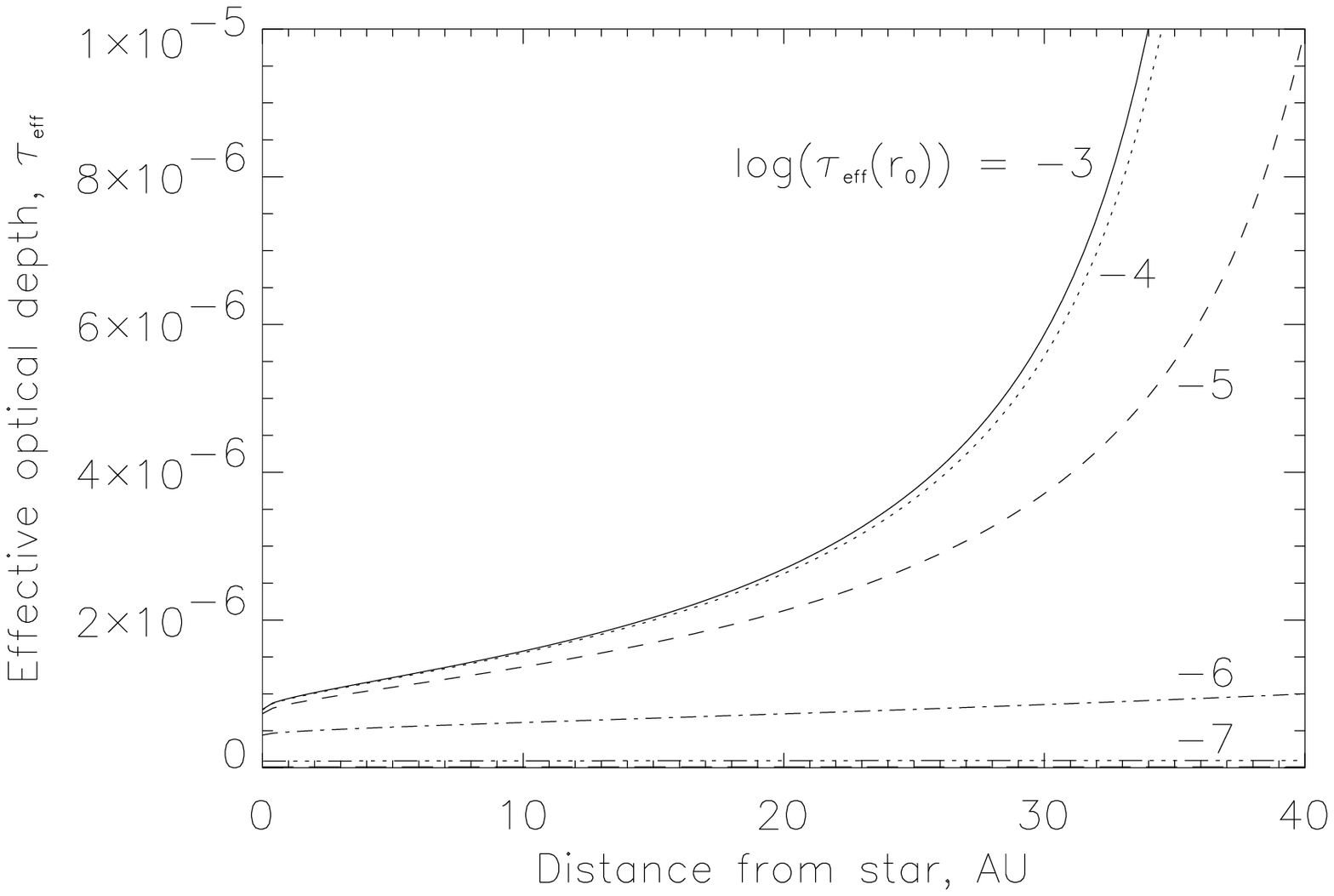,height=1.85in} & \hspace{-0.2in} \textbf{(b)} \\
    \end{tabular}
  \end{center}
  \caption{The distribution of effective optical depth in a disk resulting
  from a source of same-sized particles located at $r_0$.
  The particles evolve into the inner disk due to P-R drag and mutual
  collisions (which are assumed to be destructive).
  The steady state solution depends on the parameter
  $\eta_0$ which is defined by the source parameters.
  A value of $\eta_0 = 1$ corresponds to the situation when the
  collisional lifetime of the source particles (if they suffered no
  P-R drag evolution) is equal to their P-R drag lifetime.
  \textbf{(a)} shows the functional dependence of the distribution on
  $\eta_0$.
  \textbf{(b)} shows the distribution resulting from a source of
  particles with $\beta=0.1$ located at 40 AU from a solar mass
  star, but with different dust production rates that result in
  different effective optical depths at the source.}
  \label{fig:prcoll}
\end{figure}

Thus, the solution given by equation (\ref{eq:prwcoll}), which is shown
in Fig.~\ref{fig:prcoll}a, depends only on the value of $\eta_0$.
In disks that have high values of $\eta_0$ (or equivalently, disks that are
very dense), collisions form a central cavity region.
The particles that have been removed by collisions are presumably blown
out of the system by radiation pressure.
Disks with low values of $\eta_0$, on the other hand, suffer minimal
collisional loss, and their spatial distribution follows
equation (\ref{eq:napr}).
Note that for a disk of particles with $\beta = \beta_{pr}$ (i.e., one
with $\eta_0 = 1$), there is a significant fraction of particles interior
to the source. 

In a more realistic situation, a source produces particles of a range of
sizes at different rates.
These particles then collide with each other, resulting in either
catastrophic or non-catastrophic collisions.
If we approximate a particle's collisional lifetime as being inversely
proportional to the number density of particles of the same size as itself,
then equation (\ref{eq:prwcoll}) can be used to estimate the spatial
distribution of particles of different sizes from the same source.
Since particles of different sizes have different values of $\eta_0$,
their spatial distributions are different.
This also means that the size distribution of particles in the disk
changes with $r$ (although to know what that distribution is,
we need to understand the production rates of different sized particles
in the source).
This is, however, an over-simplified model of the collisional evolution
of particles in a disk moving under P-R drag and collisions.
A more accurate solution could be found by solving the partial
differential continuity equations (eq.~[\ref{eq:continuity}]; where
all of the parameters in this equation would now depend on $D$ as
well as $r$) numerically.
This would be a very interesting study which could be applied to studying
the collisional evolution of particles in the zodiacal cloud, but it is
not within the scope of this dissertation.

%%%%%%%%%%%%%%%%%%%%%%%%%%%%%%%%%%%%%%%%%%%%%%%%
\subsection{Evidence from the Zodiacal Cloud}
\label{sss-prinzc}

%%%% Figure of LDEF results
\begin{figure}
  \begin{center}
    \begin{tabular}{c}
      \epsfig{file=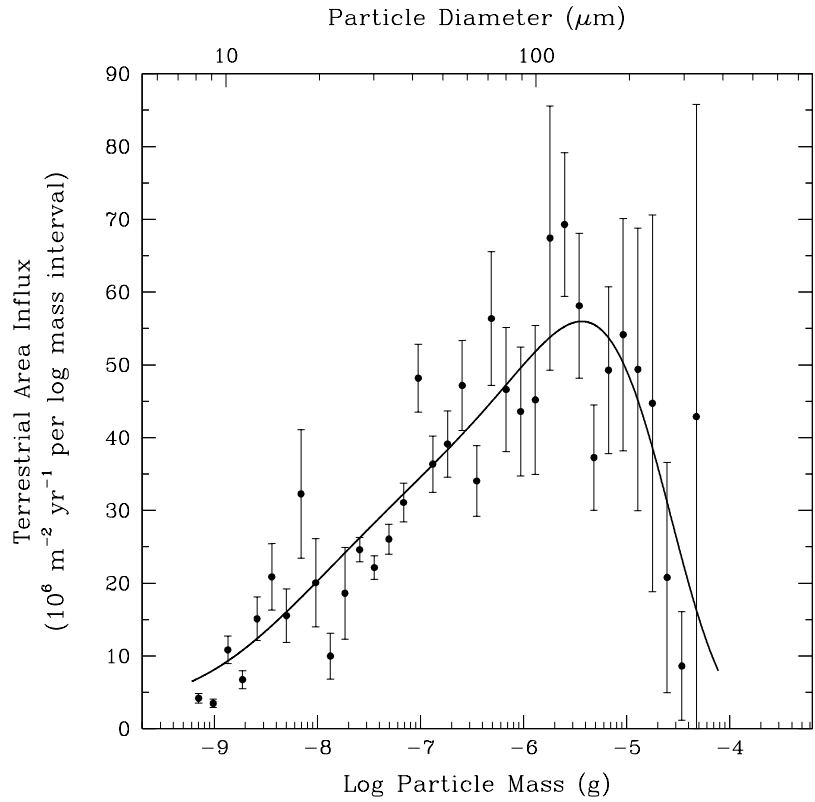,height=5in}
    \end{tabular}
  \end{center}
  \caption{Mass of micrometeorites accreted by the Earth annually per log
  differential particle mass interval (\cite{lb93}).
  This is indicative of the size-frequency distribution of material
  in the zodiacal cloud at 1 AU.
  It is, however, biased towards the type of particle that is preferentially
  accreted by the Earth (see \S \ref{s-accretion} for a further
  discussion of accretion).}
  \label{figldef}
\end{figure}

Observations of the zodiacal cloud at 1 AU show that its effective
optical depth here is $\tau_{eff} = O(10^{-7})$.
Since these particles originated in the asteroid belt at $\sim 3$ AU,
arriving at 1 AU due to the P-R drag evolution of their orbits,
the zodiacal cloud's volume density should vary $\propto 1/r$,
and its effective optical depth at 3 AU should be similar to that at 1 AU.
Assuming zodiacal cloud particles to have a density $\sim 2500$
kg/m$^3$ (\cite{lg90}), the cross-sectional area of material in the
asteroid belt should be concentrated in particles with
$D_{pr} = O$(500 $\mu$m) (eq.~[\ref{eq:dtr}]), for which both the
collisional lifetime, and the P-R drag lifetime, is $\sim 4$ Myr.
The cross-sectional area of material at 1 AU is expected to be
concentrated in particles smaller than that in the asteroid
belt, since many of the larger particles should have been
broken up by collisions before they reach the inner solar system.
This is in agreement with observations that show the
cross-sectional area distribution at 1 AU to peak for particles
with $D = 100-200$ $\mu$m (\cite{lg90}; \cite{lb93}; see
Fig.~\ref{figldef}).
The size distribution of particles evolving due to
P-R drag should have a size distribution with $q<11/6$, possibly
with $q$ as low as $5/6$ (e.g., eq.~[\ref{eq:ndpr}]).
This is supported by modeling of the brightness of the dust
bands in different wavebands which shows that the size distribution
of this material is best fitted with $q=1.43$ (\cite{gdd00};
see Fig.~\ref{figdbfit});
the deviation from $q=5/6$ is indicative of the collisional evolution
of this material.

Also, equation (\ref{eq:tcol7}) with $D_{typ} = 500$ $\mu$m,
$D_{cc}(D)/D = (10^{-4})^{1/3}$, and $q = 11/6$, predicts that the
collisional lifetime of large bodies in the asteroid belt should be:
\begin{equation}
  t_{coll} \approx 10^9\sqrt{D}, \label{eq:tcollast}
\end{equation}
where $t_{coll}$ is given in years, and $D$ in km.
Since the solar system is $\sim 4.5 \times 10^9$ years old,
this implies that asteroids larger than $\sim 20$ km should be
primordial asteroids.
This is in agreement with more accurate models of the observed size
distribution of these asteroids (\cite{dgj98}; see
Fig.~\ref{figzone1Dfit}).
Further evidence of the zodiacal cloud's collisional cascade comes from the
dust detector onboard the Ulysses spacecraft, which found evidence
of small (submicron) $\beta$ meteoroids streaming out of the
solar system on hyperbolic trajectories that were probably created in
collisions between larger bodies in the inner solar system (rather than
by sublimation near the Sun) (\cite{wm99}).

%%%%%%%%%%%%%%%%%%%%%%%%%%%%%%%%%%%%%%%%%%
\section{Disk Particle Categories}
\label{s-categories}
\begin{figure}
  \begin{center}
    \begin{tabular}{rl}
      (\textbf{a}) & \hspace{-0.4in} \psfig{file=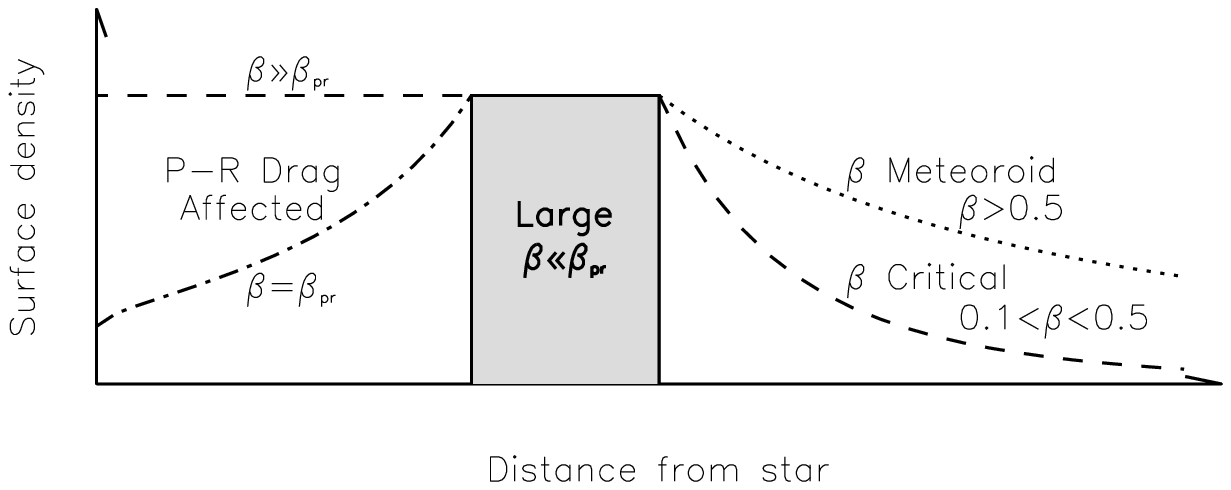,width=5.5in} \\
      \vspace{0.2in} &
    \end{tabular}
    \begin{tabular}{llr}
      \psfig{file=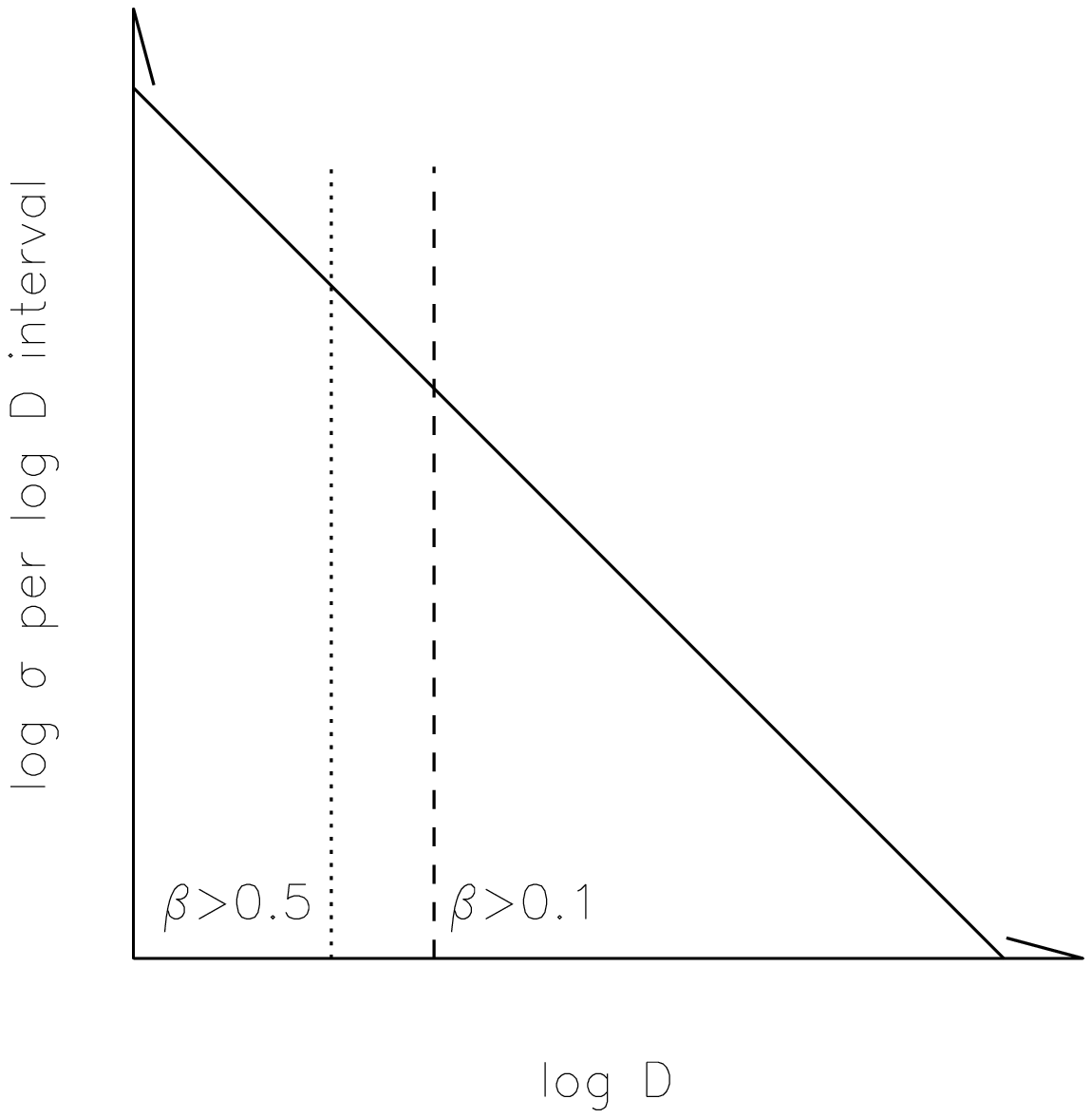,height=2.0in} &
      \psfig{file=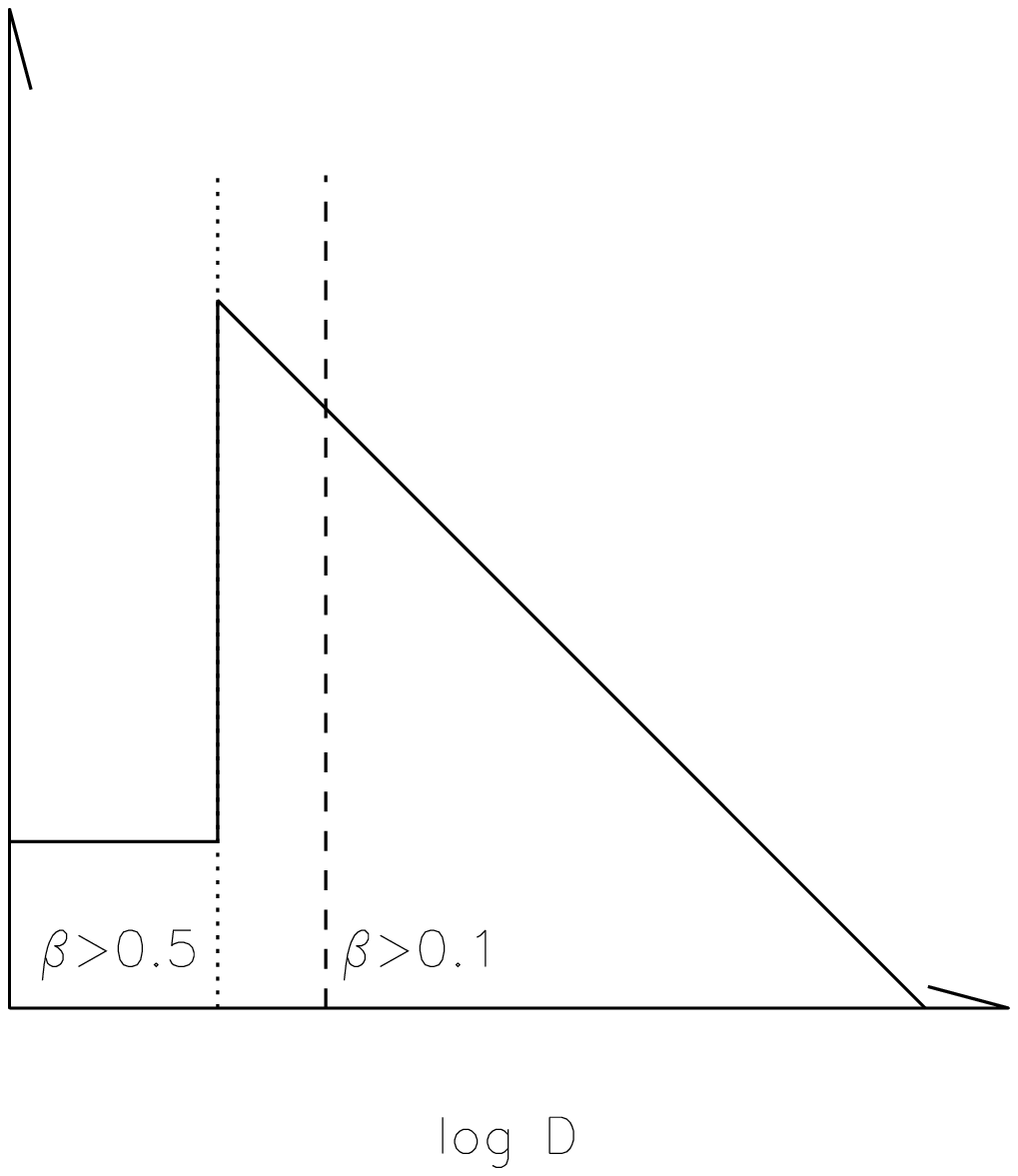,height=2.0in} &
      \psfig{file=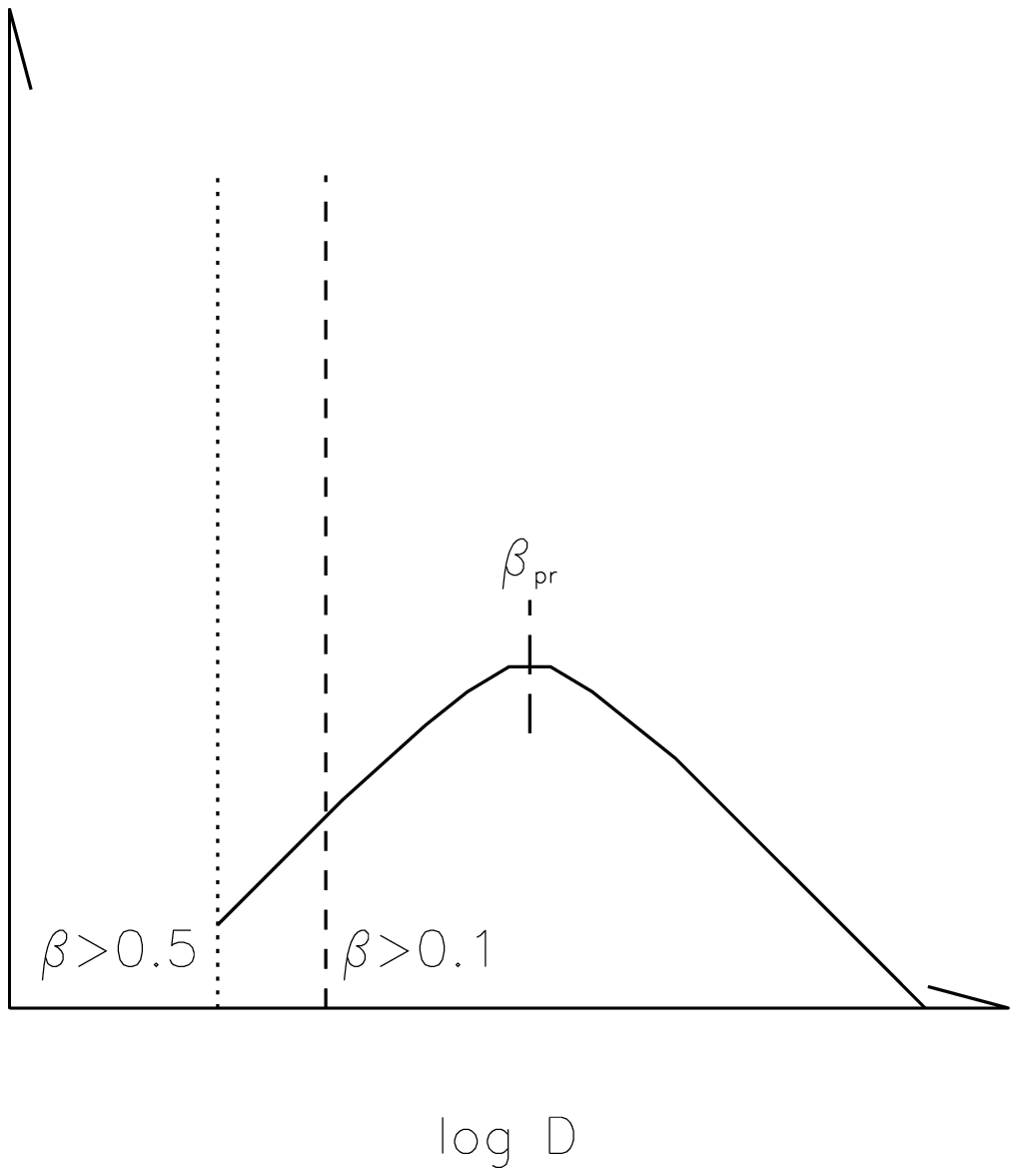,height=2.0in} \\[0.05in]
      (\textbf{b}) $\tau_{eff} > 0.1$ &
      (\textbf{c}) $\tau_{eff PR} < \tau_{eff} < 0.1$ &
      (\textbf{d}) $\tau_{eff} < \tau_{eff PR}$ 
    \end{tabular}
  \end{center}
  \caption{The spatial and size distributions of particles in a debris disk
  comprised of a narrow torus of planetesimals (such as a Kuiper belt-like
  or asteroid belt-like disk).
  (\textbf{a}) The qualitative functional form of the surface density of
  particles of different sizes (and hence different $\beta$) created by
  collisions between the planetesimals.
  The distribution of the planetesimals is shown by the shaded region, which is
  the same as the distribution of the largest collisional fragments.
  (\textbf{b}), (\textbf{c}), and (\textbf{d}) The qualitative cross-sectional
  area distribution in disks of different densities.}
  \label{figddeg}
\end{figure}

Disk particles of different sizes can be categorized according to the
dominant physical processes affecting their evolution.
Particles in the different categories have different lives;
i.e., the way they are created, their dynamical evolution,
and the way they are eventually destroyed, are all different.
Each of a disk's categories has a different spatial distribution.
Which of these categories dominates the disk's observable structure
depends on the relative contribution of each to the disk's
cross-sectional area (see \S \ref{s-los}).
The concept of particle categories is supported by the example of a
debris disk comprised of a narrow torus of planetesimals
(Fig.~\ref{figddeg}).

%%%% Figure of debris disk example

%%%%%%%%%%%%%%%%%%%%%%%%%%%%%%%%%%%%%%%%%%
\subsection{Category Definitions}
\label{ss-catdefns}
A disk's largest particles, those with $\beta < 0.1$, have orbital
elements that are initially the same as, or at least very similar to,
those of their parents.
Of these large particles, only those with $\beta < \beta_{pr}$
suffer no significant P-R drag evolution to their orbits in their
lifetime.
These truly large particles continue on the same orbits as those
of their ancestors until they collide with a particle large enough to
cause a catastrophic collision;
the resulting collisional fragments populate the collisional cascade.
The spatial distribution of these ``large'' particles in a disk is
its base distribution.
The spatial distributions of a disk's smaller particles can only be
understood in terms of how they differ from the disk's base
distribution (see Fig.~\ref{figddeg}a).

Disk particles with $\beta_{pr} < \beta < 0.1$, spiral in from their
parent's orbits due to P-R drag, so that they are closer to the star than
their parents by the time of their demise (which could be caused
either by collisions or by evaporation close to the star).
The spatial distribution of these ``P-R drag affected'' particles in
a disk differs from the disk's base distribution in that it extends
closer in to the star.
The orbits of particles with $0.1 < \beta < 0.5$ also undergo
significant P-R drag evolution before their demise, but their original
orbits are already different from those of their parents.
The spatial distribution of these ``$\beta$ critical'' particles
in a disk extends both further out, and further in, from the disk's base
distribution.
Particles with $\beta > 0.5$ leave their parents on hyperbolic
orbits and so are quickly lost from the system.
The spatial distribution of these ``$\beta$ meteoroids'' in a disk
extends further out, but not further in, from the disk's base distribution. 

Thus, a disk comprises four particle categories, each of which
has a different spatial distribution, although all are inextricably
linked to that of the large particles through the collisional
cascade.
For disks that have $\beta_{pr} > 0.5$, however,
i.e., those with $\tau_{eff}(r) > \tau_{eff PR}(r)$, where
\begin{equation}
  \tau_{eff PR}(r) = 10^{-4}\sqrt{(M_\star/M_\odot)(a_\oplus/r)},
  \label{eq:taueffnopr}
\end{equation}
there is no significant P-R drag evolution of any of its constituent
particles.
Such disks comprise just three categories, since the P-R drag affected
category is empty.

%%%%%%%%%%%%%%%%%%%%%%%%%%%%%%%%%%%%%%%%
\subsection{Category Cross-Sectional Area}
\label{ss-catsigs}
The size distributions of a disk's large particles, and its P-R drag
affected particles, were discussed in \S\S \ref{s-collisions} and
\ref{s-radiationf}.
These discussions imply that the cross-sectional area of a disk
in which there is a population of P-R drag affected particles
(see eq.~[\ref{eq:taueffnopr}]) is dominated by particles with
$D \approx D_{pr}$ (eq.~[\ref{eq:dtr}]; see Fig.~\ref{figddeg}d).
As a first-cut approximation, the size distribution of a disk in which
there are no P-R drag affected particles follows equation
(\ref{eq:nd}) from $D_{max}$ down to $D_{min} = D(\beta=0.5)$
(see, e.g., eq.~[\ref{eq:sigd}] and Fig.~\ref{figddeg}c),
although this is unlikely to be true for the smaller end of this
size distribution, since a particle's catastrophic collision rate is
affected by the size distribution of particles smaller than itself
(\cite{dd97}).
The cross-sectional area of such a disk is concentrated
in its smallest particles, and the contribution of $\beta$ critical
particles to the disk's total cross-sectional area is given by 
\begin{equation}
  d\sigma/\sigma_{tot} = [D^{5-3q}]_{D(\beta=0.1)}^{D(\beta=0.5)} /
                      [D^{5-3q}]_{D_{max}}^{D(\beta=0.5)};
\end{equation}
e.g., if $q=11/6$, and $\beta \propto 1/D$, this means that half of the
disk's cross-sectional area comes from its $\beta$ critical particles.

Since $\beta$ meteoroids have hyperbolic orbits, they are expected to
contribute little to a disk's cross-sectional area unless they are
produced at a high enough rate to replenish their rapid loss from the
system.
This could be the case if the disk was very dense, since material would
pass quickly through the cascade (see Fig.~\ref{figddeg}b);
such a disk would undergo considerable mass loss.
An estimate of how dense the disk would have to be for this to be the
case depends on the assumptions made about the physics of collisions
between small particles.
For heuristic purposes, it is assumed here that the total cross-sectional
area of $\beta$ meteoroids created by the collisional break-up of a parent
body is comparable to that of the parent itself.
This is probably an underestimate if the collision is destructive,
but an overestimate if the collision is erosive.
If this were the case, then the disk's $\beta$ meteoroids would dominate
a disk's cross-sectional area only if their lifetime,
which is of the order of the orbital period of their parents,
is longer than the lifetime of these parents,
which can be approximated by equation (\ref{eq:tcol6}),
i.e., only if
\begin{equation}
  \tau_{eff}(r) > 0.1. \label{eq:taueffnobm}
\end{equation}

In conclusion, from a theoretical stand-point, there are few solid
assumptions that can be made about a disk's size distribution.
However, it would appear that it is a disk's density that determines
the size of particles in which its cross-sectional area is concentrated,
and the denser the disk is, the smaller the diameter of particles that
its cross-sectional area is concentrated in.

%%%%%%%%%%%%%%%%%%%%%%%%%%%%%%%%%%%%%%%%%%%%%%
\section{Debris Disk Evolution}
\label{s-diskevol}
\begin{figure}
  \begin{center}
    \begin{tabular}{c}
      \epsfig{file=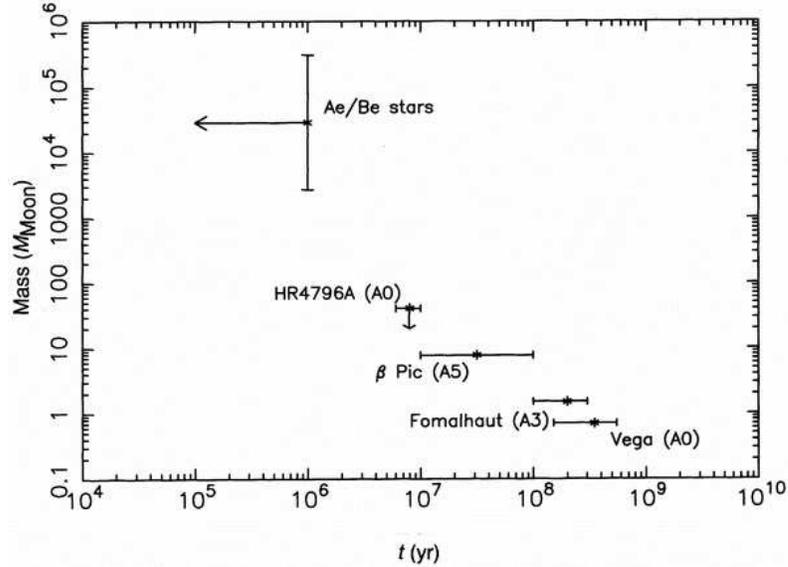,height=3in}
    \end{tabular}
  \end{center}
  \caption{Circumstellar dust masses plotted as a function of stellar
  age (Fig.~2 of \cite{hgzw98}).}
  \label{figtauvst}
\end{figure}

Collating the available observational information, we see that just
as for the protoplanetary disks (\cite{beck99}) there
is a trend for older debris disks to be less massive than younger
disks (see, e.g., Fig.~\ref{figtauvst}).
It also appears that the disk emission tends to arise from regions
analogous to the Kuiper belt in the solar system (\cite{bp93}),
with the region interior to this being relatively (but not
necessarily completely) empty, possibly due to the formation
of planets, or planetesimals, in the inner disk.
Thus, here we consider the evolution of a Kuiper belt disk
of planetesimals.

%%%% Figure of Mass vs time for circumstellar disks from Holland et al. 1998

The emission of Kuiper belt-like disks that have different densities
are dominated by the emission of different particle categories.
The physics affecting the evolution of these disks is also different.
For young disks, that have a high optical depth, their emission
is dominated by $\beta$ meteoroids that are the end products of the
grinding down of larger planetesimals (see Figs.~\ref{figddeg}a and b).
The consequently high mass loss rate results in the decay of the
planetesimal population, which causes a reduction in the disk's
optical depth, and hence a reduction in the mass loss rate.
The optical depth of young debris disks is also reduced by ongoing planet
formation.
Very young systems cannot have reached dynamical equilibrium,
and planetesimals are still lost from source regions through either
accretion onto a nascent planet or through scattering onto highly
eccentric or inclined orbits, ending up as Oort- or short period-like
comets.

The mass loss from a young disk means that eventually it becomes
tenuous enough for $\beta$ critical particles to dominate the disk
emission (eq.~[\ref{eq:taueffnobm}]; see Figs.~\ref{figddeg}a and c).
As the disk dissipates further (at an ever slowing-pace),
P-R drag becomes important, first for the evolution of the disk's
$\beta$ critical particles (eq.~[\ref{eq:taueffnopr}]),
then for the evolution of larger particles (eq.~[\ref{eq:dtr}];
see Fig.~\ref{figddeg}d).
Throughout these stages, the disk's emission (i.e., the disk's
cross-sectional area) comes from particles of increasing size.
Also, the outer edge of the disk creeps inwards, as does the
inner edge (see Fig.~\ref{figddeg}a).
If the particles' P-R drag evolution is not disrupted by a planet
at the inner edge of the disk (see Chapter \ref{c-signatures}),
the extent of migration of the inner edge can be shown by
considering the solution given by equation (\ref{eq:prwcoll})
for a disk with a constant $r_0$ and $D$ but a decreasing $n_0$;
e.g., Fig.~\ref{fig:prcoll}b shows the distribution of $\beta=0.1$
particles from a source at 40 AU from a solar mass star that
has a variable dust production rate which causes the effective optical
depth of the particles at the source to vary.
To understand these latter stages of evolution better, a more
complete analysis of the P-R drag evolution with collisions
that includes a range of particle sizes is needed
(\S \ref{sss-prcoll}).

As the disk gets older, its optical depth (and so its
brightness) decreases at all $r$.
Since a disk's collisional mass loss rate, $-\dot{M}$
(where $M$ is the disk's mass) is proportional
to the collision rate in the disk, which is proportional to the
disk's optical depth, which is roughly proportional to $M^{2/3}$,
the disk never totally disappears.
For a disk with a current mass of $M_1$ and a mass loss rate
$-\dot{M_1}$ (which can be determined from its optical depth),
its mass, $M_2$, after an elapsed time $dt$ will be:
\begin{equation}
  M_2 = M_1(1 - \dot{M_1}dt/M_1)^{-3}. \label{eq:massloss}
\end{equation}
However, a disk will eventually dissipate under the threshold when
it is no longer observable with a given detector system,
and the disks that are preferentially observed are those in
the earlier stages of their evolution, since this is when they
are brightest.

The arguments outlined in the last paragraphs apply equally for the
evolution of an inner disk, such as one associated with an
asteroid-like belt.
The zodiacal cloud can be explained as an asteroid belt disk in
the latter stages of its evolution, where its P-R drag affected
particles reach all the way into the Sun and dominate its thermal
emission (see Figs.~\ref{figddeg}a and d).
The rate of evolution of the two (inner and outer) components in a
disk depends on their stability against gravitational perturbations
from the planetary system and the initial density of the
planetesimal population.
There may also be some interaction between the two populations;
$\beta$ meteoroids and $\beta$ critical particles from the inner
disk may interact with the outer disk, and P-R drag affected
particles from the outer disk may interact with the inner disk.
Given the myriad of possible outcomes of the planetary formation
process, the wavelengths and radial location at which the emission
from a disk arises varies from disk to disk, as well as
for the same disk at different stages of its evolution.

%%%%%%%%%%%%%%%%%%%%%%%%%%%%%%%%%%%%%%%%%%%%%%
\section{The Perturbed Dynamic Disk}
\label{ss-pertns}
In addition to the physical processes described in
\S\S \ref{s-gravity}-\ref{s-radiationf}, the particles of the dynamic disk
are affected by a number of perturbing processes.
These produce subtle, but perhaps observable, and in some cases maybe
dominant, changes in a disk's structure.
The dominant perturbing processes in the zodiacal cloud are the
secular and resonant gravitational perturbations of the planets.
These are discussed in Chapter \ref{c-signatures}.

Other possible perturbing processes include:
stellar wind forces, that, at least for dust in the solar system, effectively
increase the value of $\beta$ for P-R drag (e.g., \cite{lg90});
Lorentz forces acting on charged particles, that, while negligible for
particles in the inner solar system, are increasingly important for
particles at distances further from the Sun (e.g., \cite{km98});
interactions with dust from the interstellar medium (e.g., \cite{ac97});
the sublimation of icy dust grains, which is one of the mechanisms that has
been suggested as the cause of the inner hole in the HR 4796 disk
(\cite{jmwt98});
the self gravity of a massive disk, which could have played an important
role in determining the evolution of the primordial Kuiper belt
(\cite{wh98});
the Yarkovsky force, which may be important for the evolution of
intermediate-sized meteorites (e.g., meter-sized) in the asteroid belt
(e.g., \cite{fvh98});
and gas drag, which is important for evolution in a young disk
in which the gaseous component of the protoplanetary disk has not
yet dissipated (which is thought to occur in $\sim 10^7$ years,
\cite{beck99}).

%%%%%%%%%%%%%%%%%%%%%%%%%%%%%%%%%%%%%%%%%%%%%
\chapter{SIGNATURES OF PLANETARY PERTURBATIONS}
%%%%%%%%%%%%%%%%%%%%%%%%%%%%%%%%%%%%%%%%%%%%%
\label{c-signatures}
Consider the structure of a young debris disk.
It is thought that by an age of $\sim 10^7$ years, any planets that are to
form in the disk should have accumulated most of their final mass
(\cite{liss93}).
The remainder of their mass, which they should accumulate by $\sim 10^8$
years, comes from a planetesimal population that has already been depleted
by the initial growth of the planet.
This planetesimal disk (i.e., the debris disk) already has structure that
is indicative of the nascent planetary system, since regions close to the
orbits of the planets are relatively empty, this material having been
accreted onto, or scattered by, the growing planet.
Thus the first signature of a planetary system is the large-scale
radial distribution of the reservoir of planetesimals that collide to
produce the dust that is seen by its thermal emission (or by its
scattering of the starlight).
In the solar system, the fact that planetesimals are only found
in specific regions such as the asteroid belt and the Kuiper belt is
indicative of the solar planetary system.
However, regions that are relatively empty of planetesimals are only
suggestive, rather than demonstrative, of planetary formation,
since there could be other reasons for the depletion;
i.e., we cannot rely on the planetary formation paradigm to
explain the state of a young debris disk.

What we can be certain of, is that any debris particle in a circumstellar
disk that is in a system in which there are one or more massive perturbers
inevitably has its orbit affected by the gravitational perturbations of
these bodies.
The consequent evolution of the orbits of individual particles can be
used to obtain a quantitative understanding of the effect of these
perturbations on the structure of the disk.
There are three types of gravitational perturbation from a planetary
system that act on the orbit of a disk particle:
secular perturbations, resonant perturbations, and close
encounter perturbations (such as scattering or accretion).
The signatures of planets seen in the solar system debris disk can
be attributed to these perturbations:
the offset and warp asymmetries are caused by secular perturbations;
resonant rings structures are caused by resonant perturbations;
and the large scale radial distribution of material is caused
by both resonant and close encounter perturbations.

%%%%%%%%%%%%%%%%%%%%%%%%%%%%%%%%%%%%%%%%%%%%
\section{Secular Perturbations}
\label{s-secpertns}

%%%%%%%%%%%%%%%%%%%%%%%%%%%%%%%%%%%%%%%%%%%%
\subsection{Perturbation Equations}
\label{ss-pertneqns}
The gravitational forces from a planetary system that act to perturb the
orbit of a particle in the system can be decomposed into the sum of many
terms that are described by the particle's disturbing function, $R$.
The long-term average of these forces are the system's secular
perturbations, and the terms of the disturbing function that contribute
to these secular perturbations, $R_{sec}$, can be identified as those that
do not depend on the mean longitudes of either the planets or the particle
(the other forces having periodic variations).

Consider a particle that is orbiting a star of mass $M_\star$, that also
has $N_{pl}$ massive, perturbing, bodies orbiting it.
This particle has a radiation pressure force acting on it represented by
$\beta$, and its orbit is described by the elements $a$, $e$, $I$,
$\Omega$ and $\tilde{\omega}$.
To second order in eccentricities and inclinations, the secular terms in the
particle's disturbing function are given by
(\cite{bc61}; \cite{dnbh85}; \cite{dn86}; \cite{md99}):
\begin{equation}
  R_{sec} =  na^2 \left[ \frac{1}{2}A(e^2-I^2)
    + \sum_{j=1}^{N_{pl}} \left[A_j e e_j cos(\tilde{\omega}-\tilde{\omega}_j)
    + B_j I I_j cos(\Omega-\Omega_j) \right] \right],
\label{eq:rsec}
\end{equation}
where $n = (2\pi/t_{year})\sqrt{(M_\star/M_\odot)(1-\beta)(a_\oplus/a)^3}$
is the mean motion of the particle in rad/s,
$t_{year} = 2\pi / \sqrt{GM_\odot/a_\oplus^3} = 3.156\times 10^7$ s is one year
measured in seconds, and
\begin{eqnarray}
  A   & = & + \frac{n}{4(1-\beta)}
              \sum_{j=1}^{N_{pl}}
	      \left( \frac{M_{j}}{M_\star}     \right)
	      \alpha_j\overline{\alpha}_j b_{3/2}^{1}(\alpha_j),
  \label{eq:a} \\
  A_j & = & - \frac{n}{4(1-\beta)}
	      \left( \frac{M_{j}}{M_\star}     \right)
            \alpha_j\overline{\alpha}_j b_{3/2}^{2}(\alpha_j),
  \label{eq:aj} \\
  B_j & = & + \frac{n}{4(1-\beta)}
	      \left( \frac{M_{j}}{M_\star}     \right)
            \alpha_j\overline{\alpha}_j b_{3/2}^{1}(\alpha_j),
  \label{eq:bj}
\end{eqnarray}
where $\alpha_j = a_j/a$ and $\overline{\alpha}_j = 1$ for $a_j<a$, and 
$\alpha_j = \overline{\alpha}_j = a/a_j$ for $a_j>a$,
and $b_{3/2}^s(\alpha_j) = (\pi)^{-1}\int_0^{2\pi}(1-2\alpha_j\cos{\psi}+
\alpha_j^2)^{-3/2}\cos{s\psi}d\psi$ are the Laplace coefficients ($s=1,2$).
$A$, $A_j$ and $B_j$ are in units of rad/s, and
$R_{sec}$ is in units of m$^2$/s$^2$.

The effect of these perturbations on the orbital elements of the
particle can be found using Lagrange's planetary equations
(\cite{bc61}; \cite{md99}).
The semimajor axis of the particle remains constant, 
$\dot{a}_{sec} = 0$, while the variations of its eccentricity
and inclination are best described when coupled with the
variations of its longitude of pericenter and ascending node
using the variables defined by its complex eccentricity, $z$,
and complex inclination, $y$:
\begin{eqnarray}
  z & = & e*\exp{i\tilde{\omega}}, \label{eq:zdefn} \\
  y & = & I*\exp{i\Omega}, \label{eq:ydefn}
\end{eqnarray}
where $i^2 = -1$.
Using these variables Lagrange's planetary equations give the orbital
element variations due to secular perturbations as:
\begin{eqnarray}
  \dot{z}_{sec} & = & +iAz + i\sum_{j=1}^{N_{pl}}A_j z_j, \label{eq:zdot} \\
  \dot{y}_{sec} & = & -iAy + i\sum_{j=1}^{N_{pl}}B_j y_j, \label{eq:ydot}
\end{eqnarray}
where $z_j$ and $y_j$ are the complex eccentricities and inclinations
of the perturbers, which have a slow temporal variation due to the
secular perturbations of the perturbers on each other (\cite{bc61};
\cite{md99}):
\begin{eqnarray}
  z_j(t) & = & \sum_{k=1}^{N_{pl}}e_{jk}\ast\exp{i(g_kt + \beta_k)},
    \label{eq:zj} \\
  y_j(t) & = & \sum_{k=1}^{N_{pl}}I_{jk}\ast\exp{i(f_kt + \gamma_k)},
    \label{eq:yj}
\end{eqnarray}
where $g_k$ and $f_k$ are the eigenfrequencies of the perturber system,
the coefficients $e_{jk}$ and $I_{jk}$ are the corresponding
eigenvectors, and $\beta_k$ and $\gamma_k$ are constants found from the
initial conditions of the perturber system.

%%%%%%%%%%%%%%%%%%%%%%%%%%%%%%%%%%%%%%%%%%
\subsection{Solution to Perturbation Equations}
\label{ss-pertnsoln}
\begin{figure}
  \begin{center}
    \begin{tabular}{c}
      \epsfig{file=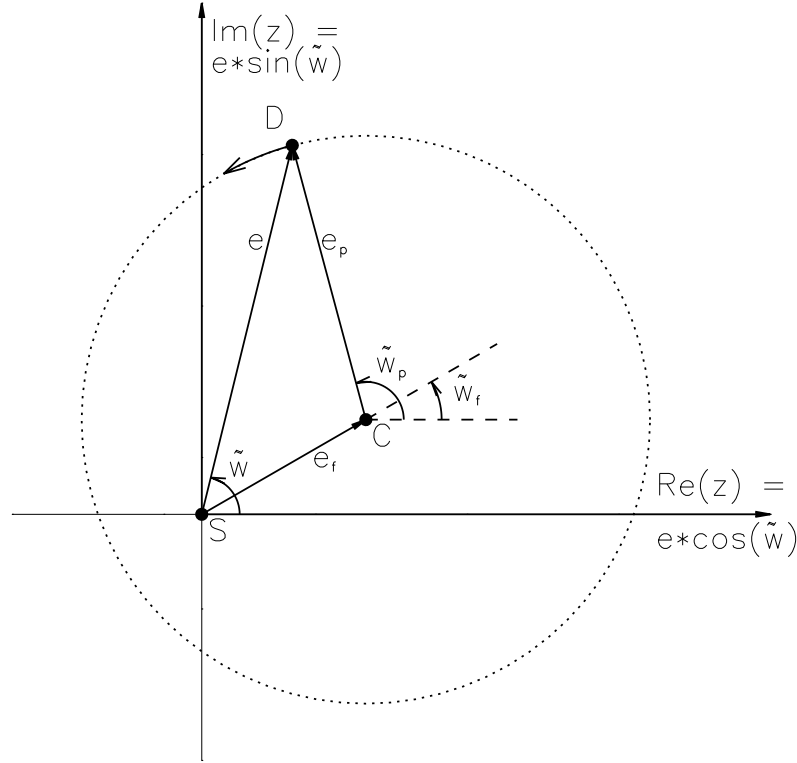,height=4in}
    \end{tabular}
  \end{center}
  \caption{The osculating (instantaneous) complex eccentricity,
  $z = e \ast \exp{i\tilde{\omega}} = SD$, of the orbit of a particle in a
  system with one or more massive perturbers can be resolved vectorially into
  two components:
  a forced eccentricity, $z_f = e_f \ast \exp{i\tilde{\omega}_f} = SC$, that is
  imposed on the particle's orbit by the perturbers;
  and a proper eccentricity, $z_p = e_p \ast \exp{i\tilde{\omega}_p} = CD$,
  that is the particle's intrinsic eccentricity.
  The secular evolution of its complex eccentricity is to precess
  counterclockwise around the dotted circle in the figure, although the
  forced eccentricity may also vary with time.
  A similar figure applies for the particle's osculating complex inclination,
  $y = I \ast \exp{i\Omega}$.}
  \label{fig2a}
\end{figure}

Ignoring the evolution of a particle's orbital elements due to P-R drag,
equations (\ref{eq:zdot}) and (\ref{eq:ydot}) can be solved to give
the secular evolution of the particle's instantaneous complex
eccentricity and inclination (a.k.a.~the particle's osculating
elements).
This secular evolution is decomposed into two distinct time-varying
elements --- the ``forced'', subscript $f$, and ``proper'', subscript $p$,
elements --- that are added vectorially in the complex planes
(see Fig.~\ref{fig2a}):
\begin{eqnarray}
  z(t) & = & z_f(t) + z_p(t),  \label{eq:zfpdef} \\
       & = & \sum_{k=1}^{N_{pl}} \left[
           \frac{ \sum_{j=1}^{N_{pl}}A_j e_{jk}}{g_k-A} \right]
           \ast\exp{i(g_kt + \beta_k)} +
           e_p\ast\exp{i(+At+\beta_0)},  \label{eq:z} \\
  y(t) & = & y_f(t) + y_p(t),  \label{eq:yfpdef} \\
       & = & \sum_{k=1}^{N_{pl}} \left[
           \frac{ \sum_{j=1}^{N_{pl}}B_j I_{jk}}{f_k+A} \right]
           \ast\exp{i(f_kt + \gamma_k)} +
           I_p\ast\exp{i(-At+\gamma_0)},  \label{eq:y}	    
\end{eqnarray}
where $e_p$, $\beta_0$, and $I_p$, $\gamma_0$ are
determined by the particle's initial conditions.

%%%% Figure 2a

These equations have simple physical and geometrical interpretations.
A particle's forced elements, $z_f$ and $y_f$, depend only on the
orbits of the perturbers in the system (that have a slow secular
evolution, eqs.~[\ref{eq:zj}] and [\ref{eq:yj}]), as well as on the
particle's semimajor axis (which has no secular evolution).
Thus, at a time $t_0$, a particle that is on an orbit with a
semimajor axis $a$, has forced elements imposed on its orbit by
the perturbers in the system that are defined by $z_f(a,t_0)$ and
$y_f(a,t_0)$.
The contribution of the particle's proper elements to its osculating
elements, $z(t_0)$ and $y(t_0)$, is then given by
$z_p(t_0) = z(t_0) - z_f(a,t_0)$ and $y_p(t_0) = y(t_0) - y_f(a,t_0)$,
thus defining the particle's proper eccentricity, $e_p$, and proper
inclination, $I_p$, which are its fundamental orbital elements
(i.e., those that the particle would have if there were no perturbers in
the system), as well as the orientation parameters $\beta_0$ and
$\gamma_0$.
Since both the forced elements, and the osculating elements, of
collisional fragments are the same as those of their parent
(apart from fragments with $\beta > 0.1$), particles from the same
family have the same proper elements, $e_p$ and $I_p$.

The evolution of a particle's proper elements is straight-forward
--- they precess around circles of fixed radius, $e_p$ and $I_p$, at a
constant rate, $A$, counterclockwise for $z_p$, clockwise for
$y_p$.
The secular precession timescale depends only on the semimajor
axis of the particle's orbit:
\begin{equation}
  t_{sec} = 2\pi /At_{year},
  \label{eq:tsec}
\end{equation}
where $t_{sec}$ is given in years, and $A$ is given in
equation (\ref{eq:a});
secular perturbations produce long period variations in a particle's
orbital elements (e.g., $t_{sec} = O$(0.1 Myrs) in the asteroid belt).
The centers of the circles that the proper elements precess
around are the forced elements (see, e.g., Fig.~\ref{fig2a}).
Actually the forced elements vary on timescales that are
comparable to the precession timescale (eq.[\ref{eq:tsec}]).
Thus, it might appear ambitious to talk of the precession of a particle's
osculating elements around circles when its real evolutionary track in
the complex eccentricity and complex inclination planes may not be
circular at all.
The reason it is presented as such is that at any given time, all of the
particles at the same semimajor axis precess (at the same rate)
around the same forced elements on circles of different radii, and this
has consequences for the global distribution of orbital elements
(see \S \ref{ss-ofwpfam}).

There are two things that are worth mentioning now about a particle's
forced elements.
If there is just one perturber in the system, $N_{pl} = 1$, its complex
eccentricity and complex inclination do not undergo any secular
evolution, and the forced elements imposed on a particle in the system
are not only constant in time, but also independent of the mass of the
perturber:
\begin{eqnarray}
  z_f & = & \left[ b_{3/2}^{2}(\alpha_j) / b_{3/2}^{1}(\alpha_j) \right]
            e_j\ast\exp{i\tilde{\omega}_j}, \label{eq:zf1} \\
  y_f & = & I_j\ast\exp{i\Omega_j}. \label{eq:yf1}
\end{eqnarray}
This implies that a body of low mass, such as an asteroid, has as
much impact on a particle's orbit as a body of high mass, such as a
Jupiter mass planet.
The perturbations from a smaller perturber, however, produce longer
secular precession timescales (eq.~[\ref{eq:tsec}]):
\begin{equation}
  t_{sec} = 4[\alpha_j \overline{\alpha}_j b_{3/2}^1(\alpha_j)
             (a_\oplus/a)^{3/2}(M_j/M_\star)\sqrt{M_\star/M_\odot}]^{-1}.
          \label{eq:tsec2}
\end{equation}
Also, for a very small perturber, the perturbations could be similar in
magnitude to those of the disk's self-gravity, which in that case could
no longer be ignored.
If there is more than one perturber in the system, $N_{pl} > 1$,
then particles on orbits for which their precession rate equals one of
the system's eigenfrequencies ($A = g_k$, or $-A = f_k$) have infinite
forced elements imposed on their orbits, and so are quickly ejected
from such a secular resonance region.

The solution given by equations (\ref{eq:z}) and (\ref{eq:y}) accounts
for the fact that small particles see a less massive star due to the
action of radiation pressure, but not for the P-R drag evolution of their
orbits.
To find the secular evolution of the orbital elements of a particle that
is affected by P-R drag, i.e., one with $\beta_{pr} < \beta < 0.1$,
the equations governing the evolution of its complex eccentricity,
$\dot{z} = \dot{z}_{sec} - 2.5(\alpha/a^2)z$
(eqs.~[\ref{eq:edot}] and [\ref{eq:zdot}]), and its complex inclination,
$\dot{y} = \dot{y}_{sec}$ (eq.~[\ref{eq:ydot}]),
must both be solved in conjunction with the P-R drag evolution of its
semimajor axis (eq.~[\ref{eq:adot}]).
While the solution given by equations (\ref{eq:z}) and (\ref{eq:y}) is
no longer applicable, the decomposition of the particle's complex
eccentricity and complex inclination into forced and proper elements
(eqs.~[\ref{eq:zfpdef}] and [\ref{eq:yfpdef}]),
and the physical meaning of these elements, is still valid.
However, each of these elements now depends on the particle's
dynamical history.

Also, the perturbation theory of \S \ref{ss-pertneqns} is only valid
for particles with small eccentricities;
i.e., it is not valid for the evolution of a disk's $\beta$ critical particles,
or its $\beta$ meteoroids.
However, if the evolution of a disk's $\beta$ critical particles is
affected by secular perturbations (i.e., if their lifetime is longer than
the secular timescale), then it is probably also affected by
P-R drag (i.e., their lifetime is probably also longer than the P-R drag
timescale), in which case the disk's $\beta$ critical particles do not
contribute much to its observable structure (\S\S \ref{s-radiationf}
and \ref{ss-catsigs}).
There is no secular evolution to the orbits of $\beta$ meteoroids because
of their short lifetimes.

%%%%%%%%%%%%%%%%%%%%%%%%%%%%%%%%%%%%%%%%%
\subsection{Offset and Plane of Symmetry of Family Material}
\label{ss-ofwpfam}
The effect of secular perturbations on the structure of a disk can be
understood by considering the effect of the secular evolution of the
constituent particles' orbits on the distribution of their orbital
elements.
The perturbation equations of \S \ref{ss-pertneqns} show that secular
pertubations affect only the distribution of disk particles' complex
eccentricities, $n(z)$, and complex inclinations, $n(y)$,
while having no effect on their size distribution (and hence the
division of the disk into its particle categories), or on their semimajor
axis distribution (and hence the disk's large-scale radial distribution).
Consider the family of collisional fragments originating from a
primordial body, the orbital elements of which were described by
$a$, $e_p$, and $I_p$.

%%%%%%%%%%%%%%%%%%%%%%%%%%%%%%%%%%%%%%%%%%
\subsubsection{Large ($\beta < 0.1$) Fragments}
\label{sss-ofwpfamlarge}

%%%% Figure of Asteroid families z and y diagrams
\begin{figure}
  \begin{center}
    \begin{tabular}{cc}
      \epsfig{file=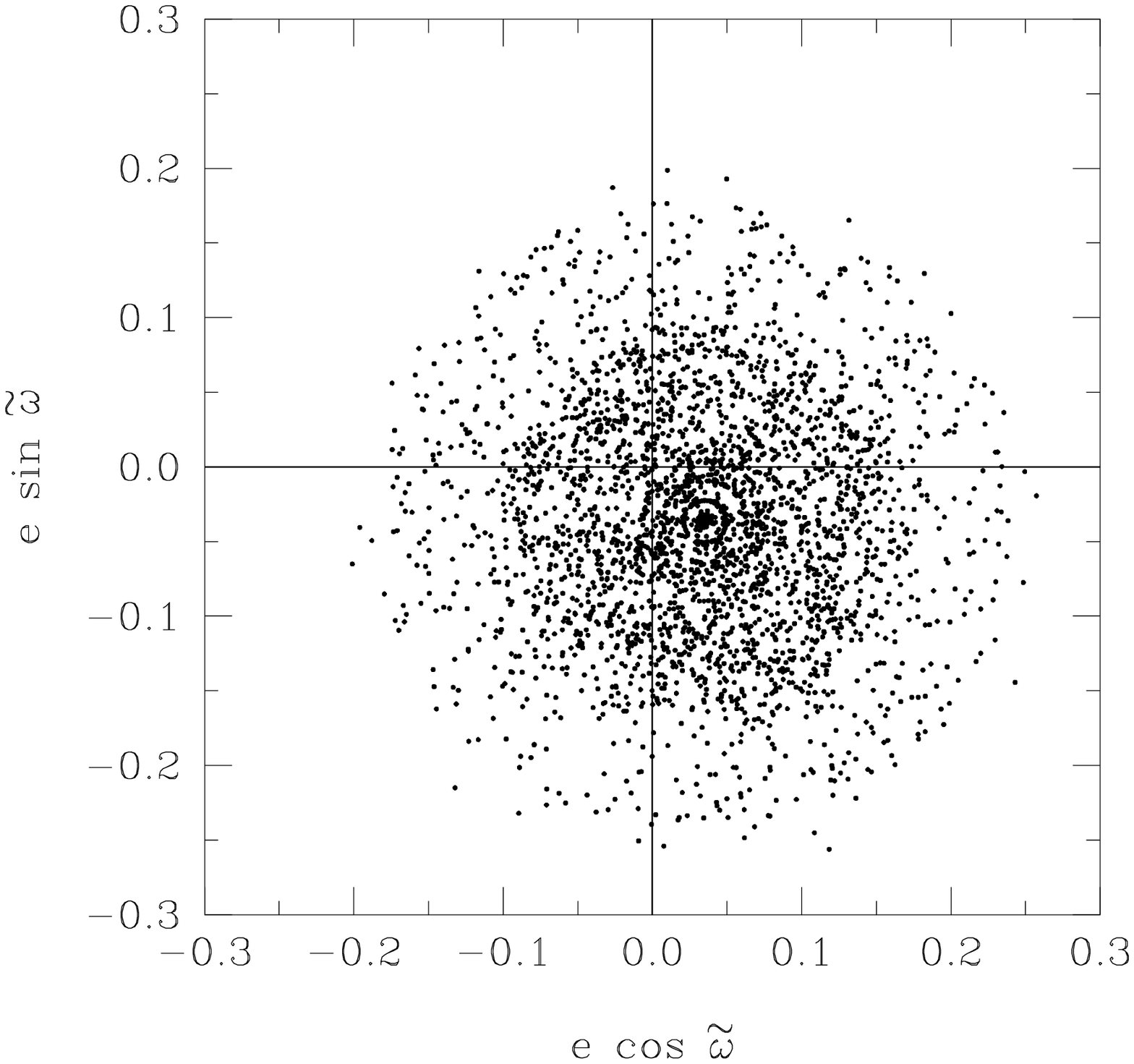,height=3in} &
      \epsfig{file=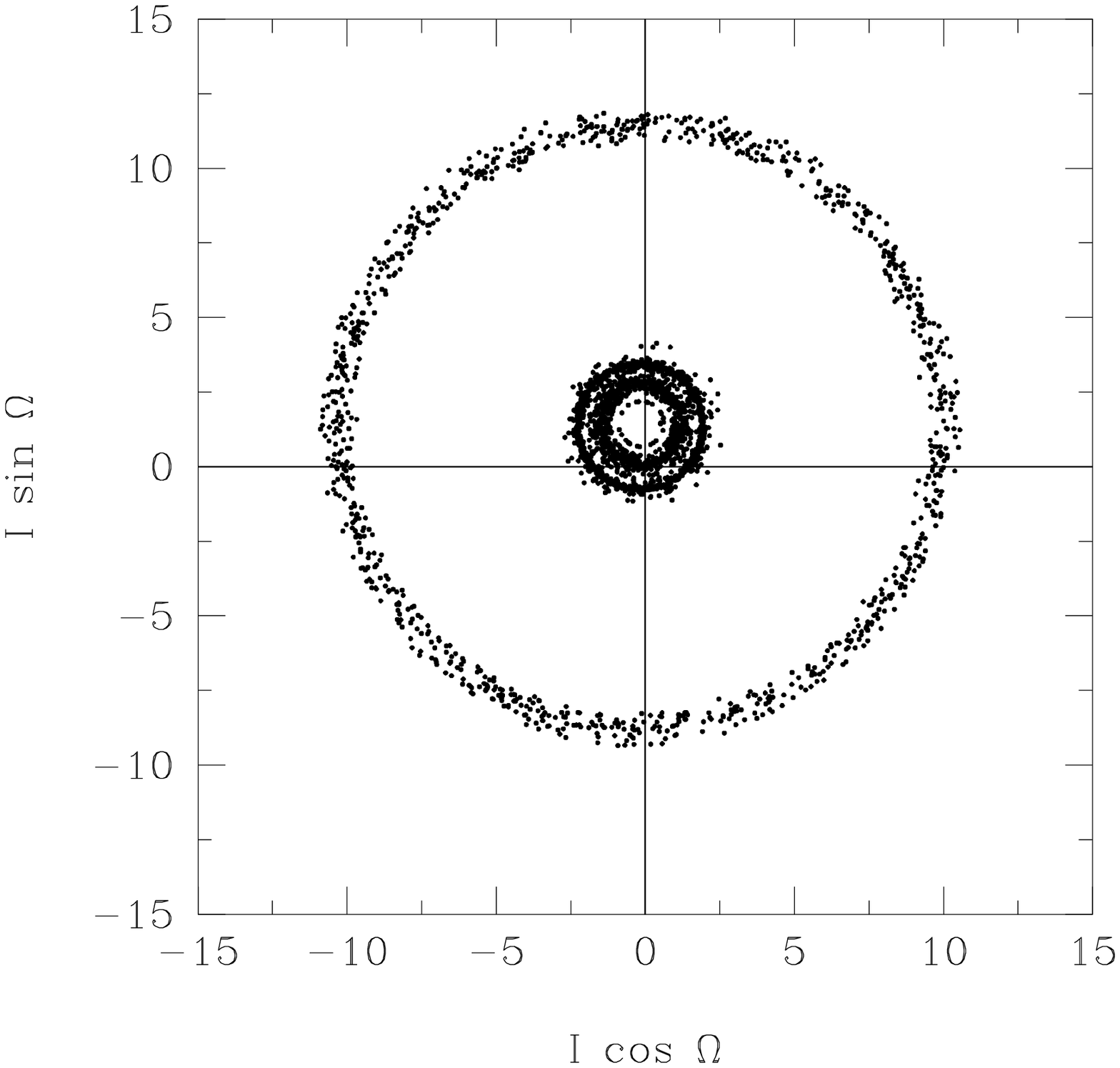,height=3in} \\
    \end{tabular}
  \end{center}
  \caption{The distribution of orbital elements of the asteroids in the
  families Eos, Themis, and Koronis (\cite{md99}).
  The asteroids in the three families have mean semimajor axes of
  3.015, 3.148, 2.876 AU, mean proper eccentricities of 0.071, 0.155, 0.047,
  and mean proper inclinations of 10.2, 1.426, $2.112^\circ$, respectively.
  The secular evolution the asteroids in each of the families 
  results in their orbital elements lying on circles in the complex
  eccentricity, $z$, and complex inclination, $y$, planes, the radii
  of which are equal to the family's mean proper elements.
  Since these asteroid families are at similar semimajor axes, the
  forced elements imposed on their orbits are almost identical, which is
  why their circles are offset from the origin by the same amount and in the
  same direction. }
  \label{figastfams}
\end{figure}

The orbital elements of the largest fragments, those with $\beta<0.1$,
created in the break-up of the primordial body are initially very
close to those of the primordial body;
they do not have identical orbits due to the velocity dispersion imparted
to the fragments in the collision.
The forced elements imposed on the orbits of all of these collisional
fragments are the same as those imposed on the primordial body.
The secular evolution of their osculating complex eccentricities
(eq.~[\ref{eq:z}]) and complex inclinations (eq.~[\ref{eq:y}]), is to precess
about the forced elements (which are also varying with time), but at
slightly different rates (due to their slightly different semimajor axes).
A similar argument applies for all particles created by the collisional
break-up of these fragments.
Thus, after a few precession timescales, the complex eccentricities
and complex inclinations of the collisional fragments of this family
lie evenly distributed around circles that are centered on $z_f(a,t)$
and $y_f(a,t)$, and that have radii of $e_p$ and $I_p$ (e.g., their
complex eccentricities lie on the circle shown in Fig.~\ref{fig2a}),
while their semimajor axes are all still close to $a$.
This is seen to be the case in the asteroid belt:
there are families of asteroids that have similar $a$, $e_p$, and $I_p$,
that are the collisional fragments resulting from the break-up
of a much larger asteroid (\cite{hira18}; see Fig.~\ref{figastfams})

%%%% Figure 2b
\begin{figure}
  \begin{center}
    \begin{tabular}{c}
      \epsfig{file=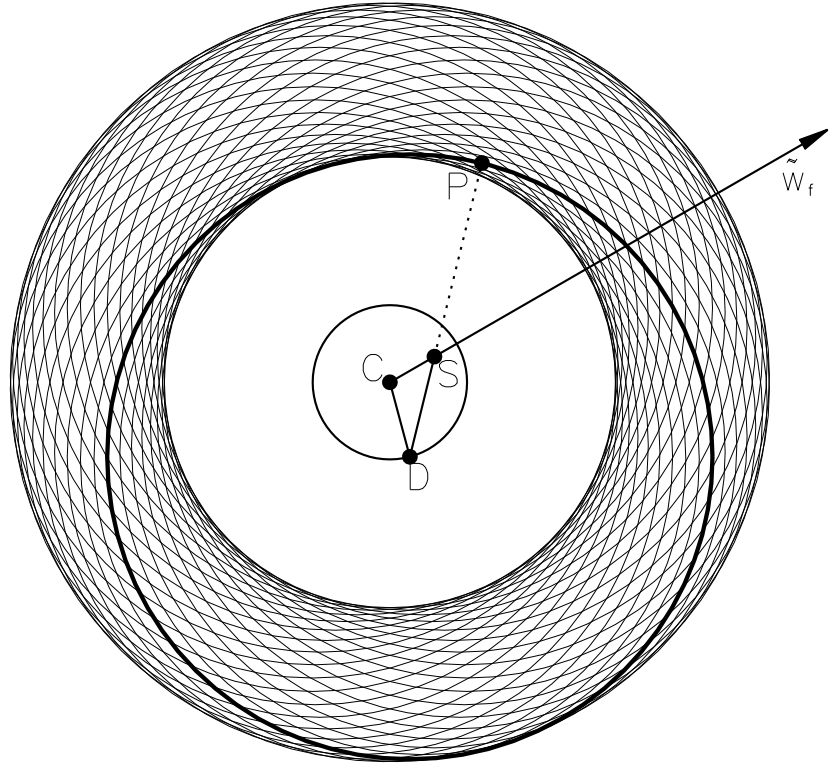,height=4in}
    \end{tabular}
  \end{center}
  \caption{Face-on view of the secular spatial distribution of large
  particles in a disk created by the break-up of one large asteroid.
  After a few secular precession timescales, these particles still have the
  same semimajor axis, $a$, and their complex eccentricities lie evenly
  distributed around a circle of radius $e_p$ that is centered on the
  complex forced eccentricity, described by $e_f$ and $\tilde{\omega}$
  (such as the dotted circle in Fig.~\ref{fig2a}).
  The contribution of each particle to the spatial distribution of material
  in the disk can be described by the elliptical ring of material coincident
  with the particle's orbit (see \S \ref{s-gravity}).
  Here, these elliptical rings have been represented by uniform
  circles of radius $a$, with centers that are offset by $ae$ in a direction
  opposite to the pericenter direction, $\tilde{\omega}$
  (this is a valid approximation to first order in the particles'
  eccentricities).
  A heavy line is used to highlight the orbital ring with a pericenter located
  at $P$, and a displaced circle center located at $D$, where $DP = a$.
  The vector $SD$ can be decomposed into its forced and proper
  components;
  this is shown by the triangle $SCD$, where $SD = ae$, $SC = ae_f$,
  and $CD = ae_p$ (there is a similar triangle in Fig.~\ref{fig2a}).
  Given that the distribution of $\tilde{\omega}_p$ is random, it follows that
  the distribution of the rings' centers, $D$, for this family disk are
  distributed on a circle of radius $ae_p$ and center $C$.
  Thus, the family forms a uniform torus of inner radius $a(1-e_p)$ and
  outer radius $a(1+e_p)$ centered on a point $C$ displaced from
  the star by a distance $ae_f$ in a direction away from the forced
  pericenter, $\tilde{\omega}_f$ (\cite{dnbh85}; \cite{dghw98}). }
  \label{fig2b}
\end{figure}

Thus, the distribution of the complex eccentricities, $n(z)$, of these
particles, has a distribution of pericenters that is biased towards the
orientation in the disk that is defined by $\tilde{\omega}_f$.
The consequence of this biased orbital element distribution on the
spatial distribution of this family material is best described with the
help of Fig.~\ref{fig2b}.
This shows how the family material forms a uniform torus of inner radius
$a(1-e_p)$ and outer radius $a(1+e_p)$ centered on a point that is
displaced from the star by a distance $ae_f$ in a direction away from the
forced pericenter, $\tilde{\omega}_f$ (\cite{dnbh85}; \cite{dghw98}).

The distribution of the complex inclinations, $n(y)$, of these particles
is also unevenly distributed.
Each particle's complex inclination describes its orbital plane,
so $n(y)$ describes the distribution of the orbital planes of the family
material;
i.e., the out-of-plane distribution of the torus described in the
last paragraph (and that shown in Fig.~\ref{fig2b}).
Changing the reference plane relative to which the particles' orbital
inclinations are defined to that described by $y_f$, shows that the
secular complex inclination distribution of this family material
leads to a torus that is symmetrical about the plane described by $y_f$,
the opening angle (half width) of which is described by $I_p$.

This picture of both the distribution of the orbital elements and the
consequent spatial distribution of the large family members is extremely
important for understanding the distribution of the smaller members
of the family, since the smaller members are created by collisions
between the larger members.

%%%%%%%%%%%%%%%%%%%%%%%%%%%%%%%%%%%%%%%%%%%%%%%%%%%%%%%%%
\subsubsection{P-R Drag Affected ($\beta_{pr} < \beta < 0.1$) Fragments}
\label{sss-ofwpfampr}

%%%% Figure of Particles in a Circle Method
\begin{figure}
  \begin{center}
    \begin{tabular}{c}
      \hspace{-0.65in} \epsfig{file=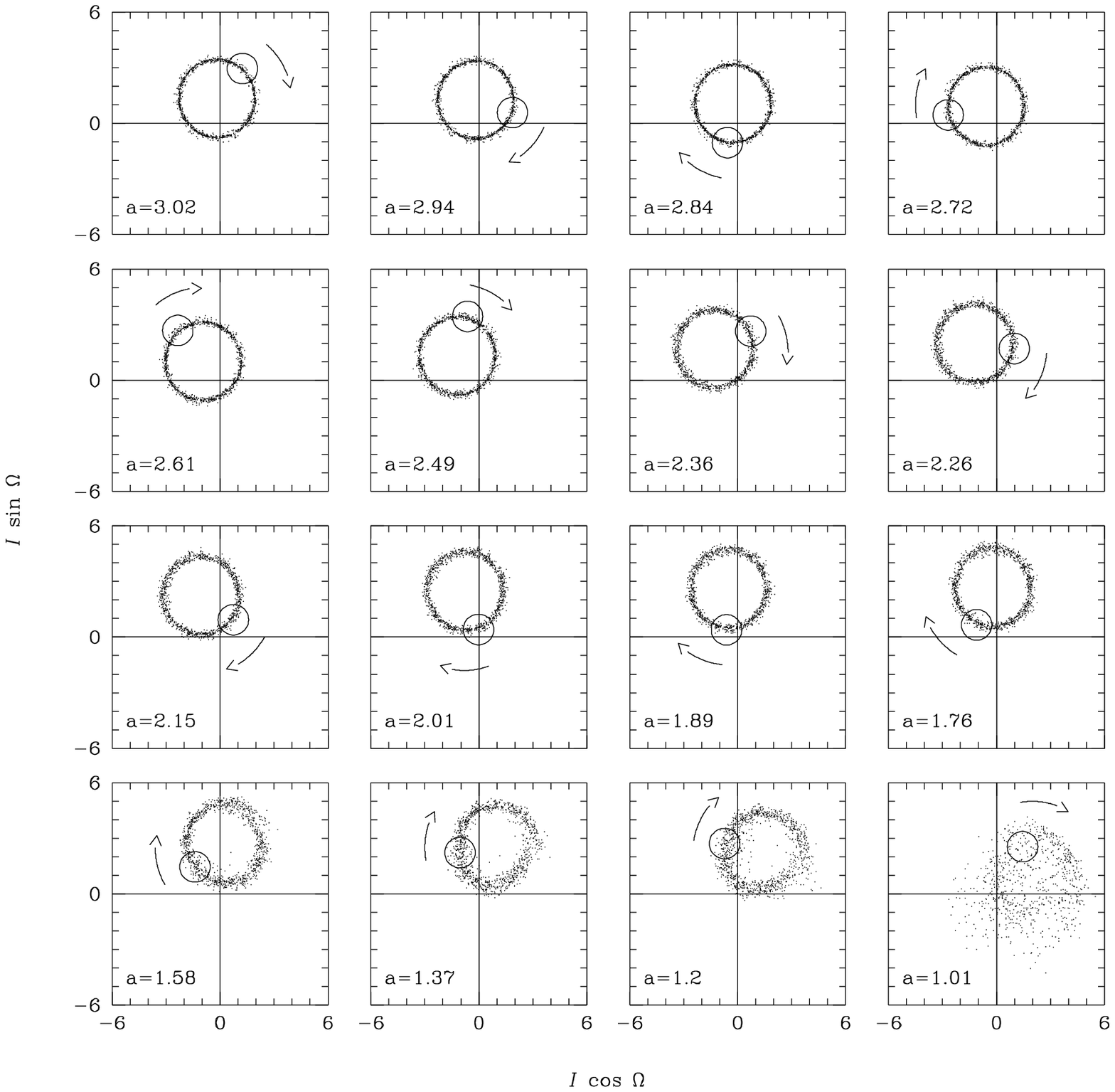,height=6.95in}
    \end{tabular}
  \end{center}
  \caption{The 52,000 year dynamical evolution of a wave of 10 $\mu$m
  particles released at the same time from material in the asteroid
  family Koronis (\cite{kd98b}).
  This evolution was found by numerical integration including the
  gravitational forces of the Sun and the solar system planets,
  radiation pressure, P-R drag, and the solar wind force.
  Each panel represents the complex inclination plot at a snapshot of time
  ($\sim 3500$ years between each panel) when the mean semimajor axis of
  the particles is shown in the panels.
  The circle highlights the dynamical evolution of one particle
  in the wave.}
  \label{figpinc}
\end{figure}

Immediately after they are created, the orbital elements of the P-R
drag affected particles are the same as the large members of the family.
The dynamical evolution of a wave of these particles, i.e., those that
were created at the same time, can be followed by numerical
integration to ascertain how the orbital elements of the particles
in the wave vary as their semimajor axes decrease due to
P-R drag;
this is the ``particles in a circle'' method (\cite{dgdg92}; see
Fig.~\ref{figpinc}).
It was found that the complex eccentricities and complex inclinations
of a wave of particles originating in the asteroid belt
remain on circles, and that as the wave's semimajor axis,
$a_{wave}$, decreases:
its effective proper eccentricity (the radius of the wave's circle
in the complex eccentricity plane) decreases
$\propto e_{p}*(a_{wave}/a)^{5/4}$;
its effective proper inclination (the radius of the wave's circle in
the complex inclination plane) remains constant at $I_p$;
the distributions of the particles' $\tilde{\omega}_p$ and $\Omega_p$
remain random;
while its effective forced elements (the centers of the circles in the
complex eccentricity and complex inclination planes) have a more
complicated variation (\cite{dgdg92}; \cite{liou93}).

Thus, the orbital element distributions, $n(z)$ and $n(y)$, of P-R drag
affected family particles are like that of the large particles,
in that they are the addition of $z_f$ and $y_f$ to symmetrical proper
element distributions, except that these distributions are different for
particles of different sizes as well as for particles at different
semimajor axes.
This means that the P-R drag affected family particles of a given P-R drag
rate that are at a given semimajor axis form a torus with an offset described
by their complex forced eccentricity and a plane of symmetry described
by their complex forced inclination.

%%%%%%%%%%%%%%%%%%%%%%%%%%%%%%%%%%%%%%%%%%%%%%%%%%%%%%%%%%
\subsubsection{Small ($\beta > 0.1$) Fragments}
\label{sss-ofwpfamsmall}
Consider the $\beta$ critical particles that are produced at the
same time from the population of family particles that have the
same $\tilde{\omega}_p$ and $\Omega_p$ at that time.
This parent population is spread out along the orbit defined by the
elements $a$, $e$, and $\tilde{\omega}$, which could be
one of the rings shown in Fig.~\ref{fig2b}.
The average pericenter orientation of these $\beta$ critical particles,
$\langle \tilde{\omega}^{'} \rangle$, is the same as that of the orbit of
the larger particles, $\tilde{\omega}$ (obvious because of the
symmetry of eq.~[\ref{eq:wprime}] with respect to $f$).
Their pericenter locations are also the same (eq.~[\ref{eq:rp}]).
Thus, the ring shown in Fig.~\ref{fig2b} defines the inner edge of
the disk of $\beta$ critical particles created in the break-up of
large particles on this ring;
i.e., their disk is offset by an amount $ae$ in the $\tilde{\omega}$
direction.
Consequently, the inner edge of the disk of $\beta$ critical particles
created in the break-up of all large particles in this family
is offset by an amount $ae_f$ in the $\tilde{\omega}_f$ direction.
This disk has the same plane of symmetry as the families' large
particles, since all particle categories from the same family have
the same distribution of orbital planes, $n(y)$.
Similar arguments apply for the families' $\beta$ meteoroids.

%%%%%%%%%%%%%%%%%%%%%%%%%%%%%%%%%%%%%%%%%%%%%%%%%%%%%%%%%
\subsection{Offset and Warp of Whole Disk}
\label{ss-ofwpdisk}
Consider first of all a disk in which there are no P-R drag affected
particles.
The tori of material from all of the families that have the same semimajor
axis, or equivalently that are at the same distance from the star, have the
same offset inner edge, and the same plane of symmetry.
This is because their large particles have the same forced elements
imposed on their orbits.
The complex eccentricities and complex inclinations of the large particles
of all of these families lie evenly distributed around circles with the same
centers, $z_f$ and $y_f$, but with a distribution of radii, $n(e_p)$ and
$n(I_p)$, that are the distributions of the proper elements of these
families (defining the radial width and opening angle of the torus
consisting of these families' material).

The whole disk is made up of families with a range of semimajor axes.
The families that are at different semimajor axes can have different
forced elements imposed on their orbits (depending on the perturbers
in the system), and in view of the proper element distributions in the
asteroid belt and in the Kuiper belt, they can also have different
distributions of proper elements.
A disk with a proper inclination distribution that increases with
semimajor axis would be flared;
protoplanetary disks are both expected (if they are thermally supported) and
observed to be flared (\cite{kh87}; \cite{bswk96}).
If the forced eccentricity imposed on the disk is non-zero, which it is if
there is at least one perturber in the system that is on a non-circular
orbit (eqs.~[\ref{eq:z}] and [\ref{eq:zf1}]), then the disk's center of
symmetry is offset from the star. 
If the forced inclination imposed on the disk is different for families
at different semimajor axes, which it is if there are two or more
perturbers in the system that are moving on orbits that are not
co-planar (eqs.~[\ref{eq:y}] and [\ref{eq:yf1}]), then the disk's plane
of symmetry varies with distance from the star;
i.e., the disk is warped.

The addition of P-R drag complicates the situation further, since $z_f$
and $y_f$ at a given distance from the star are different for particles
from different families as well as for particles of different sizes
from the same family.
However, for the same reasons as for the larger particles, this orbital
element distribution leads to a disk that is both offset and warped. 

%%%%%%%%%%%%%%%%%%%%%%%%%%%%%%%%%%%%%%%%%%%%%%
\subsection{Physical Understanding of Offset and Warp}
\label{ss-ofwpphysical}
There is also a physical explanation for the secular perturbation
asymmetries.
The secular perturbations of a massive body are equivalent to the
gravitational perturbations of the elliptical ring that contains the
mass of the perturber spread out along its orbit, the line density of
which varies inversely with the speed of the perturber in its orbit
(Gauss' averaging method, \cite{bc61}; \cite{md99}).
The ring's elliptical shape, as well as its higher line density at the
perturber's apocenter, mean that the center of mass of the star-ring
system is shifted from the star towards the perturber's apocenter.
The focus of the orbits of particles in such a system is offset
from the star;
i.e., the center of symmetry of a disk in this system is offset from
the star.
The gravitational perturbations of the ring also point to the plane
coincident with the perturber's orbital plane.
In systems with two or more perturbers, the system's plane of
symmetry (that in which the perturbing forces out of this plane
cancel) varies with distance from the star;
i.e., a disk in such a system is warped.

%%%%%%%%%%%%%%%%%%%%%%%%%%%%%%%%%%%%%%%%%%%%
\subsection{Observational Evidence of Offset and Warp in the Zodiacal Cloud}
\label{ss-ofwpinzc}

%%%% Figure 3
\begin{figure}
  \begin{center}
    \begin{tabular}{rlrl}
      (\textbf{a}) & \hspace{-0.4in}
      \epsfig{file=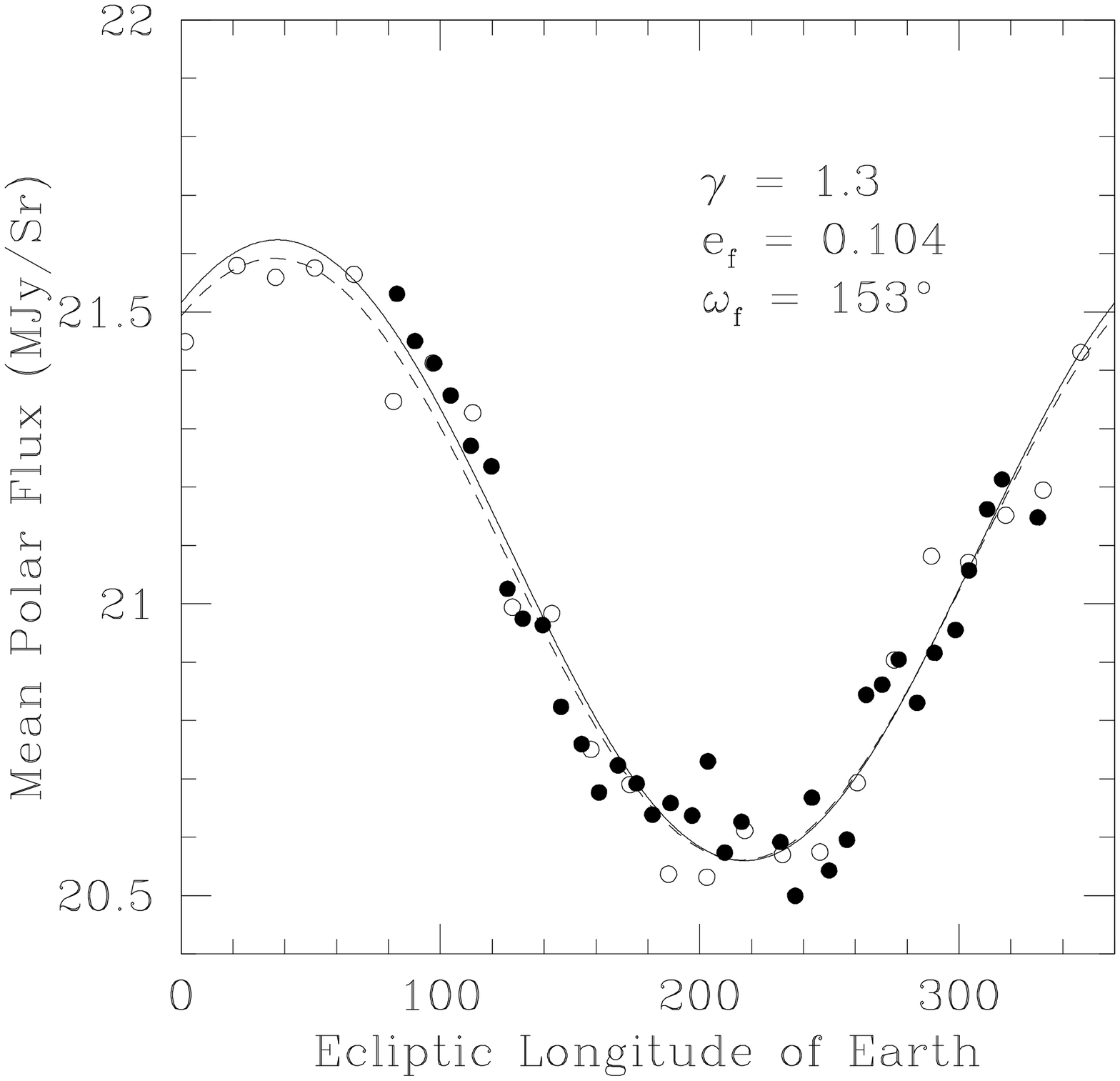,height=2.6in} &
      \hspace{0.15in} (\textbf{b}) & \hspace{-0.45in}
      \epsfig{file=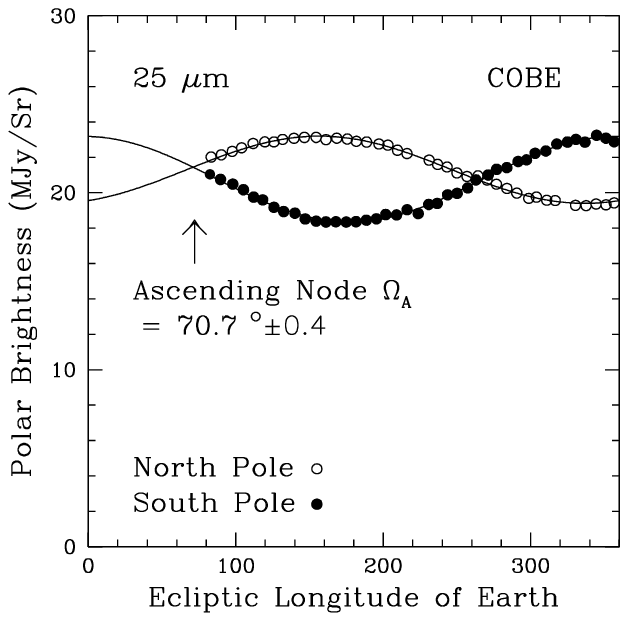,height=2.6in}
    \end{tabular}
  \end{center}
  \caption{COBE observations in the 25 $\mu$m waveband of the variation of
  brightnesses at the North, $N$, and South, $S$, ecliptic poles with
  ecliptic longitude of the Earth, $\lambda_\oplus$ (\cite{dghk99}).
  (\textbf{a}) shows the mean polar flux, $(N + S)/2$.
  The COBE observations are shown with the filled circles and the best fit
  to this observation with the solid line.
  The fact that $N+S$ is at a minimum at $\lambda_\oplus = 224 \pm 3^{\circ}$
  rather than at the Earth's aphelion, $\lambda_\oplus = 282.9^{\circ}$,
  implies that the center of symmetry of the zodiacal cloud is displaced
  from the Sun.
  Models for the zodiacal cloud with a number density fall-off
  $\propto r^{-1.3}$ show that this observation can be explained by
  the particles having $e_f = 0.104$ and $\tilde{\omega}_f = 153^\circ$
  (\cite{dggh99});
  the model observation is shown with the open circles and the dashed line.
  (\textbf{b}) $N$ and $S$ are equal when the Earth is at either
  the ascending or the descending node of the plane of symmetry of the
  cloud at 1 AU, giving an ascending node of $70.7 \pm 0.4^\circ$.
  This plane of symmetry is different from the plane of symmetry of the cloud
  at distances $>1$ AU from the Sun, implying that the cloud is
  warped. }
  \label{fig3}
\end{figure}

Mid-IR geocentric satellite observations (such as the IRAS, COBE, and
ISO observations) are dominated by the thermal emission of the
zodiacal cloud's P-R drag affected particles in all directions except
that of the Galactic plane (\cite{lg90}).
Such observations contain detailed information about the spatial
structure of the zodiacal cloud, especially since their observing
geometry changes throughout the year as the Earth moves around
its orbit.
Since there are 9 massive perturbers in the solar system,
the resulting secular perturbation asymmetries should be apparent
in the IRAS, COBE, and ISO data-sets.

Fig.~\ref{fig3}a shows COBE observations of the sum of the brightnesses
in the 25 $\mu$m waveband at the north and south ecliptic poles, $(N+S)/2$
(\cite{dghk99}), where there is no contamination from the Galactic plane.
If the zodiacal cloud was rotationally symmetric with the Sun at the
center, then the cross-sectional area density of particles in the near Earth
region would vary according to $\sigma(r,\theta,\phi) \propto
r^{-\gamma}f(\phi)$, where $\gamma$ is a constant.
Because the Earth's orbit is eccentric, geocentric observations
sample the zodiacal cloud at different radial distances from the Sun.
Thus, the minimum of the $(N+S)/2$ observation is expected to occur
either at the Earth's aphelion, $\lambda_\oplus = 282.9^{\circ}$,
or perihelion, $\lambda_\oplus = 102.9^{\circ}$, depending on whether
$\gamma > 1$ or $\gamma < 1$, which is determined by the collisional evolution
of particles in the near-Earth region (e.g., \cite{lg90} discuss the
observational evidence and conclude that $\gamma \approx 1.3$ as found
by the Helios zodiacal light experiment; see also \cite{kwfr98}).
However, the minimum in the 25 $\mu$m waveband observations occurs at
$\lambda_\oplus = 224^{\circ}$, and a similar result is found in the
12 $\mu$m waveband.
This is expected only if the Sun is not at the center of symmetry of the
zodiacal cloud.
Assuming that $\gamma = 1.3$, the observations can be explained if a
forced eccentricity of $e_f = 0.103$ with a forced pericenter of
$\tilde{\omega}_f = 153^\circ$ is imposed on the particles in the
background cloud (\cite{dggh99}).
Parametric models of the zodiacal cloud have also shown the need for an
offset to explain the observations (e.g., \cite{kwfr98}).

Fig.~\ref{fig3}b shows the variation of the brightnesses of the ecliptic
poles with ecliptic longitude of the Earth (\cite{dghk99}).
The north and south polar brightnesses are equal when the Earth is at
either the ascending or descending node of the local (at 1 AU)
plane of symmetry of the cloud, giving an ascending node of
$\Omega_{asc} = 70.7 \pm 0.4^\circ$.
However, COBE observations of the latitudes of the peak brightnesses
of the zodiacal cloud measured in the directions leading and trailing
the Earth's orbital motion give $\Omega_{asc} = 58.4^\circ$
(\cite{djxg96}).
Since such observations sample the cloud external to 1 AU,
this implies that the plane of symmetry of the zodiacal cloud varies with
heliocentric distance, i.e., that the zodiacal cloud is warped.

To observe the zodiacal cloud's offset and warp asymmetries, an
observer outside the solar system, would, at the very least, need an
observational resolution greater than the magnitude of these
asymmetries;
e.g., to observe the offset asymmetry, the observer needs a
resolution of $>(ae_f/a_\oplus)/R_\odot$ arcseconds, where the distance
from the observer to the Sun, $R_\odot$, is measured in pc.
An offset asymmetry would be more readily observable in a disk with a
central cavity, since the offset would cause a brightness asymmetry
in the emission from the inner edge of the disk (\cite{wdtf00};
see Chapter \ref{c-hrmodel}).

%%%%%%%%%%%%%%%%%%%%%%%%%%%%%%%%%%%%%%%%%%%%%%
\section{Resonant Perturbations}
\label{s-respertns}
A disk particle is subjected to resonant perturbations from a planet
either when the particle orbits the star $p$ times for every $p+q$ times
that the planet does (the external resonance case), or when the particle
orbits $p+q$ times for every $p$ planet orbits (the internal resonance
case), where both $p$ and $q$ are integers.
The nominal resonance location can be found from Kepler's third law of
planetary motion, and for an exterior resonance:
\begin{equation}
  a/a_{pl} = \left( \frac{p+q}{p} \right)^{2/3},
  \label{eq:ares}
\end{equation}
where $a$ is the semimajor axis of the particle's orbit, and $a_{pl}$ is
that of the planet.
If the particle also suffers from radiation pressure, there should be
an additional factor of $(1-\beta)^{1/3}$ on the right hand side of
equation (\ref{eq:ares}).
The physical reason for the resonance can be understood in terms of
its geometry.
Repeated encounters between the particle and the planet give the particle's
orbit a periodic kick.
These can either destabilize the particle's orbit, in which case the
particle is quickly ejected from the resonance region, or they
can stabilize its orbit, in which case the particle can be said to be
trapped in the resonance.

Just as the secular perturbations from a planetary system could be
described by the appropriate terms in the particle's disturbing function
(\S \ref{ss-pertneqns}), so can the resonant perturbations (see, e.g.,
\cite{md99}).
Quantitative analysis of the appropriate terms in the particle's
disturbing function shows how to define whether or not a particle
is in resonance, and shows whether that resonance has a stabilizing or
destabilizing effect on the particle's orbit;
e.g., a particle is formally said to be in an external $p:p+q$
eccentricity resonance when the resonant argument, $\phi$, librates
about some mean value rather than circulates through $360^\circ$, where
\begin{equation}
  \phi = (p+q)\lambda - p\lambda_{pl} - q\tilde{\omega},
  \label{eq:resarg}
\end{equation}
$\lambda$ and $\lambda_{pl}$ are the longitudes of the particle
and planet respectively, and $\tilde{\omega}$ is the particle's
longitude of pericenter.
Resonant phenomena can also be studied using numerical simulations,
since a full dynamical integration of a particle's motion as it evolves
under the gravitational influence of a star and a planetary system
automatically accounts for all types of gravitational perturbation.
It is not the aim of this section to give a quantitative analysis
of resonant phenomena (for which the reader is referred to texts such
as \cite{md99}), but rather to give a qualitative understanding
of the type of resonant phenomena we can expect to observe in
circumstellar debris disks that is based upon observations of resonant
phenomena in the solar system.

%%%%%%%%%%%%%%%%%%%%%%%%%%%%%%%%%%%%%%%%%%%%%%%%%%%
\subsection{Resonant Trapping due to Particle Migration}
\label{ss-restrapzc}

%%%% Figure of Trapping probablities into the Earth's resonant ring
\begin{figure}
  \begin{center}
    \begin{tabular}{c}
      \epsfig{file=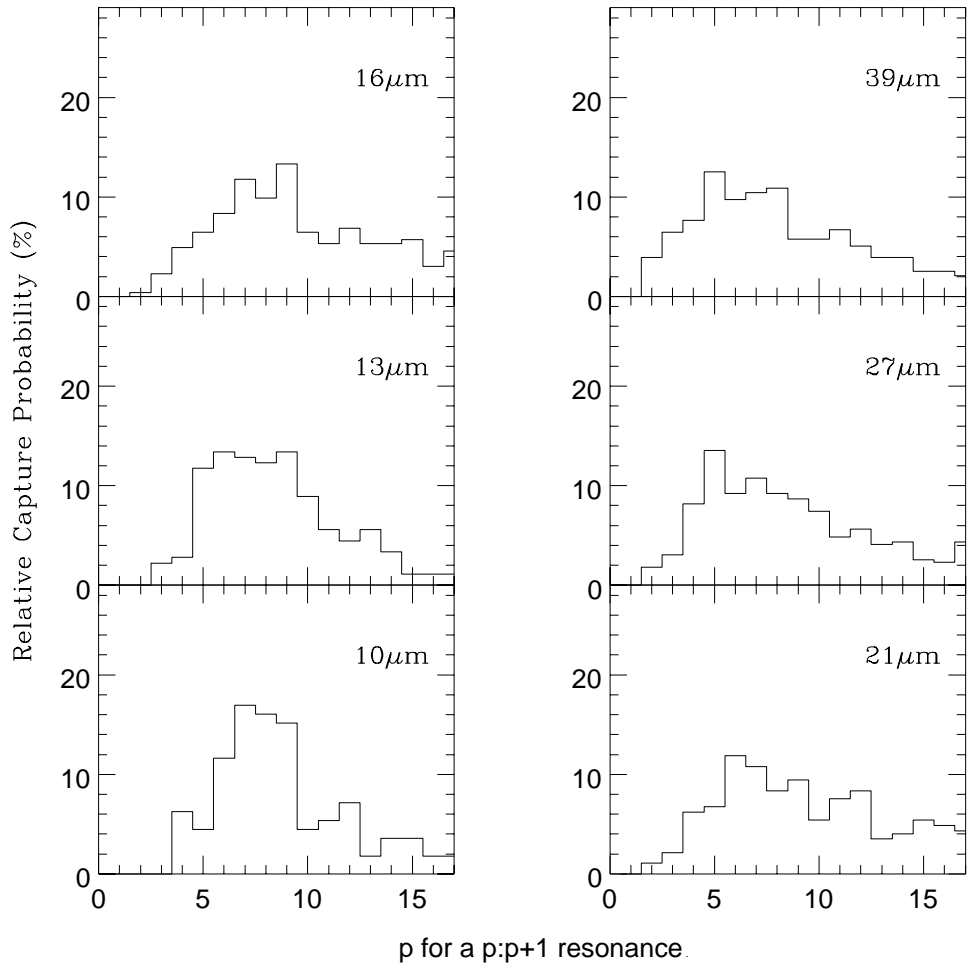,height=5in}
    \end{tabular}
  \end{center}
  \caption{The probabilities of trapping into $p:p+1$ mean motion
  resonances with the Earth for asteroidal particles of different
  sizes as they spiral in towards the Sun due to P-R drag
  (\cite{jd99}).
  The probabilities are plotted relative to the probability of
  the particle's capture into all of the first order mean motion
  resonances, which is approximately 12, 20, 29, 41, 43, and 47 \%
  for the 10, 13, 16, 21, 27, and 39 $\mu$m particles respectively.}
  \label{figcapprobs}
\end{figure}

As the particles responsible for the thermal emission of the zodiacal
cloud spiral in towards the Sun due to P-R drag,
they encounter the exterior mean motion resonances of the Earth.
A significant proportion of the particles get trapped into these
resonances.
The strongest resonances, i.e., those with the highest capture
probabilities, are the $p:p+1$ resonances.
For a particular particle, capture into a resonance is probabilistic
and depends on factors such as the particle's P-R drag rate and its
orbital parameters at the time of the encounter;
e.g., capture is more likely for a particle with a low eccentricity,
such as one originating in the asteroid belt, than for one with a high
eccentricity, such as one with a cometary origin.
Since particles of different sizes have different P-R drag rates,
they have different capture probabilities;
e.g., larger particles have a lower P-R drag rate and so have a higher
capture probability than smaller particles.
The proportion of asteroidal particles of different sizes that get
trapped into the different resonances can be ascertained by following
their dynamical evolution from the asteroid belt (\cite{djxg94}; \cite{jd99};
see Fig.~\ref{figcapprobs}).
Note that particles do not get trapped in interior mean motion resonances
with Jupiter (\cite{gff82}), because resonant trapping is far more likely
to occur when the orbits of a particle and a planet are converging rather
than diverging (\cite{dmm88}).

While a particle is trapped in resonance, the orbital decay due to P-R
drag is balanced by the resonant perturbations from the Earth;
the semimajor axis of the particle remains roughly constant and
its eccentricity increases.
Eventually the particle leaves the resonance, either when its eccentricity
is high enough to permit a close encounter with the Earth (\cite{wj93};
\cite{bf94}), or when its orbit becomes chaotic due to resonance overlap
(\cite{dmm88}; see \S \ref{ss-resolap}).
It then either continues its evolution into the inner solar system
or it gets accreted by the Earth.
The amount of time a trapped particle remains in resonance
depends on the $p$ of the resonance and the eccentricity of the particle
when it entered the resonance;
e.g., trapping times are shorter for resonances with higher $p$, since
these are closer to the Earth, and so close encounters are more likely to
occur (\cite{djxg94}).
Again, trapping time can be found by numerical integration;
e.g., it is found that the 13 $\mu$m particles that get trapped into the
Earth's 2:3 resonance remain there on average for 10,000 years (\cite{jd99}),
which is comparable to the time it would have taken them to reach the Sun
by P-R drag if they had not been trapped.

\begin{figure}
  \begin{center}
    \begin{tabular}{c}
      \epsfig{file=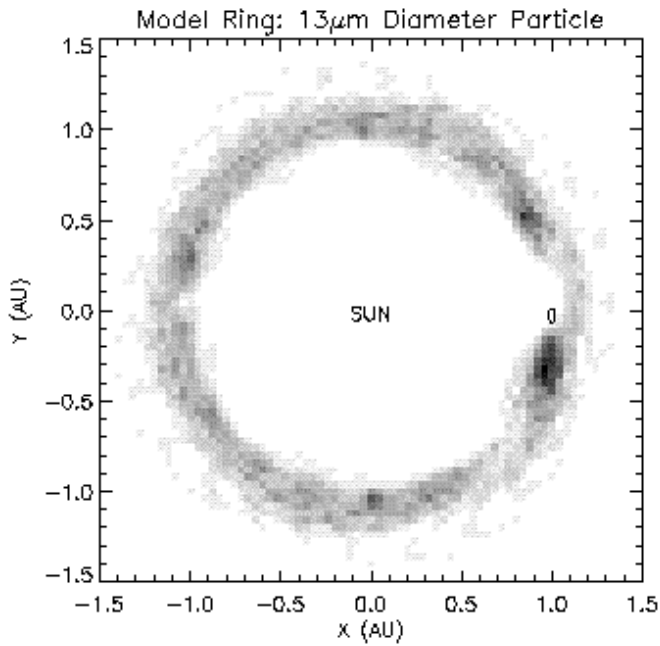,height=3.5in}
    \end{tabular}
  \end{center}
  \caption{Face-on view of a model of the 13 $\mu$m particles trapped
  in the Earth's resonant ring (\cite{jd99}).
  This is plotted in the reference frame rotating with the Earth's
  mean motion;
  i.e., this ring-like structure co-orbits with the Earth.
  The epicyclic motion of the Earth is shown by the small ellipse 1 AU
  to the right of the Sun.}
  \label{figmodelring}
\end{figure}

The density enhancement at $\sim 1$ AU caused by the particles trapped
in resonance causes the formation of a narrow ring of material that
co-orbits with the Earth;
this ring was first predicted by Gold (1975)\nocite{gold75}, and later
by Jackson \& Zook (1989)\nocite{jz89}.
Dermott et al. (1994) used numerical simulations to get a single-sized
particle model of the ring, which they later developed to include
a range of particle sizes (\cite{jd99}; see Fig.~\ref{figmodelring}).
The geometry of resonance is such that in the absence of P-R drag,
the ring would be symmetrical about the line joining the Earth and
the Sun.
Its structure would, however, be clumpy, since the ring associated
with each resonance is denser at the longitudes where the particles
in that ring are at pericenter (\cite{djxg94}).
The action of P-R drag on a resonance's geometry is to introduce a
phase lag into the equation of motion that is different for particles
with different P-R drag rates, as well as for resonances with different
$p$.
The result is that the pericenters of the particles trapped in the
ring bunch together behind the Earth, while the pericenters in front
of the Earth are dispersed.
This causes the formation of a trailing cloud that follows the Earth
in its motion (see Fig.~\ref{figmodelring}) that is estimated to have
a size of $\sim 0.2^3$ AU$^3$ and to have a number density $\sim 10\%$
above the background (\cite{djxg94}).

%%%% Figure of the Earth's Resonant Ring model

Since the trailing cloud is located permanently in the Earth's wake,
the brightness of the zodiacal cloud in the direction behind the
Earth's motion is always brighter than that of the Earth's motion.
This observational manifestation of the resonant ring was discovered
in the IRAS data-set (\cite{djxg94}).
The discovery of the Earth's resonant ring was later confirmed in the
COBE observations (\cite{rfwh95}).
In theory, the other terrestrial planets could also have resonant
rings of asteroidal material co-orbiting with them (\cite{djxg94}),
although these remain to be detected, and their expected brightness has
yet to be determined.
Dust particles created in the Kuiper belt that are spiraling in towards
the Sun due to P-R drag could also be trapped into resonance with Neptune
(\cite{lz99}).

If the P-R drag affected particles of an exosolar disk evolve past a
planet in the disk, an asymmetric resonant ring structure could result.
Observations of such a structure could be modeled using the same techniques
that were used to model the Earth's resonant ring (\cite{djxg94}; \cite{jd99}),
possibly allowing us to determine the presence, location, and even the mass of
the perturbing planet (\cite{dghw98}).
However, calculations show that when viewed from a distant point in space
normal to the ecliptic plane, the Earth's ``wake'' would only have
an IR signal $O(0.1)$ times that of the Earth (\cite{back98}).
So, regardless of the resolution requirements, if one were to observe the solar
system from outside, it would be easier to detect the Earth
directly than to infer its existence from the structure of the zodiacal
cloud.
This may not be true of dust from the Kuiper belt.
If this dust were plentiful enough to be observable, an exosolar observer
could infer the existence of Neptune and one other giant planet in the
inner solar system (\cite{lz99}).
Also, an extrasolar terrestrial planet resonant ring could be brighter
than the Earth's resonant ring if the density of the background cloud were
greater than the zodiacal cloud.

%%%%%%%%%%%%%%%%%%%%%%%%%%%%%%%%%%%%%%%%%%%%%%%%%
\subsection{Resonant Trapping due to Planetary Migration}
\label{ss-restrapkb}

%%%% Figure of Distribution of Kuiper Belt Material
\begin{figure}
  \begin{center}
    \begin{tabular}{c}
      \epsfig{file=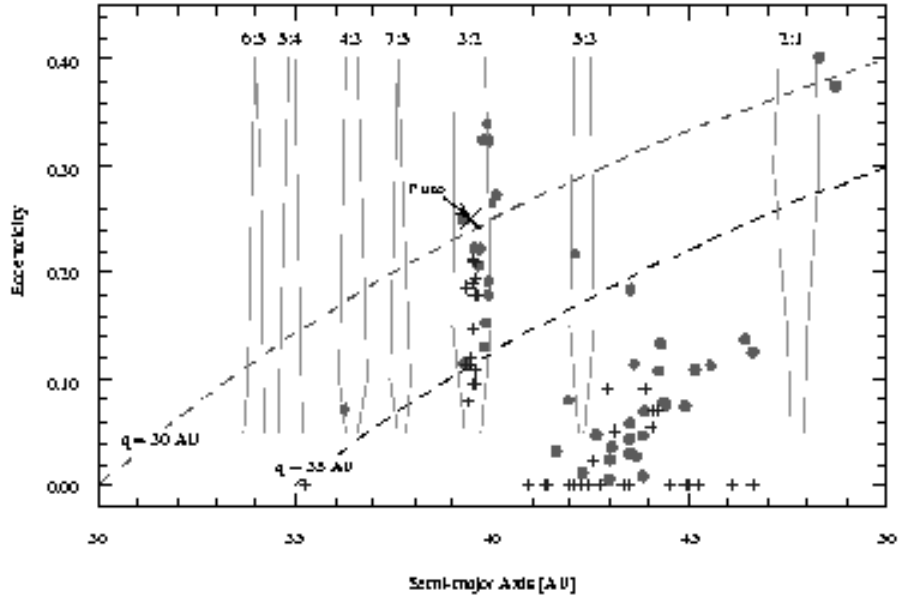,height=3.1in}
    \end{tabular}
  \end{center}
  \caption{Semimajor axis vs. orbital eccentricity plot for the known
  Kuiper belt objects (\cite{jewi99}).
  The solid circles denote multi-opposition objects and the pluses
  denote single opposition objects.
  The approximate boundaries of dominant resonances with Neptune
  are shown (\cite{malh95}).
  The dashed lines indiacte the loci of orbits that have perihelion
  distance 30 and 35 AU.}
  \label{figkbea}
\end{figure}

The distribution of known Kuiper belt objects is distinctly
non-random and shows very pronounced resonant structure (\cite{jewi99};
see Fig.~\ref{figkbea}).
About one third of the observed Kuiper belt objects are, just like Pluto,
in the 2:3 mean motion resonance with Neptune, which is located at $a=39.4$ AU;
these bodies are known as ``Plutinos''.
The Plutinos have high orbital eccentricities which puts some of their
pericenters within the orbit of Neptune.
However, close encounters with Neptune are prevented by the geometry of
the resonance;
i.e., the resonance has a stabilizing effect on their orbits.
There is almost no Kuiper belt material on orbits with semimajor axes
interior to the 2:3 resonance.
There may also be a significant fraction in the 1:2 resonance, although
these objects would be more difficult to detect because of their faintness.
The other two thirds of the Kuiper belt objects are known as ``Classical''
objects.
These are characterized by their semimajor axes, which lie between 42 and 47
AU, and their low eccentricities (which means they do not suffer close
encounters with Neptune) and inclinations.
We are also now beginning to see another population of Kuiper belt objects, the
``Scattered'' objects, which are those that are supposed to have been
scattered by Neptune.

An explanation of the origin of this detailed structure is important,
since it can provide an insight into the last stages of the planetary
formation process in the solar system (e.g., \cite{hm99}).
It is especially important, since the emission from debris disks appears
to originate from analogous Kuiper belt regions (\cite{bp93}).
In the final stages of the formation of the solar system, the residual
planetesimal population was cleared (apart from those planetesimals that
still reside in the asteroid belt) due to the gravitational perturbations
of the fully-formed giant planets.
The consequent exchange of orbital energy and angular momentum is expected
to result in the radial migration of the orbits of the giant planets
(\cite{fi84}).
The extent and rate of this migration depends on the mass of the
residual planetesimal disk, with the migration being stronger for
more massive disks.
Numerical simulations of this process show that the orbits of Saturn, Uranus,
and Neptune migrated radially outwards, while that of Jupiter migrated
slightly inwards (\cite{hm99}).

The expansion of Neptune's orbit meant that the planetesimals beyond its
orbit ended up on converging orbits with Neptune's first-order mean motion
resonances.
This allowed the possibility of them being captured into these resonances.
This resonance sweeping explains the large number of objects trapped
in the 2:3 resonance, as well as explaining the large eccentricities of
these objects (\cite{gome97b}), and predicts that a similar amount of
objects could be trapped in the 1:2 resonance (\cite{malh95}). 
It also provides a natural explanation for the other populations of
Kuiper belt objects, as well as an explanation for the formation of the
Oort cloud of comets (\cite{hm99}).
Hahn \& Malhotra (1999) showed that a mass of $\sim 50M_\oplus$ for
the residual disk is required to expand Neptune's orbit by $\sim 7$ AU
which would pump Plutino eccentricities up to $\sim 0.3$, and
they estimated that $\sim 12M_\oplus$ of this material would be deposited
in the Oort cloud, of which only $\sim 4M_\oplus$ would remain today.
However, there are still many issues to deal with when studying the
structure of the young Kuiper belt regions.
For example, if the disk was very massive, then the interaction of Neptune
with the self-gravitating disk could set up apsidal density waves, which
would both cause disk structure, and damp the eccentricity of Neptune
(\cite{wh98}).
Also, density waves launched at mean motion resonances would exert a
torque on Neptune's orbit opposing its radial migration (\cite{gt80}).

Just as for the Earth's resonant ring, the geometry of particles
trapped in Neptune's resonances is such they create clumpy
patterns that co-orbit with the planet.
In this instance, however, there is no phase lag, since the P-R drag
force acting on the Kuiper belt objects is minimal, so the resulting
resonant ring structure has mirror symmetry about the line joining
Neptune and the Sun.
The geometry of the orbits of Kuiper belt objects trapped in the 2:3
resonance means that they are expected to be most densely concentrated
$90^\circ$ in front of and behind the location of Neptune, while those
in the 1:2 resonance are expected to be concentrated $180^\circ$
from Neptune.
If the debris disks really are analogue Kuiper belts, then some of
the large particles that are colliding to create the dust seen by its
emission could have undergone the same fate as the Kuiper belt objects.
This could cause the emission to have large-scale observable
asymmetries that co-orbit with the outermost planet in the system.
A ring with all of its particles trapped in the 1:2 and 2:3 resonances
with a planet would have three lobes of emission, with the planet
residing in an empty fourth lobe.

%%%%%%%%%%%%%%%%%%%%%%%%%%%%%%%%%%%%%%%%%%%%%%%
\subsection{Resonance Removal Mechanisms}
\label{ss-resremmech}
In a planetary system, there are stable regions where large particles can
remain undisturbed over the age of the system, and unstable regions where
planetary perturbations cause any particles in these regions to be 
quickly removed.
Numerical integrations over the age of the solar system confer with observations
that show that the only stable regions in the solar system are the asteroid belt
and the Kuiper belt (e.g., \cite{dqt89}), and that the asteroid belt itself
has regions of instability.
The regions of instability in the solar system can be explained analytically
as the consequence of resonant mechanisms, and these arguments can be applied
to explain large-scale radial distribution of large particles in exosolar disks.

%%%%%%%%%%%%%%%%%%%%%%%%%%%%%%%%%%%%%%%%%%%%%%%
\subsubsection{Resonance Overlap}
\label{ss-resolap}
Close to the orbit of a planet, the planet's first-order mean motion
resonances are so tightly packed that the libration widths of the
individual resonances overlap (i.e., a particle in this region can have
two different resonant arguments, eq.[\ref{eq:resarg}], that are librating
at the same time), and this causes the particle's orbit to be chaotic.
Resonance overlap causes material on orbits within a radial width
\begin{equation}
  dr \approx 1.3a_{pl}(M_{pl}/M_\star)^{2/7}
  \label{eq:resolap}
\end{equation}
about a planet's orbit get scattered onto highly eccentric orbits within
about 1000 orbital periods (\cite{wisd80}).

%%%%%%%%%%%%%%%%%%%%%%%%%%%%%%%%%%%%%%%%%%%%%%%%%
\subsubsection{Resonance Gaps}
\label{ss-resgaps}
It was noticed that the distribution of the orbits of asteroids in
the asteroid belt contains structure;
there are significantly fewer asteroids with semimajor axes close to
some of Jupiter's interior resonances (\cite{kirk76}).
These Kirkwood gaps were later shown to be truly associated with the
resonances, since they encompass the specific regions where a particle's
resonant argument (eq.~[\ref{eq:resarg}]) is librating rather than
circulating (\cite{dm83}).
The cause of the gaps has been shown, at least for the 1:3 resonance
(e.g., \cite{wisd82}), to be due to the chaotic nature of orbits in
these resonances;
i.e., the eccentricity of a particle in the resonance is pumped up
until it is eventually removed from the resonance by a close
planetary encounter.

%%%%%%%%%%%%%%%%%%%%%%%%%%%%%%%%%%%%%%%%%%%%%%%%%
\subsubsection{Secular Resonance}
\label{ss-secres}
It was already mentioned in \S \ref{ss-pertnsoln} that material cannot
reside in regions of secular resonance (those regions which cover the range
of semimajor axes where $e_f, I_f \rightarrow \infty$).
The current asteroid belt is truncated interior to a secular resonance at 2 AU.
If the orbits of the giant planets suffered radial migration (\cite{hm99}),
then the locations of the solar planetary system's secular resonances will
have swept through a wide range of semimajor axes in the primordial
asteroid belt, pumping up the eccentricities and inclinations of material there.
This could have inhibited planetary formation in the asteroid belt
and could explain the current distribution of the asteroids' eccentricities
and inclinations (\cite{gome97a}).
Similarly, secular resonance sweeping in the primitive solar nebula has
been invoked to explain the dearth of asteroids between the 2:1 and 3:2
resonance with Jupiter (\cite{lf97}).

%%%%%%%%%%%%%%%%%%%%%%%%%%%%%%%%%%%%%%%%%%%%%%
\section{Accretion}
\label{s-accretion}
Accretion processes are very important during the initial stages of planetary
formation, since this is the method by which planetesimals grow into planets.
At the end of planetary formation, however, accretion should be minimal, since
the source of accretion material close to the planet should be much depleted.
However, material can be transported into a planet's accretion zone by
P-R drag.
While the resulting increase in the planet's mass is likely to be insignificant,
the loss of the P-R drag affected particles from the debris disk could be
important observationally.
In the absence of collisional or accretion loss, the P-R drag affected
particles from a particular source extend all the way in to the star.
Accretion could cause the formation of an inner hole in such a disk.
It is therefore relevant to question how much material that is set to evolve
past a planet due to P-R drag survives without being accreted onto the
planet.

A simple estimate for the proportion of particles lost in such a way
can be obtained by considering the P-R drag evolution of a torus of particles
with orbital elements $a$, $e$, $I$, and random $\Omega$,
$\tilde{\omega}$ and $\lambda$;
the volume of this torus is $V_{tor} = 8\pi a^3 e \sin{I}$ (\cite{syke90}).
In the time it takes for the torus to pass the planet,
\hbox{$\Delta t = (1602/\beta)(M_\odot/M_\star)(a_{pl}/a_\oplus)^2e$
yrs}, the planet accretes a volume of dust given by (\cite{kess81}):
$V_{acc} = \sigma_{cap} v_{rel} \Delta t$, where
$\sigma_{cap} = \pi R_{pl}^2(1+v_e^2/v_{rel}^2)$ is the capture
cross-sectional area, $R_{pl}$ is the radius of the planet,
$v_e = \sqrt{2GM_{pl}/R_{pl}}$ is the escape velocity of the planet,
and $v_{rel}$ is the mean relative velocity of encounter between
the planet and the particles.
Thus, the proportion of dust accreted onto the planet,
$P = V_{acc}/V_{tor}$, is given by:
\begin{equation}
  P \approx \frac{12}{\beta \sqrt{1-\beta}}
            { \left( \frac{M_\odot}{M_\star}  \right) }^{3/2}
	    { \left( \frac{M_{pl}}{M_\odot}   \right) }^{4/3}
	    { \left( \frac{a_\oplus}{a_{pl}}  \right) }^{1/2}
	    { \left( \frac{\rho_J}{\rho_{pl}} \right) }^{1/3}
	    g(e,I),
\label{eq:paccrn}
\end{equation}
where $\rho_J = 1330$ kg/m$^3$ is the mean density of Jupiter,
$g(e,I) = [(v_{rel}/v)\sin{I}]^{-1}$, and $v$ is the velocity of the particle.

For a massive enough planet, $P = 1$ for $\beta < 0.5$, implying that
no material can pass it, and an inner hole can be formed.
However, a planet that is likely to accrete much of the material that passes
it is also likely to trap the particles into a resonant ring.
This could affect the calculation of the accretion rate, although it is
not immediately obvious whether it would make accretion easier or harder.
It is thought that resonance trapping helps the accretion process,
since trapped particles may leave the resonance upon a close encounter
with the planet (e.g., \cite{kd98b}).

%%%%%%%%%%%%%%%%%%%%%%%%%%%%%%%%%%%%%%%%%%%%%%
\chapter{DISK PARTICLE OPTICAL PROPERTIES}
%%%%%%%%%%%%%%%%%%%%%%%%%%%%%%%%%%%%%%%%%%%%%%
\label{c-optprops}
In order to model circumstellar debris disk observations, we need
a model for the optical properties of the constituent particles
(\S \ref{s-modeling}).
Essentially, we need a method of finding the variation of the optical
coefficients, $Q_{abs}$ and $Q_{pr}$, with wavelength for a particle
of a given diameter, $D$.
First, we need to know the composition of the grains.
This defines the optical constants of the grain material (or materials),
i.e., the variation of the complex refractive index with wavelength:
\begin{equation}
  m(\lambda) = m^{'} + im^{"}.
\end{equation}
The optical constants for a given material can be found by some combination
of laboratory measurements and theoretical analysis;
e.g., the real part, $m^{'}$ can be calculated from the imaginary part,
$m^{"}$, using the Kramers-Kronig relation (\cite{bh83}).
Second, we need to know the morphology of the grains.
Finally, we need to use a light scattering theory (see, e.g., \cite{bh83})
to ascertain how these grains interact with incident light to determine their
optical coefficients.

To understand the composition and morphology of debris disk grains, we
need to consider their physical origin.
The grains described in Chapter \ref{c-structure} are created by the
collisional break-up of debris material (such as asteroids or comets) left
over at the end of the planetary formation process in the disk.
Thus, the grains have a long history that can be traced back to the dust
grains in the diffuse interstellar medium.
Presumably, interstellar grains can also be traced back to their creation
in events such as condensation in the outflow from cool supergiant stars.
Observations of the grain populations at different stages in this
evolutionary cycle, from diffuse interstellar medium to molecular cloud core
to protoplanetary disk to debris disk, can be pieced together to form an
understanding of the factors affecting their ultimate composition and
morphology.
Again, the solar system provides one of the best opportunities for studying
debris disk grains, since it is the only system for which we have actual
samples of the grains.

%%%%%%%%%%%%%%%%%%%%%%%%%%%%%%%%%%%%%%%%%%%%%%
\section{Interstellar Dust Grains}
\label{s-ismdust}
Information about the physical and chemical nature of interstellar dust
grains (composition, size, shape, and alignment) can be gleaned from
observations of interstellar continuum extinction and polarization from the
NIR to FUV (e.g., \cite{lg97}), as well as from dust thermal emission.
Spectral features in the interstellar extinction law identify three major
dust populations.

First, there is a population of elongated particles with a silicate core,
and an organic refractory mantle that are a few tenths of a micron in size
(e.g., \cite{lg97}).
The presence of silicates is noticeable because of strong
absorption/emission features in the extinction at about 10 and 20 $\mu$m
that are characteristic of Si-O stretching and O-Si-O bending modes.
These silicates could not be bare (i.e., have no mantle, such as
those proposed by \cite{dl84}), because their destruction rate in
the interstellar medium is too high compared with their production
rate.
An organic refractory mantle accreted onto the silicate core protects the
core from being destroyed;
this material is also responsible for the 3.4 $\mu$m extinction
feature.
The mantle material originally comprised of ices of simple chemical
compounds, but billions of years of UV photoprocessing changed the
ice mixture into a carbon-rich, oxygen-poor refractory material containing
many different organic molecules.
These core-mantle grains were shown to be a few tenths of a micron in size
and have to be elongated by a factor of about 2:1 to fit the NIR extinction
polarization (\cite{lg97}).

There is also a separate population of grains that are responsible for the hump
in the interstellar extinction at 2200 \AA.
While the exact nature of these grains is unknown, they are generally believed
to be small $< 0.01$ $\mu$m carbonaceous grains (e.g., \cite{lg97}).
It has been speculated that these grains could be made of graphite
(e.g., \cite{dl84}), however, the carbon in the interstellar medium is
produced in carbon stars that have amorphous not graphitic carbon in
their circumstellar envelopes.
Further, the precise location of the hump is very sensitive to
grain properties if graphite grains are assumed, which is inconsistent
with observations.
The final population of grains are those responsible for the FUV extinction.
These are likely to be Polycyclic Aromatic Hydrocarbons (PAHs), which
are small $< 10$ \AA~thermally spiking grains.

A unified model for interstellar dust that incorporates all three components
can simultaneously fit the observed interstellar extinction and
polarization (\cite{lg97}).
This model for interstellar grains is consistent with constraints on the
abundance of the condensable atoms (O, C, N, Mg, Si, Fe) in interstellar
dust relative to elemental hydrogen, $[X/H]_{dust}$.
This constraint states that the abundance of a particular element in the
dust cannot be greater than the difference between its cosmic abundance in
the interstellar medium, $[X/H]_{cosmic}^{ism}$, and its abundance
observed there in the gas phase, $[X/H]_{gas}^{ism}$.
Cosmic abundances can be estimated from those of the solar system.
There is, however, evidence that suggests that solar abundances of O, C,
and N are above that of the interstellar medium in general.
Such an overabundance could have occurred due to the injection of
nucleosynthetic products by a nearby supernova, or from a decreasing
rate of type II supernovae that resulted in a secular decrease in the
heavier elements in the last 5 billion years.
There is also evidence that abundances are not uniform in the galaxy, with
the inner regions of the galaxy having a greater abundance of heavier
elements.
Mixing should, however, have maintained some level of uniformity over
the local (within a few kpc) interstellar medium.
Gas phase abundances can be estimated from UV spectral observations of
background objects, which appear to give consistent results along
different lines of sight.

%%%%%%%%%%%%%%%%%%%%%%%%%%%%%%%%%%%%%%%%%%%%%%
\section{Primordial Cometary Grains}
\label{s-primordcomets}
In the process of the formation of molecular cloud cores, and subsequently
protoplanetary disks, some of the species that were in the gas phase in
the interstellar medium condense onto the refractory grains to form
an outer mantle of volatile ices (e.g., \cite{phbs94}), within which are
embedded hundreds of the small carbonaceous particles (e.g., \cite{gh90}).
The widespread occurrence of a feature near $3.0$ $\mu$m in the infrared
spectra of molecular cloud cores that could arise from the O-H stretch
fundamental indicates that H$_2$O ice is the dominant ice in the
outer mantle.
The grains in protoplanetary disks are different from those in molecular
cloud cores due to chemical reactions that process the grains as they
pass through the shock interface between the two.
In particular, these reactions increase the amount of H$_2$O and SiO$_2$
in the disks (\cite{phbs94}).
In the colder outer portions of protoplanetary disks, further condensation
of the gaseous components can occur, while in the warmer inner portions,
some of the grain species sublimate or are chemically transformed.

%%%% Figure of model of cometary grain evolution (Greenberg & Hage 90)
\begin{figure}
  \begin{center}
    \begin{tabular}{c}
      \epsfig{file=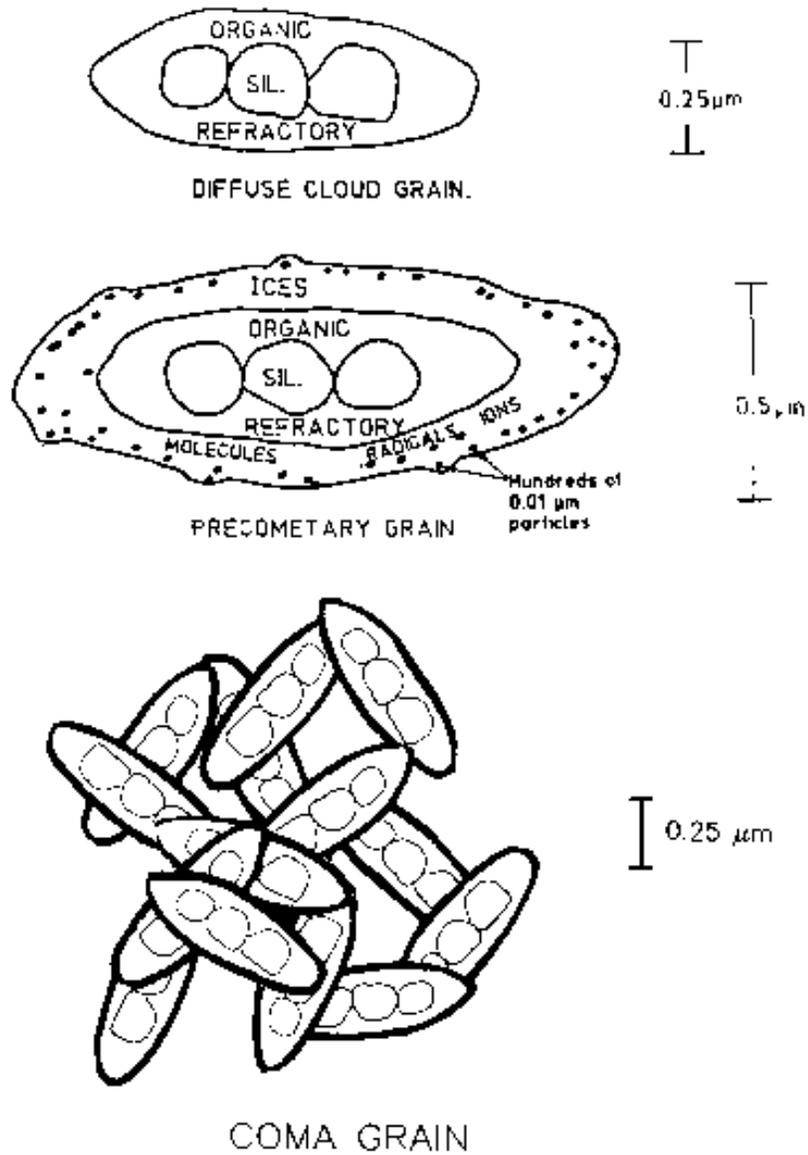,height=6in}
    \end{tabular}
  \end{center}
  \caption{The evolutionary history from interstellar grains to cometary
  grains (\cite{gh90}).
  \textit{Top}: Schematic of an interstellar dust grain which consists of
  a core of silicates and a mantle of organic refractory material.
  \textit{Middle}: An interstellar grain as it would appear in the
  protosolar dust cloud after accretion of gases onto its surface.
  These grains make up a comet.
  \textit{Bottom}: Schematic of a porous coma grain as it would appear
  according to the interstellar dust model.}
  \label{figcometgrains}
\end{figure}

Comets are widely recognized to be the most primitive bodies in the solar
system.
The currently observed coma grains can be linked to their interstellar
origin through models of how the grains were formed (\cite{gh90};
\cite{gree98}; see Fig.~\ref{figcometgrains}).
The formation of a comet nucleus starts with the aggregation of the
protosolar nebula particles.
The proposed tangled ``bird's nest'' structure of these aggregates
(\cite{gg81}) would provide the aggregates with rigidity.
These aggregates then coalesce to form the comet.
When the comet approaches the Sun, the icy mantle, and possibly part
of the organic refractory mantle, of the individual aggregates
evaporate leaving porous aggregate particles, which are lifted from the
surface of the comet to become the coma grains.
If the comet reaches close enough to the Sun, some of the silicate
core may be crystallized (\cite{gree98}).
The mass fractions of the various components of the protosolar nebula
grains, and hence the comet nucleus, can be estimated from models for
the composition of the interstellar dust (e.g., \cite{lg97}), and
observations of the solar abundance (\cite{gree98}).
These can be used to estimate the density and porosity of coma grains
which can be compared with observed meteor and nucleus densities.
Modeling the optical properties of these particles constrains the grain
properties from spectral observations of cometary comae.
The evidence supports this model of grain formation (e.g., 0.1 $\mu$m
particles have been shown to be the basic constituents of the coma
aggregates), and indicates that the coma grains have a high porosity of
between 0.93 and 0.975 (\cite{gh90}).

%%%%%%%%%%%%%%%%%%%%%%%%%%%%%%%%%%%%%%%%%%%%%
\section{Zodiacal Dust}
\label{s-zoddust}
Interplanetary dust particles (IDPs) in the zodiacal cloud originate
in both comets and asteroids (see, e.g., \cite{lg90}; \cite{gust94}).
The composition and morphology of cometary coma grains was discussed in
\S \ref{s-primordcomets}.
The composition of the solar system asteroids ranges from the comet-like
D and P types found at the outer edge of the main belt, to the less
pristine C and S types further in.
This gradation could either be a consequence of the primordial properties
of the solar nebula, or of a nearby supernova which either restructured
these asteroids or affected the solar abundance in the nebula.
Asteroids have porosities of between 0.2 and 0.3 which means that
they are much denser than comets.
They are supposed to have formed through repeated collisions between
chondrules of partially melted interstellar grains which resulted
in the observed compaction.
Asteroidal grains should therefore be more compact than cometary
grains.
If the cometary grains seen in the zodiacal cloud are those created by
the collisional break-up of a comet at the end of its active life, rather
than by their ejection in the coma, then these too could be more compact
than the cometary coma grains.

IDPs have been collected in the Earth's stratosphere, and while this
is bound to be a selective example of those in the zodiacal cloud as
a whole (e.g., due to either the preferential accretion of particles on
circular orbits or the low survival probability of loosely packed
aggregates), the collected samples do provide us with much information
about the elemental composition and structure of IDPs.
Most of the collected IDPs are chondritic, either carbonaceous or silicate,
with approximately solar composition, and often appear to be aggregates of
smaller more or less spherical silicate particles of about 0.1 $\mu$m size,
which is suggestive of an interstellar origin.
Some of the aggregates have high porosity, such as those expected to
be of cometary origin, while others are densely packed, such as would
be expected from particles created in collisions between asteroids.
The aggregates are roughly equidimensional which is consistent with the
morphology of lunar microcraters.

%%%%%%%%%%%%%%%%%%%%%%%%%%%%%%%%%%%%%%%%%%%%%%
\section{Debris Disk Dust}
\label{s-diskdust}
The most obvious place to search for an understanding of the composition
and morphology of debris disk particles is in observations of these disks.
Much information can be gleaned from spectral observations of a disk's
thermal emission.
The mid-IR spectrum of $\beta$ Pictoris was discovered to contain
silicate emission features near 10 $\mu$m very similar to those seen in the
coma of comet Halley (\cite{tk91}).
Silicate features (which are located at 9.7 and 18 $\mu$m) have since
been identified in the spectra of a broad sample of debris disks
(e.g., \cite{ftk93}; \cite{ssbm96}; \cite{sglr99}).
Careful modeling is needed for a successful interpretation of
the relative abundance of silicates in these systems, since feature strengths
depend on the temperature and size distribution of the disk particles;
e.g., the silicate emission features are only observed if most of the
disk particles are smaller than 10 $\mu$m (\cite{gh90}).
The spectra of some debris disks also show unidentified IR emission bands
which indicate oxygen and carbon rich dust species (\cite{ssbm96}) and
could be attributed to thermally spiking small grains (\cite{ssb97}).
These grains could be PAHs which are so small that on absorption of a single
UV photon, their temperature rises to between 1000 and 1500 K.
Thermally spiking grains also have emission features at $\sim 10$ $\mu$m
which means that interpretation of a 10 $\mu$m excess in terms of silicate
emission should be treated cautiously.

In fact, any inferences from spectral observations are only as strong as the
model upon which they are based;
e.g., the model must include the spatial distribution of the emitting dust
which is in general not known.
Further, the spatial distribution of the particles emitting at one
wavelength may not be the same as those at a different wavelength
(\S \ref{s-categories}).
There may also be two populations of particles that have different
compositions;
these may, or may not, be spatially related (e.g., the PAHs may originate
in the mantles of the particles responsible for the silicate emission, or
they may be an unrelated population).
However, the observations tend to concur with the theoretical arguments
that debris disk particles have their origin in interstellar dust particles.
This lead to the formation of a comet dust model for the $\beta$ Pictoris
disk (\cite{lg98}).

%%%%%%%%%%%%%%%%%%%%%%%%%%%%%%%%%%%%%%%%%%%%%%
\section{Dust Models}
\label{s-dustmodels}
When creating a model for debris disk particles, one must remember that
the methods for calculating their optical properties are limited, since
there are only rigorous solutions for light scattering from particles that
are either spheres (using Mie theory) or infinite cylinders (\cite{bh83}).
Solutions for more complicated configurations do exist (e.g., \cite{xg99}),
however at present these are prohibitively computer intensive for our
purposes.
Solutions can also be found using laboratory experiments.
I have chosen to use Mie theory throughout this dissertation, not
because I believe debris disk particles to be compact spheres, but for
its robustness (the codes are well-established in the astronomical community)
and its ability to provide a real-time solution.

\begin{figure}
  \begin{center}
    \begin{tabular}{c}
      \epsfig{file=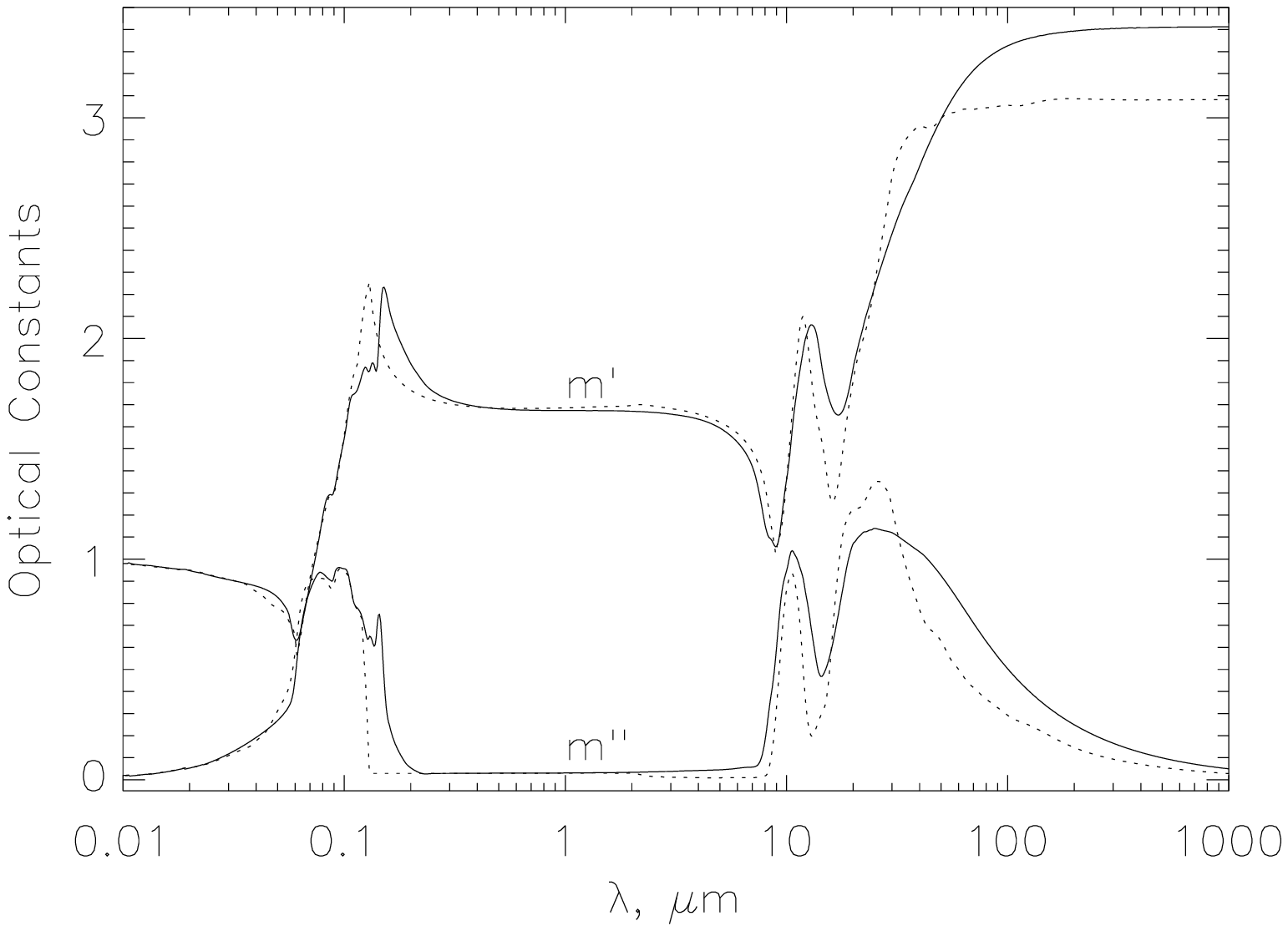,height=3.0in} \\[0.2in]
      \hspace{-0.2in} \epsfig{file=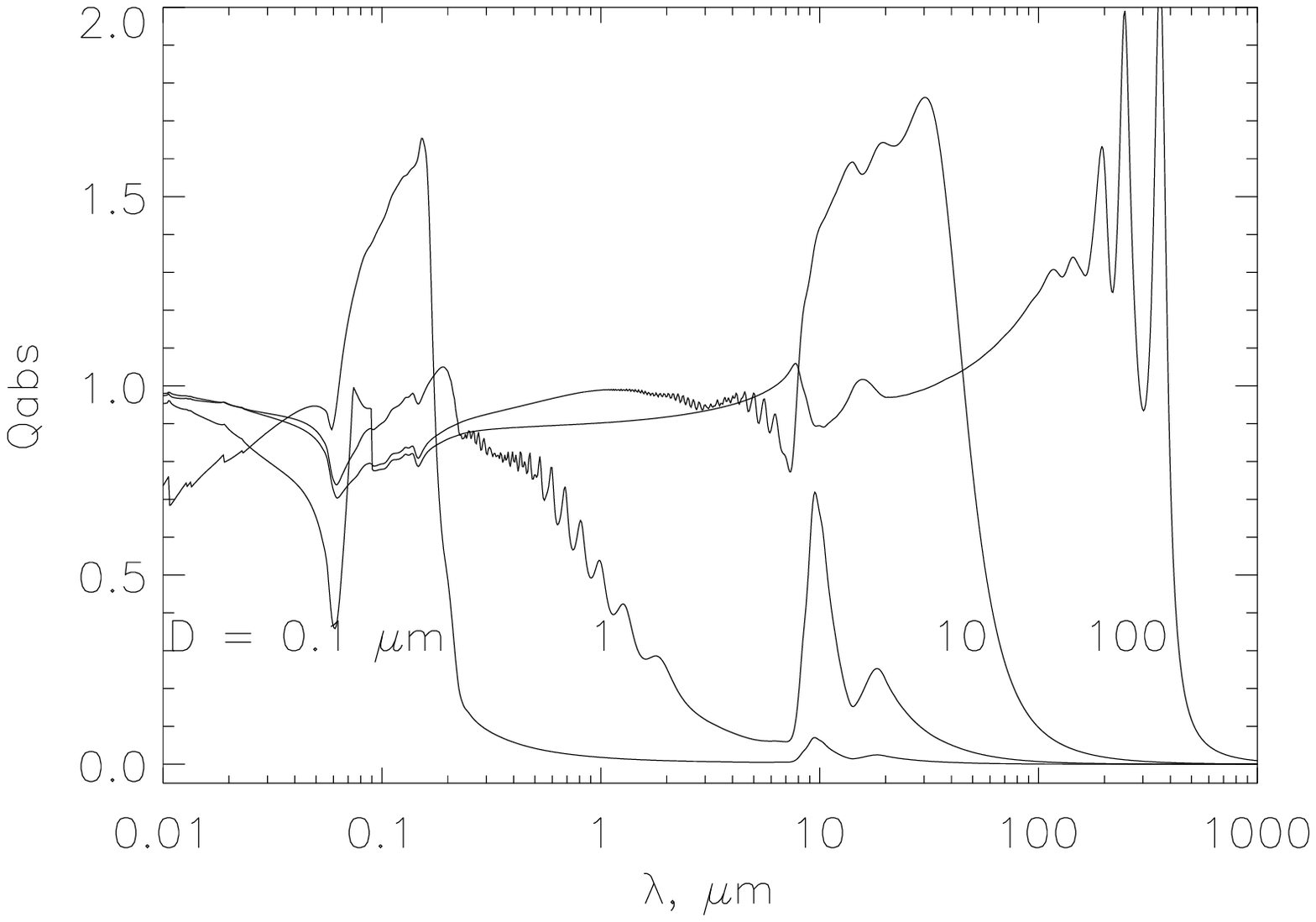,height=3.0in}
    \end{tabular}
  \end{center}
  \caption{The optical properties of particles composed of astronomical
  silicate.
  The optical constants, $m(\lambda) = m^{'} + im^{"}$, are shown in the
  top plot.
  The solid line corresponds to the astronomical silicates of Draine \&
  Lee (1984), and the dotted line corresponds to the silicates of Li \&
  Greenberg (1997).
  The absorption coefficients for solid spherical astronomical silicate
  (\cite{dl84}) particles with diameters, 0.1, 1, 10, and 100 $\mu$m,
  calulated using Mie theory, are shown in the bottom plot.}
  \label{figoptprsil}
\end{figure}

\begin{figure}
  \begin{center}
    \begin{tabular}{c}
      \epsfig{file=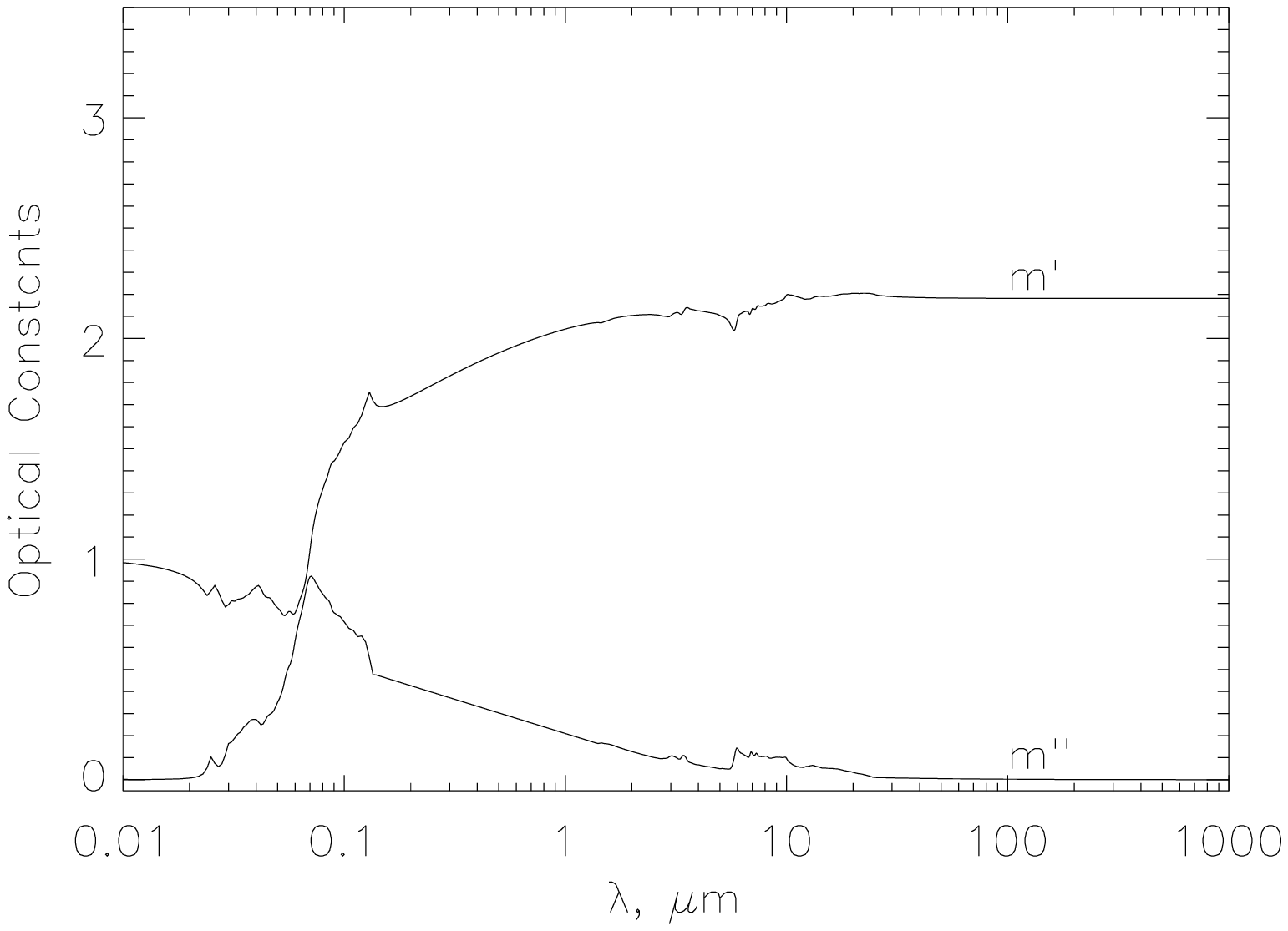,height=3.0in} \\[0.2in]
      \hspace{-0.2in} \epsfig{file=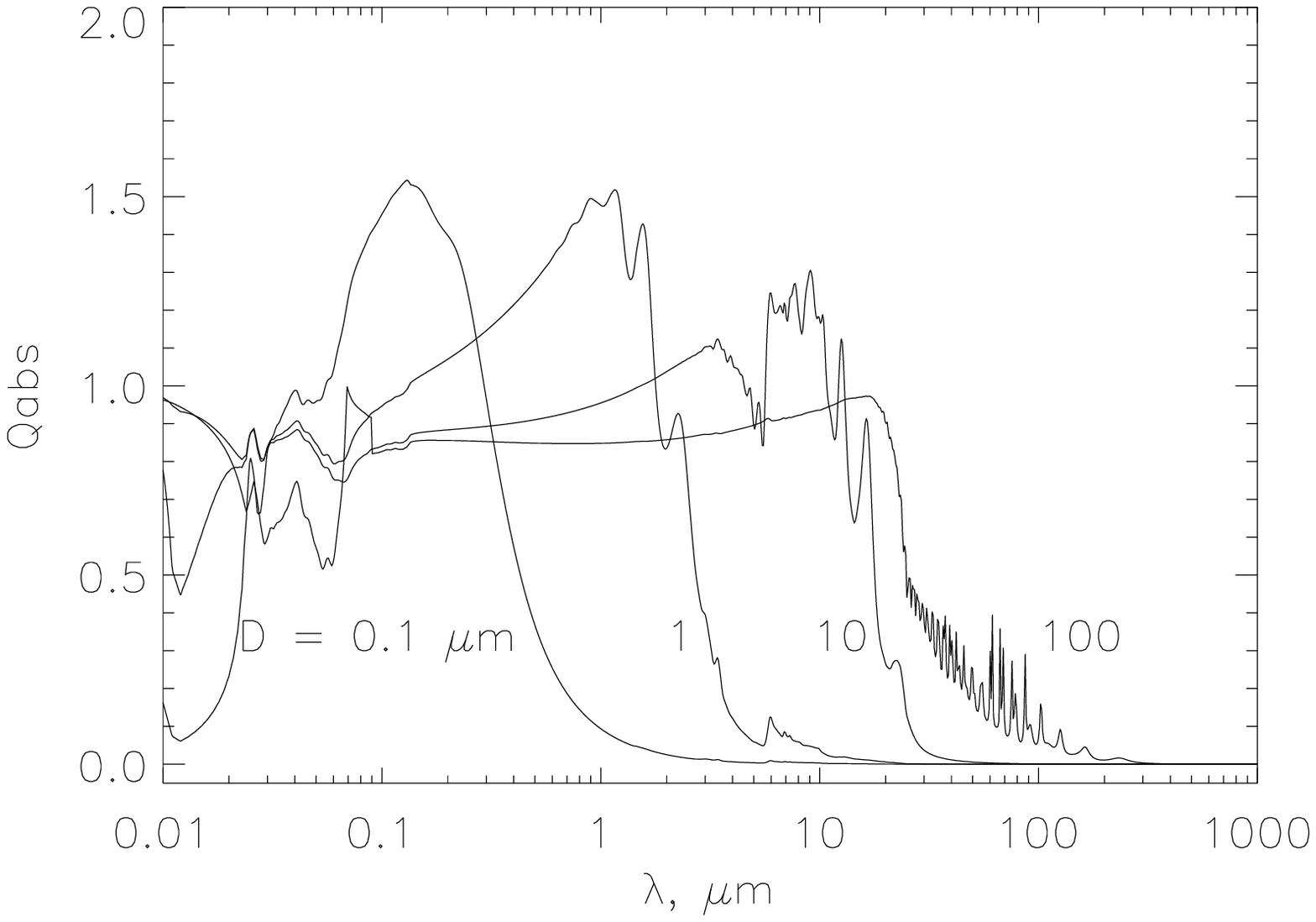,height=3.0in}
    \end{tabular}
  \end{center}
  \caption{The optical properties of particles composed of organic
  refractory (\cite{lg97}).
  The optical constants, $m(\lambda) = m^{'} + im^{"}$, are shown in the
  top plot.
  The absorption coefficients for solid spherical particles with diameters,
  0.1, 1, 10, and 100 $\mu$m, calulated using Mie theory, are shown in
  the bottom plot.}
  \label{figoptprorg}
\end{figure}

Given this constraint, the simplest model for the dust particles is that
they are spherical and composed of a single material.
The most obvious choice for this material is silicates.
Based on a combination of laboratory measurements and infrared emissivities
inferred from astronomical observations of circumstellar and interstellar
grains, Draine \& Lee (1984) obtained the optical constants of a material
they term ``astronomical silicate''.
This model has since been updated (\cite{ld93}), and the updated optical
constants for astronomical silicate are plotted in Fig.~\ref{figoptprsil}.
Since astronomical silicate is not a physical material, its density
is undefined.
Here it is taken as 2.5 g/cm$^3$, a density representative of IDPs
(\cite{gust94}).
Another set of optical constants for silicates comes from the core material
of the interstellar dust model of Li \& Greenberg (1997),
which is amorphous olivine, MgFeSiO$_4$, with a density of 3.5 g/cm$^3$.
The optical constants of this core material were found from laboratory
measurements of amorphous and crystalline olivine (and in part from
\cite{dl84}), and are plotted next to those of astronomical silicates
in Fig.~\ref{figoptprsil} for comparison.
Another choice for the dust material is the organic refractory mantle
of the interstellar grain model (\cite{lg97}), which has a density of
1.8 g/cm$^3$.
The optical constants for this material were found using a combination
of measurements of ices that have been subject to UV photoprocessing in
space, measurements of the Murchison meteorite, and theoretical reasoning
(\cite{lg97}), and are plotted in Fig.~\ref{figoptprorg}.
Other potential materials for dust particles are water ice, graphite,
or PAHs, but these are not discussed here.

The absorption coefficients, $Q_{abs}$, for Mie spheres of different
sizes composed of both astronomical silicate and organic refractory
material are plotted in Figs.~\ref{figoptprsil} and \ref{figoptprorg}.
Since the imaginary part of the optical constants, $m^{"}$, is related
to how energy is dissipated in the particle, $Q_{abs}$ is closely
related to $m^{"}$.
There is more information about how these constants lead to the temperature
and brightness dependences in a debris disk in Chapter \ref{c-hrmodel}.

%%%% Figure of optical properties of particles of astronomical silicate

%%%% Figure of optical properties of particles of organic refractory

A more complicated model of the dust particles could still be made within
the confines of Mie theory.
Using the Maxwell-Garnett effective medium theory (\cite{bh83}), we could
model the particles to be composed of a silicate core and an organic
refractory mantle (\cite{hg90}; \cite{gh90}).
The dielectric constant of the composite grain, $\epsilon_{cm} = m^2$, would be
given by:
\begin{equation}
  \epsilon_{cm} = \epsilon_m \left[
    \frac{\epsilon_c + 2\epsilon_m + 2q^3(\epsilon_c - \epsilon_m)}
         {\epsilon_c + 2\epsilon_m -  q^3(\epsilon_c - \epsilon_m)}
    \right],
    \label{eq:dielcoremantle}
\end{equation}
where $\epsilon_c$ and $\epsilon_m$ are the dielectric constants of the
core and mantle materials, respectively, and $q$ is the fractional radius
of the core to the mantle.
The same effective medium theory could also be used to model a fluffy aggregate
of core-mantle particles with a packing factor (one minus the porosity), $f$.
The resulting dielectric constant would be (\cite{hg90}; \cite{gh90}):
\begin{equation}
  \epsilon = \frac{\epsilon_{cm} + 2 - 2f(1-\epsilon_{cm})}
                  {\epsilon_{cm} + 2 +  f(1-\epsilon_{cm})},
  \label{eq:dielagg}
\end{equation}
where we have assumed that the space between the core mantle particles
is filled with a vacuum, however, this space could also be filled with
ice.

This is the essence of the cometary grain model for the $\beta$ Pictoris
disk (\cite{lg98}), which has also been used to study the HR4796 disk
(\cite{almp99}).
Given that debris disk particles are likely to originate from comet-like
bodies, this is the best available model.
There are, however, many parameters in this model, so a lot of information
about the disk's spectral energy distribution is needed in order to
constrain them.
Some of the model parameters could be estimated by constraining the abundance
of the different elements in the grain model to match that of the abundances
in the star.
Since this model is not used in this dissertation, it will not be
discussed in any further detail.

%%%%%%%%%%%%%%%%%%%%%%%%%%%%%%%%%%%%%%%%%%%%%%
\chapter{HR 4796 LITERATURE REVIEW}
%%%%%%%%%%%%%%%%%%%%%%%%%%%%%%%%%%%%%%%%%%%%%%
\label{c-hrlitrev}
The remainder of the dissertation shows the application of the
theory of the previous chapters to the modeling of a real disk
observation.
The disk that is studied is that around the A0V star HR 4796A, and the
observations that are modeled are those of Telesco et al.
(2000, herafter T2000).
Most of the modeling described here has been presented in Wyatt et al.
(1999, hereafter W99).
Before embarking on the modeling, this chapter describes the information
available about the HR 4796 system from the literature.

%%%%%%%%%%%%%%%%%%%%%%%%%%%%%%%%%%%%%%%%%%%%%%
\section{The stars HR 4796A and HR 4796B}
\label{s-hrstars}
The HR 4796 system is a binary system consisting of the A0V star,
HR 4796A (\cite{gfbd99}), and the T Tauri-like pre-main sequence M2.5
dwarf star HR 4796B (\cite{shb95}).
There is no spectroscopic evidence to suggest that these stars are also
close binaries (\cite{jgwm95}).
The position of HR 4796B relative to HR 4796A is at a projected separation of
$7.7$ arcseconds and at position angle (measured counterclockwise from north)
of $226^\circ$ (\cite{jzbs93}).
Since this separation has not changed over the last 61 years (\cite{jzbs93}),
there is strong evidence that the two stars are physically associated.
This association is supported by the similarities of their proper motions,
radial velocities, and ages (\cite{jzbs93}).
There are two stars that are located close to HR 4796A on the sky:
HR 4796C is located at 40 arcseconds separation and HR 4796D is
at 4.7 arcseconds separation.
Both of these are background stars (\cite{jzbs93}; \cite{mlbr97};
\cite{jmwt98}).

The distance to HR 4796A was measured by Hipparcos to be $67.1 \pm 3.5$ pc.
The physical parameters of HR 4796A were estimated to be (\cite{gfbd99}):
$T_{eff} = 9774 \pm 100$ K, $L_\star/L_\odot = 23.4 \pm 2.3$,
$log(g_{ev}) = 4.33 \pm 0.03$, $M_\star/M_\odot = 2.18 \pm 0.1$,
$R_\star/R_\odot = 1.68$, $t \approx 10^7$ years, and $v\sin{i} = 150$ km/s
(i.e., rapid rotation).
Like other IR excess stars, it is under luminous for its color,
which is indicative of its youth (\cite{jmwt98}).
In the modeling of W99 that is described here, the parameters
were assumed to be those of a typical A0V star (\cite{jmwt98}):
$T_{eff} = 9500$ K, $L_\star/L_\odot = 21$, $M_\star/M_\odot = 2.5$.
This discrepancy would not affect the results in any significant manner.
A typical A0 star has a main sequence lifetime of $\sim 10^9$ years,
and has low stellar winds and magnetic fields (\cite{arty97}).
The spectrum of HR 4796A is characterized by weak metal lines (\cite{gfbd99}),
however, a detailed model for the stellar atmosphere is needed to estimate
the stellar abundances.
High resolution spectroscopy of other Vega-excess stars showed that their
photospheric abundances are only slightly under the solar values (\cite{dbr97}).
Such an underabundance would be expected if some of the heavier elements
in the gaseous stellar environment had been accreted onto dust grains
before this gas was accreted onto the star.
This has been proposed as an explanation for the origin of the metal
poor A-type $\lambda$ Bootis stars (\cite{vl90}).

The parameters for HR 4796B are (\cite{jfht98}):
$T_{eff} = 3620 \pm 60$ K, $L_\star/L_\odot = 0.11 \pm 0.02$,
and $M_\star/M_\odot = 0.38 \pm 0.05$.
The spectrum of HR 4796B shows an extraordinarily strong lithium 6708
\AA~absorption line (\cite{shb95}).
From the strength of this line and from pre-main sequence isochrone
fitting, the age of HR 4796B was estimated to be 10 Myr (\cite{shb95};
\cite{jmwt98}).
This fits in with observations of the rotational velocity and
chromospheric activity of the star.
Further, the star is a source of X-rays with $L_X/L_{bol} \approx 3 \times
10^{-4}$, which is typical for a $10^7$ year old pre-main sequence star
(\cite{jmwt98}).
Assuming the two stars to be coeval, we thus have an accurate
age estimate for the system.
The current projected separation between HR 4796A and HR 4796B corresponds
to a projected distance of $\sim 517$ AU, which gives an orbital period of
$> 7000$ years (eq.~[\ref{eq:tper}]).
The relative positions of the two stars was also observed 61 years ago,
however the observational errors mean that it is not possible to derive
the binary orbit from the change in their relative positions (\cite{jzbs93}).
Based on samples of other binary orbits, the long period of the binary
orbit implies that it could be highly eccentric ($e>0.5$; \cite{blac97}).

%%%% Figure showing relation of HR 4796 to the CLA, the TWA, and Loop I
\begin{figure}
  \begin{center}
    \begin{tabular}{c}
      \epsfig{file=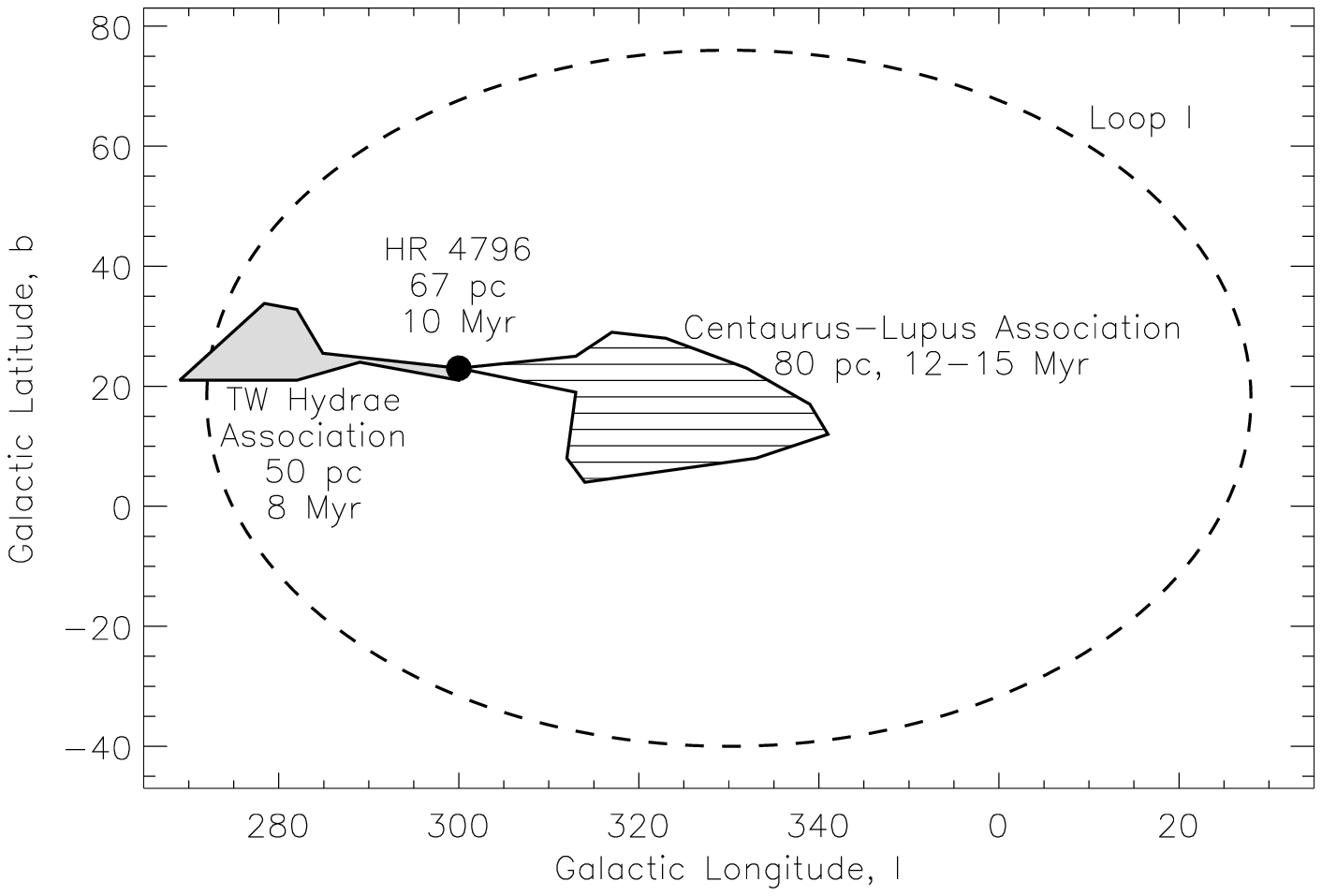,height=3.8in}
    \end{tabular}
  \end{center}
  \caption{The positions on the sky of HR 4796, the Centaurus-Lupus
  association (\cite{ddl89}; \cite{jgwm95}), the TW Hydrae association
  (\cite{wzpp99}; \cite{jhff99}), and the Galactic giant radio loop,
  Loop I (\cite{bhs71}; \cite{ea95}).}
  \label{fighr4796assoc}
\end{figure}

The HR 4796 system is located at an R.A. and Dec. of 12h 36m, $-40^\circ$,
or in Galactic coordinates, at $l=300^\circ$, $b=23^\circ$
(see Fig.~\ref{fighr4796assoc}).
It was first speculated that HR 4796 might be a very young and close
member of the Centaurus-Lupus region (\cite{jzbs93}; \cite{jgwm95}),
where stars are at $80 \pm 20$ pc from the Sun, and have an age of 12-15 Myr
(\cite{ddl89}), since its proper motion and radial velocity are similar to
those of the rest of the group.
HR 4796 lies at the edge of this region, which is centered
$\sim 25^\circ$ away close to $l=325^\circ$, $b=18^\circ$.
More recently, HR 4796 has been associated with the TW Hydrae
association (TWA) (\cite{wzpp99}; \cite{jhff99}).
At $\sim 50$ pc and an age of $\sim 8$ Myr (\cite{wzpp99}), the TWA is
the closest known region of recent star formation.
If HR 4796 is indeed a member of the TWA, then it is the most distant
member of the group.
Also, HR 4796 lies on the edge of the group, the center
of which is $\sim 18^\circ$ away, close to TW Hydrae itself, which is
at $l=279^\circ$, $b=23^\circ$.
The two groups Centaurus-Lupus and the TWA do not overlap,
rather, HR 4796 lies in between them (see Fig.~\ref{fighr4796assoc}).
All members of the TWA have the same signatures of youth:
high X ray fluxes, large lithium abundances, and strong chromospheric
activity (as does HR 4796B);
the age of the TWA is $\sim 8$ Myr (\cite{wzpp99}) and
most members are T Tauri stars.
They also have similar proper motions and radial velocities.
There has been much speculation about the origin of the TWA, because
there appears to be very little nearby molecular gas, and most 
young stars are found in regions of molecular clouds (\cite{lbt89});
the natal clouds of the TWA stars appear to have been dispersed or destroyed.
Both the nearest dark cloud and the nearest CO emission are $>15^\circ$
from HR 4796.
Based on its motion through space, HD 98800, which is a
quadruple system in the TWA, was shown to be within the Scorpio-Centaurus
complex at the time of its birth $\sim 10$ Myr ago if its velocity
relative to this complex was $\sim 15$ km/s (\cite{sksn98}). 

%%%% Figure 5 of Egger & Aschenbach 1995
\begin{figure}
  \begin{center}
    \begin{tabular}{c}
      \epsfig{file=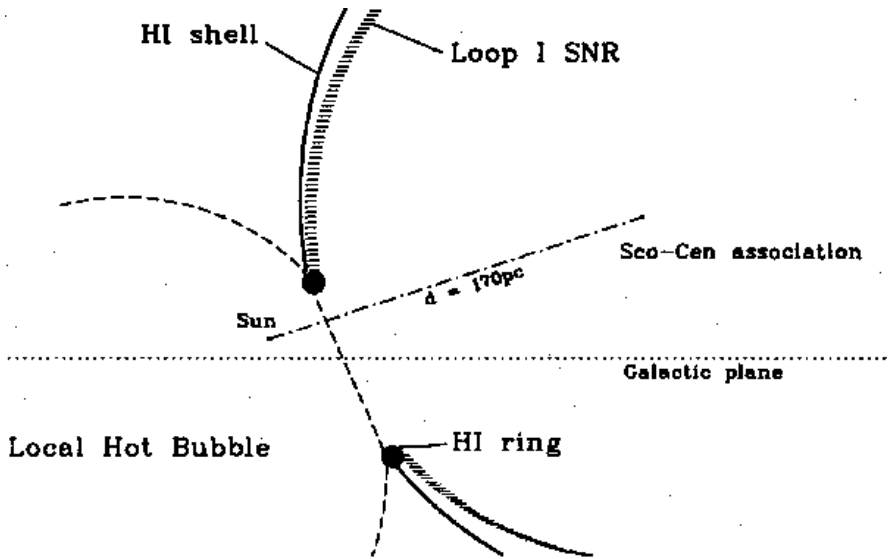,height=3in}
    \end{tabular}
  \end{center}
  \caption{The possible spatial arrangement of the Sun, the Local Hot
  Bubble, the Scorpio-Centaurus association, the HI shell, and Loop I
  (\cite{ea95}).
  This is a vertical cut normal to the Galactic plane through the
  interaction area of Loop I and the LHB.}
  \label{figea95fig5}
\end{figure}

Since HR 4796 is nearby, we should know something about the interstellar
medium within which it resides (for a review of the local interstellar
medium, see \cite{ferl99}).
Loop I is a Galactic giant radio loop of $\sim 58^\circ$ radius centered
at $l=330^\circ$, $b=18^\circ$ (\cite{bhs71}; see Fig.~\ref{fighr4796assoc}),
which is caused by the interaction of a shock wave with the
interstellar medium.
It can be modeled as a superbubble of radius $\sim 160$ pc produced
by collective stellar winds and several consecutive SN events in the
Scorpio-Centaurus OB association which is at a distance of
$\sim 170$ pc (\cite{ea95}; see Fig.~\ref{figea95fig5}). 
The terminal shock of the bubble is still expanding outwards at a velocity
of $\sim 20$ km/s;
the most recent SNe was $\sim 2 \times 10^5$ years ago (\cite{ea95}).
It is interacting with the Local Hot Bubble (LHB), the superbubble
within which the Sun is embedded (although not at the center),
which has a radius of $\sim 100$ pc (\cite{ferl99}). 
At the shock wall, a ring of neutral matter is expected to form that
is denser by a factor of 20-30 than the ambient medium (\cite{yi90}).
The HI column density in this direction does indeed increase by this
factor at about 70 pc, however, the exact shape and distance to the
ring is not uniform, since other estimates put the neutral gas wall
at $\sim 40 \pm 25$ pc (\cite{ea95}).

Thus, it appears that both HR 4796 and the TWA sit either at the edge
of or within the shock front neutral gas wall.
Could the TWA be associated with the terminal shock, which when it
passed a molecular cloud some $\sim 10^7$ years ago, triggered both star
formation and the dispersal of this cloud?
This would explain the proper motion of HD 98800, the natal cloud of which
would have attained a proper motion relative to the shock center.
If this were true, it could explain the origin of the proper motion of
the whole TWA.
The shock could also have been responsible for star formation in the
Centaurus-Lupus association, since at a distance of 80 pc, it is closer
to the shock center, thus the shock would have passed it longer ago
than the TWA, explaining the older age of these stars.
An estimate of the speed of the shock front in this scenario depends on
the geometry of the two associations with respect to the plane of
the shock front, which may be significantly deformed due to its
interaction with the LHB (\cite{ea95}).
Without studying this geometry further, I note that a distance of
30 pc traveled in 5.5 Myr corresponds to a velocity of $\sim 5$ km/s,
which is comparable to the velocity of the expanding shock front.
So far, the evidence is purely speculative.
More detailed analysis of the spatial geometry of the shock front
with respect to the associations from radio and X-Ray observations
is warranted.
There may be other observational consequences of this scenario which
either support it or rule it out.

%%%%%%%%%%%%%%%%%%%%%%%%%%%%%%%%%%%%%%%%%%%%%%%
\section{The Pre-discovery Disk}
\label{s-prediscovery}

%%%% Figure of S.E.D. of HR 4796
\begin{figure}
  \begin{center}
    \begin{tabular}{c}
      \epsfig{file=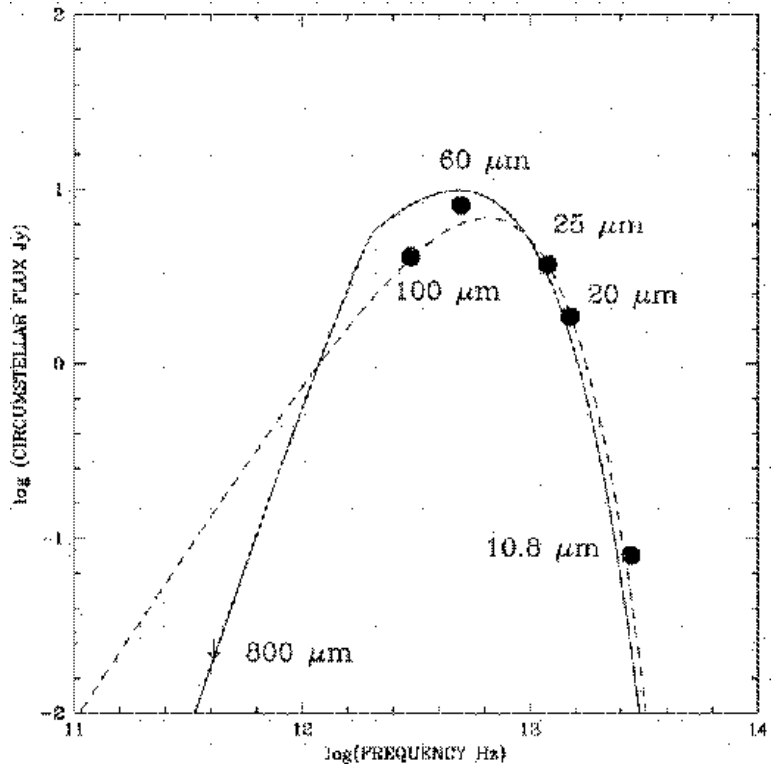,height=4in}
    \end{tabular}
  \end{center}
  \caption{The excess dust emission from the HR 4796 disk above that
  of the photosphere of HR 4796A (\cite{jmwt98}).
  The solid circles are the IRAS measured excesses (\cite{jura91}),
  and the limit to the sub-mm excess comes from observations described
  in Jura et al. (1995).
  The dashed line indicates emission from a black body at a temperature
  of 110 K.
  The solid line is the emission from a model of a disk composed
  of icy grains (\cite{jmwt98}).}
  \label{fighr4796SED}
\end{figure}

HR 4796A was one of the stars observed by IRAS to have an IR flux
above that expected from the photospheric emission from the star that
was seen to be indicative of debris disks.
Until the disk was resolved, interpretation of the IR excess had to be
based on the shape of the spectral energy distribution (\cite{jura91};
Fig.~\ref{fighr4796SED}).
The mean optical depth of the dust debris is very high at
$\tau = L_{dust}/L_\star = 5 \times 10^{-3}$ (\cite{jura91}).
The shape of the SED implied that the dust in the disk is at a temperature
of 110 K and therefore has an inner hole $\approx$ 40 AU
in radius from the star (\cite{jzbs93}).
Limits from the non-detection of reflection nebulosity put an outer
limit of 130 AU to the dust location (\cite{mlbr97}; \cite{jmwt98}).

Thus, the disk emission arises from analogue Kuiper belt regions.
Since the age of the HR 4796 system (\cite{shb95}) places it at a stage
in its evolution when the formation of any planets is expected to be
almost complete (e.g., \cite{liss93}), many authors speculated that
the central cavity could be indicative of planetary formation in
this inner region.
It was thought that a sweeper companion orbiting HR 4796A at a
distance of half the hole radius (\cite{al94}) could be the cause
of the inner hole (\cite{jgwm95}; \cite{jmwt98}).
Speckle imaging at 2 $\mu$m showed that any companion between 11 and 120
AU from HR 4796A must be less than 0.2 $M_\odot$ (\cite{jgwm95};
\cite{jmwt98}).
Another origin that was proposed for the inner hole was that the dust
grains are icy.
Icy grains would sublimate in the inner hole region due to the high
temperatures attained by the dust there (\cite{jmwt98}).
A model that gave the spatial distribution of icy grains in the
disk based on their sublimation times was shown to provide a good fit to
the observed excess (\cite{jmwt98}; see Fig.~\ref{fighr4796SED}).

A lower limit to the disk's mass was obtained from its optical depth,
since this arises from the disk's small emitting particles that contain
the disk's surface area, but not its mass.
An upper limit to its mass was obtained from the sub-mm flux, since this
arises from the larger particles in the disk ($\sim 1$ mm) which are more
indicative of the disk mass.
These gave the constraint that $0.1M_\oplus < M_{disk} < 1.0M_\oplus$
(\cite{jgwm95}; \cite{jmwt98}), which is $O(10^{-3})$ less than that
expected to have been around the star in its pre-main sequence stage
(\cite{jzbs93}).
Study of the nature of the HR 4796 disk was understood to be particularly
important, since it could yield clues to the scenario of disk formation
in binary systems (\cite{cp91}).

%%%%%%%%%%%%%%%%%%%%%%%%%%%%%%%%%%%%%%%%%%%%%%%
\section{Disk Discovery Images}
\label{s-discovery}

%%%% Figure of discovery images taken by UF and Mirlin teams
\begin{figure}
  \begin{center}
    \begin{tabular}{rccl}
      \textbf{(a)} & \hspace{-0.1in} \epsfig{file=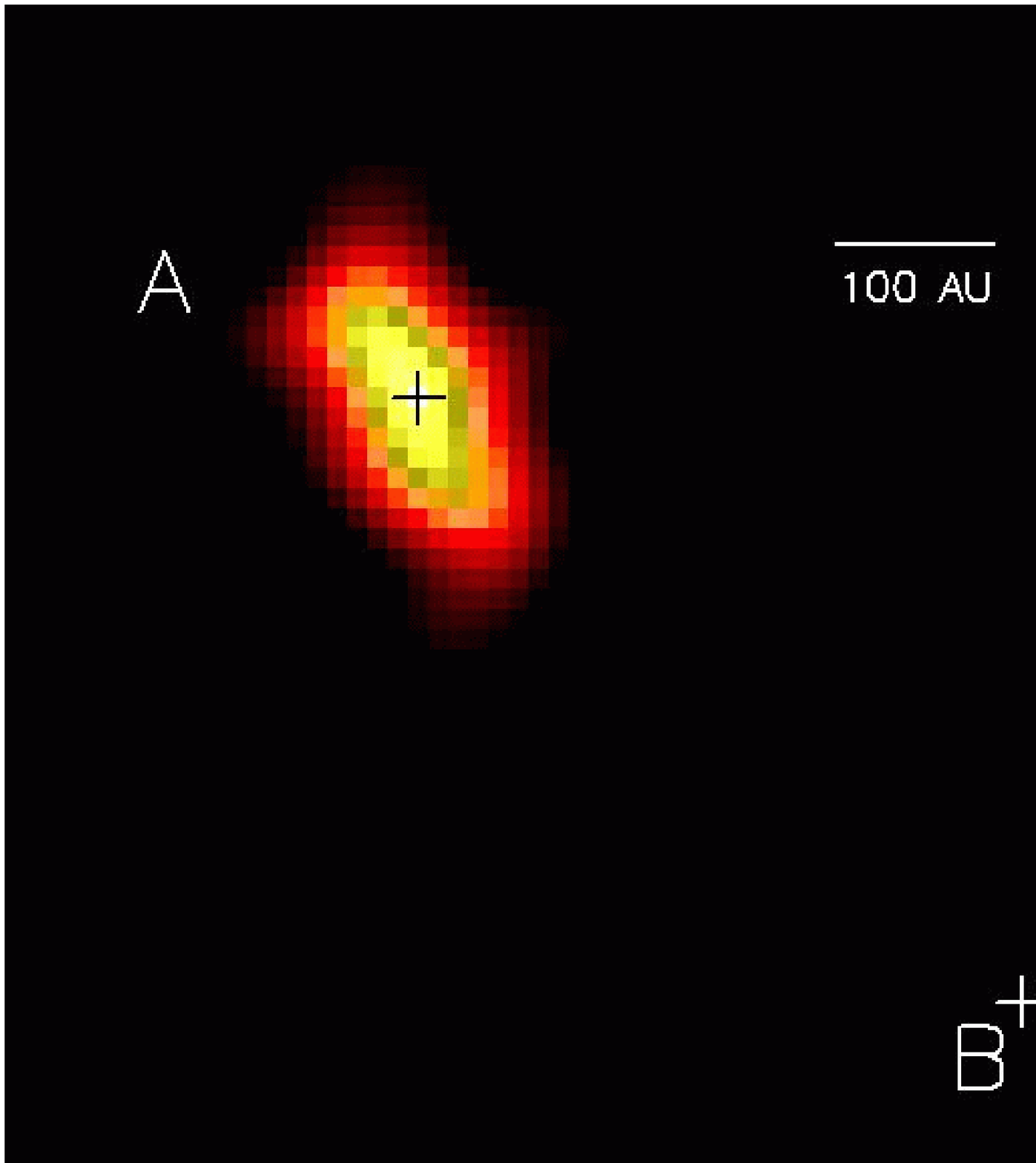,height=2.5in} &
      \epsfig{file=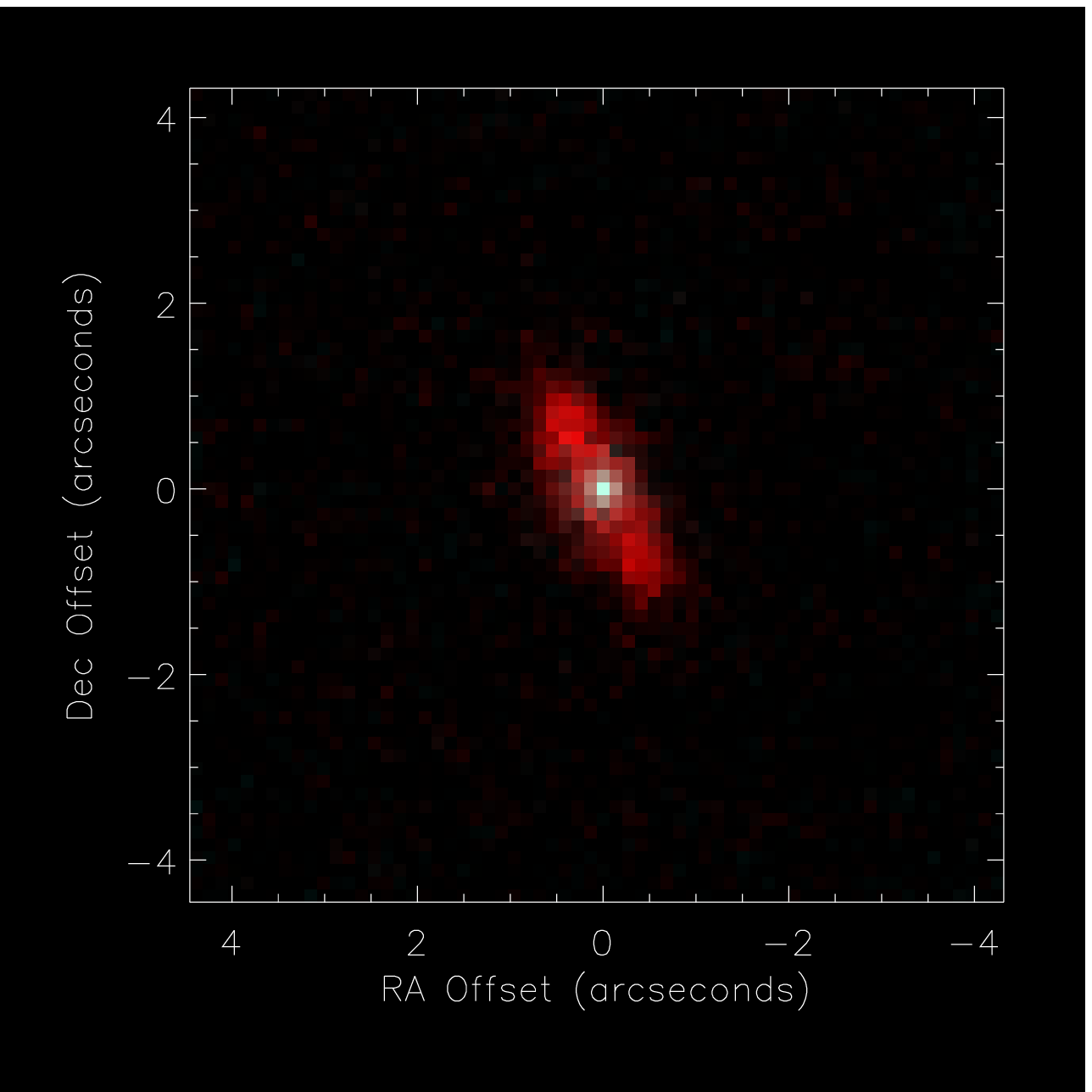,height=2.5in} & \hspace{-0.1in} \textbf{(b)} \\
    \end{tabular}
  \end{center}
  \caption{Discovery images of the mid-IR emission from the HR 4796
  disk.
  \textbf{(a)} False color image of the HR 4796 disk in the IHW18
  (18.2 $\mu$m) waveband (\cite{jfht98}).
  The positions of star A and star B are marked with crosses.
  \textbf{(b)} Composite two-color image of the emission from HR 4796 at
  $\lambda = 12.5$ (light) and 20.8 $\mu$m (dark) (\cite{krwb98}).}
  \label{figdischr4796}
\end{figure}

With the new generation of astronomical instrumentation making it possible
to image the thermal emission from the debris disks, observers had to make
a choice of which objects to observe, and HR 4796 was high on the list.
Two teams reported the simultaneous discovery of the spatially
resolved disk (\cite{jfht98}; \cite{krwb98}).
Their images, shown in Fig.~\ref{figdischr4796}, show the disk's 18
and 21 $\mu$m emission to be concentrated in two lobes, one on either
side of the star.
The emission is extended along a position angle of $210 \pm 10^\circ$,
which is similar to the direction of HR 4796B.
These observations indicate that the disk is being observed nearly edge-on,
and that its inner region is almost completely devoid of dust.
The outer edge of the disk at 18 $\mu$m was observed to be at $\sim 110$ AU
(\cite{jfht98}).
Comparison of parametric models for the disk morphology with these
observations gave the radial location of the inner hole at $55 \pm 15$ AU
from the star, and an inclination of the disk plane of $\sim 18^\circ$
to the line of sight (\cite{krwb98}).
In contrast to the 21 $\mu$m emission, the $12.5$ $\mu$m emission was observed
to be faint and concentrated at the stellar position, and was
concluded to arise from a tenuous population of hot grains close to the
star (\cite{krwb98}).

%%%%%%%%%%%%%%%%%%%%%%%%%%%%%%%%%%%%%%%%%%%%%%
\section{The Post-discovery Disk}
\label{s-postdiscovery}

%%%% Figures of Telesco et al. IHW18 and N images of HR 4796
\begin{figure}
  \begin{center}
    \begin{tabular}{c}
      \epsfig{file=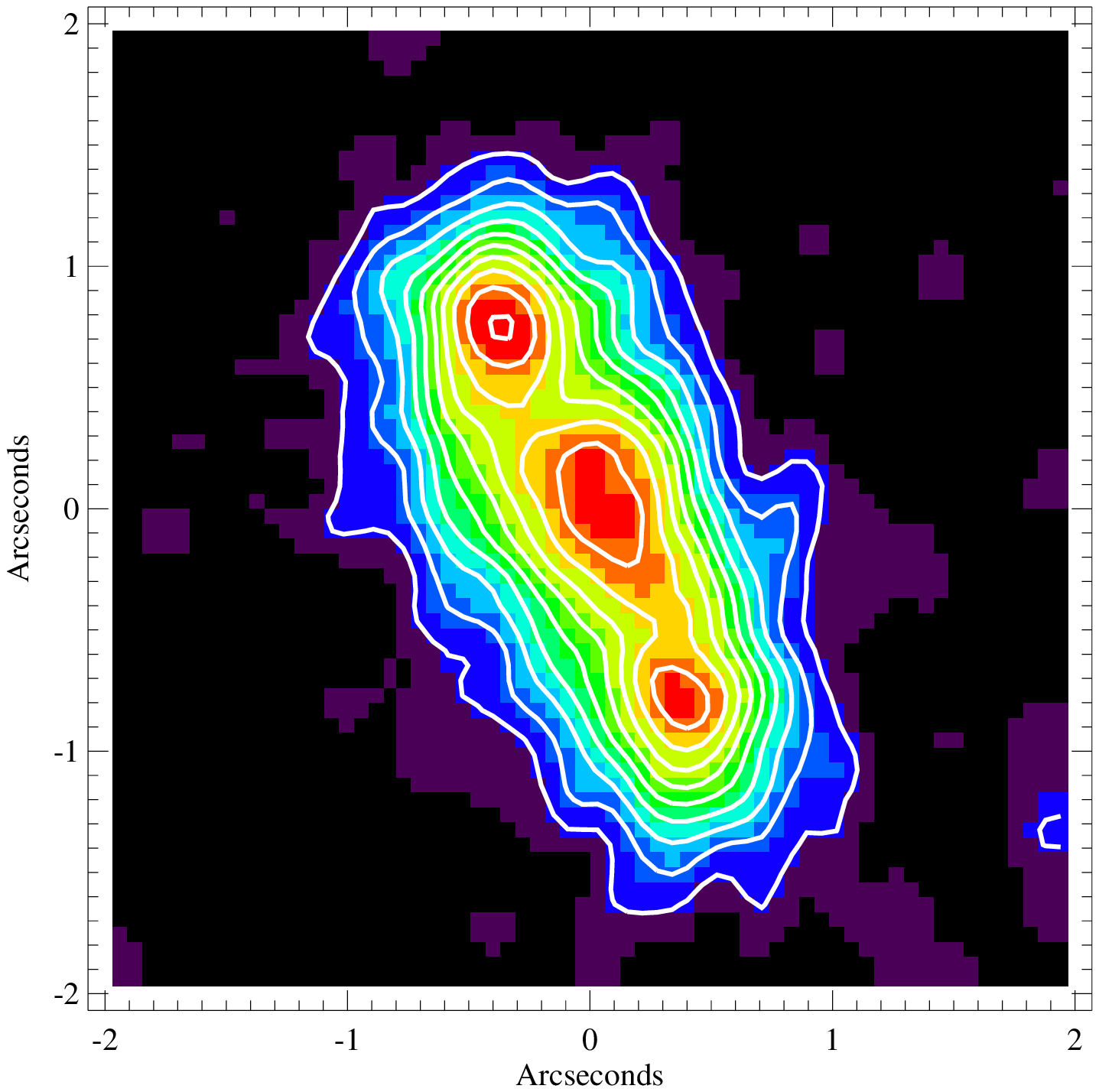,width=3.5in} \\
      \epsfig{file=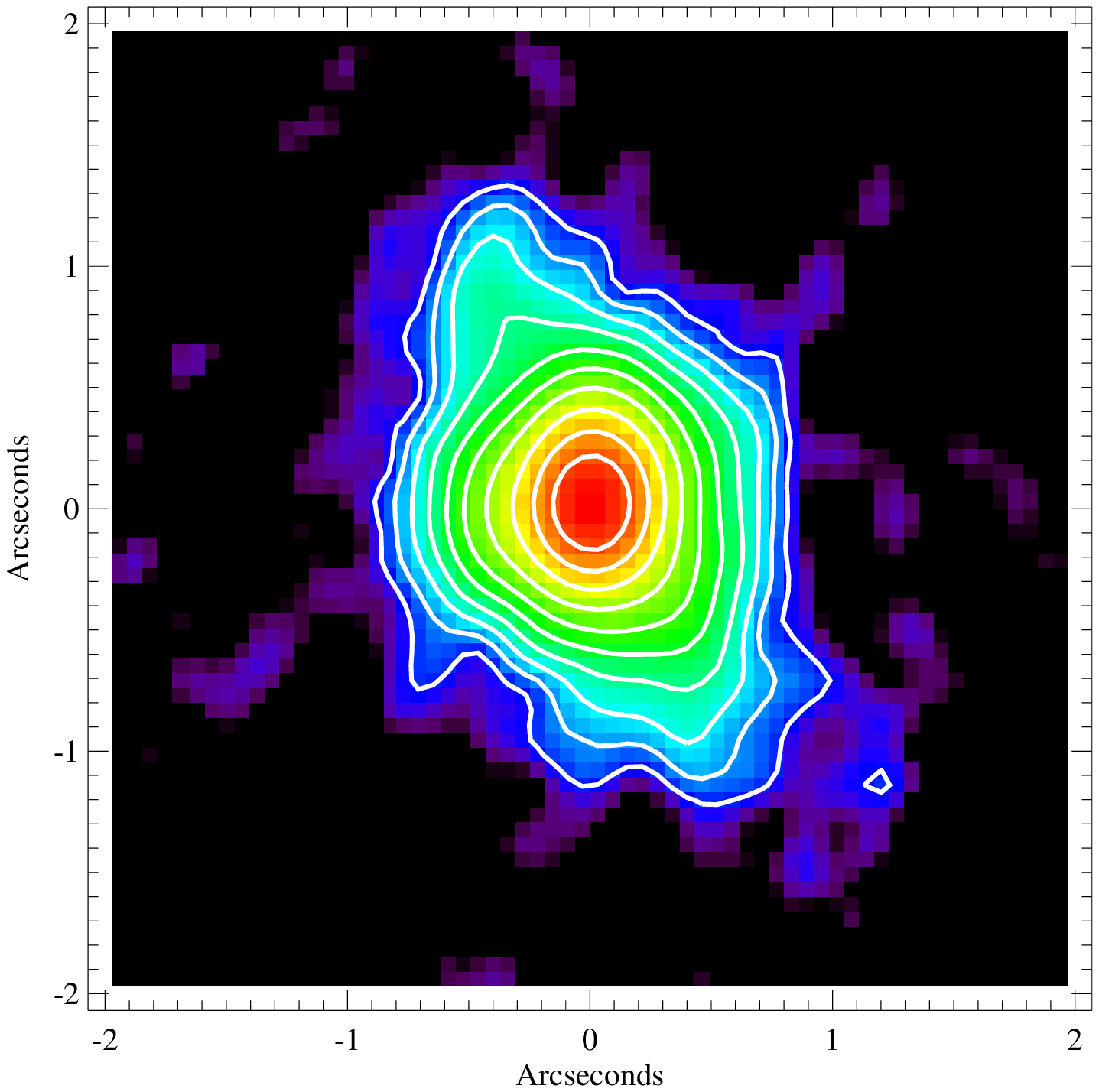,width=3.5in}
    \end{tabular}
  \end{center}
  \caption{Images of HR 4796A made with OSCIR at Keck (\cite{tfpk00}).
  North is up, east is to the left, and HR 4796A is located at the center
  of the images.
  \textit{Top}: IHW18 (18.2 $\mu$m) waveband image with contours
  (mJy/arcsec$^2$) spaced linearly at 58, 93, 128, 163, 198, 234,
  269, 304, 339, 374.
  \textit{Bottom}: N ($10.8$ $\mu$m) waveband image with contours
  (mJy/arcsec$^2$) spaced logarithmically at 6.4, 9.9, 15, 23, 36,
  55, 85, 132, 203, 312.}
  \label{figt99hr4796}
\end{figure}

The discovery images (\cite{jfht98}; \cite{krwb98}) attracted a lot of
attention and there has consequently been a significant amount of work
done on this disk.
One of the teams involved in the discovery made follow-up observations of
the disk in the N (10.8 $\mu$m) and IHW18 (18.2 $\mu$m) wavebands from
Keck (T2000).
These observations, which are shown in Fig.~\ref{figt99hr4796},
are the best (in terms of sensitivity and resolution) mid-IR images
of the disk to date.
These not only confirmed the 18 $\mu$m double-lobed structure in more detail
than before, but also showed a further interesting feature:
the disk's lobes appear to be of unequal brightness, although this observed
asymmetry is of low statistical significance ($\sim 1.8\sigma$).
Their observation suggests that the NE lobe (on the upper left of the image in
Fig.~\ref{figt99hr4796}a) is $\sim 5$\% brighter than the SW lobe.
The same asymmetry is suggested in the 21 $\mu$m images
of Koerner et al. (1998).
The modeling of these images is described in W99, and is also described
in the next two chapters.

The premise of the W99 modeling was that we know that if there is at least
one massive perturber in the HR 4796 system that is on an eccentric orbit,
then the system's secular perturbations would have caused the
disk's center of symmetry to be offset from the star (\cite{dghw98};
\S \ref{s-secpertns}).
This offset would mean that the material in one of the disk's observed
lobes is closer to the star than that in the other lobe;
consequently this lobe would be hotter and brighter.
The aim of the modeling was to ascertain how large the perturbations would
have to be to cause the observed 5\% asymmetry,
and to discuss whether perturbations of this magnitude are physically
realistic, or even to be expected, in this system.
As well as providing an explanation for the asymmetry, the W99 modeling
confirmed the large scale disk morphology and orientation.
Combining the IHW18 and N band observations, the modeling also showed
that the particles in the outer disk are likely to be small 2-3 $\mu$m
particles that are observed in the process of being blown out of the
system.

%%%% Figure of scattered light images of HR 4796
\begin{figure}
  \begin{center}
    \begin{tabular}{c}
      \epsfig{file=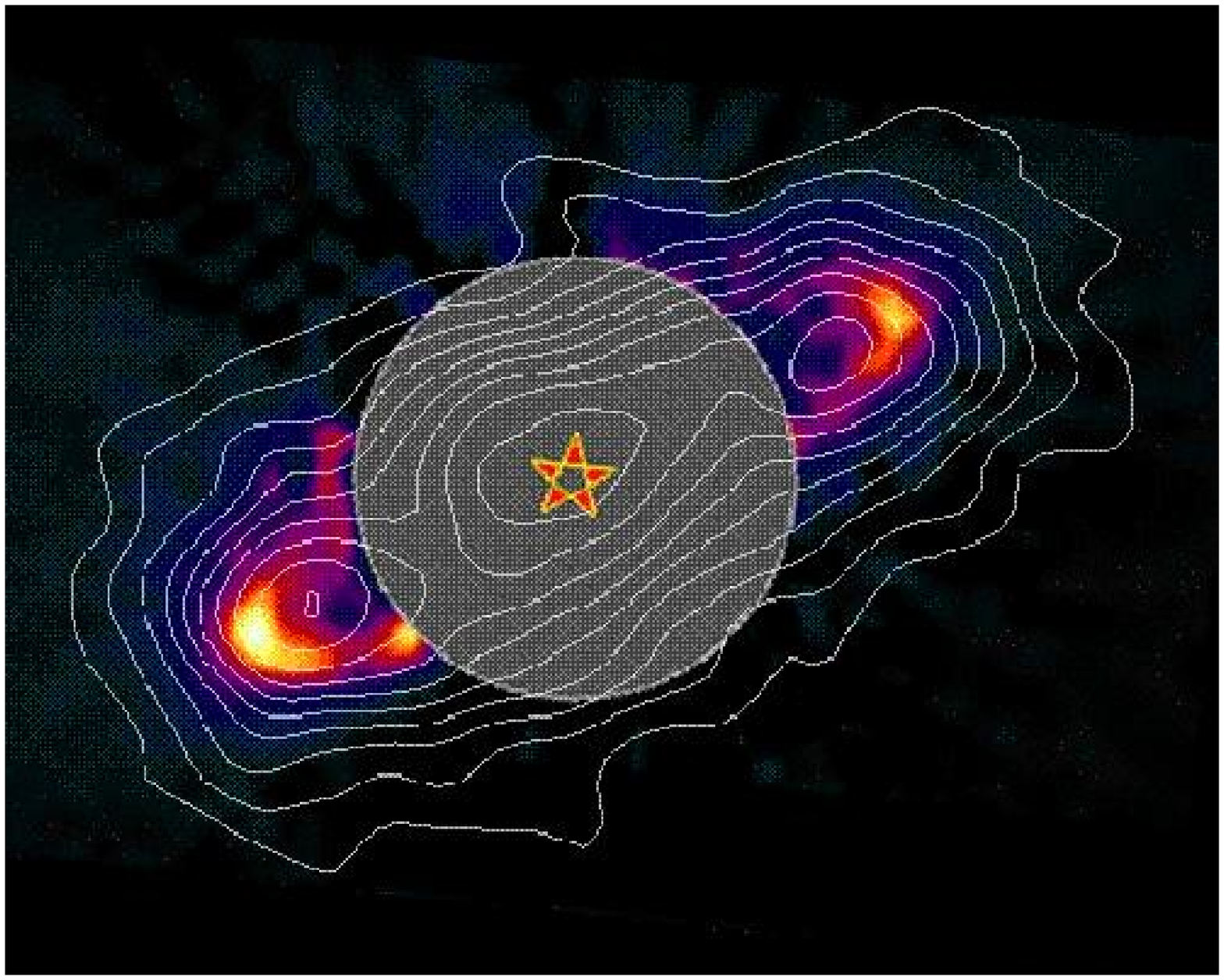,height=3in}
    \end{tabular}
  \end{center}
  \caption{NICMOS 1.1 $\mu$m image of the scattered light from the
  HR 4796A circumstellar ring (\cite{ssbk99}).
  East is approximately down.
  The unusable area circumscribing the coronographic hole is indicated by the
  gray circle.
  For comparison, the 18.2 $\mu$m contours from the observation of T2000
  (shown in Fig.~\ref{figt99hr4796}) have been overlaid onto the scattered
  light image.}
  \label{fighr4796scat}
\end{figure}

The disk was also imaged in the near-IR, $1.1$ and $1.6$ $\mu$m,
by NICMOS (\cite{ssbk99}).
These images, one of which is reproduced in Fig.~\ref{fighr4796scat},
show the disk in scattered light.
These showed that the morphology of the disk at 1.1 and 18.2 $\mu$m
and are almost identical (T2000).
They also appear to show the same lobe asymmetry as the T2000 images.
Asymmetries in images of a disk's scattered light can, however, arise
from purely geometrical causes (\cite{kj95}).
From these observations, the NICMOS team inferred that the disk must
be confined to a 17 AU wide ring, although this is not based on any
detailed modeling of the disk morphology.
Since the ring appears somewhat red in reflection, they also inferred
that the mean particle sizes are in excess of several $\mu$m.

A model of the HR 4796 disk was produced that was based on a compilation
of all previous observations of its spectral energy distribution, and using
the NICMOS images for information about the spatial distribution of
material in the disk (\cite{almp99}).
The model used a simple parametric distribution of grains, the optical
properties of which were modeled in the same way as proposed by
Li \& Greenberg (1998) for the grains in $\beta$ Pictoris.
It was found that the disk observations could be best fitted by
two independent grain populations:
a cold annulus peaking at 70 AU from the star with a width of
$\approx 14$ AU comprised of amorphous interstellar-like grains
that have a porosity of 0.6, a small
proportion of ices, and which have sizes that range from 10 $\mu$m to
$\sim 1$ m, giving a mass of this component of a few $M_\oplus$;
and a tenuous warm inner component at 9-10 AU from the star made of
crystalline comet-like grains with a porosity of 0.97.
The inner population of this model contributes almost all of the 10 $\mu$m
flux, but little at longer wavelengths.
This is a contentious issue, since the model did not have access to the
results of T2000 that indicate that much of the 10 $\mu$m emission does
indeed come from the outer disk.
It would be interesting to see what the results of this modeling would
have been had it included the observations of T2000.

Kenyon et al. (1999) modeled the planetesimal accretion in the HR 4796
disk, since it was not expected that planet-sized objects would form
on short timescales at large distances from the star.\nocite{kwww99}
They showed that a dusty ring with a width of 7-15 AU and a height
of 0.3-0.6 AU at 70 AU from HR 4796A would indeed form in $\sim 10$ Myr
as a natural outcome of the planetary formation process if the initial
mass of the protoplanetary disk was 10-20 times the minimum mass solar
nebula.
The modeling process they used was the same as that used to estimate
that the formation of Pluto would have taken $10-40$ Myr at 35 AU from
the Sun (\cite{kl99}).
They treat the planetesimals as a statistical ensemble of bodies with
a distribution of horizontal and vertical velocities about keplerian
orbits.
Collisions between the planetesimals is assumed to result in either
mergers (with potential cratering debris), catastrophic disruption,
or inelastic rebound.
Gas drag and mutual gravitational interactions are also included in the
model.
The planetesimal growth is simulated from when the planetesimals are
80 m in size.
Thus, this planetesimal accretion model is based on assumptions about the
outcome of the early stages of planet formation (i.e., the origin of the
swarm of 80 m planetesimals) which are not as well understood as the latter
stages.
They find that there is initially a period of slow growth when planetesimals
grow from 80 to 1000 m.
In this stage collisions damp the velocity dispersion.
After this there is runaway growth of the largest bodies in the disk from
1 to 100 km in several Myr.
This increases the velocity dispersion, thus ending runaway growth.
Finally there is slow growth from 100 to 1000 km.
 
Liseau (1999) observed HR 4796A in the CO (1-0) and (2-1) lines, but
neither continuum or molecular line emission was detected.\nocite{lise99}
They concluded that this could either be a consequence of a dirth of
gas in the system, possibly lost in the planetary formation process,
or of photodissociation of the molecules from the intense stellar
UV field. 
They did not put any quantifiable constraint on the gas in the system.
Independent observations put the gas mass at $< 1 - 7M_\oplus$
(\cite{gmh99}), which means that it would probably not play an
important role in the dynamics of the disk.

%%%%%%%%%%%%%%%%%%%%%%%%%%%%%%%%%%%%%%%%%%%%%%
\chapter{THE DYNAMIC HR 4796 DISK MODEL}
%%%%%%%%%%%%%%%%%%%%%%%%%%%%%%%%%%%%%%%%%%%%%%
\label{c-hrmodel}
This chapter describes a model of the HR 4796 disk that accounts for the
brightness distribution seen in the IHW18 waveband observation of T2000
(Fig.~\ref{figt99hr4796}; also reproduced in Fig.~\ref{fig8}a).
This model is based on the theory of Chapters \ref{c-obsn} to
\ref{c-optprops}, and is also presented in W99.
While the modeling techniques that are used are new to the study of
circumstellar disks, they have already been widely used to study the
observed structure of the zodiacal cloud (see, e.g.,
\cite{djxg94}; \cite{gdjx97}), and so can be used with a certain degree
of confidence.
There are three components of the observation that had to be included
in the model (\S \ref{s-modeling}):
the disk's structure, $\sigma(r,\theta,\phi)$, defined in \S \ref{s-structure};
the combination of the optical properties and the size distribution of the
disk's particles given by $P(\lambda,r)$, defined in \S \ref{s-los};
and the disk's orientation.

%%%%%%%%%%%%%%%%%%%%%%%%%%%%%%%%%%%%%%%%%%%%
\section{Model of Offset Disk Structure, $\sigma(r,\theta,\phi)$}
\label{ssec-mds}
The disk's structure was modeled as that of a secularly perturbed dynamic
disk.
A combination of the theory of Chapters \ref{c-structure} and
\ref{c-signatures}, and inferences from the observation (Fig.~\ref{fig8}a),
was used to parameterize the distribution of the orbital elements of the
disk's large particles (\S \ref{sssec-doe}).
This was used to create parameterized models of the spatial distribution
of these large particles (\S \ref{sssec-simul}), that could then be
compared with the observed spatial distribution.
Since the observed spatial distribution is that of the emitting
particles, the underlying assumption is that these emitting particles
either are the disk's large particles, or that their spatial distribution is
the same as that of the disk's large particles (whether they are large or
not).
The implications of this assumption are discussed in the interpretation
of the model (Chapter \ref{c-interpretation}).

%%%%%%%%%%%%%%%%%%%%%%%%%%%%%%%%%%%%%%%%%%%%
\subsection{Distribution of Orbital Elements,
$\sigma(a,e,I,\Omega,\tilde{\omega})$}
\label{sssec-doe}
The quintessentially secular part of the distribution of the orbital
elements of the large particles in a secularly perturbed disk is the
distributions of their complex eccentricities, $n(z)$, and complex
inclinations, $n(y)$.
For a particle with an orbit of a given semimajor axis, $a$, its complex
eccentricity, $z$, and complex inclination, $y$, are the addition of
forced elements, $z_f(a)$ and $y_f(a)$, to proper elements
that have $\tilde{\omega}_p$ and $\Omega_p$ chosen at random,
while $e_p$ and $I_p$ are chosen from the distributions $n(e_p)$
and $n(I_p)$.

%%%% Figure of distribution of ep and Ip in the asteroid belt
\begin{figure}
  \begin{center}
    \begin{tabular}{cc}
      \epsfig{file=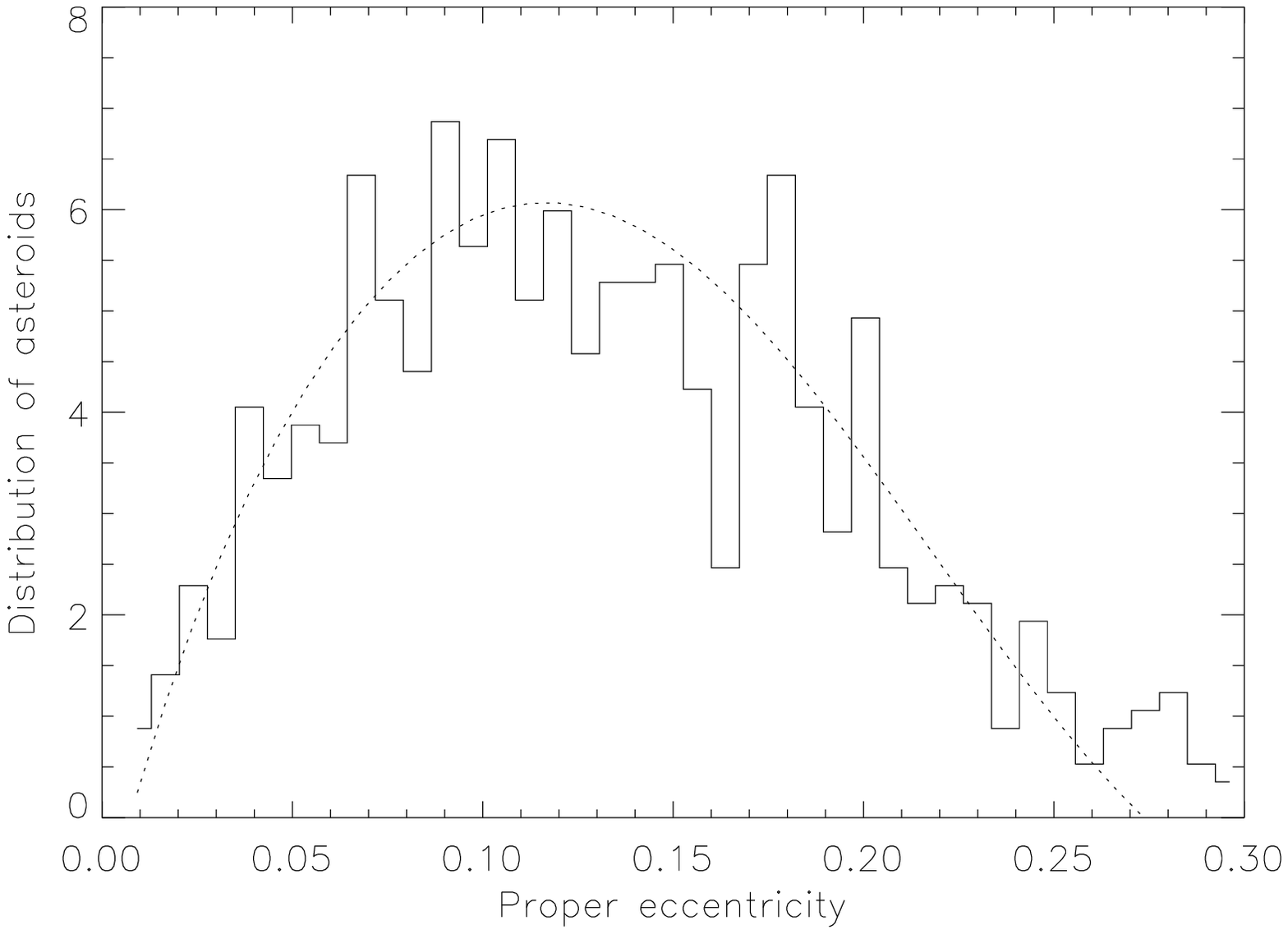,height=2in} &
      \epsfig{file=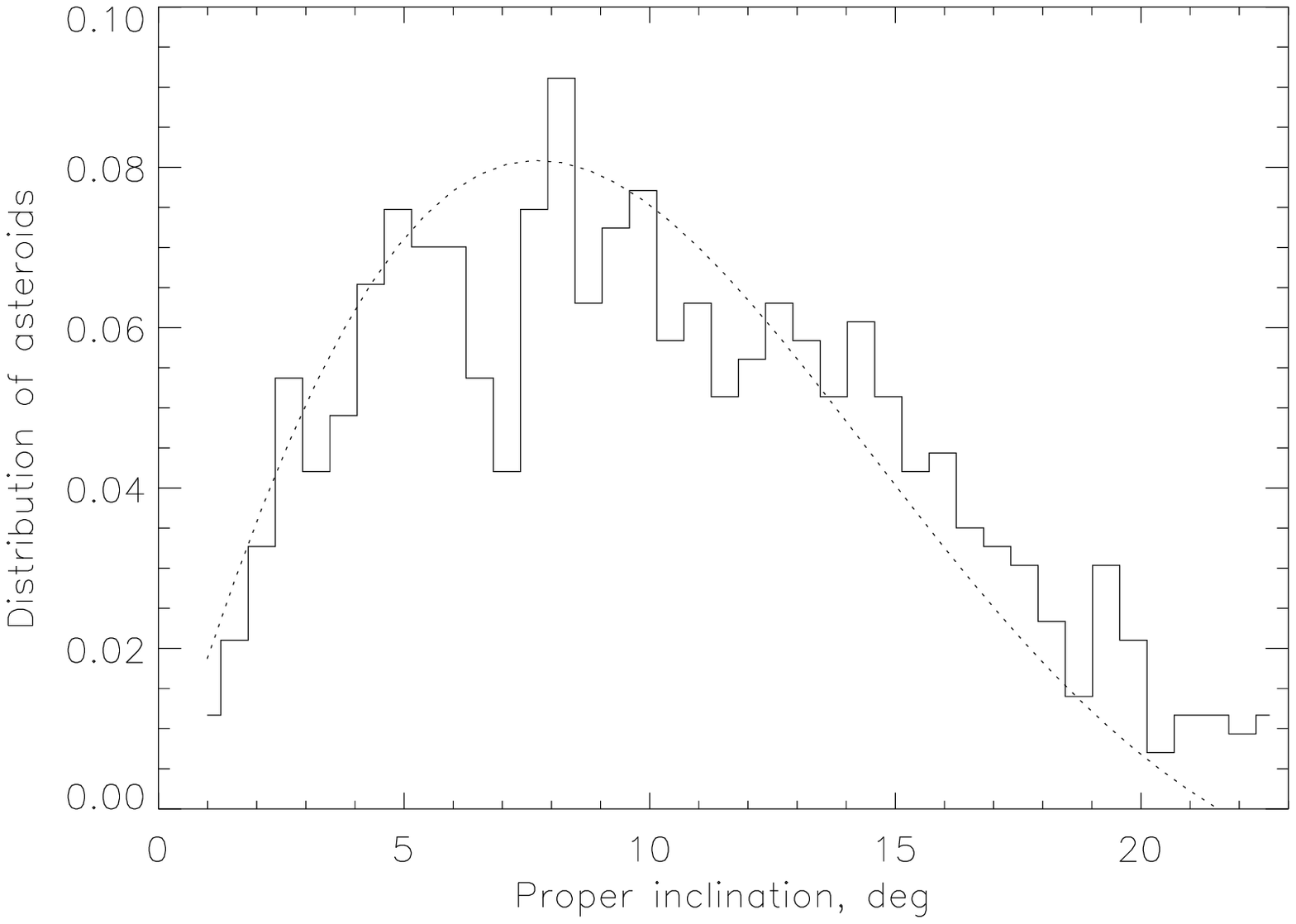,height=2in} \\
    \end{tabular}
  \end{center}
  \caption{Distribution of the proper eccentricities and inclinations
  of the mainbelt asteroids.
  The histogram indicates the observed distribution and the dotted
  line is the parameterized fit to this distribution that was used
  in the model.}
  \label{figepipast}
\end{figure}

Since there is insufficient information available to determine the
variation of the forced and proper elements with semimajor axis
in this disk, they were assumed to be constant across the disk.
The forced elements were left as model variables:
the forced eccentricity, $e_f$, defines the magnitude of the offset
asymmetry in the disk model;
the forced pericenter orientation, $\tilde{\omega}_f$, defines the
orientation of this asymmetry;
and the forced inclination, $y_f$, defines the plane of symmetry of
the disk model.
When creating a disk model, both $\tilde{\omega}_f$ and $y_f$ were
set to zero;
these were incorporated later into the description of the disk's
orientation to the line of sight of the observation (see \S \ref{ssec-do}).
The distributions of the proper eccentricities, $n(e_p)$, and proper
inclinations, $n(I_p)$, of particles in the disk model were taken to be
like those of the main-belt asteroids with absolute magnitudes
$H < 11$ (\cite{bowe96}; see Fig.~\ref{figepipast}).
These large asteroids constitute a bias-free set (\cite{bowe96}) and have
mean proper eccentricities and proper inclinations of
$\langle e_p \rangle = 0.130$ and $\langle I_p \rangle = 10.2^\circ$.
Not enough Kuiper belt objects have been discovered yet to infer a
bias-free distribution for their orbital elements.

The distribution of the semimajor axes, $n(a)$,  of particles in the
disk defines its radial distribution.
There is no way of guessing this distribution from theoretical
considerations, since it depends on the outcome of the system's
planetary formation process, which varies from system to system
(compare the distribution of the solar system's planets,
and its disk material, with those found in exosolar systems, e.g.,
\cite{bp93}; \cite{mb98}).
Thus, it had to be deduced purely observationally.
Fig.~\ref{fig8}a shows that the disk has an inner edge, inside of
which there is a negligible amount of dust;
this was modeled as a sharp cut-off in the distribution of
semimajor axes at $a_{min}$, a model variable.
The observation also shows that the disk has an outer edge at
$\sim 130$ AU;
this was modeled as a sharp cut-off in the distribution of
semimajor axes at $a_{max} = 130$ AU (this is a non-critical
parameter, since particles near the outer edge of the disk contribute
little to the observation, see \S \ref{sssec-pg}).
The distribution between $a_{min}$ and 130 AU was taken as
$n(a) \propto a^\gamma$, where $n(a)da$ is the number of particles
on orbits with semimajor axes in the range $a \pm da/2$, and
$\gamma$ is a model variable.
To get an idea of the radial distribution resulting from this semimajor
axis distribution, consider that if the particles had zero eccentricity,
this distribution would result in a volume density (number of particles
per unit volume) distribution that is $\propto r^{\gamma-2}$
(since the number of particles in a spherical shell of width $dr$, the
volume of which is $\propto r^2dr$, would contain a number of particles
that is $\propto r^\gamma dr$).

%%%%%%%%%%%%%%%%%%%%%%%%%%%%%%%%%%%%%%%%%%%%%
\subsection{Conversion to Spatial Distribution, $\sigma(r,\theta,\phi)$}
\label{sssec-simul}
Disk models were created from the orbital element distribution of
\S \ref{sssec-doe} using the ``SIMUL'' program;
SIMUL was developed by the solar system dynamics group at the
University of Florida (\cite{dgdg92}).
A disk model is a large three-dimensional array,
$\sigma(r,\theta,\phi)$, that describes the spatial distribution of the
cross-sectional area of material in the disk model per unit volume binned
in: $r$, the radial distance from the star;
$\theta$, the longitude relative to an arbitrary direction
(set here as the forced pericenter direction, $\tilde{\omega}_f$);
and $\phi$, the latitude relative to an arbitrary plane
(set here as the forced inclination, $y_f$, or symmetry, plane).
SIMUL creates a disk model by taking the total cross-sectional
area of material in the disk (specified by the model variable
$\sigma_{tot}$), and dividing it equally among a large number of
orbits (5 million in this case), the elements of each of which are
chosen randomly from the specified distribution (\S \ref{sssec-doe}).
The disk model is populated by considering the contribution of each
orbit to the cross-sectional area density in each of the cells it crosses.

%%%% Figure 4
\begin{figure}
  \begin{center}
    \begin{tabular}{c}
      \epsfig{file=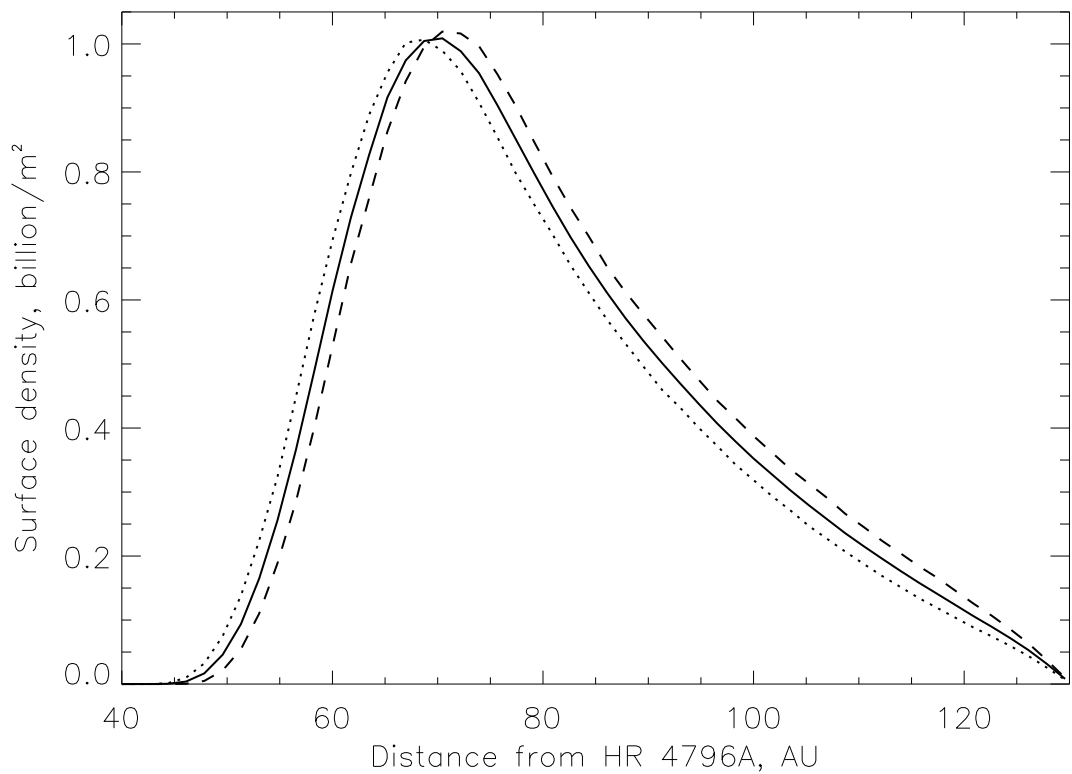,height=4in}
    \end{tabular}
  \end{center}
  \caption{ The surface number density of the 2.5 $\mu$m dust grains
  in the HR 4796 disk derived from the 18.2 $\mu$m brightness
  distribution (T2000).
  The solid curve is the azimuthal average of the surface number density,
  and the dotted and dashed lines indicate the density through the
  disk towards and away from the forced pericenter direction of the model,
  respectively.
  The offset is a result of the forced eccentricity imposed on the disk
  model;
  the inner edge of each side of the disk is offset by
  $\sim a_{min}e_f \approx 1$ AU.
  The disk's surface density peaks at $\sim 1.02 \times 10^9$ m$^{-2}$
  at $\sim 70$ AU.
  Interior to this, the surface density falls to zero by 45 AU;
  the sloping cut-off is due to the eccentricities of the disk model
  particles' orbits.
  Exterior to 70 AU, the surface density falls off $\propto r^{-3}$;
  this is due to the distribution of the disk model particles' semimajor
  axes, $n(a) \propto a^{-2}$.}
  \label{fig4}
\end{figure}

The spatial distribution of material in one such model of the HR 4796
disk can be described by the three variables $a_{min}$, $\gamma$,
and $e_f$;
$\sigma_{tot}$ simply scales the amount of material in the model,
and $y_f$ and $\tilde{\omega}_f$ describe the reference plane and the
reference direction of the model relative to the line of sight of
the observation.
Fig.~\ref{fig4} is a plot of the surface density of material in a disk
model with $a_{min} = 62$ AU, $e_f = 0.02$, and $\gamma = -2$
(this is the final model of \S \ref{ssec-mpr}).
This illustrates how the specified distribution of orbital elements
affects the spatial distribution of material in the disk model:
the sharp cut-off in semimajor axes at $a_{min}$ determines the radial
location of the inner hole, which has a sloping cut-off in $r$ due
to the particles' eccentricities;
as predicted in \S \ref{ss-ofwpfam}, particles at the inner edge of
the disk in the forced pericenter direction are closer to the star than
those in the forced apocenter direction by $\sim 2a_{min}e_f$;
the distribution of semimajor axes has produced a surface density
distribution that is $\propto r^{\gamma-1}$, but only exterior to 70 AU.

%%%%%%%%%%%%%%%%%%%%%%%%%%%%%%%%%%%%%%%%%%%%%%%%%%
\section{Model of $P(\lambda,r)$}
\label{ssec-mpp}

%%%%%%%%%%%%%%%%%%%%%%%%%%%%%%%%%%%%%%%%%%%%%%%%
\subsection{Optical Properties of Disk Particles}
\label{sssec-pop}
The optical properties of the disk particles were found assuming them
to be made of astronomical silicate (\cite{dl84}; \cite{ld93}),
since this is a common component of interplanetary dust found in both
the zodiacal cloud (\cite{lg90}) and exosolar systems (e.g., \cite{tk91};
\cite{ftk93}; \cite{sglr99}).
The presence of silicates in the disk can be tested at a later date using
spectroscopy to look for silicate features in the disk emission.
Furthermore, the particles were assumed to be solid, spherical, and have a
density of $\rho = 2500\textrm{kg}/\textrm{m}^3$.
Their optical properties were calculated using Mie theory, assuming that
HR 4796A has a luminosity and temperature of $L_\star = 21L_\odot$,
$T_\star = 9500^{\circ}$K (\cite{jmwt98}), and using for its spectrum,
that of the A0V star Vega.
In other words, the model for the optical properties of the disk particles
is a very simple one.
The consequences of these simplifications are discussed in the next chapter,
but since the modeling aims to reproduce an image of the disk in just one
waveband, this should not affect the interpretation of the disk's observed
spatial structure (\S \ref{s-modeling}).

%%%% Figure 5
\begin{figure}
  \begin{center}
    \begin{tabular}{cccc}
      & \epsfig{file=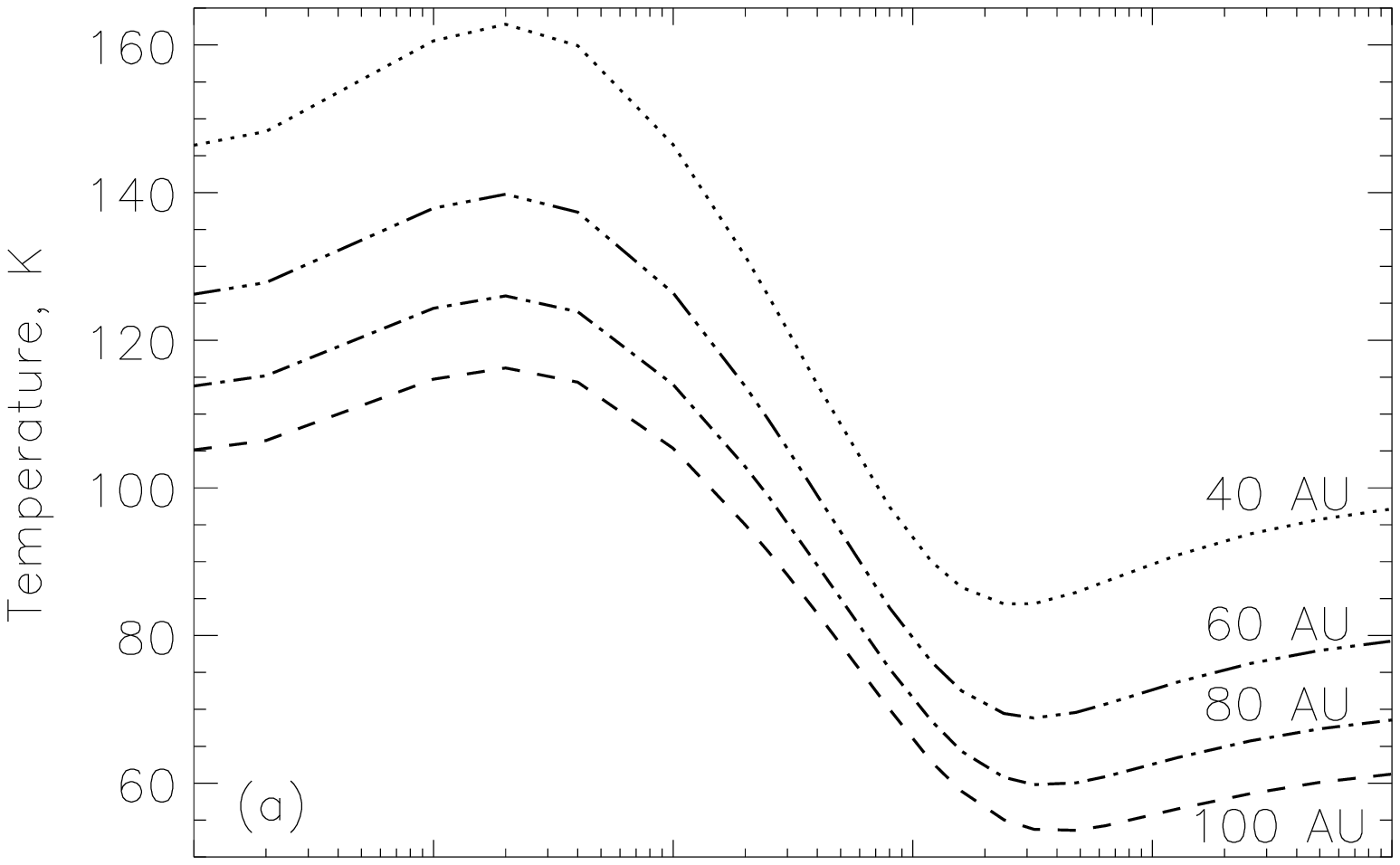,height=1.87in} & \hspace{-0.2in}
        \epsfig{file=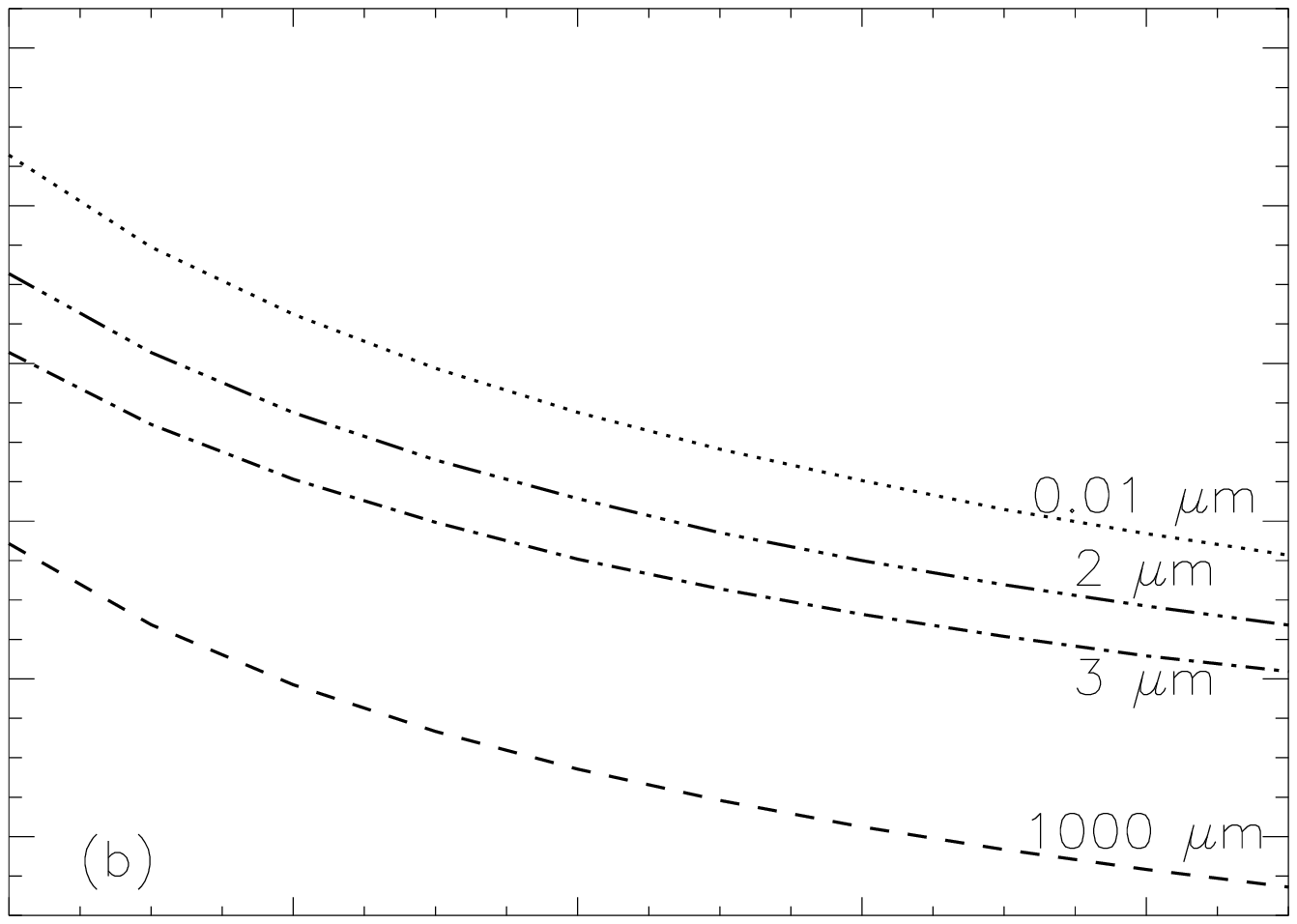,height=1.87in} & \vspace{-0.08in} \\
      & \hspace{-0.15in} \epsfig{file=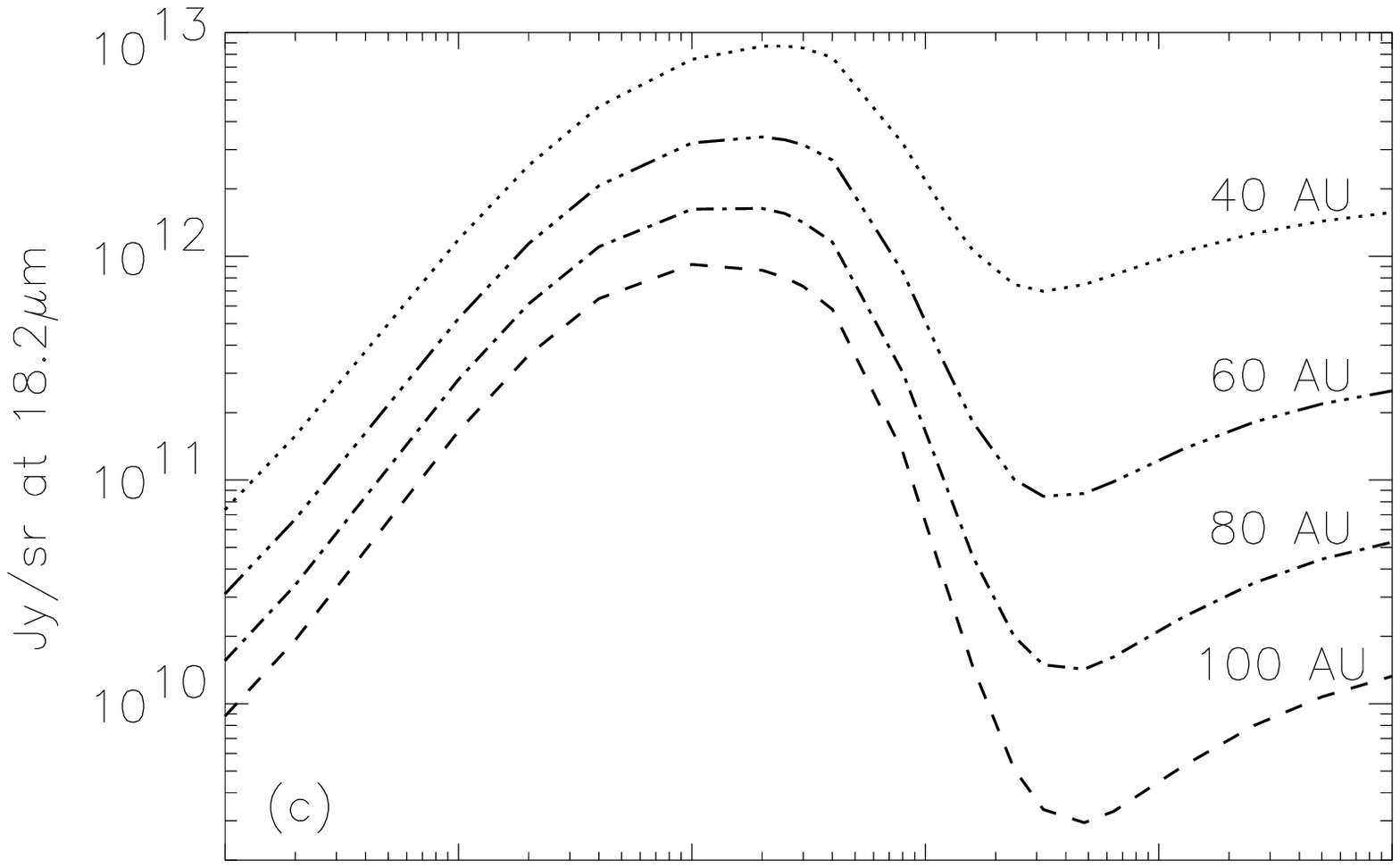,height=1.925in} & \hspace{-0.2in}
        \epsfig{file=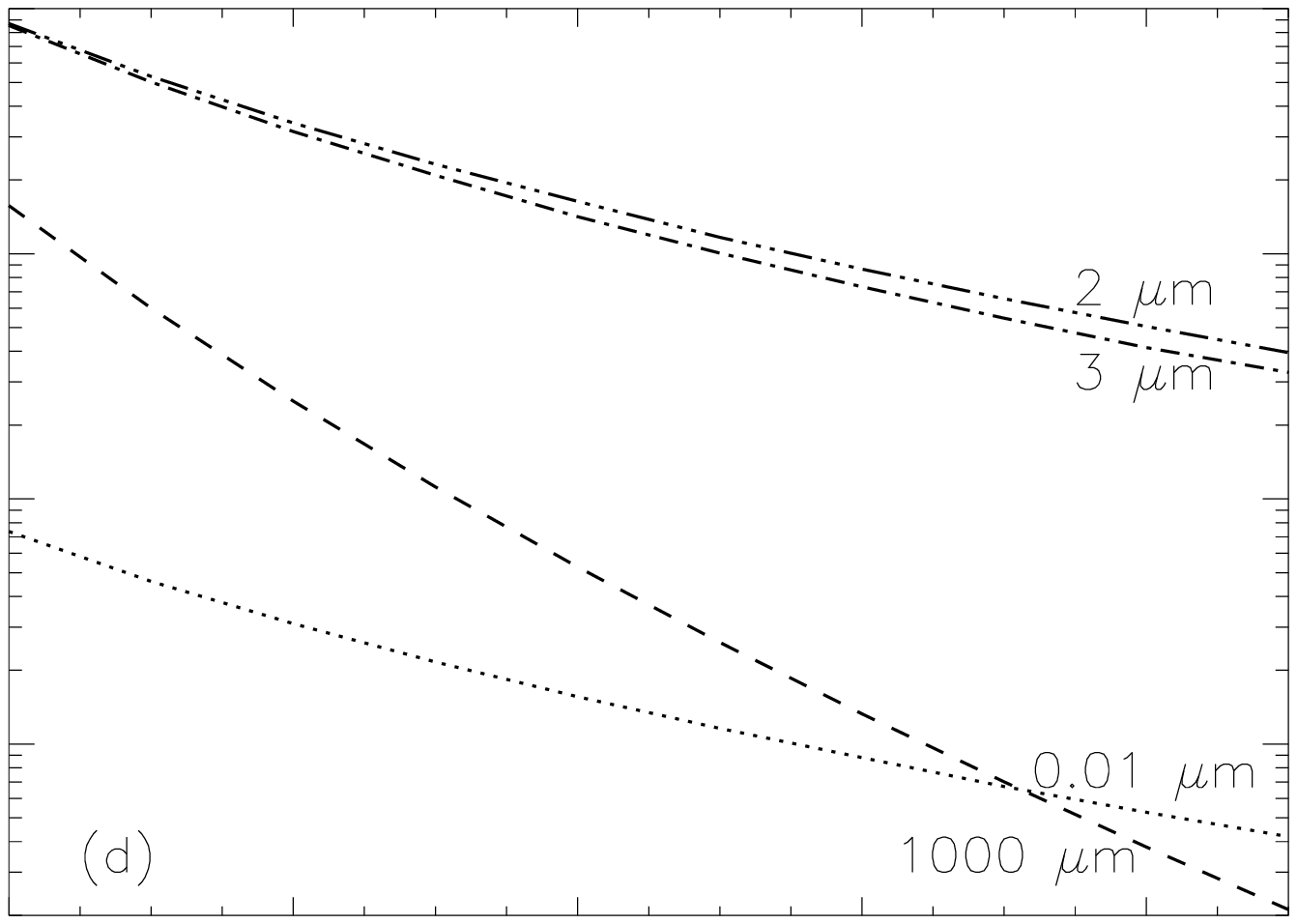,height=1.87in} & \vspace{-0.03in} \\
      & \hspace{-0.15in} \epsfig{file=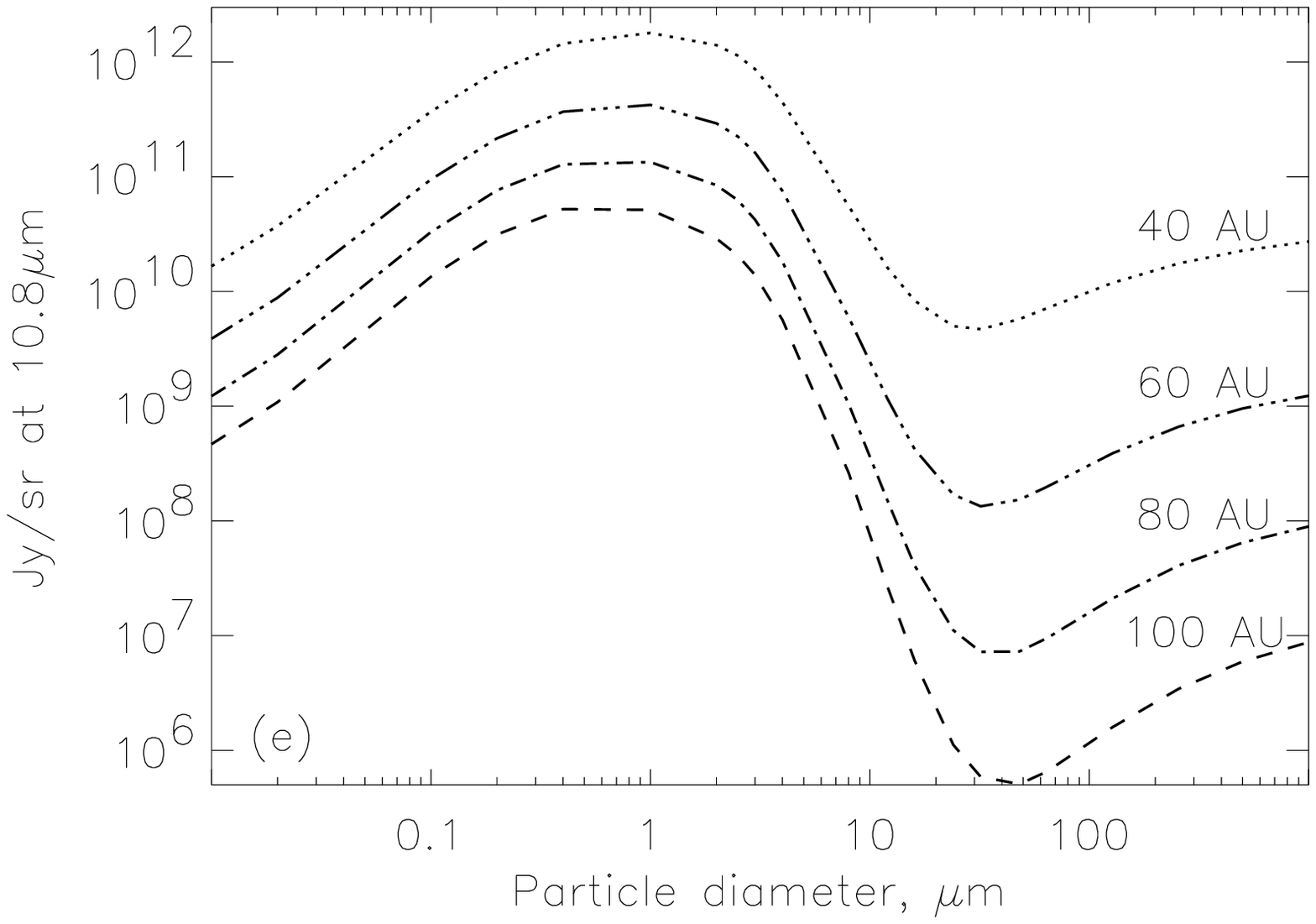,height=2.18in} &
        \hspace{-0.2in}  \epsfig{file=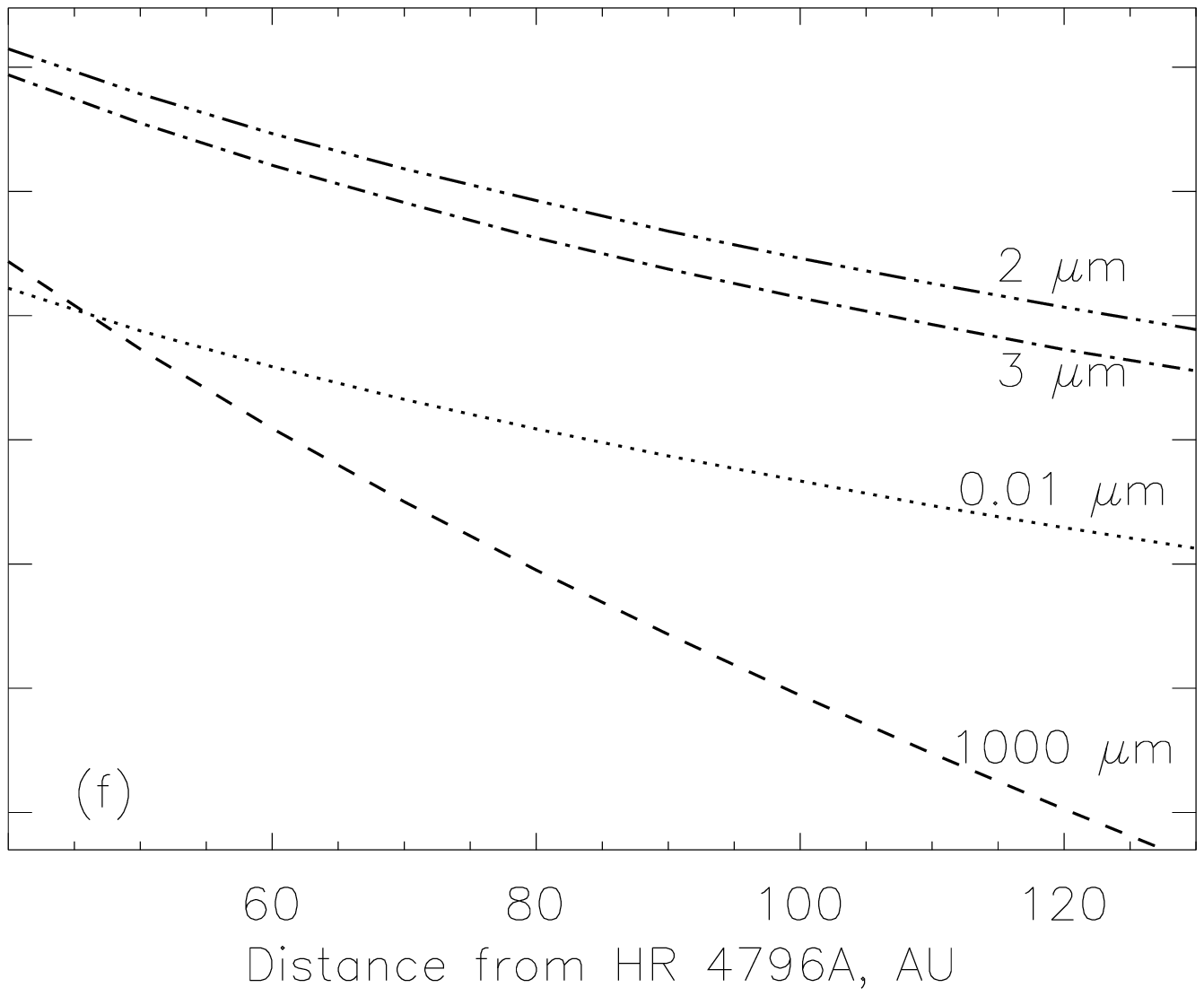,height=2.18in} & 
    \end{tabular}
  \end{center}
  \caption{ The thermal properties of astronomical silicate Mie spheres
  in the HR 4796 disk, plotted for
  particles of different sizes at 40, 60, 80, and 100 AU from
  HR 4797A (\textbf{a}), (\textbf{c}), and (\textbf{e}),
  and for 0.01, 2, 3, and 1000 $\mu$m diameter particles at different
  distances from HR 4796A (\textbf{b}), (\textbf{d}), and (\textbf{f}).
  The temperatures that these particles attain is plotted in (\textbf{a})
  and (\textbf{b}).
  The contribution of a particle's thermal emission to the flux density
  received at the Earth per solid angle that its cross-sectional area
  subtends there, $Q_{abs}(D,\lambda)B_\nu[T(D,r),\lambda]$
  (eq.~[\ref{eq:flux}]), is plotted for observations in the IHW18,
  18.2 $\mu$m, (\textbf{c}) and (\textbf{d}), and N, 10.8 $\mu$m,
  (\textbf{e}) and (\textbf{f}) wavebands.
  The brightnesses of disk models in these two wavebands were
  calculated by taking $P(\lambda,r)$ from the lines on (\textbf{d})
  and (\textbf{f}) corresponding to particles of diameter $D_{typ}$. }
  \label{fig5}
\end{figure}

The properties of particles of different sizes, and at different distances
from HR 4796A, are shown in Fig.~\ref{fig5}.
The temperatures of the particles are plotted in Figs.~\ref{fig5}a
and \ref{fig5}b.
The form of Fig.~\ref{fig5}a can be understood by consideration of
equation (\ref{eq:tdr}), and the wavelengths at which the star and a
particle, if it was a black body, emit most of their energy:
$\lambda_\star \approx 2898/T_\star$ $\mu$m, and
$\lambda_{bb} \approx 2898/T_{bb} =
10\sqrt{r/a_\oplus}(L_\odot/L_\star)^{0.25}$ $\mu$m.
As a crude approximation, a particle with diameter $D$ has
$Q_{abs} \approx 1$ for $\lambda \ll \pi D$ and $Q_{abs} \rightarrow 0$ for
$\lambda \gg \pi D$.
This is in qualitative agreement with Fig.~\ref{figoptprsil}.
Thus, in terms of their thermal properties, disk particles can be divided
into four categories:
the largest particles, $D \gg \lambda_{bb}/\pi$, are efficient absorbers and
emitters at all relevant wavelengths and so achieve nearly black body
temperatures, $T_{bb}$;
particles with $D \ll \lambda_{bb}/\pi$, are inefficient emitters at
their black body temperature, and so need temperatures higher
than $T_{bb}$ to re-radiate all of the incident energy;
the smallest particles, $D \ll \lambda_\star/\pi$, are also
inefficient absorbers at the stellar temperature, and so do not need as high
temperatures as slightly larger particles to re-radiate the absorbed energy;
and particles with $D \approx 20$ $\mu$m have temperatures below that of
a black body --- this is because these particles are super-efficient
emitters at their black body temperatures (due to silicate resonances,
$Q_{abs}$ can go up as high as 2), and so need lower temperatures to
re-radiate the incident energy.
The form of Fig.~\ref{fig5}b can be understood in the same way.
The fall-off of a large (e.g., $D = 1000$ $\mu$m) particle's temperature with
distance from HR 4796A is like that of a black body, i.e.,
$T \propto 1/\sqrt{r}$.
The fall-off for smaller particles, however, is not that steep because
these particles emit less efficiently the further they are from the star
(due to their lower temperatures, and consequently higher $\lambda_{bb}$);
e.g., the fall-off for $D = 2.5$ $\mu$m particles is $T \propto 1/r^{0.34}$,
which is close to the $1/r^{1/3}$ fall-off expected for particles with an
emission efficiency that decreases $\propto 1/\lambda^2$ (e.g., \cite{bp93}).

More important observationally is the variation of
\mbox{$Q_{abs}(\lambda,D)B_\nu[\lambda,T(D,r)]$}, since this determines
the contribution of a particle's thermal emission to the flux density
received at the Earth (eqs.~[\ref{eq:flux}],[\ref{eq:fnu}] and [\ref{eq:p}]).
This is plotted for $\lambda = 18.2$ $\mu$m in Figs.~\ref{fig5}c and
\ref{fig5}d, and for $\lambda = 10.8$ $\mu$m in Figs.~\ref{fig5}e
and \ref{fig5}f.
The form of Figs.~\ref{fig5}c and \ref{fig5}e can be explained in the
same way that Fig.~\ref{fig5}a was explained:
all three figures have similar forms, which is to be expected since
the particles' temperature also appears in $B_\nu$;
Figs.~\ref{fig5}c and \ref{fig5}e are, however, attenuated for
$D \ll \lambda/\pi$, since these particles are inefficient emitters at that
wavelength.
The fall-off with distance of the different particles shown in
Figs.~\ref{fig5}d and \ref{fig5}f is due solely to their different temperature
fall-offs (Fig.~\ref{fig5}b);
e.g., for $\lambda = 18.2$ $\mu$m, the approximate fall-off for
$D = 1000$ $\mu$m particles is $\propto 1/r^{5.4}$, while that for
$D = 2.5$ $\mu$m particles is $\propto 1/r^{2.6}$.

%%%%%%%%%%%%%%%%%%%%%%%%%%%%%%%%%%%%%%%%%%%%%
\subsection{Cross-sectional Area Distribution}
\label{sssec-dtyp}
The definition of $P(\lambda,r)$ (eqs.~[\ref{eq:p}] and [\ref{eq:p2}])
shows that it is the convolution of $Q_{abs}B_\nu$
(Fig.~\ref{fig5}c - \ref{fig5}f) with the cross-sectional area distribution,
$\bar{\sigma}(D,r)$.
Since theoretical arguments cannot supply an accurate size distribution
(\S \ref{ss-catsigs}), this model uses the assumption of
equation (\ref{eq:pdtyp}),
which is that $P(\lambda,r)$ is equal to the $Q_{abs}B_\nu$ of particles
in the disk with characteristic size, $D_{typ}$.
The model further assumes that this $D_{typ}$ is constant across the disk
in the IHW18 waveband;
i.e., the brightness of a disk model observation in the IHW18 waveband
is calculated using $P(\lambda,r)$ from the line on Fig.~\ref{fig5}c
corresponding to $D_{typ}$, where $D_{typ}$ is a model variable.
A qualitative understanding of $D_{typ}$ comes from Fig.~\ref{fig5}c:
since $Q_{abs}B_\nu$ is fairly flat for particles larger than $\sim 8$
$\mu$m, an IHW18 waveband observation is dominated by those
particles with the most cross-sectional area, unless there is a
significant amount of particles smaller than 8 $\mu$m;
Fig.~\ref{fig5}c also shows that the observation is unlikely to be dominated
by particles smaller than $O$(0.01 $\mu$m), unless their contribution to
the disk's cross-sectional area is much higher than that of
larger particles.

In addition to the IHW18 waveband, T2000 observed HR 4796 in the
N ($\lambda = 10.2$ $\mu$m) waveband.
The N band observations also show the double-lobed feature,
but the inaccuracy of the subtraction of the image of
HR 4796A from the observations means that they cannot be used to
constrain the disk's structure.
The N band observations can, however, be used to constrain the
disk's brightness in this waveband.
The brightness of a disk in different wavebands differs only in
the factor $P(\lambda,r)$ (eq.~[\ref{eq:fnu}]), and so the N band
observation can be used to obtain information about the disk's
size distribution.
Since Figs.~\ref{fig5}c and \ref{fig5}e have similar forms, the two
observations should be dominated by the emission of similarly sized
particles;
i.e., they should have the same $D_{typ}$, and show similar
structures.
The same characteristic particle size, $D_{typ}$, was used to
calculate the brightness of a disk model in both the IHW18 and N
wavebands, using $P(\lambda,r)$ from the appropriate
lines on the plots of Figs.~\ref{fig5}d and \ref{fig5}f.
There is just one $D_{typ}$ that simultaneously matches the disk's
observed brightnesses in both wavebands.

%%%% Figure 6a
\begin{figure}
  \begin{center}
    \begin{tabular}{c}
      \epsfig{file=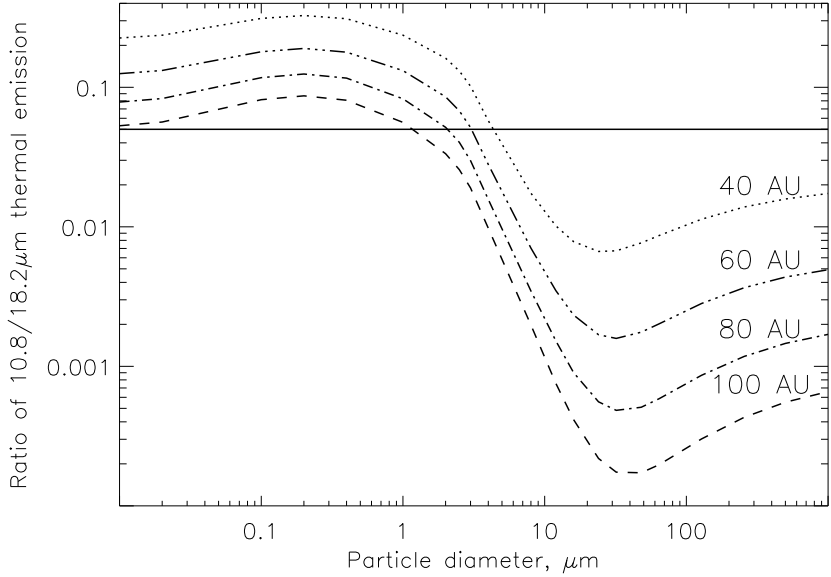,height=3in}
    \end{tabular}
  \end{center}
  \caption{The ratio of the thermal emission in the N, 10.8
  $\mu$m, and IHW18, 18.2 $\mu$m, wavebands, of astronomical
  silicate Mie spheres of different sizes at 40, 60, 80, and 100 AU
  from HR 4796A (i.e., Fig.~\ref{fig5}e divided by Fig.~\ref{fig5}c).
  Assuming the disk's flux densities in the two wavebands to be
  dominated by the emission of particles at 60-80 AU, the observed
  ratio of flux densities, $O(0.05)$ (T2000; \S \ref{sssec-dtyp}), can
  be used to estimate that the disk's emitting particles have
  $D_{typ} = 2-3$ $\mu$m.}
  \label{fig6a}
\end{figure}

An initial estimate for $D_{typ}$ was made based on the flux densities
observed in the two wavebands:
the flux densities of the disk in the IHW18 and N wavebands are
857 and 40 mJy, respectively (T2000);
i.e., the observed flux density ratio (N/IHW18) is $O(0.05)$.
The expected flux density ratio of the two wavebands (the ratio of
$P(\lambda,r)$ for different $D_{typ}$) is plotted in Fig.~\ref{fig6a}.
Assuming the emission to arise mostly from particles near the inner
edge of the disk, $r =$ 60-80 AU, Fig.~\ref{fig6a} shows that the
observed emission can be fairly well-constrained to come from
particles with $D_{typ} = 2-3$ $\mu$m.

%%%%%%%%%%%%%%%%%%%%%%%%%%%%%%%%%%%%%%%%%%%%%%
\subsection{Pericenter Glow}
\label{sssec-pg}

%%%% Figure 7a
\begin{figure}
  \begin{center}
    \begin{tabular}{c}
      \epsfig{file=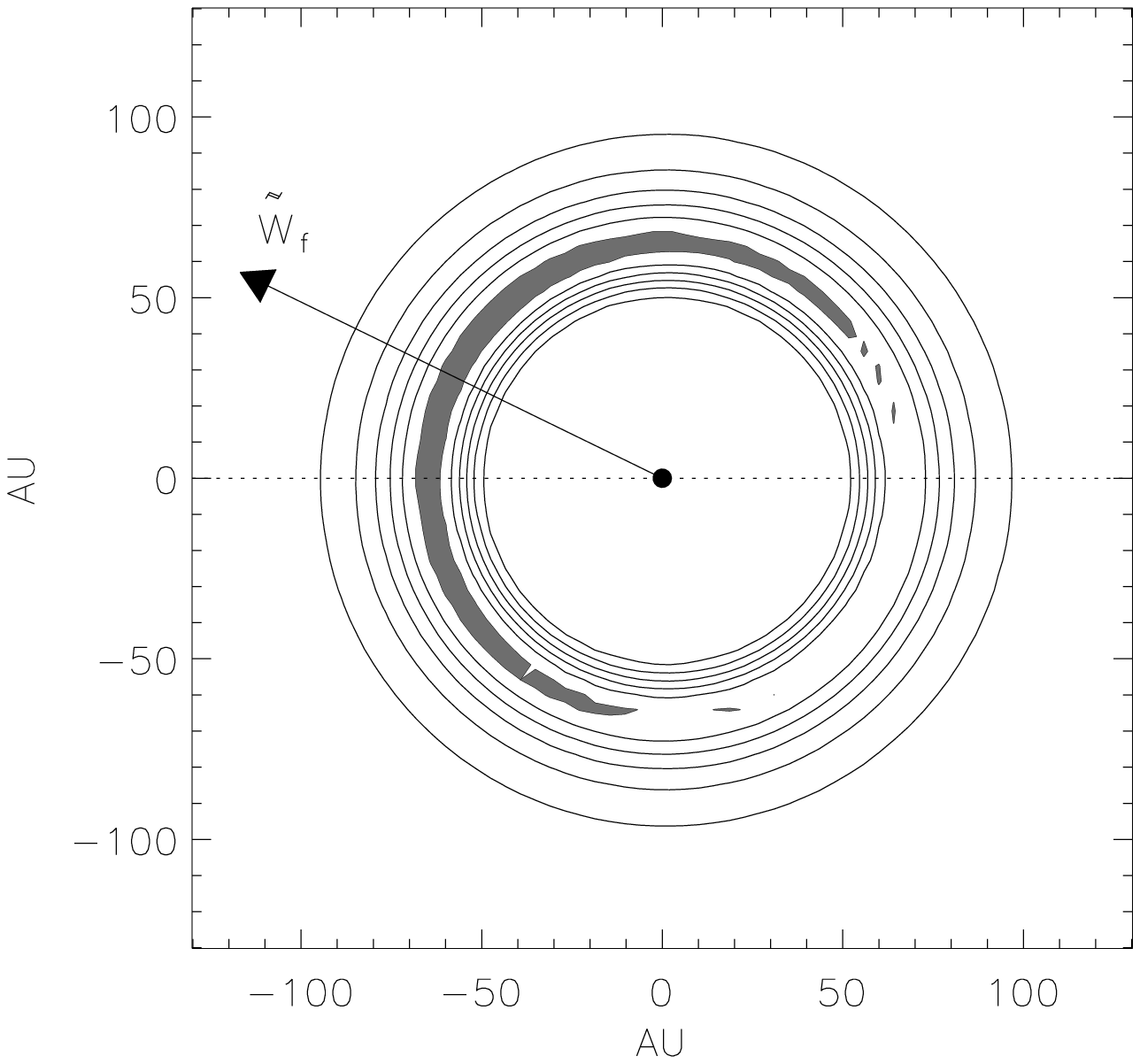,height=3.5in}
    \end{tabular}
  \end{center}
  \caption{Contour plot of an unsmoothed face-on view of the
  HR 4796 disk model (shown also in Fig.~\ref{fig4}) seen in the IHW18
  (18.2 $\mu$m) waveband.
  The contours are spaced linearly at 0.17, 0.34, 0.51, 0.68, 0.85, and
  1.02 mJy/pixel.
  The disk's offset causes particles in the forced pericenter direction,  
  located at a position angle of $90^\circ - \tilde{\omega}_f$, where
  $\tilde{\omega}_f = 26^\circ$, as measured from North in a
  counterclockwise direction, to be hotter, and hence brighter, than
  those in the forced apocenter direction;
  this is the ``pericenter glow'' phenomenon, evident in this figure
  by the shape of the brightest (filled-in) contour.
  The geometry of the observation is defined such that the
  disk as it is shown here is rotated by a further $90^\circ - I_{obs}$
  about the dotted line, where $I_{obs} = 13^\circ$.}
  \label{fig7a}
\end{figure}

Fig.~\ref{fig7a} shows a contour plot of an unsmoothed IHW18 waveband
observation of the disk model of Fig.~\ref{fig4} viewed face-on
(i.e., perpendicular to the disk's plane of symmetry, $y_f$).
This shows the observational consequence of the offset center of symmetry
of the disk model.
Because the particles at the inner edge of the disk (those that contribute most
to the disk's brightness) are closer to the star in the forced pericenter
direction, $\tilde{\omega}_f$, than those in the forced apocenter direction
(Fig.~\ref{fig4}), they are hotter and so contribute more to the disk's thermal
emission than those at the forced apocenter.
This is the ``pericenter glow'' phenomenon, which leads to the
horseshoe-shaped highest contour line (the filled-in 1.02 mJy/pixel line),
which is pointed in the $\tilde{\omega}_f$ direction.
This asymmetry is a consequence of $a_{min}$ and $e_f$ only,
and its magnitude is determined by $e_f$ only.
In particular, if there is a gradient of $e_f$ across the disk, then it is
$e_f$ at the inner edge of the disk that controls the magnitude of the
asymmetry.
The outermost contour plotted on Fig.~\ref{fig7a}, which is an
offset circle with a radius of 95 AU, is that corresponding to 0.17 mJy/pixel.
Thus, there is little emission from the outer edge of the disk,
justifying the arbitrary use of $a_{max} = 130$ AU in the modeling.

%%%%%%%%%%%%%%%%%%%%%%%%%%%%%%%%%%%%%%%%%%%%%%%%
\section{Disk Model Orientation}
\label{ssec-do}
The two variables that define the orientation of the HR 4796 disk to the
line of sight of the observation are $\tilde{\omega}_f$ and $I_{obs}$;
how they define this orientation is best explained using Fig.~\ref{fig7a}.
Imagine that the disk starts face-on with the forced pericenter
direction pointing to the left.
It is then rotated clockwise by $\tilde{\omega}_f$ (this is shown in
Fig.~\ref{fig7a}, where $\tilde{\omega}_f = 26^{\circ}$),
and then tilted by $90^{\circ} - I_{obs}$ about the dotted line on
Fig.~\ref{fig7a}.
The direction of this tilt, whether the top or bottom of the disk ends
up closer to the observer, is not constrained in the modeling,
since no account was made for either the extinction of the disk's
emission by the disk itself, or for the disk's scattered light (e.g., an
observer would see forward-scattered starlight from the closest
part of the disk and back-scattered starlight from the farthest part,
a phenomenon that could produce an apparent asymmetry in a
symmetric disk, \cite{kj95}).
If the resulting inclination of the disk's symmetry plane to the
line of sight, $I_{obs}$, is small, then the resulting nearly edge-on
observation shows two lobes, one either side of the star.
Since the hotter, brighter, pericenter glow material is predominantly
in one of the lobes (unless $\tilde{\omega}_f = 90^\circ$),
the lobes have asymmetric brightnesses.

%%%%%%%%%%%%%%%%%%%%%%%%%%%%%%%%%%%%%%%%%%%%%
\section{Modeling Process and Results}
\label{ssec-mpr}

%%%% Figure 8
\begin{figure}
  \begin{center}
    \begin{tabular}{rlll}
      (\textbf{a}) & \hspace{-0.5in} \epsfig{file=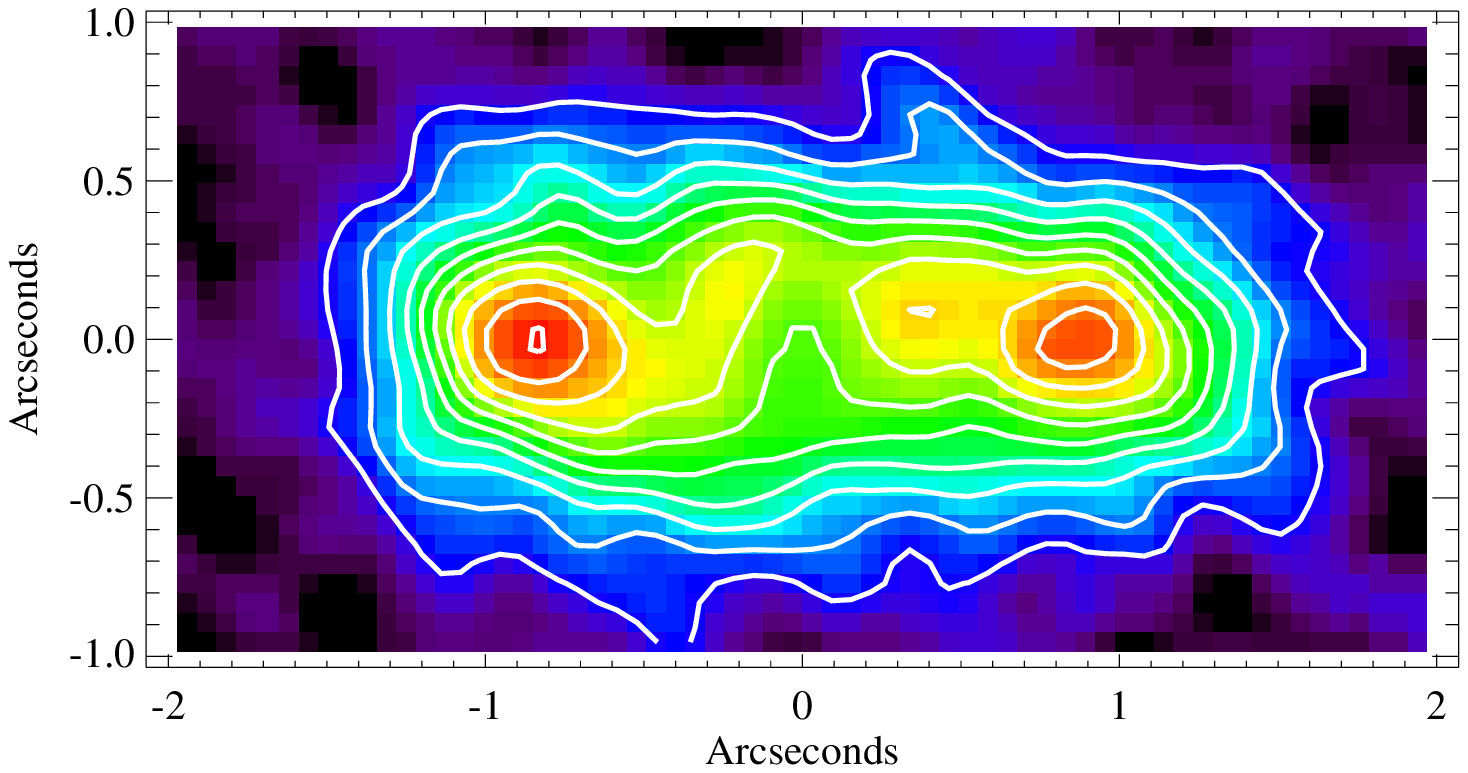,height=1.55in} &
      \hspace{-0.22in} \epsfig{file=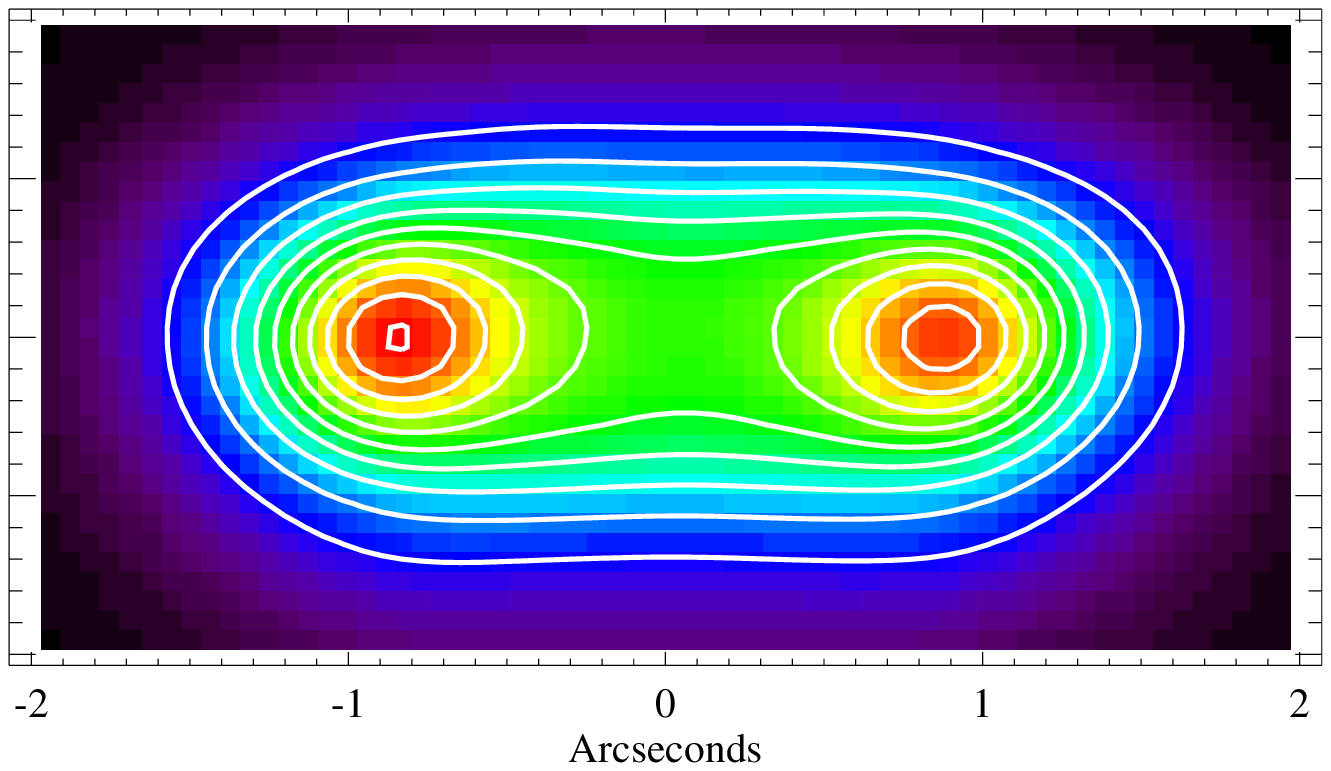,height=1.55in} &
      \hspace{-0.2in} (\textbf{b}) \vspace{0.1in}
    \end{tabular}
    \begin{tabular}{ccc}
      \epsfig{file=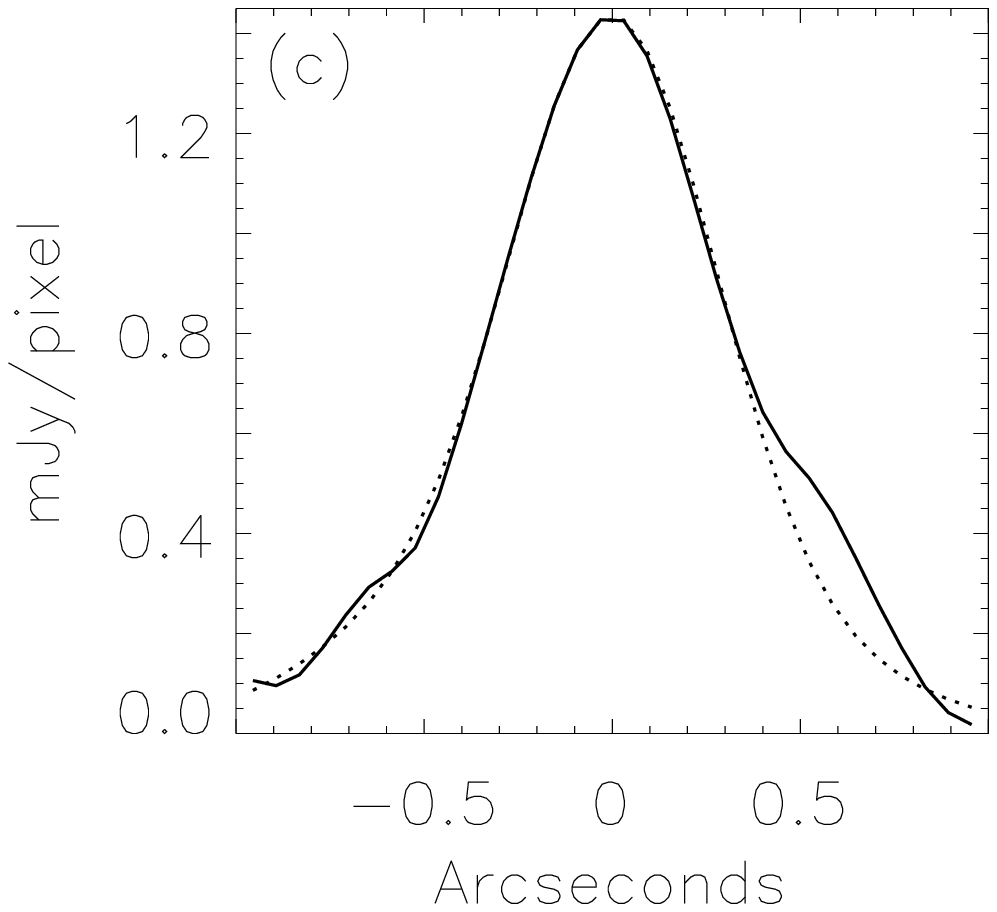,height=1.6in} &
      \hspace{-0.22in} \epsfig{file=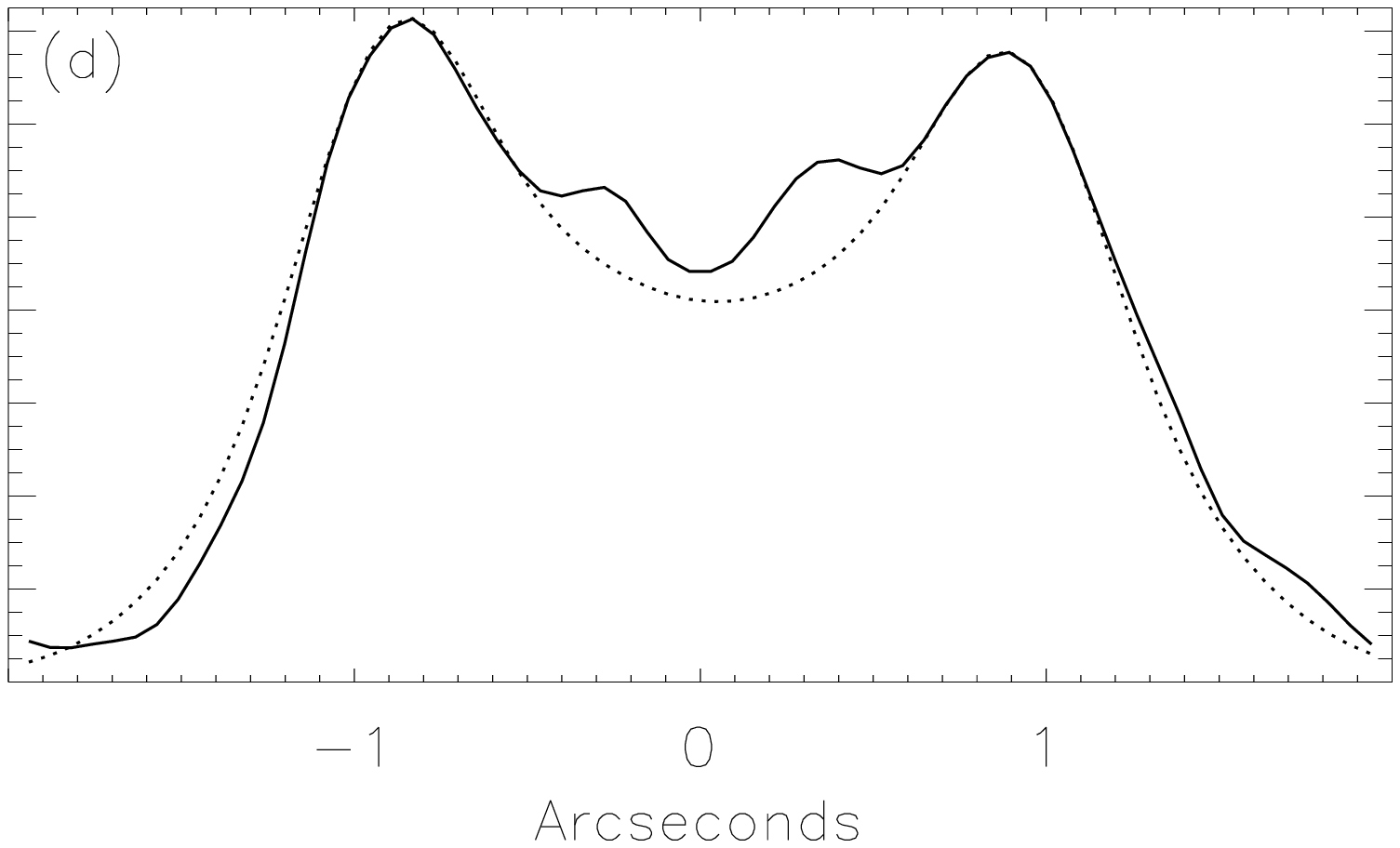,height=1.6in} &
      \hspace{-0.22in} \epsfig{file=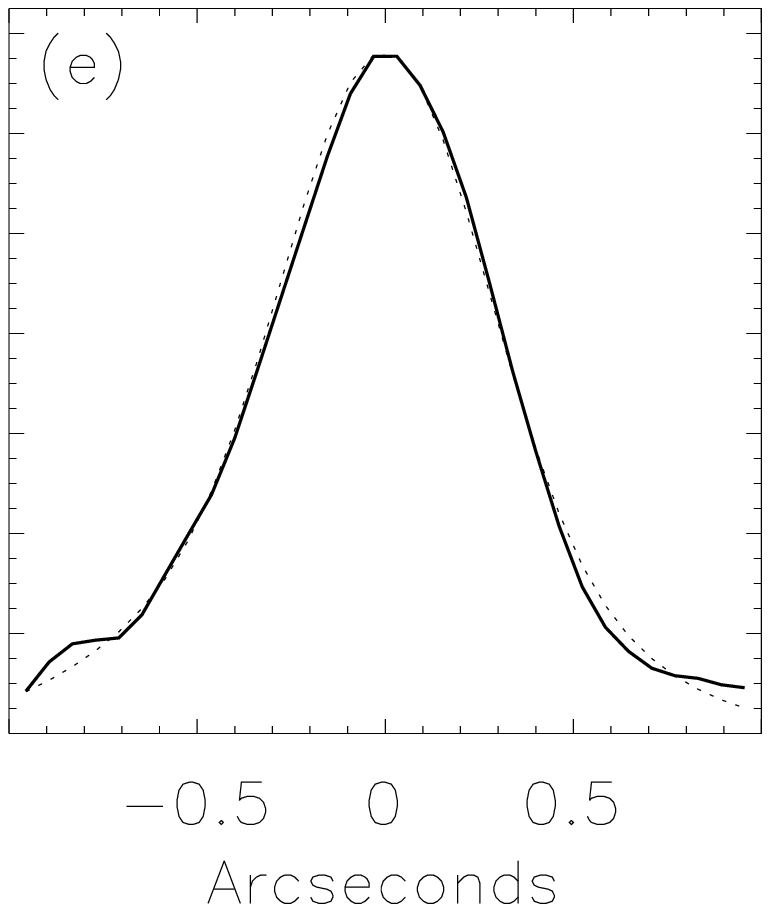,height=1.6in}
    \end{tabular}
  \end{center}
  \newpage
  \caption{Comparison of the HR 4796 disk model observation with the
  real observation.
  \textit{Top} --- False color images of HR 4796 in the IHW18
  (18.2 $\mu$m) waveband.
  Both the observation (\textbf{a}), on the left, and the model (\textbf{b}),
  on the right, have been rotated to horizontal with the NE lobe on the left.
  The contours are spaced linearly at 0.22, 0.35, 0.49, 0.62, 0.75, 0.89,
  1.02, 1.15, 1.29, and 1.42 mJy/pixel.
  The observation has had the photospheric emission of HR 4796A
  subtracted, and a 3 pixel FWHM gaussian smoothing applied.
  The image of the model mimics the observation both in pixel
  size (1 pixel $= 0\farcs0616 = 4.133$ AU) and
  smoothing (using an observed PSF and including the 3 pixel
  post-observational smoothing).
  \textit{Bottom} --- Line-cuts in the vertical direction through the NE
  (\textbf{c}) and SW (\textbf{e}) lobes, and in the horizontal
  direction through the center of both lobes (\textbf{d}).
  The observations are shown with a solid line and the
  model with a dotted line. }
  \label{fig8}
\end{figure}

Pseudo-observations of disk models in the IHW18 and N wavebands were
produced that mimicked the real OSCIR (the University of Florida mid-IR
imager) observations in both pixel size,
\hbox{1 pixel $= 0\farcs0616 = 4.133$ AU} at 67 pc,
and smoothing, using the observed PSFs (which are asymmetric and slightly
fatter than diffraction limited, T2000), and including the post-observational
gaussian smoothing of FWHM = 3 pixels.
The model variables: $a_{min}$, $\gamma$, $I_{obs}$, $e_f$,
$\tilde{\omega}_f$, and $\sigma_{tot}$, were optimized so that the
modeled IHW18 observation correctly predicts the observed IHW18 brightness
distribution;
at the same time, the variable, $D_{typ}$, was optimized so that the modeled
N band observation correctly predicts the observed N band brightness.

The model observations were compared with the real observations using the
following diagnostics:
the lobe brightnesses, $F_{ne,sw}$, and their projected radial offsets from
HR 4796A, $R_{ne,sw}$, that were found by fitting a quintic polynomial surface
to a $10 \times 10$ pixel region around each of the lobes with 0.1 pixel
resolution (the location of HR 4796A in the real observations was also found
in this way);
and line-cuts through the disk both parallel (e.g., Figs.~\ref{fig8}d and
\ref{fig9}), and perpendicular (e.g., Figs.~\ref{fig8}c and \ref{fig8}e)
to the line joining the two lobes in the IHW18 observations.
An understanding of how the different model variables affect the
different diagnostics allowed the modeling process to be decoupled
into solving for:
the disk's symmetrical structure, defined by $I_{obs}$, $a_{min}$, and
$\gamma$;
the particle size $D_{typ}$;
and the disk's asymmetrical structure, defined by $e_f$ and
$\tilde{\omega}_f$.
Throughout the modeling, the amount of material in a model,
$\sigma_{tot}$, was scaled so that the model observation predicted
the correct observed mean brightness of the lobes in the IHW18
waveband, $F_{mean} = (F_{ne}+F_{sw})/2 = 1.40 \pm 0.02$ mJy/pixel;
its final value, $\sigma_{tot} = 2.03\times 10^{24}$ m$^2$, was calculated
once the other variables had been constrained.

%%%%%%%%%%%%%%%%%%%%%%%%%%%%%%%%%%%%%%%%%%%%%%%
\subsection{Symmetrical Disk Structure}
\label{sssec-dss}
Since $e_f$ and $\tilde{\omega}_f$ only pertain to the disk's asymmetrical
structure, then for a given $D_{typ}$, the variables pertaining to the
disk's symmetrical structure, $I_{obs}$, $a_{min}$, and $\gamma$, could
be solved using a model with $e_f = 0$;
such a model is axisymmetric, and so the variable $\tilde{\omega}_f$ is
redundant.
The inclination of the disk's plane of symmetry to the line of sight,
$I_{obs} = 13 \pm 1^{\circ}$, was constrained to give the best fit to the
line-cuts perpendicular to the lobes (Figs.~\ref{fig8}c and \ref{fig8}e);
this inclination agrees with that found by previous models of HR 4796
disk observations (\cite{krwb98}; \cite{ssbk99}).
Since the model is of a fat disk, a different proper inclination
distribution would result in a different $I_{obs}$;
e.g., if the disk is actually thinner than modeled here,
$\langle I_p \rangle < 10.2^\circ$, then the inferred $I_{obs}$ is an
underestimate, and vice versa.
The inner edge of the disk, \hbox{$a_{min} = 62 \pm 2$ AU},
was constrained such that the model observation reproduces the
observed mean radial offset of the lobes from HR 4796A,
\hbox{$R_{mean} = (R_{ne}+R_{sw})/2 = 58.1 \pm 1.3$ AU}.
The semimajor axis distribution, $\gamma = -2 \pm 1$, was constrained
to give the best fit to the cut along the line joining the two lobes
(Fig.~\ref{fig8}d).

%%%%%%%%%%%%%%%%%%%%%%%%%%%%%%%%%%%%%%%%%%%%%%%%
\subsection{Particle Size}
\label{sssec-tps}

%%%% Figure 9
\begin{figure}
  \begin{center}
    \epsfig{file=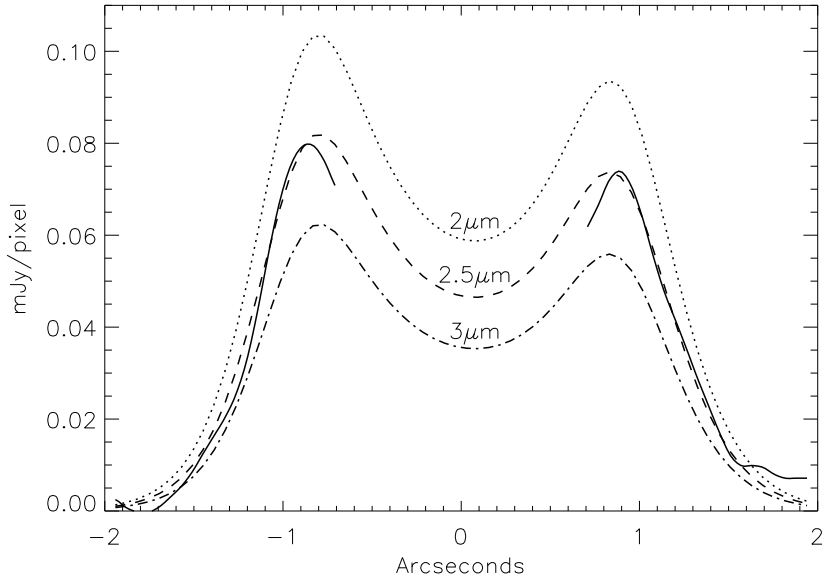,height=3in}
  \end{center}
  \caption{Horizontal line-cuts along the plane of the lobes in the
  N (10.8 $\mu$m) band.
  The observation is shown with a solid line and models with particle
  diameters of $D_{typ}$ = 2, 2.5, and 3 $\mu$m are shown with dotted,
  dashed, and dash-dot lines.
  The total amount of cross-sectional area in the models, $\sigma_{tot}$,
  has been scaled to fit the observed mean brightness of the lobes in the
  IHW18 (18.2 $\mu$m) waveband;
  the model with $D_{typ} = 2.5$ $\mu$m gives the best fit to the observed
  lobe brightnesses in the N band.
  The observed N band flux density is not well constrained within
  0\farcs8 of HR 4796A due to imperfect subtraction of the stellar
  photosphere from the image (T2000), and so it is not shown here. }
  \label{fig9}
\end{figure}

Since $D_{typ}$ was already estimated to be about 2-3 $\mu$m
(\S \ref{sssec-dtyp}), the modeling of the disk's symmetrical structure
was repeated for $D_{typ} =$ 2, 2.5, and 3 $\mu$m.
Adjusting $D_{typ}$ by such a small amount did not affect the inferred
symmetrical structure parameters.
This was expected, since the $P$(18.2 $\mu$m$,r)$ for each of
these $D_{typ}$ are very similar (Fig.~\ref{fig5}d).
Remembering that the model is always normalized to predict the
observed mean IHW18 lobe brightnesses, the predicted N band lobe
brightnesses are compared for the three values of $D_{typ}$ in
Fig.~\ref{fig9}.
This shows that the particle size can be constrained to be
$D_{typ} = 2.5 \pm 0.5$ $\mu$m;
i.e., the crude method of calculating the particle size of
\S \ref{sssec-dtyp} gives a very good estimate of this size.
This particle size means that the total mass of emitting particles in
the disk model is $\sim 1.4 \times 10^{-3} M_\oplus$, where
$M_\oplus = 3 \times 10^{-6}M_\odot$ is the mass of the Earth.
However, this is not a useful constraint on the disk's mass, since
the disk's mass is expected to be concentrated in its largest particles. 
The best estimate of the mass of the HR 4796 disk, from
submillimeter observations, is that it is between $0.1M_\oplus$ and
$1.0M_\oplus$ (\cite{jgwm95}; \cite{jmwt98}),
which, as expected, is well above the mass of this disk model.

%%%%%%%%%%%%%%%%%%%%%%%%%%%%%%%%%%%%%%%%%%%
\subsection{Asymmetrical Disk Structure}
\label{sssec-das}

%%%% Figure 7b
\begin{figure}
  \begin{center}
    \begin{tabular}{c}
      \epsfig{file=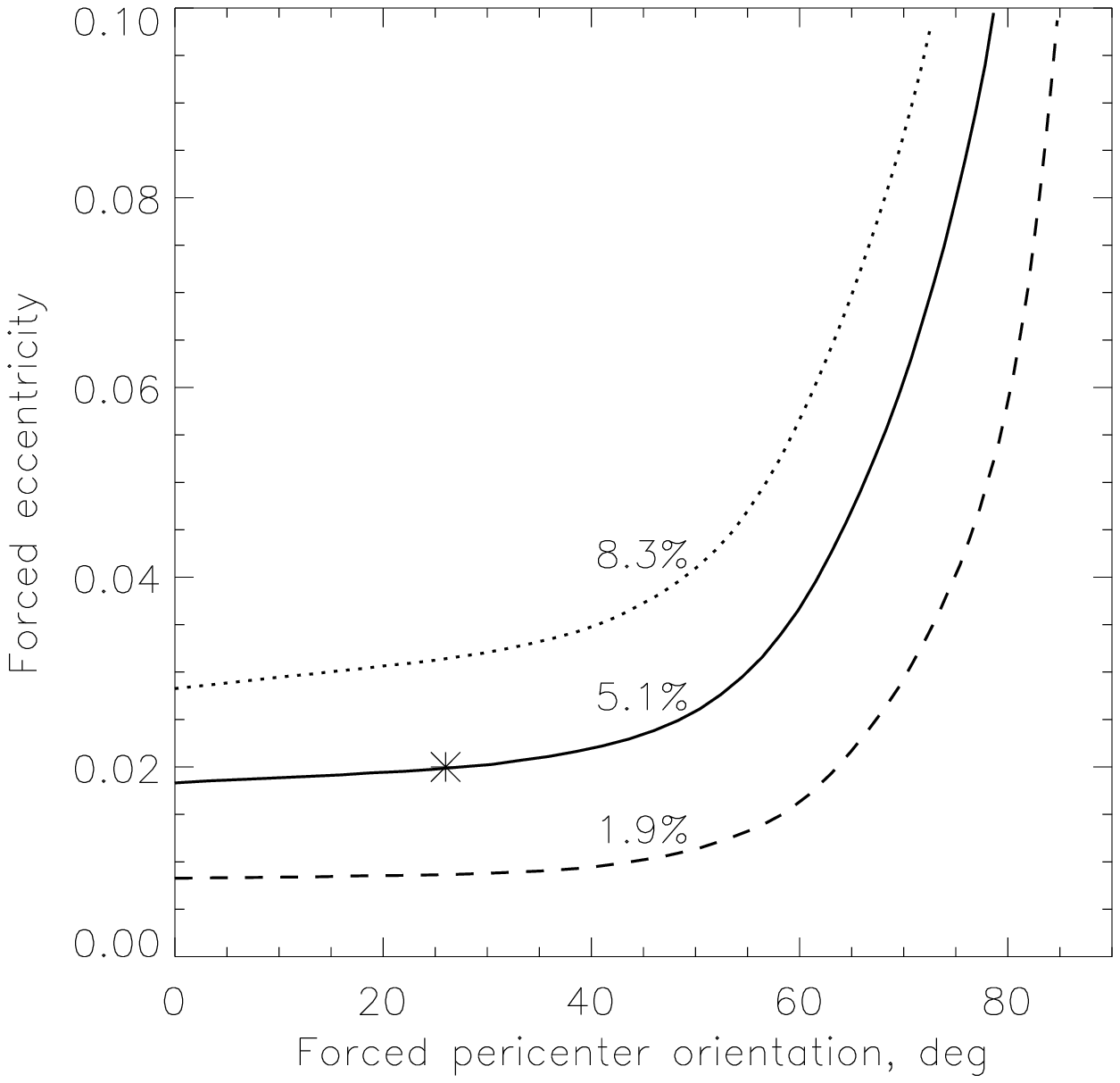,height=3.5in}
    \end{tabular}
  \end{center}
  \caption{Relation of the forced eccentricity, $e_f$, to the orientation
  of the longitude of forced pericenter, $\tilde{\omega}_f$, in the model
  needed to achieve the observed lobe brightness asymmetry of $5.1 \pm 3.2\%$
  (a similar relationship is required to achieve the observed radial
  offset).
  A forced eccentricity as small as 0.02 would suffice to achieve
  the observed asymmetry, but a higher forced eccentricity could be
  necessary if the forced pericenter is aligned in an unfavorable
  direction.
  The final model, shown in Figs.~\ref{fig4}, \ref{fig7a}, \ref{fig8},
  and \ref{fig9}, has $e_f = 0.02$ and $\tilde{\omega}_f = 26^\circ$,
  and this point is shown with an asterisk on this plot.}
  \label{fig7b}
\end{figure}

The disk's observed asymmetries are defined by
the lobe brightness asymmetry,
$(F_{ne} - F_{sw})/F_{mean} = 5.1 \pm 3.2\%$,
and the radial offset asymmetry,
$(R_{sw} - R_{ne})/R_{mean} = 6.4 \pm 4.6\%$.
These asymmetries are also apparent in the disk model, and their
magnitudes are determined by both $e_f$ and $\tilde{\omega}_f$.
The lobe brightness asymmetry was used to constrain $e_f$ and
$\tilde{\omega}_f$, and it was found that the forced eccentricity
needed to cause the $5.1 \pm 3.2\%$ asymmetry depends on the geometry
of the observation according to the relation shown in Fig.~\ref{fig7b}.
Thus, for the majority of the geometries, a forced eccentricity of between
0.02 and 0.03 is sufficient to cause the observed brightness asymmetry.
In the context of this modeling, the observed brightness asymmetry
implies a radial offset asymmetry of $\sim 5\%$, which is within
the limits of the observation.
The final model shown in Figs.~\ref{fig4}, \ref{fig7a}, \ref{fig8}, and
\ref{fig9} assumes a modest value of $e_f = 0.02$, which corresponds to a
disk orientation described by $\tilde{\omega}_f = 26^\circ$ (Fig.~\ref{fig7b}).
Whether this rotation puts the pericenter glow material above or below
the horizontal, $\tilde{\omega}_f = 26^\circ$ or $-26^\circ$, is not
constrained here, since it has a minimal effect on the observation:
in the model observation of Fig.~\ref{fig8}, the top of the disk is
brighter than the bottom of the disk by a fraction that would be
undetectable in the observation due to noise and the disk's unknown
residual structure.
The line-cuts of Fig.~\ref{fig8} show how well the model fits all aspects
of the observation --- the vertical structure, the horizontal structure,
and the lobe location and asymmetry.

%%%%%%%%%%%%%%%%%%%%%%%%%%%%%%%%%%%%%%%%%%%%%
\subsection{Statistical Significance}
\label{sssec-ss}
The standard deviations of the OSCIR lobe observations quoted in this
chapter were found using model observations that mimicked the noise
present in the OSCIR observations.
The background sky noise in the IHW18 observation was found to be
approximately gaussian with zero mean and a $1\sigma$ noise per pixel
of $0.15$ mJy (T2000).
In a noisy model observation, such a random noise field was included
after the PSF smoothing, but before the post-observational smoothing.
Observations of noisy models were repeated for 50,000 different noise
fields to obtain the quoted standard deviations.
Since the observed PSF was asymmetric (T2000), this introduces
an apparent lobe asymmetry of $-0.8\%$ in an observation of a
symmetric disk (i.e., one with $e_f=0$), and so the observed lobe
asymmetry is $5.9 \pm 3.2\%$ from the mean, and its statistical
significance is $1.8\sigma$.
While this is small, it does show that the pericenter glow phenomenon is
observable with current technology:
HR 4796 was observed with the infrared imager OSCIR for only one hour
on Keck II (T2000);
in the background-limited regime, the significance level of any asymmetry
increases at a rate $\propto \sqrt{t}$;
thus, one good night on a 10 meter telescope should be enough to get a
definitive observation of the HR 4796 lobe asymmetry.
The real significance of the HR 4796 asymmetry may also be higher
than quoted above, since it also seems to be apparent in other
observations of this disk (\cite{krwb98}; \cite{ssbk99}).
Even if subsequent observations happen to disprove the existence of
the asymmetry, this is still significant, since, as \S \ref{ssec-ias}
shows, an asymmetry is to be expected if the companion
star, HR 4796B, is on an eccentric orbit.

%%%%%%%%%%%%%%%%%%%%%%%%%%%%%%%%%%%%%%%%%%%%%
\chapter{INTERPRETATION OF THE DYNAMIC HR 4796 DISK MODEL}
%%%%%%%%%%%%%%%%%%%%%%%%%%%%%%%%%%%%%%%%%%%%%
\label{c-interpretation}
The interpretation of the HR 4796 disk model is broken down into
sections that cover discussions of:
the dynamics of the disk particles, \S \ref{ssec-ddch};
the lobe asymmetry, \S \ref{ssec-ias};
the emitting particle category, \S \ref{ssec-deps};
the origin of the inner hole, \S \ref{ssec-lsd};
and the residual structure of the observation once the model has been
subtracted, \S \ref{ssec-rs}.

%%%%%%%%%%%%%%%%%%%%%%%%%%%%%%%%%%%%%%%%%%%%%%%%%%%
\section{The Dynamic HR 4796 Disk}
\label{ssec-ddch}
An interpretation of the observed structure of the HR 4796 disk starts
with a discussion of the dynamics of the particles in the disk,
and where the emitting particles fit into our understanding of the
dynamic disk (Chapter \ref{c-structure}).
In all calculations, the mass of HR 4796A was assumed to be
$M_\star = 2.5M_\odot$ (\cite{jfht98}).

%%%%%%%%%%%%%%%%%%%%%%%%%%%%%%%%%%%%%%%%%%%%%%%%%
\subsection{Radiation Forces, $\beta$}
\label{sssec-rfh}

%%%% Figure 6b
\begin{figure}
  \begin{center}
    \begin{tabular}{c}
      \epsfig{file=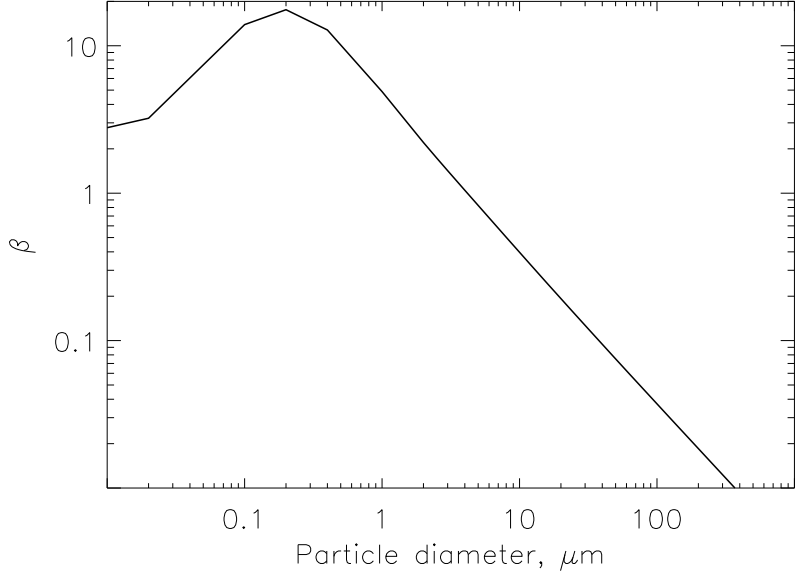,height=3in}
    \end{tabular}
  \end{center}
  \caption{The ratio, $\beta$, of the radiation pressure force to
  the gravitational force acting on astronomical silicate Mie spheres
  of different sizes in the HR 4796 disk.}
  \label{fig6b}
\end{figure}

The radiation forces, defined by $\beta$ (eq.~[\ref{eq:beta}]), acting on
particles in the HR 4796 disk can be found from their optical properties.
Fig.~\ref{fig6b} shows the $\beta$ of particles with the optical properties
assumed in this model (\S \ref{sssec-pop}).
Thus, particles in the disk with $D<8$ $\mu$m are $\beta$ meteoroids
(even the submicron particles have $\beta>0.5$), and a good approximation for
the non-$\beta$ meteoroids is:
\begin{equation}
  \beta \approx 4/D, \label{eq:betah}
\end{equation}
where $D$ is measured in $\mu$m (in agreement with eq.~[\ref{eq:beta2}]).
Fig.~\ref{fig9} shows that the emitting particles have
$D_{typ} \approx 2.5$ $\mu$m; these particles have $\beta = 1.7$.
So, the modeling implies that the emitting particles are $\beta$ meteoroids,
i.e., that they are blown out of the system on hyperbolic orbits.
Since the lifetime of $\beta$ meteoroids is $O$(370 years) for those
created at $\sim 70$ AU, which is much shorter than the age of the
HR 4796 system, any $\beta$ meteoroids that are currently in the disk
cannot be primordial particles, rather they must be continuously
created from a reservoir of larger particles;
i.e., the existence of $\beta$ meteoroids implies the existence of a
dynamically stable population of larger particles.

If the emitting particles are on hyperbolic orbits, we can use the mass of
the disk's emitting particles, $\sim 1.4 \times 10^{-3}M_\oplus$, and the
emitting lifetime of these particles, $\sim 370$ years, to estimate the mass
loss rate of the disk to be $\sim 4 \times 10^{-6}M_\oplus/$year.
If this mass loss rate has been sustained over the age of the system,
the original disk must have been $\sim 40M_\oplus$ more massive than
it is today.
In fact, a more massive disk would have had a higher mass loss rate, since
this rate increases proportionally with the total cross-sectional
area in the disk (i.e., $\propto m_{tot}^{2/3}$).
This means that if the current disk has a mass $1.0M_\oplus$, the original
disk must have had a mass $\sim 7 \times 10^4M_\oplus$
(see eq.~[\ref{eq:massloss}]);
i.e., the HR 4796 disk may provide evidence for the type of collisional
mass loss that may have happened in the early Kuiper belt (\cite{sc97}). 

%%%%%%%%%%%%%%%%%%%%%%%%%%%%%%%%%%%%%%%%%%%
\subsection{Collisional Processes}
\label{sssec-collh}
The collisional lifetime of the disk's emitting particles can be calculated
directly from the disk model using equations (\ref{eq:tcol3}) and
(\ref{eq:fccdtyp}):
$t_{coll}(D_{typ}) < 10^{4}$ years across most of the disk (55-85 AU),
with a minimum at $\sim 70$ AU of $\sim 4500$ years.
This collisional lifetime is much less than the age of the HR 4796
system.
Thus, the emitting particles cannot be primordial particles (irrespective
of whether they are $\beta$ meteoroids).
The collisional lifetime of the emitting particles can be corroborated
using equation (\ref{eq:tcol6}).
The model's effective optical depth (eq.~[\ref{eq:taueffdefn}]) is its
surface density, plotted in Fig.~\ref{fig4}, multiplied by the cross-sectional
area of a particle in the model, $\sigma = \pi D_{typ}^2/4$.
This peaks at 70 AU, where $\tau_{eff}$(70 AU) = $5 \times 10^{-3}$, and
$t_{per} = O$(370 years), giving a collisional lifetime that has a minimum
of $t_{coll}(D_{typ}) \approx 6000$ years.
The disk's effective optical depth at 70 AU, $\tau_{eff}$(70 AU), can
also be corroborated directly from the observation using equation
(\ref{eq:taueff}).
The observed edge-on, smoothed lobe brightness is $\sim 1.40$ mJy/pixel;
this can be scaled to the unsmoothed face-on brightness by the factor
1.10/1.40 (see Figs.~\ref{fig7a} and \ref{fig8}b), which, since each pixel
subtends $(0.0616*\pi/648000)^2$ sr, gives a brightness of
$F_\nu$(18.2 $\mu$m,70 AU)$/\Omega_{obs} = 12 \times 10^9$ Jy/sr.
For 2.5 $\mu$m particles, Figs.~\ref{fig5}c and \ref{fig5}d give
$P$(18.2 $\mu$m,70 AU) $\approx 2.2 \times 10^{12}$ Jy/sr, thus
confirming that $\tau_{eff}$(70 AU) $\approx 5 \times 10^{-3}$;
this is also in agreement with $\tau \approx 5 \times 10^{-3}$
found by Jura et al.~(1995).
In fact, assuming that the disk's total IHW18 flux density, 857 mJy (T2000),
comes from particles between $70 \pm 15$ AU, the unsmoothed face-on
brightness of the disk can also be corroborated;
equation (\ref{eq:fnuobs}) with $R_\star = 67$ pc gives
$F_\nu$(18.2 $\mu$m, 70 AU)$/\Omega_{obs} = O(10^{10}$ Jy/sr).

Since particles are only broken up by collisions with particles that
have diameters more than a tenth of their own (eq.~[\ref{eq:dimp}]),
the collisional lifetime of the disk's large particles must be longer
than that of the smaller emitting particles.
Assuming the cross-sectional area distribution to follow equation
(\ref{eq:nd}) with $q = 11/6$ down to particles of size $D_{typ}$, the
collisional lifetime of particles with $D \gg D_{typ}$ can be estimated
to be (eq.~[\ref{eq:tcol7}]):
\begin{equation}
  t_{coll} \approx 4 \times 10^7 \sqrt{D},
\end{equation}
where $t_{coll}$ is measured in years, and D in km.
Since particles for which $t_{coll} < t_{sys}$ cannot be primordial,
this implies that particles currently in the HR 4796 disk that are smaller
than 60 m cannot be original, rather they must have been created in the
break-up of larger particles;
i.e., these disk particles form a collisional cascade, through
which the spatial distributions of the smaller particles are related
to that of the larger particles.
Particles in the disk that are larger than 60 m could be primordial.

%%%%%%%%%%%%%%%%%%%%%%%%%%%%%%%%%%%%%%%%%%%
\subsection{P-R Drag}
\label{sssec-prh}
The disk's high effective optical depth means that $\beta_{pr} = 132$
at 70 AU (eq.~[\ref{eq:btr}]), which in turn means that there is not
expected to be significant P-R drag evolution for any of the disk's
particles.
Analysis of the P-R drag evolution of disk particles shows that even the
most affected particles, those with $\beta=0.5$, have a P-R drag lifetime
of $t_{pr} = O$(1.6 Myr) at 70 AU (eq.~[\ref{eq:tpr}]), and that these
particles would only make it to $\sim 69.8$ AU before they are broken up by
collisions and blown out of the system by radiation pressure.

But, P-R drag could still be important for disks in which the individual
particles of that disk are not affected significantly by P-R drag in their
lifetime.
While the inner edge of a disk's largest particles may be at $a_{min}$,
the cumulative effect of P-R drag on all of the stages of the collisional
cascade that a primordial particle goes through before the fragments are
small enough to be blown out of the system by radiation pressure
could mean that the inner edge of these blow-out particles is further
in than $a_{min}$.
To assess whether the cumulative effect of P-R drag has an impact
on the inner edge of the emitting particles in the HR 4796 disk,
consider its size distribution to follow that assumed in equation
(\ref{eq:sigd}), where $D_{min} = D_{typ} = 2.5$ $\mu$m, and $q = 11/6$.
The lifetime of an intermediate particle can be found using
equations (\ref{eq:tcol5}) and (\ref{eq:fcc2}),
and the amount of P-R drag evolution in its lifetime is given by
(eqs.~[\ref{eq:adot}] and [\ref{eq:betah}]):
\begin{equation}
  da \approx -0.0125 t_{coll}(D,r)/Da, \label{eq:dapr}
\end{equation}
where $da$ and $a$ are measured in AU, $D$ in $\mu$m,
and $t_{coll}(D,r)$ in years.
Assuming that the largest fragment created in a collision has a
diameter half that of the original particle, a $\beta$ meteoroid particle
(e.g., a 2.5 $\mu$m particle) at 62 AU, assuming it to have originated from
a gravitationally bound particle (i.e., one with $\beta < 0.5$), can at the
very most be removed by 27 generations from its primordial ancestor, which
would at the very most have been originally at 62.1 AU.

%%%%%%%%%%%%%%%%%%%%%%%%%%%%%%%%%%%%%%%%%%%
\subsection{The Dynamic HR 4796 Disk}
\label{sssec-ddch}
Since P-R drag is not an important process in the HR 4796 disk's evolution,
there are just three categories of particles in the disk:
large particles, $\beta$ critical particles, and $\beta$ meteoroids.
If the modeled emitting particle size is to be believed (discussed further
in \S \ref{ssec-deps}), the particles that are seen in both the IHW18 and
the N band observations are the disk's $\beta$ meteoroids.
Since the disk was modeled as if the emitting particles are large particles,
this inconsistency needs to be borne in mind in the interpretation of
the model.

While the modeling used a distribution of orbital elements that is
only appropriate for the disk's large particles, it was the spatial
distribution of the disk's emitting material, $\sigma(r,\theta,\phi)$,
that was constrained by the modeling, not the distribution of orbital
elements, $\sigma(a,e,I,\Omega,\tilde{\omega})$.
Therefore, the inferred distribution, $\sigma(r,\theta,\phi)$, which is
that shown in Fig.~\ref{fig4}, is an accurate description of the spatial
distribution of the disk's emitting particles, whatever their size.
Indeed, it is in excellent agreement with that inferred from other
observations of the HR 4796 disk
(\cite{jfht98}; \cite{krwb98}; \cite{ssbk99}).
If the emitting particles are the disk's large particles, then the
inferred model variables have physical interpretations
for the distribution of the orbits of the disk's large particles.
If, as appears to be the case, the emitting particles are $\beta$
meteoroids, then further modeling of $\sigma(r,\theta,\phi)$ needs
to be done to infer the distribution of the orbits of these particles.

However, since the disk's $\beta$ critical and $\beta$ meteoroid particles
are created from its large particles, the spatial distributions of all of these
particles share a great deal in common (see \S\S \ref{s-categories} and
\ref{ss-ofwpfam}):
they all have the same plane of symmetry, the same flaring,
and the same offset and warp asymmetries, but the radial distributions
of the smaller particles are more extended than that of the large particles.
This means that if we are seeing the disk's $\beta$ meteoroids, then
the spatial distribution of its large particles has a plane of symmetry that
is defined by $I_{obs}$, and an inner edge that is at the same radial
location, and that is offset by the same amount and in
the same direction, as that shown in Fig.~\ref{fig4};
their radial distribution, however, would not be as extended as that of
Fig.~\ref{fig4}.
Thus, the model parameters $a_{min}$, $e_f$, and $\tilde{\omega}_f$
have physical interpretations for the distribution of the orbits of the
disk's large particles, irrespective of the size of the emitting particles.

%%%%%%%%%%%%%%%%%%%%%%%%%%%%%%%%%%%%%%%%%%%%%%%
\section{Interpretation of Lobe Asymmetry: HR 4796's Secular Perturbations}
\label{ssec-ias}
The lobe asymmetry in the model of Chapter \ref{c-hrmodel} is due solely
to the offset inner edge of the disk.
The model shows that secular perturbations amounting to a comparatively
small forced eccentricity, $e_f = 0.02$, imposed on the orbits of large
particles at the inner edge of the disk, $a = 62$ AU, would cause the disk's
inner edge to be offset by a sufficient amount to cause the observed
5\% lobe asymmetry.
This section considers what kind of a perturber system would impose such
a forced eccentricity on the inner edge of the disk, and whether such a
system is physically realistic.
A discussion of the system's secular perturbations also allows
interpretation of the disk's inferred orientation, defined by the
parameters $\tilde{\omega}_f$ and $I_{obs}$.

\begin{figure}
  \begin{center}
    \begin{tabular}{cccc}
      & \epsfig{file=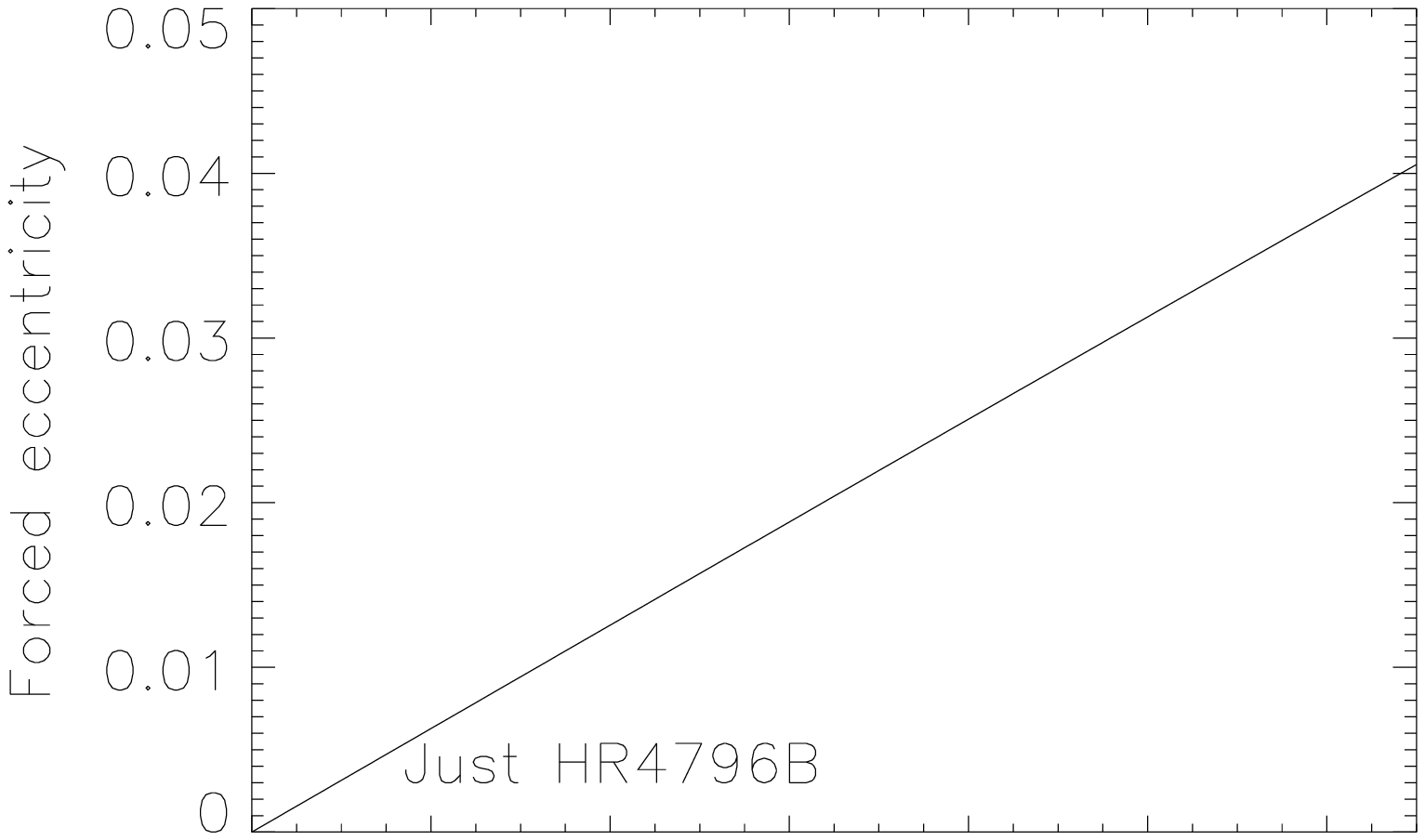,height=1.64in} &
        \epsfig{file=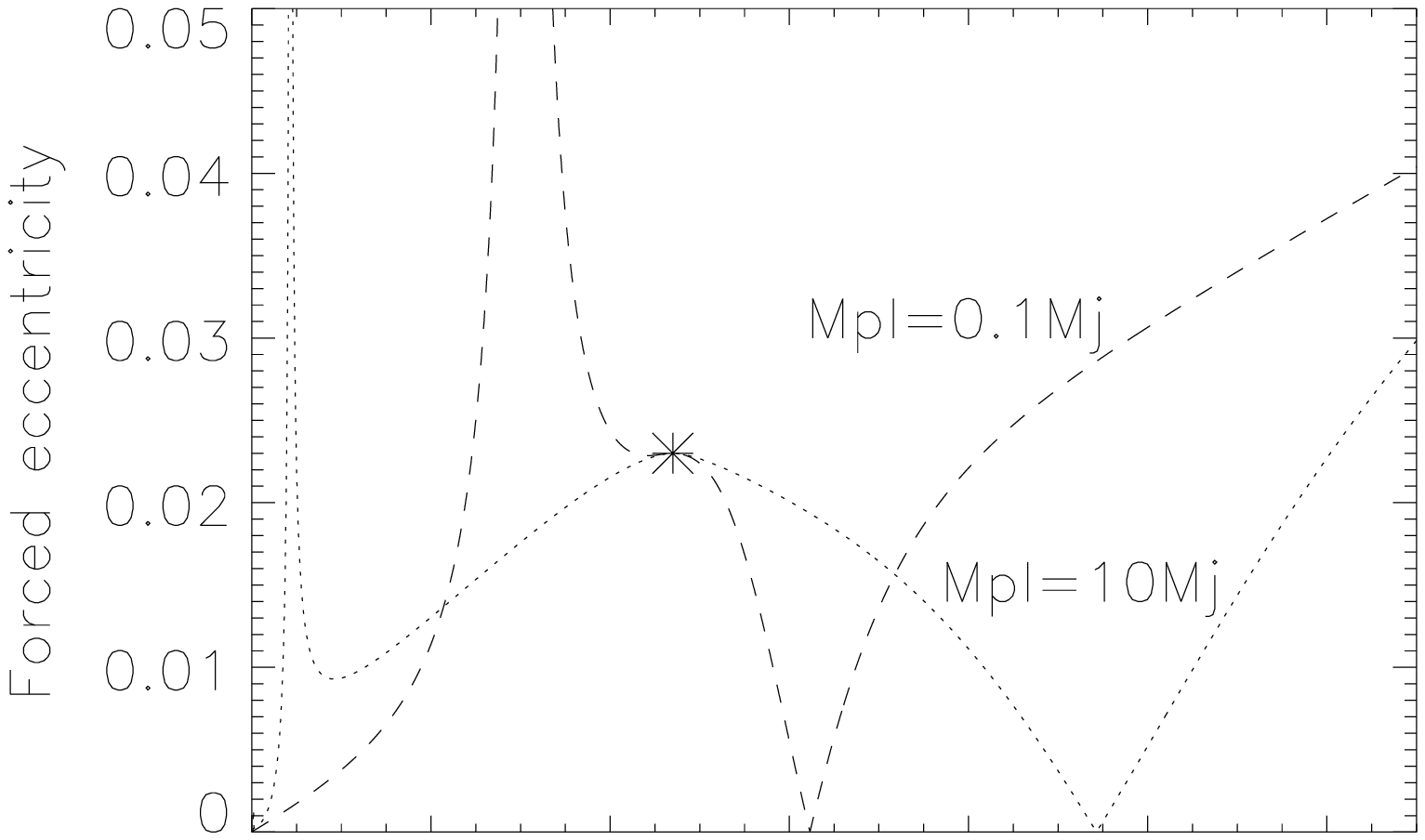,height=1.64in} & \\ \vspace{-0.02in}
      & \epsfig{file=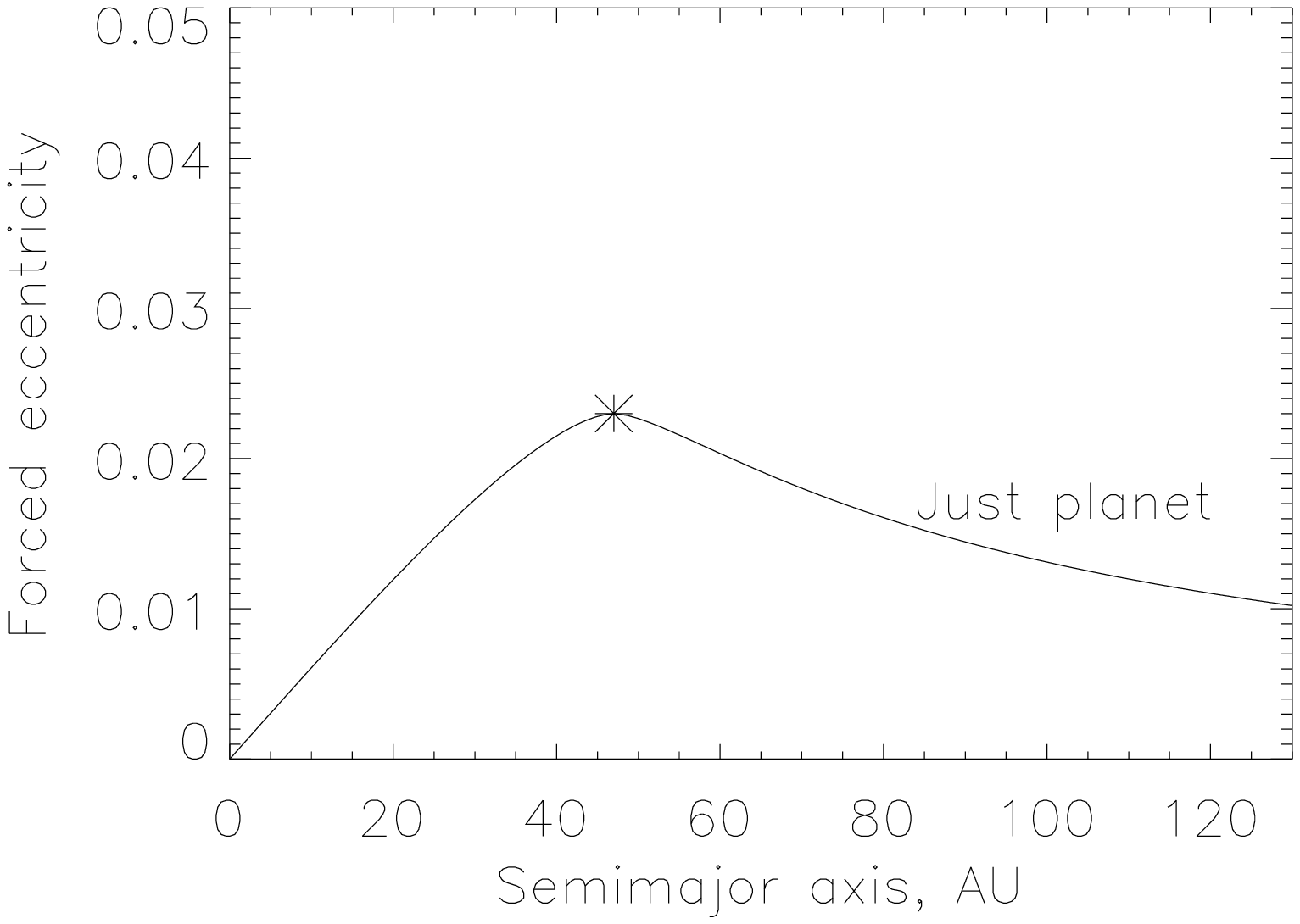,height=1.976in} &
        \hspace{0.04in} \epsfig{file=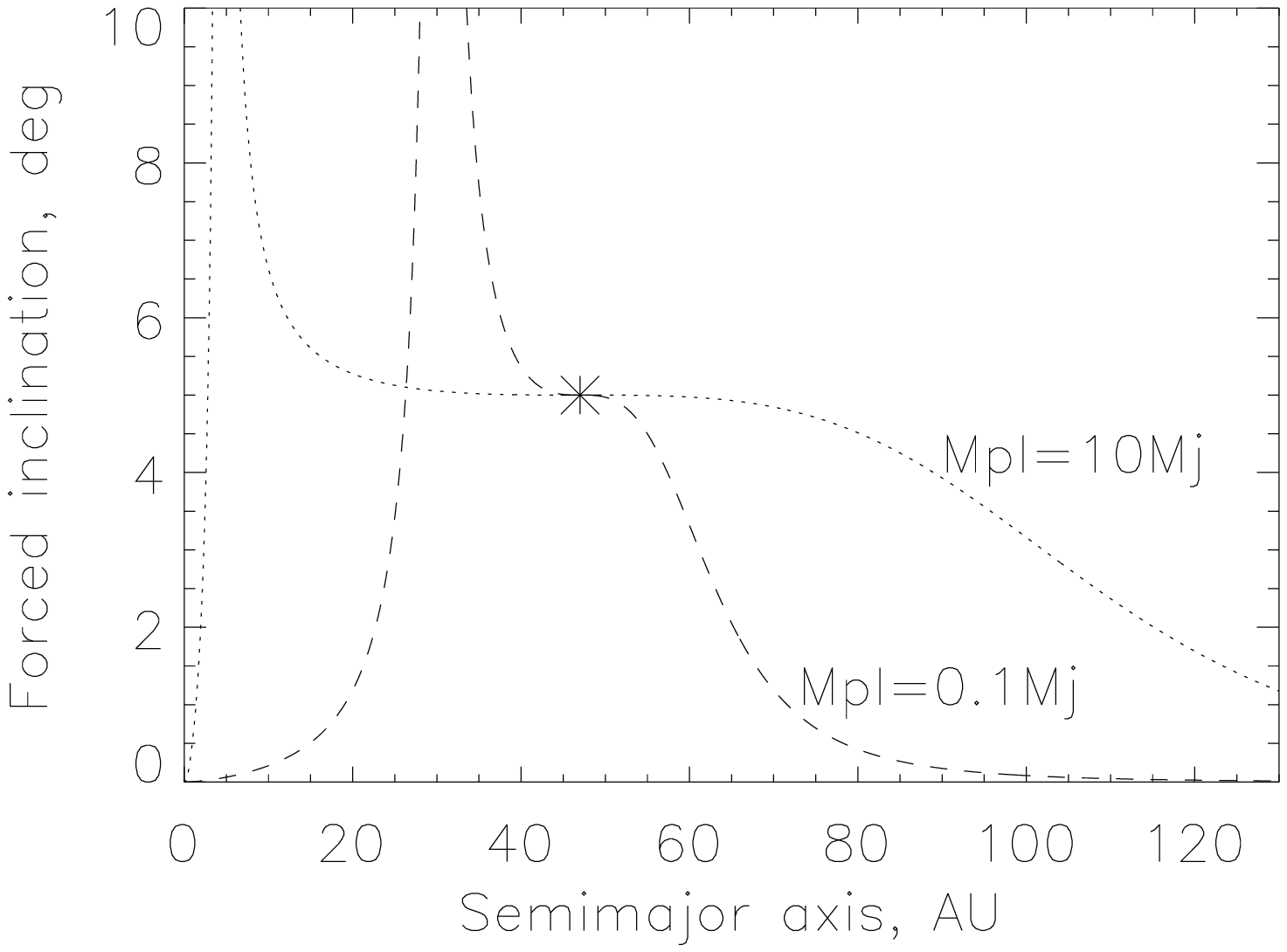,height=1.976in} & \\
    \end{tabular}
  \end{center}
  \caption{Plots of the forced eccentricities imposed on the orbits of
  particles in the HR 4796 system as a function of their semimajor axes,
  assuming different combinations of perturbers in the system:
  (\textbf{a}) HR 4796B only;
  (\textbf{b}) HR 4796B and a planet at the inner edge of the disk;
  (\textbf{c}) a planet only.
  The forced inclinations imposed by the two perturber system are
  shown in (\textbf{d}).
  The inner edge of the disk is at $a = 62$ AU.
  HR 4796B and the planet are assumed to have:
  $M_B = 0.38M_\odot$, $a_B = 517$ AU, $e_B = 0.13$, and $I_B = 0$;
  $M_{pl} = 0.1$ and 10 $M_J$, where $M_J$ is the mass of Jupiter,
  $a_{pl} = 47$ AU, $e_{pl} = 0.023$, and $I_{pl} = 5^\circ$.
  The orbital elements of the planet are marked by an asterisk on
  the forced elements plots.
  In a one perturber system the forced eccentricity is independent
  of the mass of the perturber, and the forced inclination is the plane of
  the perturber's orbit.
  In a two perturber system, the shapes of the forced element plots depend
  on the mass of the planet, and the forced eccentricity also
  depends on the orientations of the perturbers' orbits.
  The forced eccentricity is plotted in (\textbf{b}) assuming that
  $\tilde{\omega}_{pl} = \tilde{\omega}_{B} + 180^\circ$.
  This means that $\tilde{\omega}_f = \tilde{\omega}_{pl}$ for $a < a_{crit}$,
  and $\tilde{\omega}_f = \tilde{\omega}_{B}$ for $a > a_{crit}$, where
  $a_{crit}$ is the semimajor axis of a particle's orbit for which $e_f = 0$;
  a similar alignment with the perturbers' orbital planes seen in
  the plot of the particles' forced inclinations.
  Thus, the lobes have both their asymmetries and their plane of symmetry
  aligned with the orbit of the planet if $M_{pl} > 0.1M_J$, and with
  the orbit of HR 4796B if $M_{pl} < 0.1M_J$. }
  \label{fig10}
\end{figure}

If HR 4796A's binary companion, HR 4796B, is on an eccentric orbit,
it would have imposed a forced eccentricity on the disk particles.
However, a forced eccentricity could also have been imposed on the disk
by an unseen planet close to the inner edge of the disk, a planet which
could be responsible for clearing the inner region
(e.g., \cite{al94}; \cite{rsss94}).
The secular perturbations imposed on the HR 4796 disk by a perturber
system that includes HR 4796B and a putative planet located at the inner
edge of the disk are shown in Fig.~\ref{fig10} for the four cases:
$M_{pl} = 0$, $M_{pl} = 0.1M_J$, $M_{pl} = 10M_J$, and just planet.
The parameters of the two perturbers are assumed to be:
\begin{description}
\item{\textbf{HR 4796B}}
  $M_B = 0.38M_\odot$(\cite{jfht98});
  the orbit of HR 4796B is unknown at present (\cite{jzbs93}), so the
  semimajor axis of its orbit is arbitrarily taken as its projected distance,
  $a_B = 517$ AU (\cite{jmwt98}; note that this is not the same as assuming
  that this is the semimajor axis of the ellipse that the star's orbit traces
  on the sky) --- this gives an orbital period of $\sim 7000$ years
  (eq.~[\ref{eq:tper}]), and a timescale for secular perturbations from
  HR 4796B to have built up at 62 AU of $O$(1 Myr) (eq.~[\ref{eq:tsec2}]);
  $e_B = 0.13$, the eccentricity necessary to cause $e_f = 0.02$ at
  $a = 62$ AU if there were no unseen perturbers (eq.~[\ref{eq:zf1}]);
  $I_B = 0^{\circ}$, defining the reference plane for the analysis.
\item{\textbf{Planet}}
  $M_{pl}$ is a variable measured in Jupiter masses,
  where $M_J = 10^{-3}M_\odot$ (current observations have limited
  the size of a planet in the system to $M_{pl}<200M_J$, \cite{jmwt98});
  $a_{pl} = 47$ AU (see \S \ref{ssec-lsd});
  $e_{pl} = 0.023$, the eccentricity necessary to cause $e_f = 0.02$ at
  $a = 62$ AU if the planet was the only perturber (eq.~[\ref{eq:zf1}]);
  $I_{pl} = 5^{\circ}$, an arbitrary choice that represents the fact that
  the orbital plane of the planet is not necessarily be aligned with
  that of HR 4796B.
\end{description}

%%%% Figure 10

%%%%%%%%%%%%%%%%%%%%%%%%%%%%%%%%%%%%%%%
\subsection{Just HR 4796B}
\label{sssec-jb}
For the cases when there is just one perturber in the system,
the forced elements in the system can be found from equations (\ref{eq:zf1})
and (\ref{eq:yf1}):
the forced eccentricity, $e_f $, is determined by the ratio of the
semimajor axes of the perturber and the particle, and by the eccentricity
of the perturber's orbit, but is independent of the perturber's mass;
the forced pericenter, $\tilde{\omega}_f$, is aligned with the pericenter of
the perturber;
and the plane of symmetry of the disk, $y_f$, is constant across the disk,
and is the orbital plane of the perturber.

So, if the only perturber is HR 4796B, then to impose $e_f = 0.02$
at $a = 62$ AU, the eccentricity of its orbit would have to be $e_B = 0.13$;
the consequent forced eccentricity imposed on the disk is plotted in
Fig.~\ref{fig10}a.
This also means that if $e_B > 0.1$, then a brightness asymmetry in this
disk of $>5\%$ would be expected unless adverse geometrical conditions
prevented it.
The position angle from north of HR 4796B relative to HR 4796A is
$225^\circ$ (\cite{jfht98}), while that of the SW lobe (i.e., the least
bright lobe) is $206^\circ$ (T2000).
For the lobe asymmetry to be the consequence of perturbations from
HR 4796B only, HR 4796B must currently be close to its apastron, and
its orbital plane must be the plane of symmetry of the disk,
i.e., inclined at $13^{\circ}$ to the line of sight.
All of these conclusions are consistent with the initial estimate that
the semimajor axis of the star's orbit is equal to its observed projected
distance, 517 AU.

%%%%%%%%%%%%%%%%%%%%%%%%%%%%%%%%%%%%%%%%
\subsection{HR 4796B and a Planet}
\label{sssec-bp}
Consider the effect of adding a planet at the inner edge of the disk into
the HR 4796 system\footnote{The orbital elements of a low-mass planet
in the system would, just like the disk particles, have forced and
proper components, while a high-mass planet would perturb the orbit of
HR 4796B.}.
If there are two perturbers in the system, then the forced element
variation with semimajor axis depends both on the masses of the
perturbers and on the orientations of their orbits.
In the plot of Fig.~\ref{fig10}b, $\tilde{\omega}_{pl} = \tilde{\omega}_B +
180^{\circ}$ was chosen so that the forced eccentricity (and hence the
lobe asymmetry) is aligned with the planet's pericenter for $a < a_{crit}$,
and aligned with HR 4796B's pericenter for $a > a_{crit}$, where
$a_{crit}$ is the semimajor axis for which $e_f = 0$.
Since the two perturbers were also chosen to have different orbital planes,
a similar change in the disk's alignment is seen in the plot of $I_f$
(Fig.~\ref{fig10}d):
the disk's plane of symmetry is aligned with the planet's orbital plane at
its inner edge, and with the orbital plane of HR 4796B at its outer edge;
this could cause an image of the disk to appear warped.
Such a warp could be modeled using the same modeling techniques
that are described in this dissertation, and would provide further constraints
on the perturbers in the system (even if no warp was observed).

So, it is possible that the brightness asymmetry, and the symmetry plane of
the lobes, are determined by a planet close to the edge of the disk
that has $M_{pl} > 0.1M_J$, rather than by HR 4796B.
Using such an analysis, the pericenter glow phenomenon could be used to
test for the existence of a planet in the HR 4796 system, but only after
the orbit of HR 4796B has been determined;
e.g., if $\tilde{\omega}_B$, $e_B$ or the plane of HR 4796B's orbit
contradicted the observed asymmetry orientation, brightness asymmetry
magnitude, or the plane of symmetry of the lobes, then the existence of a
planet at the inner edge of the disk with $M_{pl} > 0.1 M_J$ could be
inferred.

%%%%%%%%%%%%%%%%%%%%%%%%%%%%%%%%%%%%%%%%%%%
\subsection{Just Planet}
\label{sssec-jp}
A double-lobed disk structure could also be observed in a system with no
observable companion.
The only possible perturber in such a system is an unseen planet,
the secular perturbations of which warrant the same kind of discussion
as for the case when HR 4796B was the only perturber (\S \ref{sssec-jb}).
Depending on the planet's mass, radial location and eccentricity,
it too could give rise to a detectable pericenter glow.
The only constraint on the planet's mass is that the disk must be old enough
for its secular perturbations to have affected the distribution of orbital
elements of the disk particles over the age of the system.
Since it takes of the order of one precession timescale to distribute the
complex eccentricities of collisional fragments around the circle centered
on the forced eccentricity, the constraint on the planet's mass can be
approximated as that for which is that the age of the system, $t_{sys}$, is
greater than the secular timescale, $t_{sec} \propto 1/M_{pl}$
(eq.~[\ref{eq:tsec2}]);
for the HR 4796 system this limit is $M_{pl} > 10M_\oplus$.
The constraint on the planet's eccentricity is even less stringent than for
the binary companion because the planet is closer to the edge of the disk:
a planet in the HR 4796 system would only need an eccentricity of
$e_{pl} > 0.02$ to produce the observed $5\%$ lobe asymmetry,
and the forced eccentricity imposed on the disk by a planet with
$e_{pl} = 0.023$ is plotted in Fig.~\ref{fig10}c.
So, the signature of even a low-mass planet would not escape detection
and symmetrical double-lobed features are unlikely to be observed in
systems that contain planets.

%%%%%%%%%%%%%%%%%%%%%%%%%%%%%%%%%%%%%%%%%%
\subsection{Other Considerations}
\label{sssec-oc}
If the disk itself is massive enough to cause significant gravitational
perturbations to the orbits of the disk particles, then the mass of the
disk should be incorporated into the analysis of the secular perturbations
in the system.
A massive disk could dampen the eccentricity of a planet at the inner
edge of the disk (\cite{wh98}), thus reducing the offset asymmetry.

%%%%%%%%%%%%%%%%%%%%%%%%%%%%%%%%%%%%%%%%%%%%%%%%
\section{Discussion of Emitting Particle Category}
\label{ssec-deps}
The emitting lifetime of the disk's $\beta$ meteoroids, $O$(370 years),
is less than the emitting lifetime of their parents, $O(10^4$ years);
equivalently, $\tau_{eff} < 0.1$.
Thus, our understanding of the dynamic disk
implies that the disk's cross-sectional area distribution should not
contain a significant amount of $\beta$ meteoroids.
Rather, since there are no P-R drag affected particles in the disk, the
disk's emission is expected to come from its $\beta$ critical
particles and its smallest large particles (\S \ref{ss-catsigs}).
Not all disk particles have the same composition and morphology;
even if this were a close approximation, there is, as yet, no evidence
to suggest whether the particle properties chosen in this model
(\S \ref{sssec-pop}) are correct.
Are we indeed seeing the disk's $\beta$ meteoroid particles, or did
the assumptions of the modeling lead us to this conclusion?

Consider the initial crude estimate of the particle size (\S \ref{sssec-dtyp});
this proved to be an accurate method for estimating the particle size
(\S \ref{sssec-tps}).
If different assumptions had been made about the particles'
properties (e.g., if the particles had been assumed to have
silicate cores with organic refractory mantles)
or morphologies (e.g., if the particles had been assumed to be like the
bird's nest structures of \cite{gust94}), both Figs.~\ref{fig6a} and
\ref{fig6b} would have been different, and a different conclusion might have
been drawn about the $\beta$ of the emitting particles.
If a size distribution had been included in the modeling, this would also
have affected the conclusion.
These are considerations that need to be modeled before any firm
conclusion about the dynamics of the emitting particles can be reached.
However, since irrespective of their assumed properties large particles
have black body temperatures and brightness ratios similar to those
of the $D > 100$ $\mu$m particles in Fig.~\ref{fig6a}, a flexible interpretation
of the observed brightness ratio is that the emitting particles must have
temperatures that are hotter than black body.
Thus, the emitting particles are either small (e.g., the simple analysis of
\S \ref{sssec-pop} implies that $D < 10$ $\mu$m), in which case
they are likely to be $\beta$ meteoroids (e.g., 10 $\mu$m particles have
$\beta > 0.5$ unless they have densities $> 3000$ kg/m$^3$,
eq.~[\ref{eq:beta2}]), or they are large particles that are made up of
smaller hotter particles.

%%%% Figure 5 but for Organic refractory particles
\begin{figure}
  \begin{center}
    \begin{tabular}{cccc}
      & \epsfig{file=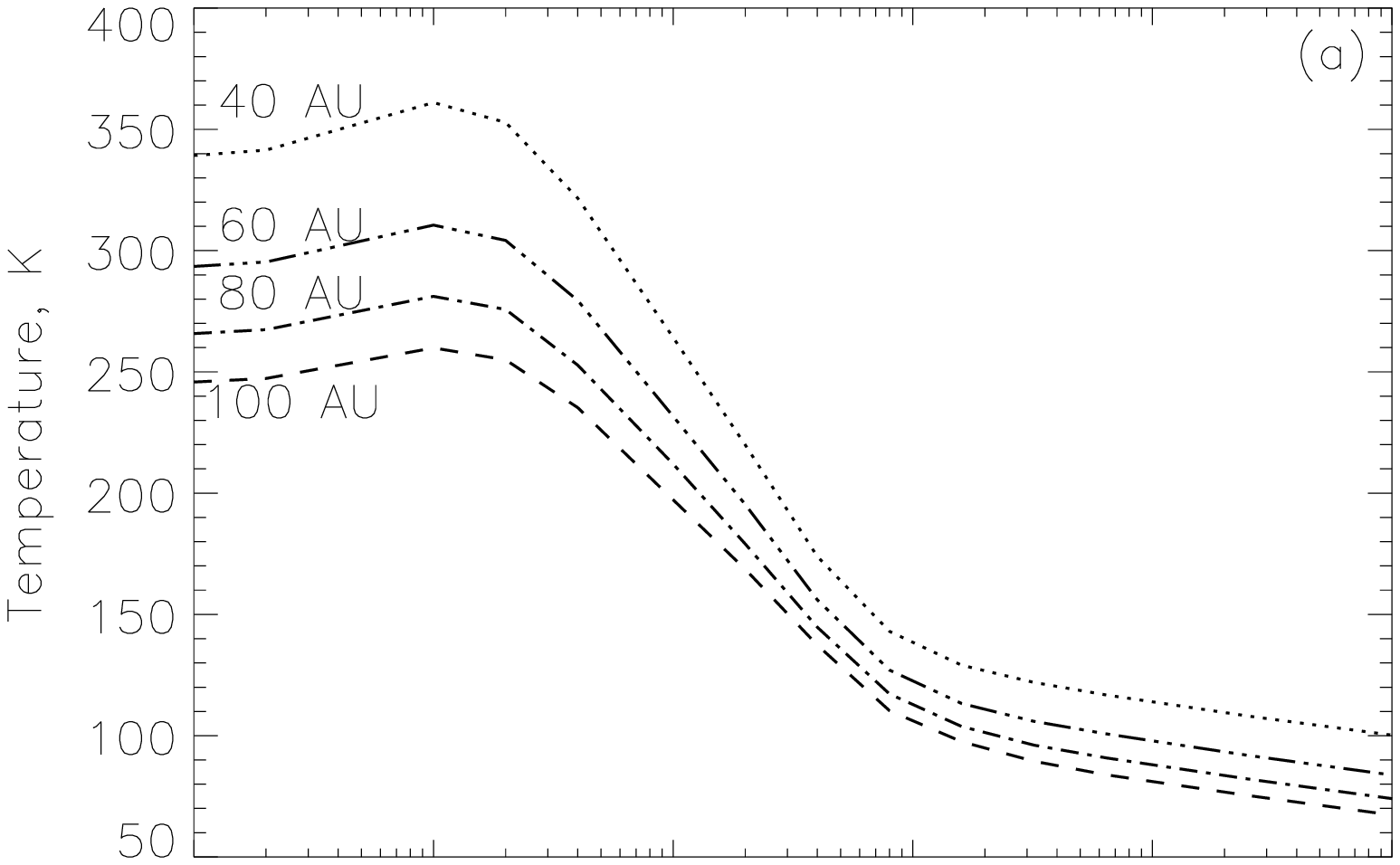,height=1.87in} & \hspace{-0.2in}
        \epsfig{file=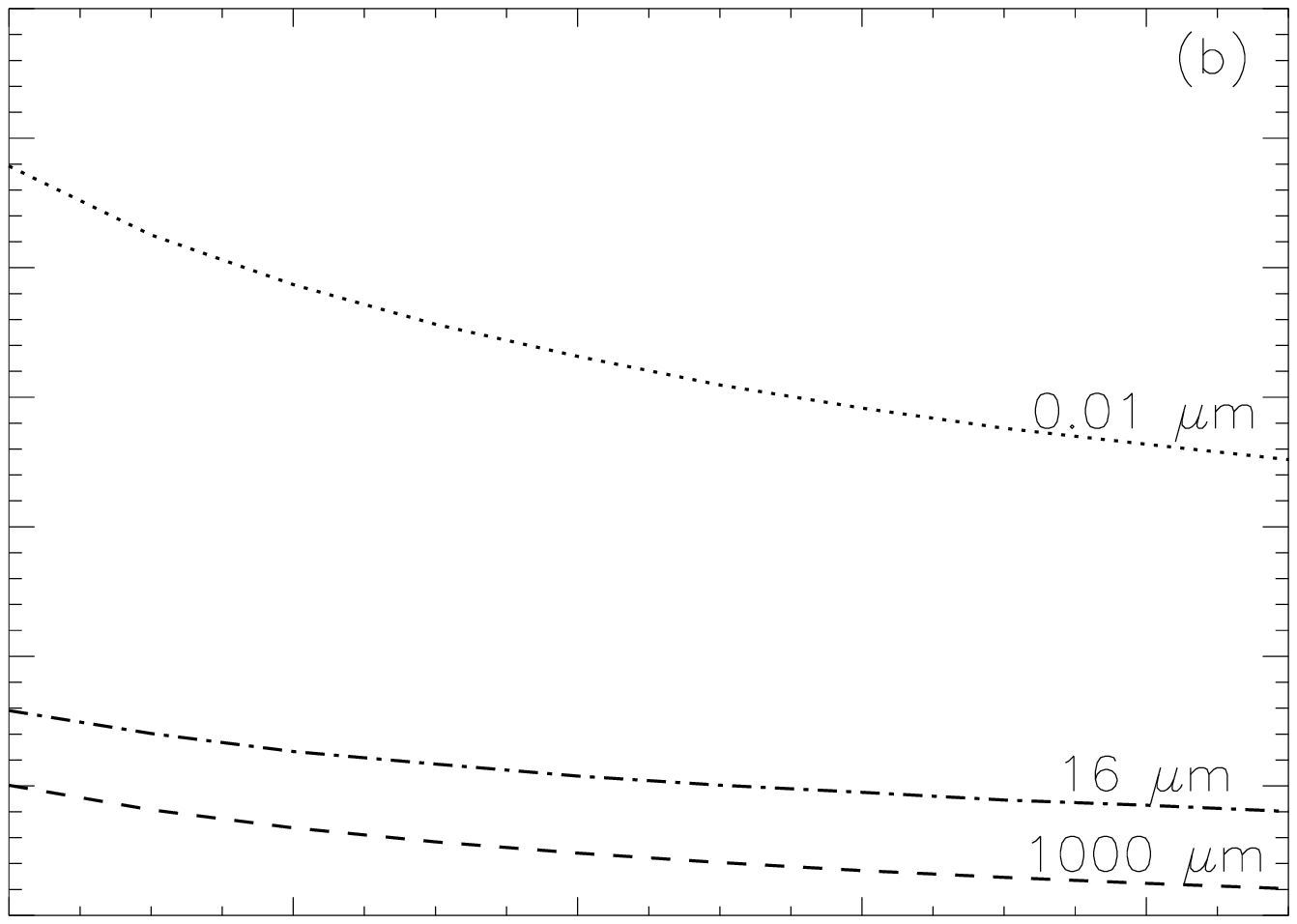,height=1.87in} & \vspace{-0.04in} \\
      & \hspace{-0.15in} \epsfig{file=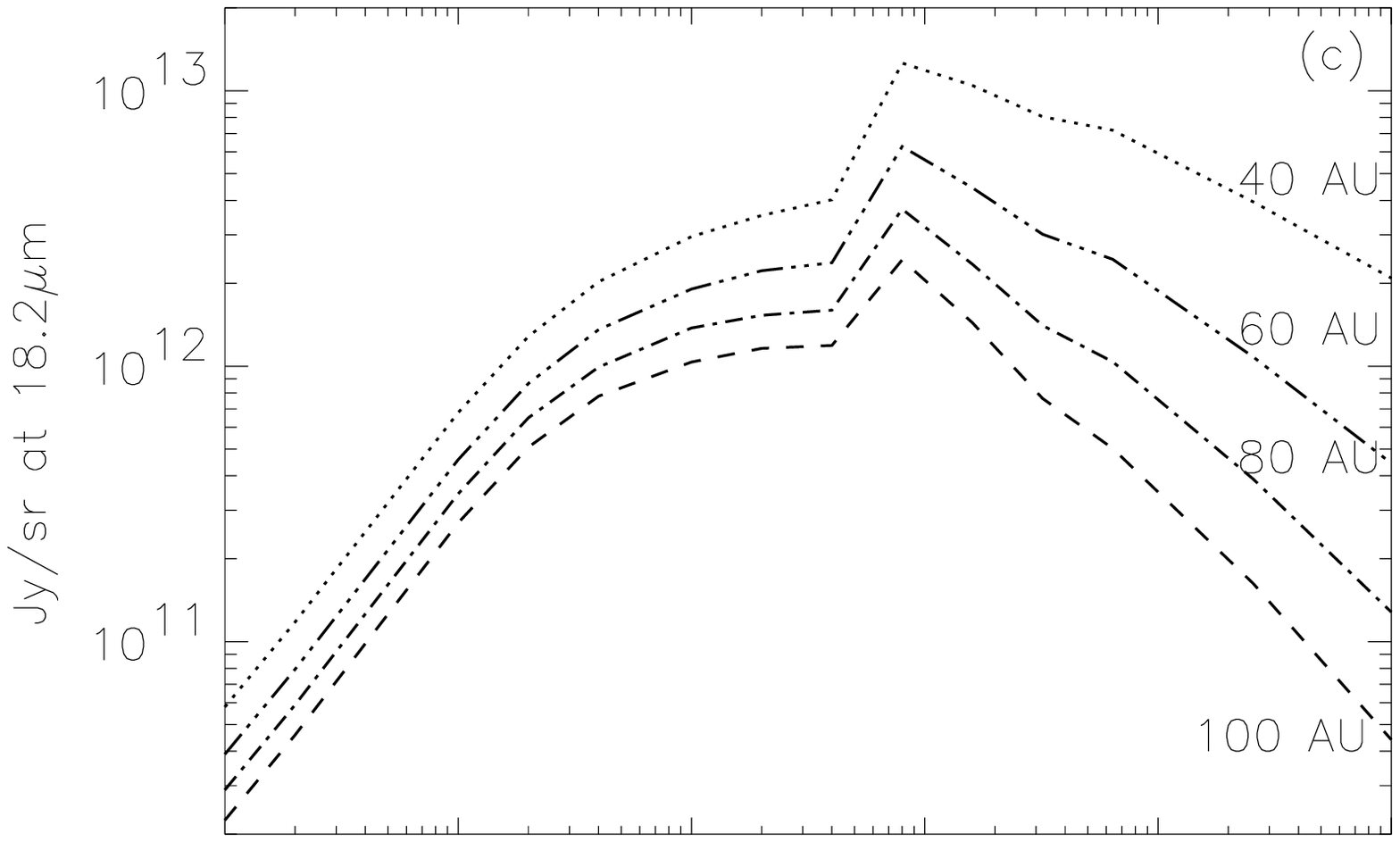,height=1.87in} &
        \hspace{-0.2in}  \epsfig{file=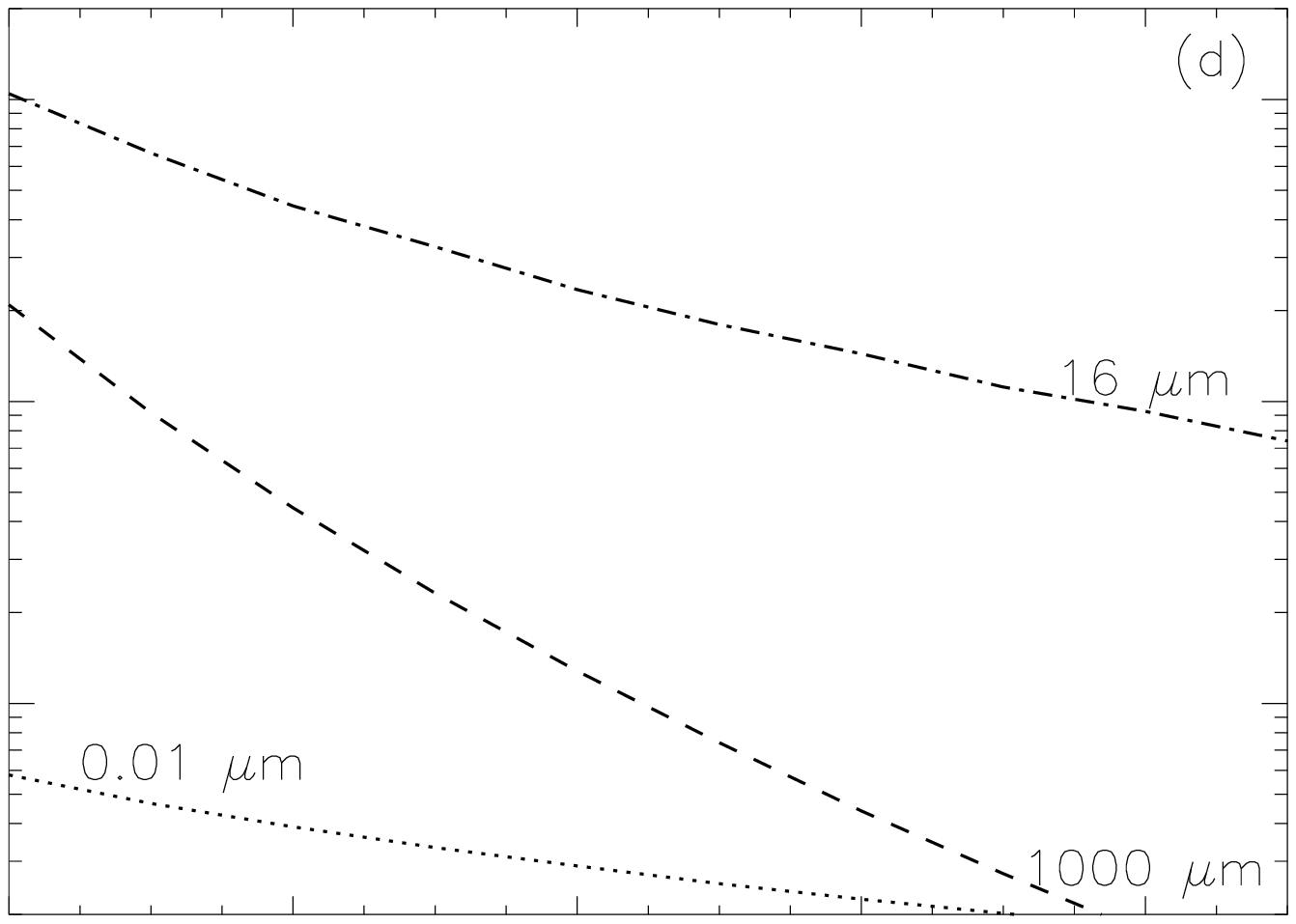,height=1.87in} &
        \vspace{-0.10in} \\
      & \hspace{-0.15in} \epsfig{file=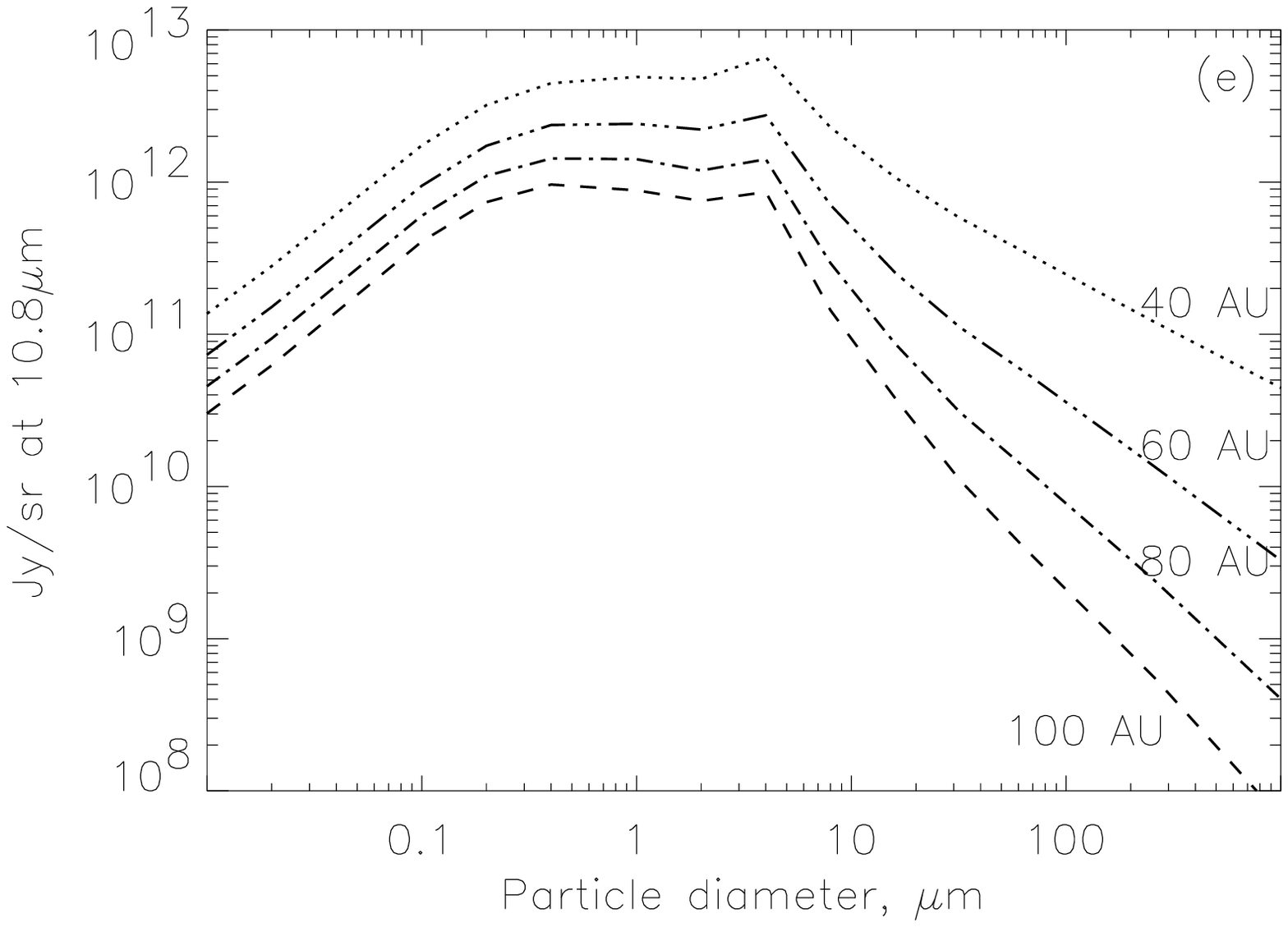,height=2.235in} &
        \hspace{-0.2in}  \epsfig{file=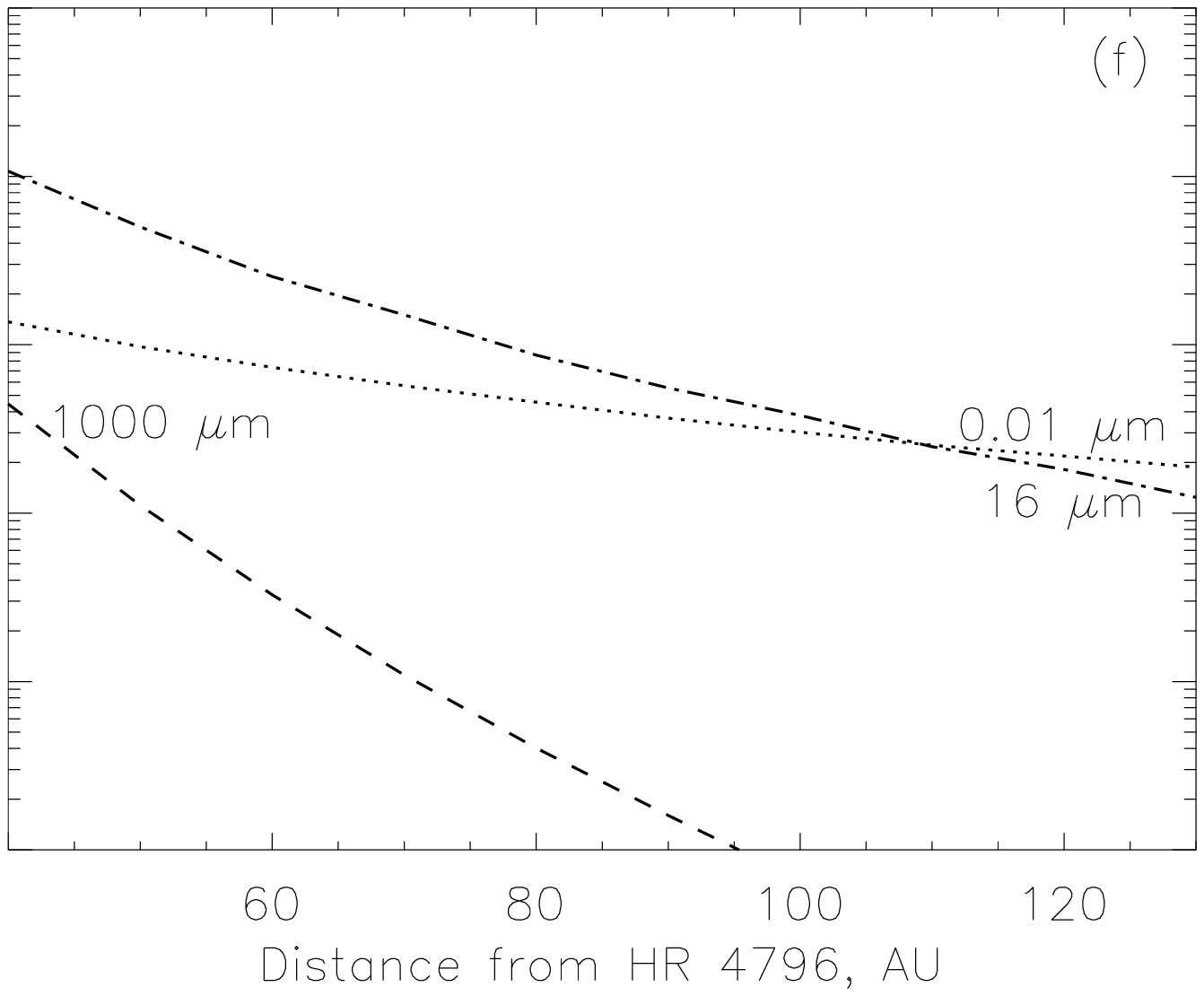,height=2.18in} & 
    \end{tabular}
  \end{center}
  \caption{ The thermal properties of organic refractory Mie spheres
  in the HR 4796 disk, plotted for particles of different sizes at
  40, 60, 80, and 100 AU from HR 4797A (\textbf{a}), (\textbf{c}), and
  (\textbf{e}), and for 0.01, 16, and 1000 $\mu$m diameter particles at
  different distances from HR 4796A (\textbf{b}), (\textbf{d}), and
  (\textbf{f}).
  The temperatures that these particles attain is plotted in (\textbf{a})
  and (\textbf{b}).
  The contribution of a particle's thermal emission to the flux density
  received at the Earth per solid angle that its cross-sectional area
  subtends there, $Q_{abs}(D,\lambda)B_\nu[T(D,r),\lambda]$
  (eq.~[\ref{eq:flux}]), is plotted for observations in the IHW18,
  18.2 $\mu$m, (\textbf{c}) and (\textbf{d}), and N, 10.8 $\mu$m,
  (\textbf{e}) and (\textbf{f}) wavebands. }
  \label{fig5org}
\end{figure}

%%%% Figure of dynamical category of particles if they are organic refractory
\begin{figure}
  \begin{center}
    \begin{tabular}{c}
      \hspace{0.035in} \epsfig{file=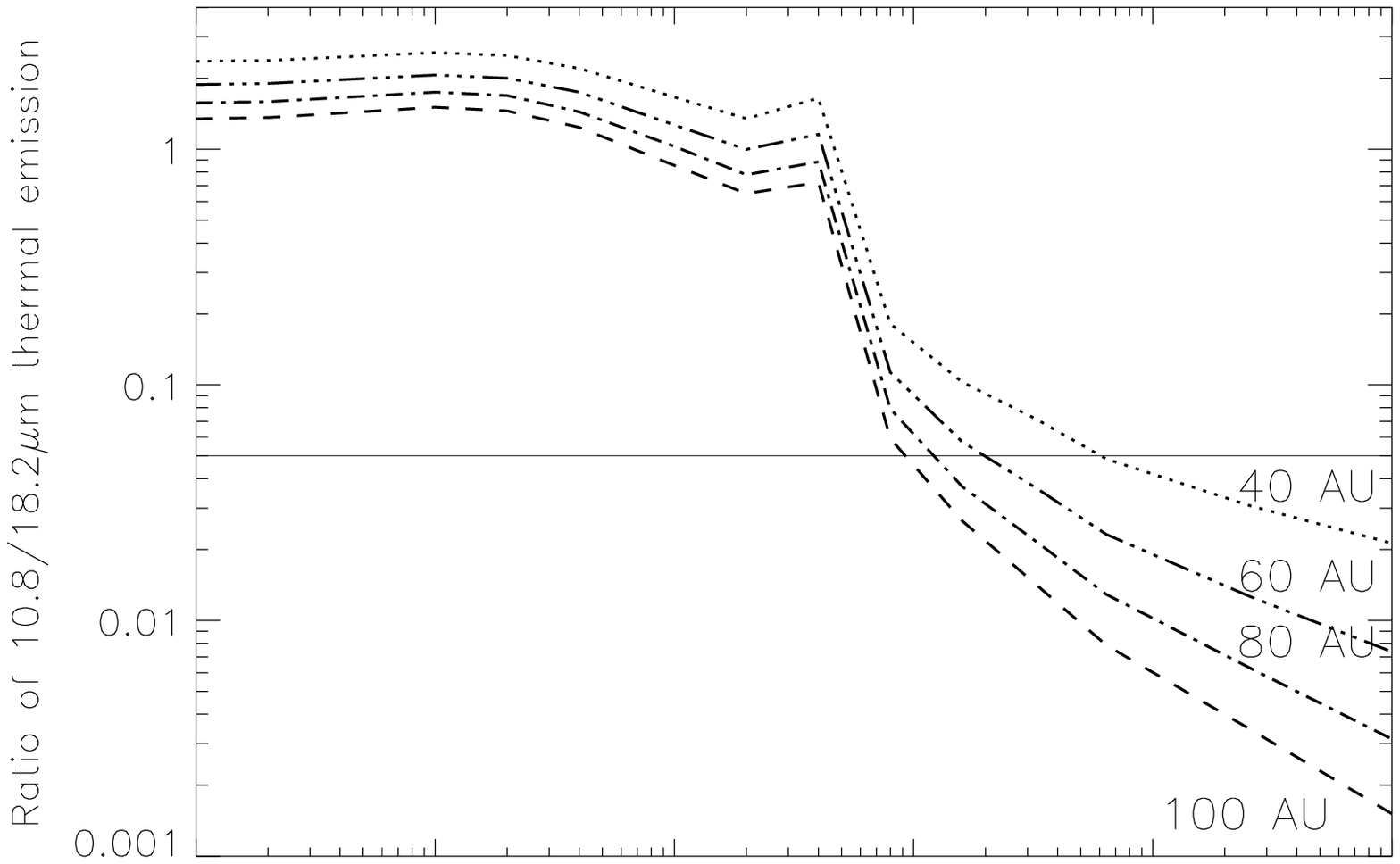,height=2.58in} \vspace{-0.05in}\\
      \epsfig{file=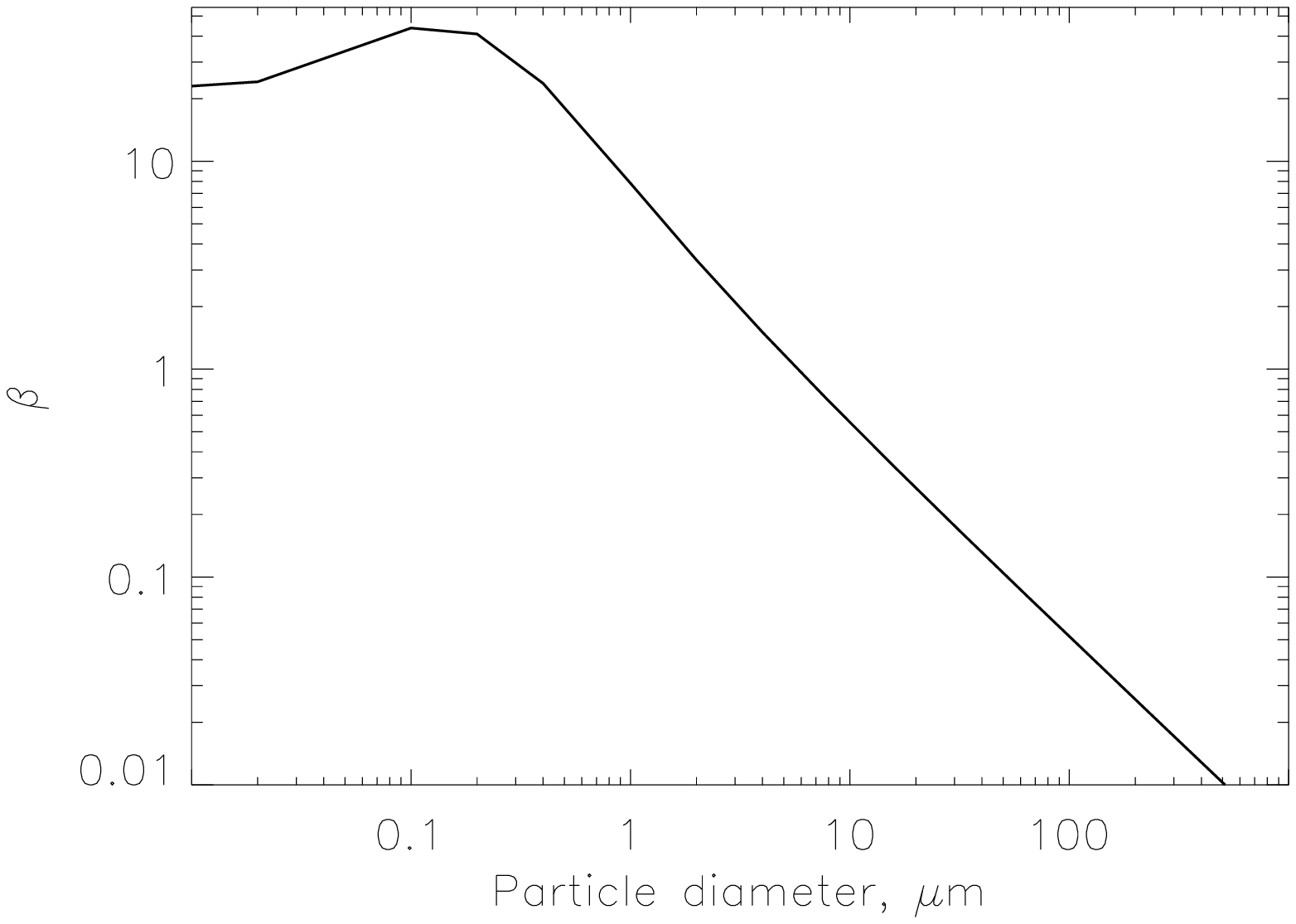,height=3in}
    \end{tabular}
  \end{center}
  \caption{Estimate of the dynamical category of the mid-infrared
  emitting particles in the HR 4796 disk if they are made of organic
  refractory material.
  \textit{Top} - The ratio of the thermal emission in the N, 10.8
  $\mu$m, and IHW18, 18.2 $\mu$m, wavebands, of organic refractory
  Mie spheres of different sizes at 40, 60, 80, and 100 AU
  from HR 4796A (cf.~Fig.~\ref{fig6a}).
  Assuming the disk's flux densities in the two wavebands to be
  dominated by the emission of particles at 60-80 AU, the observed
  ratio of flux densities, $O(0.05)$ (T2000; \S \ref{sssec-dtyp}), can
  be used to estimate that the disk's emitting particles have
  $D_{typ} = 12-20$ $\mu$m.
  \textit{Bottom} - The ratio, $\beta$, of the radiation pressure force to
  the gravitational force acting on organic refractory Mie spheres
  of different sizes in the HR 4796 disk (cf.~Fig.~\ref{fig6b}).
  The size estimate implies that the particles have a $\beta$ between
  0.2 and 0.5, i.e., that they are the disk's $\beta$ critical particles.}
  \label{figorginhr}
\end{figure}

%%%% Figure 9 for organic refractory material
\begin{figure}
  \begin{center}
    \epsfig{file=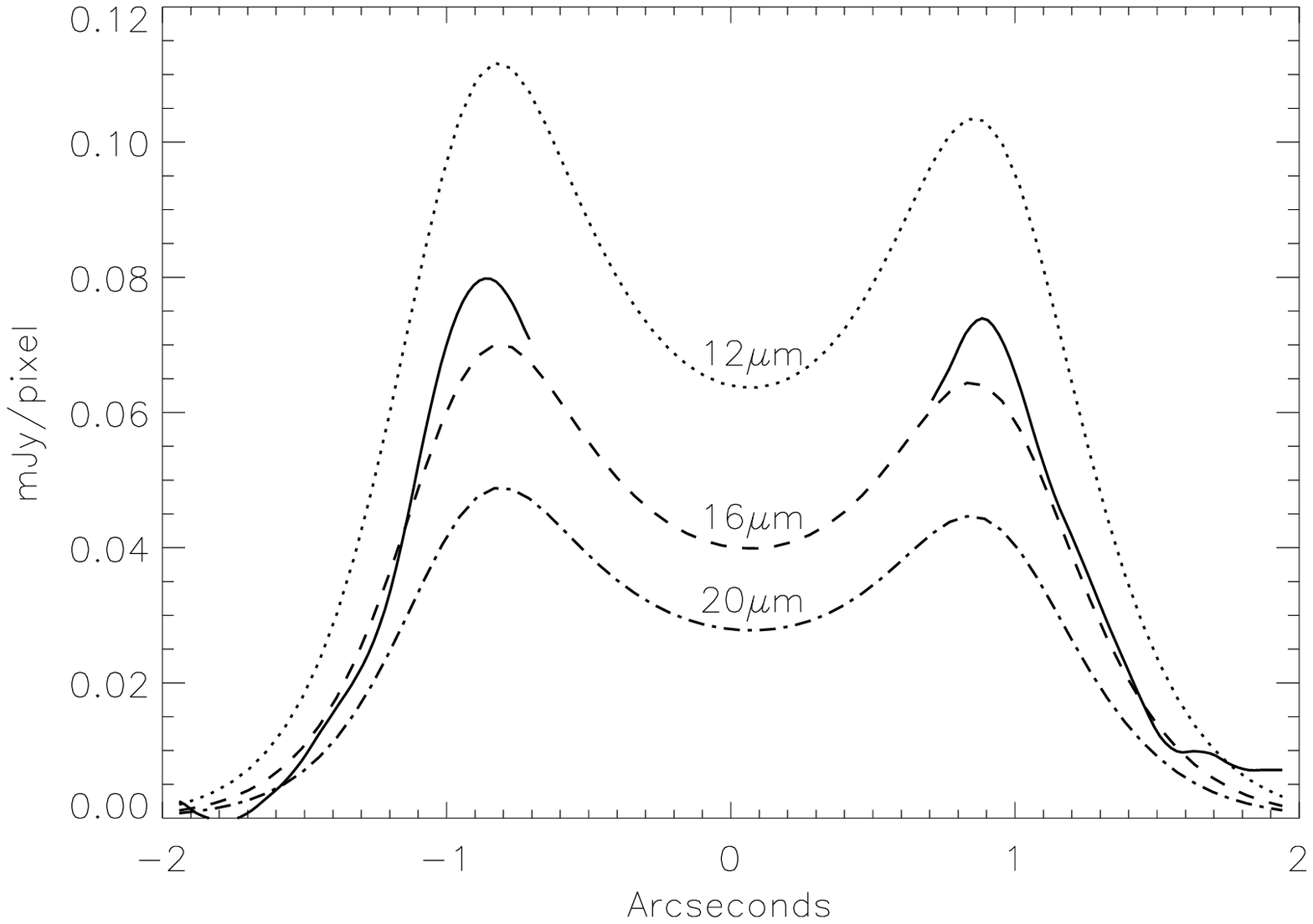,height=3in}
  \end{center}
  \caption{Horizontal line-cuts along the plane of the lobes in the
  N (10.8 $\mu$m) band.
  The observation is shown with a solid line and models with organic
  refractory Mie spheres with
  diameters of $D_{typ}$ = 12, 16, and 20 $\mu$m are shown with dotted,
  dashed, and dash-dot lines.
  This is analagous to Fig.~\ref{fig9} which shows the same plot, but
  for models with astronomical silicate Mie spheres of different sizes.}
  \label{fig9org}
\end{figure}

A preliminary investigation was carried out to determine the effects of
the assumptions about the particles' optical properties on the conclusion
about their dynamical category.
It was worried that the choice of astronomical silicates for the particle
composition might have biased the particle size estimate since the
wavebands that were used to estimate that size lie on the silicate
emission features, and so this size was re-estimated assuming the particles
to be made of organic refractory material (\cite{lg97}).
The optical properties of organic refractory grains in HR 4796 are shown
in Fig.~\ref{fig5org}, which is analogous to Fig.~\ref{fig5} for
astronomical silicate grains;
an explanation of the shape of the plots in Fig.~\ref{fig5org} is
similar to that for Fig.~\ref{fig5} (see \S \ref{sssec-pop}).
A crude estimate for the emitting particle size was found using
Fig.~\ref{figorginhr}, which is analogous to Fig.~\ref{fig6a}.
Fig.~\ref{figorginhr} shows that if the particles in the HR 4796 disk had
been modeled as organic refractory Mie spheres, then $D_{typ}$
would have been estimated to lie in the range 12-20 $\mu$m.
This was confirmed when the grain size was estimated more accurately using
Fig.~\ref{fig9org}, which is analogous to Fig.~\ref{fig9}.
This would have implied that the emitting particles have a $\beta$ in the
range 0.2-0.5, i.e., that the emitting particles are the disk's $\beta$
critical particles, as expected from dynamical considerations.
This appears to be inconsistent with the statement in the last paragraph
that because the emitting particles are hotter than black body, they should
be smaller than 10 $\mu$m.
It is not, however, since Fig.~\ref{figorginhr}a shows that even large
(up to 1000 $\mu$m) organic refractory particles have temperatures that
are higher than black body, and this is because these large particles
do not emit efficiently at their black body temperatures;
e.g., Fig.~\ref{figoptprorg} shows that 100 $\mu$m particles do not
have $Q_{abs} \approx 1$ for $\lambda < \pi D$, rather for
$\lambda < D/\pi$.
Note that the temperature of 2.5 $\mu$m astronomical silicate grains at
70 AU from HR 4796A is 103.5 K, while that of 16 $\mu$m organic refractory
grains is 108.4 K. 

The assumptions about the particle properties in the model could also
have affected the conclusion about the disk's effective optical depth and
consequently that for the collisional lifetime of its emitting particles
(see eq.~[\ref{eq:tcol5}]).
Consider the estimate of $\tau_{eff}$(70 AU) derived from the IHW18
lobe brightness (\S \ref{sssec-collh}).
Changing the properties of particles in the model would change the
estimate of $\tau_{eff}$(70 AU), because of the resultant changes in
$P(18.2$ $\mu$m,70 AU);
e.g., if the disk had been modeled using 30-50 $\mu$m astronomical silicate
particles, more cross-sectional area would have had to have been put in the
model for it to give the observed lobe brightness, resulting in a higher
effective optical depth.
Since for 16 $\mu$m organic refractory particles,
$P(18.2$ $\mu$m,70 AU)$ = 3.3 \times 10^{12}$ Jy/sr, which is just
$\sim 1.4$ times that of the 2.5 $\mu$m diameter astronomical silicates
(Fig.~\ref{fig5}c), if organic refractory material had been used in the
modeling rather than astronomical silicate, the inferred effective optical
depth of the disk would have been smaller by a factor of $\sim 1.4$, and
the collisional lifetime of the emitting particles would have been longer
by the same factor (i.e., not much different).
It is possible that different optical properties would affect this estimate
of $\tau_{eff}$(70 AU) by a greater amount, since $P(18.2$ $\mu$m,70 AU)
has the potential to cover a wide range of values;
e.g., astronomical silicate Mie spheres that are larger than 0.01 $\mu$m
imply a possible range of $0.034-2.3 \times 10^{12}$ Jy/sr
(Fig.~\ref{fig5}c), while the same range for organic refractory Mie spheres
is $0.033 - 4.8 \times 10^{12}$ Jy/sr (Fig.~\ref{fig5org}c).
However, given the temperature of the emitting particles inferred from
the observed brightness ratio, $\sim 5 \times 10^{-3}$ remains the
best estimate for the disk's effective optical depth at 70 AU, and
$O(10^4$ years) the best estimate for the emitting particles'
collisional lifetime there.
A further note of caution is necessary about the inferred collisional
lifetime:
\S \ref{ss-tcoll} assumes collisions between disk particles
to be either catastrophic or irrelevant.
While this may be appropriate for the disk's larger particles, since
these are likely to be solid bodies, collisions between its smaller
particles, which may have fluffy bird's nest structures (\cite{gust94}),
could be more erosive than destructive.

%%%% Figure of thermal emission spectrum of silicates and organics in HR 4796
\begin{figure}
  \begin{center}
    \begin{tabular}{c}
      \epsfig{file=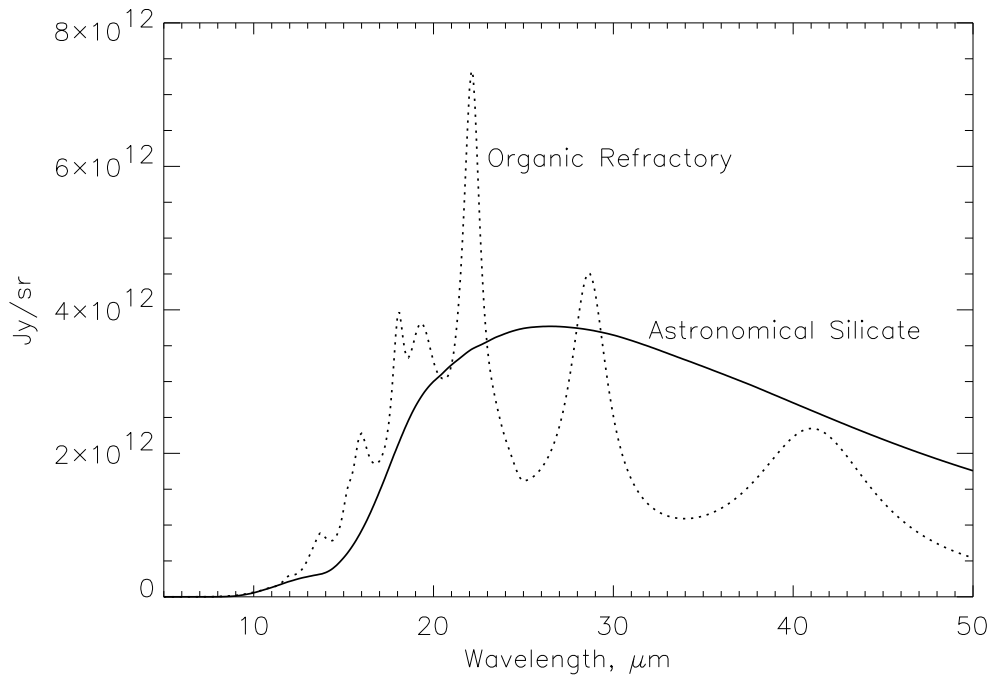,height=3.5in}
    \end{tabular}
  \end{center}
  \caption{Mid-infrared thermal emission spectra of particles
  at 70 AU from HR 4796A.
  The solid line depicts 2.5 $\mu$m diameter astronomical silicate
  particles, and the dotted line depicts 16 $\mu$m diameter organic
  refractory particles.}
  \label{figspectra}
\end{figure}

Thus, both the conclusion about the dynamical category
of the disk's emitting particles and, to some extent, the conclusion
about their collisional lifetime are sensitive to the assumptions about
the particles' composition.
Since the particles' composition cannot be determined from just two
points on the disk's SED, this really just emphasizes the ambiguity of
conclusions drawn in such a way.
However, interpretation of the spectrum of the HR 4796 disk emission
could be used to reveal properties of its particles.
Particles with different properties have different spectral signatures in
the mid-IR because the wavelengths at which they emit most efficiently
are different (see Figs.~\ref{figoptprsil} and \ref{figoptprorg});
e.g., the spectrum of the thermal emission of 2.5 $\mu$m astronomical
silicate particles and that of 16 $\mu$m organic refractory particles
at 70 AU from HR 4796A is shown in Fig.~\ref{figspectra}.
A spectrum with such well-defined features as those of
Fig.~\ref{figspectra} is not expected, since the observed spectrum
contains the emission from a range of particle sizes, and from particles
at different distances from the star;
however, such a spectrum could be modeled, and this would reveal
information not only about the particles' properties, but also about
their size distribution.
While the spectrum of HR 4796 is not yet available, its SED has been
observed (see, e.g., Fig.~\ref{fighr4796SED}), and this would be the
starting point of such modeling.

In conclusion, without further study, neither observational nor
theoretical considerations can provide a definitive answer as to the
dynamics of the mid-IR emitting particles in the HR 4796 disk.
However, because of the short collisional lifetime of these particles,
it is certain that they are not primordial,
and that there are no P-R drag affected particles in the disk.
Because of the inferred temperature of the particles, 
it is most likely that these particles are either $\beta$ meteoroids or
$\beta$ critical particles.
$\beta$ critical particles are preferred from theoretical considerations,
however a disk with mid-IR emitting particles that are on hyperbolic orbits is
not an outlandish possibility.
Such particles have been inferred from observations of the disks around
both $\beta$ Pictoris (\cite{tdbw88}) and, possibly, HD 141569
(\cite{ftpk99}).
The disk around BD$+31^\circ643$ has also been interpreted as a radiation
pressure outflow (\cite{lvf98});
however this disk is quite unlike the HR 4796 disk, since it is around a
binary star, there is a significant amount of gas in the system, and the
disk particles are thought to be created $\sim 2300$ AU from the star.

%%%%%%%%%%%%%%%%%%%%%%%%%%%%%%%%%%%%%%%%%%
\section{Origin of the Inner Hole}
\label{ssec-lsd}
Whatever the size of the emitting particles, analysis of the optical depth
of the disk's inner region shows that it is a few hundred times less
than that of the outer disk (T2000), and so there must be very few emitting
particles in this region.
Because this central hole is necessary for the secular perturbation offset
asymmetry to be observed (without the hole only the radial offset
could be observed), its physical origin requires attention.
Since the existence of small emitting particles in the disk requires the
existence of large particles, the question to answer is why there
are so many large particles in the outer disk, but so few in the inner
disk?
Either the physical conditions were such that they were able to form in the
outer region, but not in the inner region, or they formed across the whole disk,
but those formed in the inner region have since been removed.
Rather than discussing the planetary formation process and the stage of the
system's evolution (although these are of utmost importance in determining
the physics of the disk), this section offers possible dynamical
explanations for the removal of the large particles.

If a planetary system did form in the inner region, then a lack of debris
material in this region is to be expected, most obviously because much of
the debris material that did form there would have been swept up by the
growing planets.
Also, a lot of the material in the inner region that did not end up accreted
onto a planet by the end of the planetary formation process would be ejected
from this region due to resonant perturbations from the planetary system
(see discussions of resonance overlap and secular resonance in
\S \ref{s-respertns}).
For resonant mechanisms to remove all of the remaining material, the system
would have to consist of either many planets, or just one planet that is
either very large, or on a very eccentric orbit.
Radial migration by the planets, however, would allow them to affect a
larger fraction of the disk than equations such as equation
(\ref{eq:resolap}) imply.
Note that none of these removal mechanisms would cause the inner edge
of the disk to have a sharp cut-off in radial distance from the star,
rather, since all of these removal mechanisms take longer than a particle's
orbital period to take effect, they would cause an inner cut-off in the
disk particles' semimajor axes, which causes a sloping radial cut-off
(see \S \ref{sssec-simul}).

To estimate the orbit of a putative planet at the inner edge of the HR 4796
disk that is causing the cut-off, consider the inner edge of the Kuiper belt,
which is located at the \textbf{2:3} resonance with Neptune
(\S \ref{ss-restrapkb}).
By analogy, assuming that the inner cut-off of the disk's large particles
occurs at the planet's \textbf{2:3} resonance location, and that this cut-off
can be described by $a_{min} = 62$ AU, equation (\ref{eq:ares}) can be used
to estimate that the orbit of the planet has a semimajor axis of
\begin{equation}
  a_{pl} = a_{min}[\textbf{2/3}]^{2/3},
  \label{eq:apl}
\end{equation}
giving $a_{pl} = 47$ AU, and an orbital period of $\sim 200$ years
(eq.~[\ref{eq:tper}]).

%%%%%%%%%%%%%%%%%%%%%%%%%%%%%%%%%%%%%%%%%%%%%%%
\section{Interpretation of the Residual Structure}
\label{ssec-rs}

%%%% Figure of residuals
\begin{figure}
  \begin{center}
    \begin{tabular}{c}
      \epsfig{file=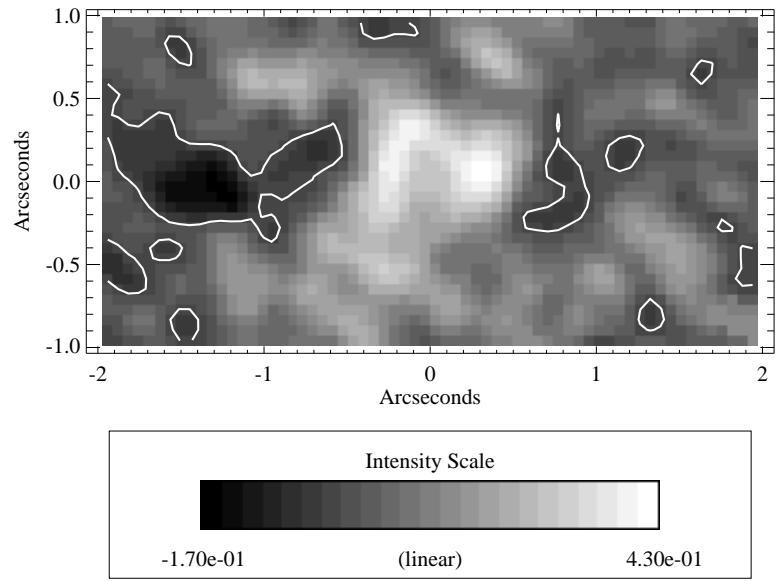,height=3.0in}
    \end{tabular}
  \end{center}
  \caption{Image of the residual IHW18 (18.2 $\mu$m) waveband emission
  from the HR 4796 disk.
  These residuals are the observed image (\cite{tfpk00}) minus the model
  image (\cite{wdtf00}; see Fig.~\ref{fig8}).
  The intensity scale is in mJy/pixel, and the contour line is that of
  zero residual emission.}
  \label{figres}
\end{figure}

So far, no explanation has been offered for the structure of the
residuals (what is left after subtracting the model from the observation;
see Fig.~\ref{figres}).
Analogy with the zodiacal cloud implies that there could be a population
of warm dust in the inner region that may be unrelated to the cold dust in
the outer disk.
Indeed, many authors have argued for the existence of such a population
(\cite{krwb98}; \cite{almp99}), although their claims do not appear
to concur with the observations of T2000 (see Chapter \ref{c-hrlitrev}).
Depending on the perturbers in the inner region, such a population could
contain considerable structure.
Analysis of the emission from such regions would reveal a great deal about
the system's perturbers, however, the detection of this emission would
not be easy, since the resolution required to map such small-scale structure
is at the limit of current technological capabilities.
In addition, such emission is masked by that of both the stellar
photosphere and the outer disk, the subtraction of which is unlikely
to be perfectly accurate (see Chapter \ref{c-obsn}).

There may also be residual structure associated with the outer regions of
the disk.
If there is a planet orbiting HR 4796A close to the inner edge of the
disk, then the distribution of large particles in the outer disk would contain
structure associated with the planet's gravitational perturbations in
addition to the secular perturbations already discussed
(\cite{djxg94}; \cite{malh96}; \cite{dghw98}; \cite{wh98}).
Some of the emitting particles might be trapped in its resonant ring,
which would be analogous to the Kuiper belt resonant ring
(\S \ref{ss-restrapkb}) rather than the Earth's resonant ring
(\S \ref{ss-restrapzc}).
Such a ring could be responsible for some of the observed lobe asymmetry.
The existence of such a ring would give the inner edge of the disk structure
that co-orbits with the planet;
i.e., observations of this structure would vary on timescales of
$O$(200 years), offering a method of distinguishing between this
structure and the large-scale background structure, which would vary on
secular timescales of $O$(1 Myr).
The amount of disk material that is trapped in an anolgous Kuiper belt
ring, and the structure of that ring, depends on the extent of radial
migration of the planet's orbit.
The edge-on viewing geometry of the HR 4796 disk makes it more difficult
to detect such resonant structure.

%%%%%%%%%%%%%%%%%%%%%%%%%%%%%%%%%%%%%%%%%%%%%%
\chapter{CONCLUSION}
%%%%%%%%%%%%%%%%%%%%%%%%%%%%%%%%%%%%%%%%%%%%%%
\label{c-conclusions}

%%%%%%%%%%%%%%%%%%%%%%%%%%%%%%%%%%%%%%%%%%%%%%
\section{Conclusion}
\label{s-con}
This dissertation provides a physical model for the evolution of
circumstellar debris disks that is based on our understanding of the
physics affecting the evolution of the particles in such disks.
The physical processes that are included in this model are those that
affect the evolution of the debris disk in the solar system, the
zodiacal cloud, but generalized for disks that may be at different
evolutionary stages.
It was shown how disk particles can be grouped into one of
four categories (large, P-R drag affected, $\beta$ critical, and $\beta$
meteoroid) depending on the dominant physical processes affecting
their evolution, and how the spatial distribution of material in each of
these categories is related.
The relative proportion of each of these categories in a disk depends on
its evolutionary status (and in particular on the disk's density);
e.g., the thermal emission from a relatively old (tenuous) disk such as
the zodiacal cloud is dominated by emission from its P-R drag affected
particles.

This model allows predictions to be made about the evolution of a
particular system, but it is always constrained by the initial conditions
of that system (i.e., the outcome of the planetary formation process).
The outcome of the planetary formation process is not generally known,
and it is one of the primary motives behind circumstellar disk
observations to infer that outcome.
The effect of the gravitational perturbations from a planetary system
on the dynamical evolution of a disk's particles was included in the model,
and it was shown how the planets cause certain signatures to be imprinted
on the disk's structure.
The signatures that were discussed are the same as those that are
observed in the structure of the zodiacal cloud:
an offset center of symmetry, a warp, and asymmetric resonant ring
structures co-orbiting with individual planets.

The model can be used to interpret debris disk observations in a
manner that can provide quantitative information about the physics
of that disk, and in particular about the outcome of the planetary
formation process in that system.
The necessary modeling techniques were demonstrated by their application
to observations of the HR 4796 disk (W99; T2000).
This disk provides an important opportunity for studying the planetary
formation process, since the system's age, $\sim 10$ Myr, places it at an
evolutionary epoch when any planetary system should be substantially formed.
Both the mid-IR and the near-IR emission from this disk has been shown to
be concentrated in two lobes, one either side of the star.
This indicates that the disk is being observed nearly edge-on, and that its
inner region ($\sim 40$ AU in radius, the same size as the solar system)
is almost completely devoid of dust.
The observations of T2000 show another interesting
feature: the mid-IR lobes appear to be of unequal brightness.
This dissertation presented a model of the T2000 observations that not
only reproduced the large-scale structure of the disk's
18.2 $\mu$m emission, but also showed that the lobe brightness asymmetry
is to be expected if there is another body orbiting HR 4796A that is
on an eccentric orbit, since the long-term effect of the gravitational
perturbations from this body would force the center of symmetry of the disk
to be offset from the star in a direction away from the pericenter of
the body's orbit, thus causing the dust near this forced pericenter to glow;
this model was previously presented in W99.

The modeling showed that a forced eccentricity as small as 0.02 is all that
is necessary to cause the observed 5\% lobe brightness asymmetry.
This forced eccentricity could have been caused by perturbations from either
the binary companion, HR 4796B, or from an unseen planet orbiting close to the
inner edge of the disk.
If the eccentricity of the orbit of HR 4796B is larger than 0.13, then a
forced eccentricity of 0.02 is to be expected.
However, if there is a planet of mass $> 0.1M_J$ located close
to the inner edge of the disk, then the forced eccentricity, and hence the
asymmetry, imposed on material in the disk's lobes
is controlled by the planet rather than the binary companion.
If a forced eccentricity is indeed the cause of the observed lobe
asymmetry then observations that constrain the orbit of HR 4796B
would help to clarify whether such a planet exists.
If the HR 4796 system had no binary companion, a forced eccentricity
of 0.02 could have been imposed on the disk by a lone planet with a
mass of $>10M_\oplus$, and an eccentricity of $> 0.02$ orbiting at 47 AU
from HR 4796A.

The statistical significance of the HR 4796 disk's lobe asymmetry in the
observations of T2000 is only at the $1.8 \sigma$ level, however, it is also
apparent in the observations of other authors (\cite{krwb98}; \cite{ssbk99}).
Also, it would take one good night on a 10 meter telescope to get a clear
observation of the HR 4796 asymmetry.
Thus, the signatures of even small mass planets orbiting far from
the star that are hiding in circumstellar disks are observable with current
technology, and these modeling techniques have the potential to be used to
infer the existence of otherwise undetectable planetary systems.
This is particularly important, since the direct detection of planets
around even nearby stars is well beyond current capabilities
(\cite{back98}), and indirect detection techniques permit detection
only of very massive planets that are close to the star
(\S \ref{s-search}; see Fig.~\ref{figmarcyplanets}).
This could be the only currently available method of detecting terrestrial-mass
extrasolar planets.

If there is a massive (e.g., $>0.1M_J$) planet close to the inner edge
of the disk, then the offset asymmetry would not be the only planetary
signature that we would expect to observe in the disk's structure.
Many of the disk's particles could be trapped in resonance with that
planet, thus forming a resonant ring, and the disk could be warped.
Such a resonant ring would give the inner edge of the disk structure
that orbits the star with the same orbital period as that of the planet
(i.e., $\sim 200$ years).
This structure could be contributing to the observed lobe asymmetry,
and may also be present in the residuals of the observation.
These possibilities could be explored with further observations and
modeling.

Because of the physical nature of the modeling process, it revealed
much information about the disk system, and in particular about the dynamic
properties of its emitting particles.
Assuming the particles to be astronomical silicate Mie spheres, the
diameter of the mid-IR emitting particles was estimated to be
$D_{typ} = 2-3$ $\mu$m.
Particles this small have radiation forces that are characterized by
$\beta \approx 2$, and so are blown out of the system on hyperbolic orbits on
timescales of $\sim 370$ years (i.e., they are the disk's $\beta$ meteoroids).
However, if the particles had been assumed to be organic refractory Mie
spheres, their diameter would have been estimated to be $12-20$ $\mu$m, with
a $\beta$ of between 0.2 and 0.5, implying that these particles are the
disk's $\beta$ critical particles.
From a theoretical point of view, the disk's inferred optical depth,
$\tau_{eff}$(70 AU) $\approx 5 \times 10^{-3}$, supports the view that the
emitting particles are the disk's $\beta$ critical particles,
however, no firm conclusion about whether the emitting particles are
$\beta$ meteoroids or $\beta$ critical particles can be reached without
further observations and modeling.
Other conclusions about the dynamic HR 4796 disk were that, because
the HR 4796 disk is very dense, and the collisional lifetime of its
emitting particles is $\sim 10^4$ years, these particles cannot be
primordial, rather they must be continuously created from a reservoir
of larger particles.
Also, because the collisional lifetimes of all of the disk's particles
are shorter than their P-R drag lifetimes, none of the disk's particles
are affected by P-R drag.

Thus, the HR 4796 disk modeling demonstrates the current status of our
interpretation of debris disk observations:
the complexity of these systems far outstretches the scope of the
observations;
i.e., there is too little information about these disks to be able to
completely infer their physics, rather conclusions are necessarily
speculative.
This physical model is a step forward, however, since it puts the conclusions
on a firm physical basis.

%%%%%%%%%%%%%%%%%%%%%%%%%%%%%%%%%%%%%%%%%%
\section{Further Work}
\label{s-furtherwork}
Since the field of imaging debris disk structure is so young, many more
disks with resolved structure are likely to be discovered in the coming
years.
With 10 meter class telescopes such as Gemini becoming widely available,
the quality of these disk observations will be similar to that of the
observations described in T2000.
Such observations are bound to test the limits of the model for
the evolution of debris disks that is described in this dissertation.
Many ways have already been discussed to develop the model for
the HR 4796 disk, and these are summarized in \S \ref{ss-hr4796}.
Since each disk is unique in that it describes the outcome of the planetary
formation process at a different epoch and with different starting conditions,
it will test the model in a unique way.
Thus, the developments that will be necessary to the theoretical model
cannot be predicted, since they will follow the observations.
\S \ref{ss-otherdisks} describes just two debris disks
to which this modeling technique can be applied and
against which it can be tested.
With a better understanding of the evolution of debris disks, and the
possibility of indirect detections of terrestrial mass planets, this
modeling technique should allow a greater understanding of
the formation and evolution of planetary systems.

%%%%%%%%%%%%%%%%%%%%%%%%%%%%%%%%%%%%%%%%%%
\subsection{HR 4796}
\label{ss-hr4796}
Several observational tests and model improvements have already been
discussed that would provide a clearer understanding of the
HR 4796 disk:
\begin{itemize}
  \item First, more accurate observations of the disk emission at 10 and 18
    $\mu$m are needed.
    One good night on a 10 meter telescope (preferably one without an
    asymmetric PSF like the Keck telescope) would suffice for this.
    This would allow the following:
    \begin{itemize}
      \item to unconditionally confirm the lobe asymmetry and its magnitude;
      \item to confirm a radial offset asymmetry for the lobes that agrees with
        the lobe asymmetry;
      \item to test for the presence of a disk warp, which could
        be modeled using techniques similar to those described in this
        disseration (and W99);
      \item and to determine the residuals with sufficient accuracy for them
        to be modeled, possibly as resonant structure associated with a
        planet at the inner edge of the disk.
    \end{itemize}
    Residual structure associated with a planet at the inner edge
    of the disk would co-orbit with the planet.
    If the lobe asymmetry, even in part, was caused by such resonant
    structure, its magnitude would vary with a period of $\sim 200$ years.
    Subsequent observations at intervals of $\sim 3$ years would
    be able to ascertain whether the residual structure is truly associated
    with an orbiting planet, since the structure would have moved by one
    pixel on the image in this timescale. 
  \item Either new observations, or careful study of archived observations,
    of the separation and position angle of HR 4796B relative to HR 4796A
    are needed to determine the orbit of this binary system.
    This would allow a much more unambiguous interpretation of the
    lobe asymmetry, and of any potential warp.
  \item The model needs to include a distribution of particle sizes,
    and a better model for the optical properties of the particles
    (including their scattering properties).
    Possible physical models for these were discussed in this dissertation.
    These can be constrained using the disk's SED (see, e.g., \cite{almp99})
    and its mid-IR spectrum.
    This will allow a better understanding of the dynamics of the emitting
    particles, and is vital if we are to constrain the model for the
    evolution of debris disks.
    Ideally, a disk model should be able to explain images and
    spectra at all wavelengths simultaneously.
  \item With a better understanding of the dynamics of the emitting particles,
    it may be necessary to expand the theoretical model to model the
    dynamics of particles that are on hyperbolic or highly eccentric orbits.
    The theory for this has already been described in the dissertation, however,
    it has not yet been translated into the realm of the modeling.
\end{itemize}

%%%%%%%%%%%%%%%%%%%%%%%%%%%%%%%%%%%%%%%%%%
\subsection{Other Disks}
\label{ss-otherdisks}

%%%%%%%%%%%%%%%%%%%%%%%%%%%%%%%%%%%%%%%%%%
\subsubsection{Fomalhaut}
\label{ss-fomalhaut}

%%%% Figure of SCUBA images of Fomalhaut and E Eri
\begin{figure}
  \begin{center}
    \begin{tabular}{rccl}
      \textbf{(a)} & \hspace{-0.45in}
      \epsfig{file=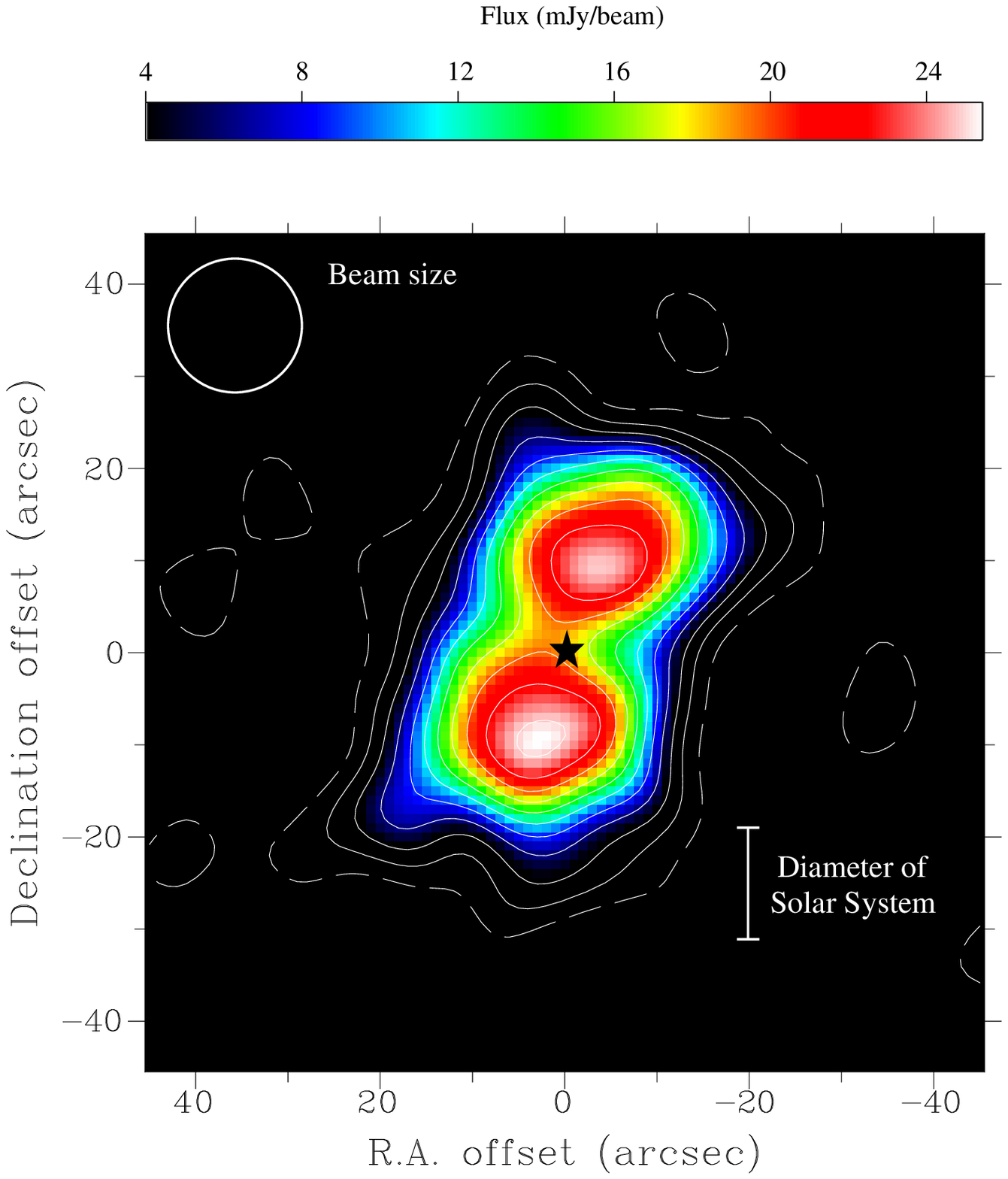,height=3.2in} &
      \epsfig{file=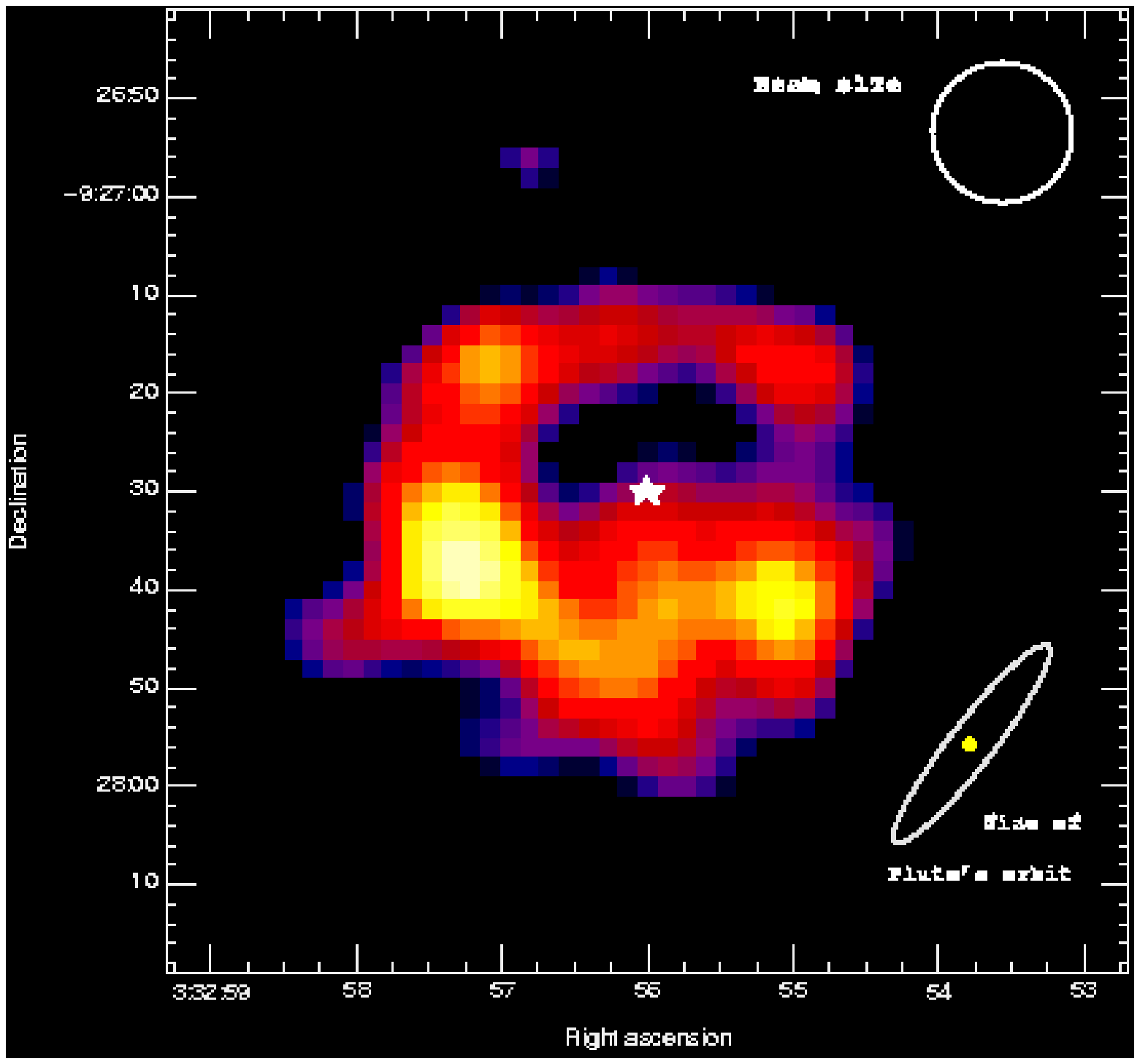,height=2.667in} &
      \textbf{(b)} \\
    \end{tabular}
  \end{center}
  \caption{Submillimeter, 850 $\mu$m, images of the disks of dust
  around (\textbf{a}) Fomalhaut (\cite{hgzw98}) and (\textbf{b})
  $\epsilon$ Eridani (\cite{ghmj98}).}
  \label{figfomeeri}
\end{figure}

A double-lobed feature similar to that observed in the HR 4796 disk
has also been observed in the disk around the star Fomalhaut (\cite{hgzw98};
see Fig.~\ref{figfomeeri}a).
A similar explanation (planetary formation) has been proposed for its origin
(\cite{hgzw98}).
The brightnesses of the Fomalhaut disk's lobes also appear to be
asymmetric (\cite{hgzw98}), but the statistical significance of this
asymmetry is low.
Fomalhaut is a wide visual binary system.
Gliese 879 is Fomalhaut's common proper motion companion
(\cite{bshb97});
the two stars are separated by $\sim 2^\circ$, which corresponds to a
projected separation of $O$(55,000 AU) at $7.7$ pc.
At such a distance, the forced eccentricity imposed on the disk
by the binary star is insignificant (see eq.~[\ref{eq:zf1}]).
A secular perturbation offset asymmetry in this disk would be expected
only if there is a planet in the disk that has a non-circular orbit.
This disk could be modeled using the same techniques that were used
to model the HR 4796 disk.

%%%%%%%%%%%%%%%%%%%%%%%%%%%%%%%%%%%%%%%%%%%%
\subsubsection{$\epsilon$ Eridani}
\label{ss-eeri}
$\epsilon$ Eridani is one of the prototype debris disks.
It was imaged at 850 $\mu$m by SCUBA at JCMT, and
these observations show that the disk's structure is clumpy (\cite{ghmj98};
see Fig.~\ref{figfomeeri}b).
In fact, the observations seem to show the tri-lobed structure
expected from an analogue Kuiper belt the particles of which are
trapped in the 1:2 and 2:3 resonances of a planet located at the
inner edge of the disk (see \S \ref{ss-restrapkb}).
Modeling of this structure will reveal whether this is a plausible
explanation for the observed structure.
It would be very exciting if resonant rings were the predominantly
observable structures of some exosolar systems, since such rings
would be good indirect indications of extrasolar planets.

%%%%%%%%%%%%%%%%%%%%%%%%%%%%%%%%%%%%%%%%%%%%%
%\subsubsection{HD 141569}
%\label{ss-hd141569}
%The disk around the star HD 141569 has been imaged by NICMOS
%(\cite{wbss99}) and resolved in the mid-IR (\cite{ftpk99}).

%%%% Appendixes (optional)
\ifx\IncludeApp\True
  \DoubleSpace
  \addcontentsline{toc}{startchapter}{\ShiftHack APPENDIXES}
  \appendix
  \input{appendixes.tex}
\fi

%%%% References
%\pagebreak[4]
\SingleSpace
\addcontentsline{toc}{chapter}{REFERENCES}
\bibliography{thesis}

\begin{thebibliography}{}

\bibitem[Artymowicz \& Lubow 1994]{al94} Artymowicz, P., \& Lubow, S. H.
  1994, \apj, 421, 651
\bibitem[Artymowicz 1997]{arty97} Artymowicz, P. 1997,
  Annu. Rev. Earth Planet. Sci., 25, 175
\bibitem[Artymowicz \& Clampin 1997]{ac97} Artymowicz, P., \& Clampin, M.
  1997, \apj, 490, 863
\bibitem[Augereau et al. 1999]{almp99} Augereau, J. C., Lagrange, A. M.,
  Mouillet, D., Papaloizou, J. C. B., \& Grorod, P. A. 1999, \aap, 348, 557
\bibitem[Aumann 1985]{auma85} Aumann, H. H. 1985, \pasp, 97, 885 
\bibitem[Aumann et al. 1984]{agbj84} Aumann, H. H., Gillett, F. C.,
  Beichman, C. A., De Jong, T., Houck, J. R., Low, F. J.,
  Neugebauer, G., Walker, R. G., \& Wesselius, P. R. 1984, \apj, 278, L23
\bibitem[Backman 1998]{back98} Backman, D. E. 1998, in Exozodiacal
  Dust Workshop Conference Proceedings, eds. D. E. Backman, L. J. Caroff,
  S. A. Sandford, \& D. H. Wooden (NASA/CP-1998-10155), 13
\bibitem[Backman \& Paresce 1993]{bp93} Backman, D. E., \& Paresce, F.
  1993, in Protostars \& Planets III,
  eds E. H. Levy, \& J. Lunine (Tucson: Univ. Ariz. Press), 1253
\bibitem[Backman et al. 1995]{bds95} Backman, D. E.,
  Dasgupta, A., \& Stencel, R. E. 1995, \apjl, 450, L35
\bibitem[Barrado Y Navascu\'{e}s et al. 1997]{bshb97} Barrado Y
  Navascu\'{e}s, D., Stauffer, J. R., Hartmann, L., \& Balachandran, S. C.
  1997, \apj, 475, 313
\bibitem[Barrado Y Navascu\'{e}s et al. 1999]{bssc99} Barrado Y Navascu\'{e}s,
  D., Stauffer, J. R., Song, I., \& Caillault, J.-P. 1999, \apj, 520, L123
\bibitem[Beaug\'{e} \& Ferraz-Mello 1994]{bf94} Beaug\'{e}, C., \&
  Ferraz-Mello, S. 1994, Icarus, 110, 239
\bibitem[Beckwith 1999]{beck99} Beckwith, S. V. W. 1999,
  to appear in The Physics of Star Formation and Early Evolution II,
  eds. C. J. Lada, \& N. D. Kylafis (astro-ph/9905003)
\bibitem[Beckwith et al. 1999]{bhn99} Beckwith, S. V. W., Henning, T.,
  \& Nakagawa, Y. 1999,
  to appear in Protostars \& Planets IV (astro-ph/9902241)
\bibitem[Beckwith \& Sargent 1996]{bs96} Beckwith, S. V. W., \& Sargent,
  A. I. 1996, \nat, 383, 139
\bibitem[Beckwith et al. 1990]{bscg90} Beckwith, S. V. W., Sargent, A. I.,
  Chini, R. S., \& Gusten, R. 1990,\aj, 99, 924
\bibitem[Beichman 1996]{beic96} Beichman, C. A. 1996, A Road Map for the
  Exploration of Neighboring Planetary Systems (ExNPS) (JPL Publ.
  96-22, Tech. Rep., Pasadena, CA)
\bibitem[Berkhuijsen et al. 1971]{bhs71} Berkhuijsen, E. M., Halsam, C. G. T.,
  \& Salter, C. J. 1971, \aap, 14, 252
\bibitem[Black 1997]{blac97} Black, D. C. 1997, \apjl, 490, L171
\bibitem[Blum et al. 1999]{bwph98} Blum, J., Wurm, G., Poppe, T., \& Heim,
  L.-O. 1999, in Laboratory Astrophysics and Space Research, Astrophysics and
  Space Science Library, eds. P. Ehrenfreund, K. Krafft, H. Kochan, \& V.
  Pirronello (Dordrecht: Kluwer), 236, 399
\bibitem[Bohren \& Huffman 1983]{bh83} Bohren, C. F., \& Huffman, D. R. 1983,
  Absorption and Scattering of Light by Small Particles (New York: Wiley)
\bibitem[Boss 1995]{boss95} Boss, A. P. 1995, Science, 267, 360 
\bibitem[Bowell 1996]{bowe96} Bowell, E. 1996, Asteroid Orbital
  Elements Database (Lowell Observatory)
\bibitem[Brouwer \& Clemence 1961]{bc61} Brouwer, D., \& Clemence, G. M.
  1961, Methods of Celestial Mechanics (New York: Academic Press)
\bibitem[Burns et al. 1979]{bls79} Burns, J. A., Lamy, P. L.,
  \& Soter, S. 1979, Icarus, 40, 1
\bibitem[Burrows et al. 1996]{bswk96} Burrows, C. J., Stapelfeldt, K. R.,
  Watson, A. M., Krist, J. E., Ballester, G. E., Clarke, J. T., Crisp,
  D., Gallagher, J. S. III, Griffiths, R. E., Hester, J. J., Hoessel,
  J. G., Holtzman, J. A., Mould, J. R., Scowen, P. A., Trauger, J. T.,
  Westphal, J. A. 1996, \apj, 473, 437
\bibitem[Cameron 1997]{came97} Cameron, A. G. W. 1997, Icarus, 126, 126
\bibitem[Clarke \& Pringle 1991]{cp91} Clarke, C. J., \& Pringle, J. E.
  1991, \mnras, 249, 588
\bibitem[de Geus et al. 1989]{ddl89} de Geus, E. J., de Zeeuw, P. T., \&
  Lub, J. 1989, \aap, 216, 44
\bibitem[Dent et al. 1998]{dmw98} Dent, W. R. F., Matthews, H. E., \&
  Ward-Thompson, D. 1998, \mnras, 301, 1049 
\bibitem[Dermott et al. 1992]{dgdg92} Dermott, S. F., Gomes, R. S.,
  Durda, D. D., Gustafson, B. \AA. S., Jayaraman, S., Xu, Y. L., 
  \& Nicholson, P. D. 1992, in Chaos, Resonance and Collective
  Dynamical Phenomena in the Solar System, ed. S. Ferraz-Mello
  (Dordrecht: Kluwer), IAU Symposium, 152, 333
\bibitem[Dermott et al. 2000]{dggh99} Dermott, S. F., Grogan, K.,
  Gustafson, B. \AA. S., Holmes, E. K., Jayaraman, S., \& Wyatt, M. C.
  2000, in Interplanetary Dust, eds. E. Grun, B. \AA. S. Gustafson,
  S. F. Dermott, \& H. Fechtig (Tucson: Univ. Ariz. Press),
  in press
\bibitem[Dermott et al. 1999]{dghk99} Dermott, S. F., Grogan, K.,
  Holmes, E. K., \& Kortenkamp, S. 1999, in Formation and Evolution of
  Solids in Space, eds. J. M. Greenberg, \& A. Li (Dordrecht: Kluwer), 565
\bibitem[Dermott et al. 1998]{dghw98} Dermott, S. F., Grogan, K.,
  Holmes, E. K., \& Wyatt, M. C. 1998, in Exozodiacal Dust Workshop
  Conference Proceedings, eds. D. E. Backman, L. J. Caroff, S. A. Sandford,
  \& D. H. Wooden (NASA/CP-1998-10155), 59
\bibitem[Dermott et al. 1996]{djxg96} Dermott, S. F., Jayaraman, S.,
  Xu, Y. L., Grogan, K., \& Gustafson, B. \AA. S. 1996, in Unveiling the
  Cosmic Infrared Background, ed. E. Dwek (New York: AIP), 25
\bibitem[Dermott et al. 1994]{djxg94} Dermott, S. F., Jayaraman, S.,
  Xu, Y. L., Gustafson, B. \AA. S., \& Liou, J. C. 1994, \nat, 369, 719
\bibitem[Dermott et al. 1988]{dmm88} Dermott, S. F., Malhotra, R., \&
  Murray, C. D. 1988, Icarus, 76, 295
\bibitem[Dermott \& Murray 1983]{dm83} Dermott, S. F., \& Murray, C. D.
  1983, \nat, 301, 201
\bibitem[Dermott \& Nicholson 1986]{dn86} Dermott, S. F., \& Nicholson, P. D.
  1986, \nat, 319, 115
\bibitem[Dermott et al. 1984]{dnbh84} Dermott, S. F., Nicholson, P. D.,
  Burns, J. A., \& Houck, J. R. 1984, \nat, 312, 505
\bibitem[Dermott et al. 1985]{dnbh85} Dermott, S. F., Nicholson, P. D.,
  Burns, J. A., \& Houck, J. R. 1985, in Properties and Interactions
  of Interplanetary Dust, eds. R. H. Giese, \& P. Lamy (Dordrecht: Reidel),
  395
\bibitem[Dohnanyi 1969]{dohn69} Dohnanyi, J. S. 1969, J. Geophys. Res.,
  74, 2531
\bibitem[Dominik \& Tielens 1997]{dt97} Dominik, C., \& Tielens, A. G. G.
  M. 1997, \apj, 480, 647
\bibitem[Draine \& Lee 1984]{dl84} Draine, B. T., \& Lee, H. M. 1984,
  \apj, 285, 89
\bibitem[Duncan et al. 1989]{dqt89} Duncan, M., Quinn, T., \&
  Tremaine, S. 1989, Icarus, 82, 402
\bibitem[Dunkin et al. 1997]{dbr97} Dunkin, S. K., Barlow, M. J., \& Ryan, S.
  G. 1997, \mnras, 286, 604
\bibitem[Durda \& Dermott 1997]{dd97} Durda, D. D., \& Dermott, S. F. 1997,
  Icarus, 130, 140
\bibitem[Durda et al. 1998]{dgj98} Durda, D. D., Greenberg, R., \& Jedicke,
  R. 1998, Icarus, 135, 431
\bibitem[Egger \& Aschenbach 1995]{ea95} Egger, R. J., \& Aschenbach, B. 1995,
  \aap, 294, L25
\bibitem[Fajardo-Acosta et al. 1993]{ftk93} Fajardo-Acosta,
  S. B., Telesco, C. M., \& Knacke, R. F. 1993, \apj, 417, L33
\bibitem[Farinella et al. 1998]{fvh98} Farinella, P., Vokrouhlicky, D.,
  \& Hartmann, W. K. 1998, Icarus, 132, 378
\bibitem[Ferlet 1999]{ferl99} Ferlet, R. 1999, \aapr, 9, 153
\bibitem[Fern\'{a}ndez \& Ip 1984]{fi84} Fern\'{a}ndez, J. A., \&
  Ip, W.-H. 1984, Icarus, 58, 109
\bibitem[Fisher et al. 2000]{ftpk99} Fisher, R. S., Telesco, C. M.,
  Pi\~{n}a, R. K., Knacke, R. F., \& Wyatt, M. C. submitted, \apj, 2000
\bibitem[Gatewood 1987]{gate87} Gatewood, G. 1987, \aj, 94, 213
\bibitem[Gerbaldi et al. 1999]{gfbd99} Gerbaldi, M., Faraggiana, R., Burnage,
  R., Delmas, F., G\'{o}mez, A. E., \& Grenier, S. 1999, \aaps, 137, 273
\bibitem[Gold 1975]{gold75} Gold, T. 1975, Icarus, 25, 489
\bibitem[Goldreich \& Tremaine 1980]{gt80} Goldreich, P., \& Tremaine, S.
  1980, \apj, 241, 425
\bibitem[Gomes 1997a]{gome97a} Gomes, R. S. 1997a, \aj, 114, 396
\bibitem[Gomes 1997b]{gome97b} Gomes, R. S. 1997b, \aj, 114, 2166
\bibitem[Gonczi et al. 1982]{gff82} Gonczi, R., Froeschle, Ch., \&
  Froeschle, C. 1982, Icarus, 51, 633
\bibitem[Gorkavyi et al. 1997]{gomt97} Gorkavyi, N. N., Ozernoy, L. M.,
  Mather, J. C., \& Taidakova, T. 1997, \apj, 488, 268
\bibitem[Greaves et al. 1998]{ghmj98} Greaves, J. S., Holland, W. S.,
  Moriarty-Schieven, G., Jenness, T., Dent, W. R. F., Zuckerman, B.,
  McCarthy, C., Webb, R. A., Butner, H. M., Gear, W. K., \& Walker, H. J.
  1998, \apj, 506, L133
\bibitem[Greaves et al. 1999]{gmh99} Greaves, J. S., Mannings, V., \&
  Holland, W. S. submitted, Icarus, 1999
\bibitem[Greenberg 1998]{gree98} Greenberg, J. M. 1998, \aap, 330, 375
\bibitem[Greenberg \& Gustafson 1981]{gg81} Greenberg, J. M., \&
  Gustafson, B. \AA. S. 1981, \aap, 93, 35
\bibitem[Greenberg \& Hage 1990]{gh90} Greenberg, J. M., \& Hage, J. I.
  1990, \apj, 361, 260
\bibitem[Grogan et al. 2000]{gdd00} Grogan, K., Dermott, S. F.,
  Durda, D. D. submitted, Icarus, 2000
\bibitem[Grogan et al. 1996]{gdg96} Grogan, K., Dermott, S. F., \&
  Gustafson, B. \AA. S. 1996, \apj, 472, 812
\bibitem[Grogan et al. 1997]{gdjx97} Grogan, K., Dermott, S. F.,
  Jayaraman, S., \& Xu, Y. L. 1997, \planss, 45, 1657
\bibitem[Gr\"{u}n et al. 1985]{gzfg85} Gr\"{u}n, E., Zook, H. A.,
  Fechtig, H., \& Giese, R. H. 1985, Icarus, 62, 244
\bibitem[Gurnett et al. 1997]{gakg97} Gurnett, D. A., Ansher, J. A., Kurth,
  W. S., \& Granroth, L. J. 1997, Geo. Res. Lett., 24, 3125
\bibitem[Gustafson 1994]{gust94} Gustafson, B. \AA. S. 1994,
  Annu. Rev. Earth Planet. Sci., 22, 553
\bibitem[Hage \& Greenberg 1990]{hg90} Hage, J. I., \& Greenberg, J. M.
  1990, \apj, 361, 251
\bibitem[Hahn \& Malhotra 1999]{hm99} Hahn, J. M., \& Malhotra, R. 1999,
  \aj, 117, 3041
\bibitem[Hayashi 1981]{haya81} Hayashi, C. 1981, Prog. Theor. Phys. Suppl.,
  70, 35
\bibitem[Henning \& Stognienko 1996]{hs96} Henning, Th., \& Stognienko,
  R. 1996, \aap, 311, 291
\bibitem[Henry et al. 2000]{hmbv00} Henry, G. W., Marcy, G. W., Butler,
  R. P., \& Vogt, S. S. 2000, \apjl, in press
\bibitem[Hirayama 1918]{hira18} Hirayama, K. 1918, \aj, 31, 185
\bibitem[Holland et al. 1998]{hgzw98} Holland, W. S., Greaves, J. S.,
  Zuckerman, B., Webb, R. A., McCarthy, C., Coulson, I. M., Walther, D. M.,
  Dent, W. R. F., Gear, W. K., \& Robson, I. 1998, \nat, 392, 788
\bibitem[Ishiguro 1999]{ishi99} Ishiguro, M. 1999, December 8, Zodiacal
  Light, http://zodi.planet.sci.kobe-u.ac.jp/$\sim$ishiguro
\bibitem[Jackson \& Zook 1989]{jz89} Jackson, A. A., \& Zook, H. A. 1989,
  \nat, 337, 629
\bibitem[Jayaraman \& Dermott 2000]{jd99} Jayaraman, S., \& Dermott, S. F.
  submitted, Icarus, 2000
\bibitem[Jayawardhana et al. 1998]{jfht98} Jayawardhana, R., Fisher, R. S.,
  Hartmann, L., Telesco, C. M., Pi\~{n}a, R. K., \& Fazio, G. 1998,
  \apj, 503, L79
\bibitem[Jayawardhana et al. 1999]{jhff99} Jayawardhana, R., Hartmann, L.,
  Fazio, G., Fisher, R. S., Telesco, C. M., \& Pi\~{n}a, R. K. 1999,
  \apjl, 521, L129
\bibitem[Jewitt 1999]{jewi99} Jewitt, D. C. 1999, AREPS, in press
\bibitem[Jura 1991]{jura91} Jura, M. 1991, \apjl, 383, L79
\bibitem[Jura et al. 1995]{jgwm95} Jura, M., Ghez, A. M., White, R. J.,
  McCarthy, D. W., Smith, R. C., \& Martin, P. G. 1995, \apj, 445, 451 
\bibitem[Jura et al. 1998]{jmwt98} Jura, M., Malkan, M., White, R.,
  Telesco, C. M., Fisher, R. S., \& Pi\~{n}a, R. K. 1998, \apj, 505, 897
\bibitem[Jura et al. 1993]{jzbs93} Jura, M., Zuckerman, B., Becklin, E. E.,
  \& Smith, R. C. 1993, \apj, 418, L37
\bibitem[Kalas 1998]{kala98} Kalas, P. 1998, Science, 281, 182
\bibitem[Kalas \& Jewitt 1995]{kj95} Kalas, P., \& Jewitt, D. C. 1995, \aj,
  110, 794
\bibitem[Kelsall et al. 1998]{kwfr98} Kelsall, T., Weiland, J. L.,
  Franz, B. A., Reach, W. T., Arendt, R. G., Dwek, R., Freudenreich, H. T.,
  Hauser, M. G., Moseley, S. H., Odegard, N. P., Silverberg, R. F., \&
  Wright, E. L. 1998, \apj, 508, 44
\bibitem[Kenyon \& Hartmann 1987]{kh87} Kenyon, S. J., \& Hartmann, L. W. 1987,
  \apj, 323, 714
\bibitem[Kenyon \& Luu 1999]{kl99} Kenyon, S. J., \& Luu, J. X. 1999, \aj,
  118, 1101
\bibitem[Kenyon et al. 1999]{kwww99} Kenyon, S. J., Wood, K., Whitney, B. A.,
  \& Wolff, M. J. 1999, \apjl, 524, L119
\bibitem[Kessler 1981]{kess81} Kessler, D. J. 1981, Icarus, 48, 39
\bibitem[Kimura \& Mann 1998]{km98} Kimura, H., \& Mann, I. 1998, \apj,
  499, 454
\bibitem[Kirkwood 1876]{kirk76} Kirkwood, D. 1876, On the Distribution
  of the Asteroids (Salem, MA: Salem Press)
\bibitem[Koerner et al. 1998]{krwb98} Koerner, D. W., Ressler, M. E.,
  Werner, M. W., \& Backman, D. E. 1998, \apj, 503, L83
\bibitem[Kortenkamp \& Dermott 1998]{kd98b} Kortenkamp, S. J.,
  \& Dermott, S. F. 1998, Icarus, 135, 469
\bibitem[Lachaume et al. 1999]{ldlh99} Lachaume, R., Dominik, C., Lanz, T.,
  \& Habing, H. J. 1999, \aap, 348, 897
\bibitem[Lagrange et al. 1987]{lfv87} Lagrange, A.-M., Ferlet, R., \&
  Vidal-Madjar, A. 1987, \aap, 173, 289
\bibitem[Laor \& Draine 1993]{ld93} Laor, A., \& Draine, B. T. 1993, \apj,
  402, 441
\bibitem[Lazzaro et al. 1994]{lsrg94} Lazzaro, D., Sicardy, B., Roques, F.,
  Greenberg, R. 1994, Icarus, 108, 59
\bibitem[Lecar \& Franklin 1997]{lf97} Lecar, M., \& Franklin, F. 1997,
  Icarus, 129, 134
\bibitem[Lecavelier des Etangs et al. 1995]{ldvf95} Lecavelier des Etangs,
  A., Deleuil, M., Vidal-Madjar, A., Ferlet, R., Nitschelm, C., 
  1995, \aap, 299, 557
\bibitem[Lecavelier des Etangs et al. 1998]{lvf98} Lecavelier des Etangs,
  A., Vidal-Madjar, \& Ferlet, R. 1998, \aap, 339, 477
\bibitem[Leinert \& Gr\"{u}n 1990]{lg90} Leinert, C., \& Gr\"{u}n, E. 1990,
  in Space and Solar Physics, Vol. 20, Physics and Chemistry in Space:
  Physics of the Inner Heliosphere I, eds. R. Schween, \& E. Marsch
  (Berlin: Springer), 207
\bibitem[Leisawitz et al. 1989]{lbt89} Leisawitz, D., Bash, F. N., \&
  Thaddeus, P. 1989, \apjs, 70, 731
\bibitem[Levison \& Duncan 1997]{ld97} Levison, H. F., \& Duncan, M. J. 1997,
  Icarus, 127, 13
\bibitem[Li \& Greenberg 1997]{lg97} Li, A., \& Greenberg, J. M. 1997, \aap,
  323, 566
\bibitem[Li \& Greenberg 1998]{lg98} Li, A., \& Greenberg, J. M. 1998, \aap,
  331, 291
\bibitem[Lin \& Papaloizou 1979]{lp79} Li, D. N. C., \& Papaloizou, J. C. B.
  1979, \mnras, 186, 799
\bibitem[Liou 1993]{liou93} Liou, J. C. 1993, Ph.D. Thesis, Univ. Florida
\bibitem[Liou et al. 1995]{ldx95} Liou, J. C., Dermott, S. F., \& Xu, Y. L.
  1995, \planss, 43, 717
\bibitem[Liou \& Zook 1999]{lz99} Liou, J. C., \& Zook, H. A. 1999, \aj,
  118, 580
\bibitem[Liou et al. 1996]{lzd96} Liou, J. C., Zook, H. A., \& Dermott,
  S. F. 1996, Icarus, 124, 429
\bibitem[Liseau 1999]{lise99} Liseau, R. 1999, \aap, 348, 133
\bibitem[Lissauer 1993]{liss93} Lissauer, J. J. 1993, \araa, 31, 129
\bibitem[Love \& Brownlee 1993]{lb93} Love, S. G., \& Brownlee, D. E. 1993,
  Science, 262, 550
\bibitem[Low et al. 1984]{lbgg84} Low, F. J., Beintema, D. A., Gautier, T. N.,
  Gillet, F. C., Beichman, C. A., Neugebauer, G., Young, E., Aumann, H. H.,
  Rowan-Robinson, M., Soifer, B. T., Walker, R. G., \&  Wesselius, P. R.
  1984, \apjl, 278, L19
\bibitem[Malhotra 1995]{malh95} Malhotra, R. 1995, \aj, 110, 420
\bibitem[Malhotra 1996]{malh96} Malhotra, R. 1996, \aj, 111, 504
\bibitem[Mannings \& Barlow 1998]{mb98b} Mannings, V., \& Barlow, M. J. 1998,
  \apj, 497, 330
\bibitem[Marcy \& Butler 1998]{mb98} Marcy, G. W., \& Butler, R. P. 1998,
  \araa, 36, 57
\bibitem[Marcy et al. 1999]{mbvf99} Marcy, G. W., Butler, R. P., Vogt,
  S. S., \& Fischer, D. 1999, December 8, Discovery of Extrasolar Planets,
  http://cannon.sfsu.edu/$\sim$gmarcy/planetsearch
\bibitem[Mayor \& Queloz 1995]{mq95} Mayor, M., \& Queloz, D. 1995,
  \nat, 378, 355
\bibitem[Mouillet et al. 1997a]{mlbr97} Mouillet, D., Lagrange, A.-M.,
  Beuzit, J.-L., \& Renaud, N. 1997, \aap, 324, 1083
\bibitem[Mouillet et al. 1997b]{mlpl97} Mouillet, D., Larwood, J. D.,
  Papaloizou, J. C. B., \& Lagrange, A. M. 1997, \mnras, 292, 896
\bibitem[Murray \& Dermott 1999]{md99} Murray, C. D., \& Dermott, S. F.
  1999, Solar System Dynamics (Cambridge: Cambridge University Press)
\bibitem[Padgett et al. 1999]{pbss99} Padgett, D. L., Brandner, W., Stapelfeldt,
  K. R., Strom, S. E., Terebey, S., \& Koerner, D. 1999, \aj, 117, 1490
\bibitem[Pantin et al. 1997]{pla97} Pantin, E., Lagage, P. O., \& Artymowicz,
  P. 1997, \aap, 327, 1123
\bibitem[Papaloizou \& Terquem 1999]{pt99} Papaloizou, J. C. B., \& Terquem,
  C. 1999, \apj, 521, 823
\bibitem[Pollack et al. 1994]{phbs94} Pollack, J. B., Hollenbach, D., Beckwith,
  S. V. W., Simonelli, D. P., Roush, T., \& Fong, W. 1994, \apj, 421, 615
\bibitem[Reach et al. 1995]{rfwh95} Reach, W. T., Franz, B. A.,
  Weiland, J. L., Hauser, M. G., Kelsall, T. N., Wright, E. L., Rawley, G.,
  Stemwedel, S. W., \& Spiesman, W. J. 1995, Nature, 374, 521
\bibitem[Roques et al. 1994]{rsss94} Roques, F., Scholl, H.,
  Sicardy, B., \& Smith, B. A. 1994, Icarus, 108, 37
\bibitem[Schneider et al. 1999]{ssbk99} Schneider, G., Smith, B. A.,
  Becklin, E. E., Koerner, D. W., Meier, R., Hines, D. C., Lowrance, P. J.,
  Terrile, R. J., Thompson, R. I., \& Rieke, M. 1999, \apjl, 513, L127
\bibitem[Schultz \& Heap 1998]{sh98} Schultz, A., \& Heap, S. 1998, January,
  3, Astronomers Have Found a New Twist in a Suspected Protoplanetary Disk,
  http://oposite.stsci.edu/pubinfo/pr/1998/03/
\bibitem[Shu et al. 1987]{sal87} Shu, F. H., Adams, F. C., \&
  Lizano, S. 1987, \araa, 25, 23
\bibitem[Shu et al. 1993]{sngo93} Shu, F. H., Najita, J., Galli, D.,
  Ostricker, E., \& Lizano, S. 1993, in Protostars \& Planets III,
  eds E. H. Levy, \& J. Lunine (Tucson: Univ. Ariz. Press), 3
\bibitem[Sitko et al. 1999]{sglr99} Sitko, M. L., Grady, C. A., Lynch, D. K.,
  Russell, R. W., \& Hanner, M. S. 1999, \apj, 510, 408
\bibitem[Skinner et al. 1995]{ssgb95} Skinner, C. J., Sylvester, R. J., Graham,
  J. R., Barlow, M. J., Meixner, M., Keto, E., Arens, J. F., \& Jernigan, J. G.
  1995, \apj, 444, 861
\bibitem[Smith \& Terrile 1984]{st84} Smith, B. A., \& Terrile, R. J. 1984,
  Science, 226, 1421
\bibitem[Soderblom et al. 1998]{sksn98} Soderblom, D. R., King, J. R., Siess,
  L., Noll, K. S., Gilmore, D. M., Henry, T. J., Nelan, E., Burrows, C. J.,
  Brown, R. A., Perryman, M. A. C., Benedict, G. F., McArthur, B. J., Franz,
  O. G., Wasserman, L. H., Jones, B. F., Latham, D. W., Torres, G., \&
  Stefanik, R. P. 1998, \apj, 498, 385 
\bibitem[Stauffer et al. 1995]{shb95}
  Stauffer, J. R., Hartmann, L. W., \& Barrado Y Navascu\'{e}s, B. 1995,
  \apj, 454, 910
\bibitem[Stern 1995]{ster95} Stern, S. A. 1995, \aj, 110, 856
\bibitem[Stern \& Colwell 1997]{sc97} Stern, S. A., \& Colwell, J. E. 1997,
  \apj, 490, 879
\bibitem[Strom et al. 1989]{sns89} Strom, K. M., Newton, G.,\& Strom,
  S. E. 1989, \apss, 71, 183S
\bibitem[Sykes et al. 1986]{slhl86} Sykes, M. V., Lebofsky, L. A.,
  Hunten, D. M., \& Low, F. J. 1986, Science, 232, 1115
\bibitem[Sykes 1990]{syke90} Sykes, M. V. 1990, Icarus, 84, 267
\bibitem[Sylvester et al. 1997]{ssb97} Sylvester, R. J., Skinner, C. J.,
  \& Barlow, M. J. 1997, \mnras, 289, 831
\bibitem[Sylvester et al. 1996]{ssbm96} Sylvester, R., J., Skinner, C. J.,
  Barlow, M. J., \& Mannings, V. 1996, \mnras, 279, 915
\bibitem[Telesco et al. 1988]{tdbw88} Telesco, C. M., Decher, R.,
  Becklin, E. E., \& Wolstencroft, R. D. 1988, \nat, 335, 51
\bibitem[Telesco et al. 2000]{tfpk00} Telesco, C. M., Fisher, R. S.,
  Pi\~{n}a, R. K., Knacke, R. F., Dermott, S. F., Wyatt, M. C., Grogan, K.,
  Holmes, E. K., Ghez, A. M., Prato, L. A., Hartmann, L. W., \&
  Jayawardhana, R. 2000, \apj, in press
\bibitem[Telesco \& Knacke 1991]{tk91} Telesco, C. M., \& Knacke, R. F.
  1991, \apj, 372, L29
\bibitem[Trilling et al. 1998]{tbgl98} Trilling, D. E., Benz, W., Guillot, T.,
  Lunin, J. I., Hubbard, W. B., \& Burrows, A. 1998, \apj, 500, 428
\bibitem[Trilling \& Brown 1998]{tb98} Trilling, D. E., \& Brown, R. H. 1998,
  \nat, 395, 775
\bibitem[Vedder 1998]{vedd98} Vedder, J. D. 1998, Icarus, 131, 283
\bibitem[Venn \& Lambert 1990]{vl90} Venn, K. A., \& Lambert, D. L. 1990,
  \apj, 363, 234
\bibitem[Walker \& Wolstencroft 1988]{ww88} Walker, H. J., \& Wolstencroft,
  R. D. 1988, \pasp, 100, 1509
\bibitem[Ward \& Hahn 1998]{wh98} Ward, W. R., \& Hahn, J. M. 1998, \apj,
  116, 489
\bibitem[Webb et al. 1999]{wzpp99} Webb, R. A., Zuckerman, B., Platais, I.,
  Patience, J., White, R. J., Schwartz, M. J., \& McCarthy, C. 1999, \apjl,
  512, L63
\bibitem[Wehry \& Mann 1999]{wm99} Wehry, A., \& Mann, I. 1999, \aap, 341, 296
\bibitem[Weidenschilling \& Cuzzi 1993]{wc93} Weidenschilling, S. J., \&
  Cuzzi, J. N. 1993, in Protostars \& Planets III, eds. E. H. Levy, \&
  J. I. Lunine (Tucson: Univ. Ariz. Press), 1031
\bibitem[Weidenschilling \& Jackson 1993]{wj93} Weidenschilling, S. J.,
  \& Jackson, A. A. 1993, Icarus, 104, 244
%\bibitem[Weinberger et al. 1999]{wbss99} Weinberger, A. J., Becklin, E. E.,
%  Schneider, G., Smith, B. A., Lowrance, P. J., Silverstone, M. D.,
%  Zuckerman, B., \& Terrile, R. J. 1999, \apjl, 525, L53
\bibitem[Wetherill 1980]{weth80} Wetherill, G. W. 1980, \araa, 18, 77
\bibitem[Wisdom 1980]{wisd80} Wisdom, J. 1980, \aj, 85, 1122
\bibitem[Wisdom 1982]{wisd82} Wisdom, J. 1982, \aj, 87, 577
\bibitem[Wolszczan \& Frail 1992]{wf92} Wolszczan, A., \& Frail, D. A. 1992,
  \nat, 355, 145
\bibitem[Woolf \& Angel 1998]{wa98} Woolf, N., \& Angel, J. R. 1998, \araa,
  36, 507 
\bibitem[Wyatt et al. 1999]{wdtf00} Wyatt, M. C., Dermott, S. F., Telesco,
  C. M., Fisher, R. S., Grogan, K., Holmes, E. K., \& Pi\~{n}a, R. K. 1999,
  \apj, in press
\bibitem[Xu \& Gustafson 1999]{xg99} Xu, Y. L., \& Gustafson, B. \AA. S.
  1999, \apj, 513, 894
\bibitem[Yoshioka \& Ikeuchi 1990]{yi90} Yoshioka, S., \& Ikeuchi, S. 1990,
  \apj, 360, 352

\end{thebibliography}

%%%% Biography page
\ifx\IncludeBio\True
\clearpage
%  \pagebreak
  \thispagestyle{plain}
  \addcontentsline{toc}{chapter}{BIOGRAPHICAL SKETCH}
  \vspace*{65pt}
\centerline{BIOGRAPHICAL SKETCH}
\vspace{.5in}

\DoubleSpace

Mark Wyatt was born 15 November 1972 in Rainham, a town in Kent,
England.
He grew up mostly in Bath, where he moved in 1979, and
where he attended school from age 7 to 18.
In 1991 he went up to Peterhouse College, Cambridge University, to read
Engineering, and graduated with a first class degree in 1994.
After spending a year in Barranquilla, Colombia, he went to Queen
Mary and Westfield College, University of London, to study for an M.Sc.
in Astrophysics, which he was awarded with distinction.
He then spent four months working as a Human Resources Analyst for
J. P. Morgan in London before moving to Gainesville, Florida, to study for
a Ph.D. in Astronomy under the tutelage of Dr. Stan Dermott.
Once the Ph.D. is completed he will return to the UK to continue with
his research as a postdoctoral fellow at the University of Edinburgh.
He is getting married in July 2000 to Maxine Neill, whom he met while
at Peterhouse.

\SingleSpace

\fi

%%%% Signature page(s)
\ifx\IncludeSig\True
  \SingleSpace
  \pagebreak
  \pagestyle{empty} % WRM changed from thispagestyle 5-13-94
  \input{signaturepage.tex}
\fi

%%%% General audience abstract
\ifx\IncludeGen\True
  \SingleSpace
  \pagebreak
  \pagestyle{empty}
  \input{genaudabspage.tex}
\fi

\end{document}